\documentclass[12pt,a4paper]{article}
\usepackage[latin1]{inputenc}
\usepackage{amsmath}
\usepackage{amssymb,mathrsfs}
\usepackage{graphics}
\usepackage{epsfig}
\newtheorem{definition}{Definition}
\newtheorem{proposition}{Proposition}
\newtheorem{theorem}{Theorem}
\newtheorem{remark}{Remark}
\newtheorem{lemma}{Lemma}
\newtheorem{corollary}{Corollary}
\newcounter{figurecounter}
\newcounter{stepcounter}
\newenvironment{step}{\refstepcounter{stepcounter}\it Step \thestepcounter:}{\medskip}

\newcommand{\Bild}[4]
{\begin{figure}[h] \centering \setlength{\epsfxsize}{#1}
\epsfbox{#2} \caption{#3}\end{figure}\refstepcounter{figurecounter}\label{#4}}

\newcommand{\ignore}[1]{}
\newcommand{\qed}{\hfill $\Box$}
\newcommand{\e}{{\rm e}}
\newcommand{\I}{{\rm i}}

\newcommand{\veps}{\varepsilon}

\newcommand{\NNN}{\mathbb{N}}
\newcommand{\RRR}{\mathbb{R}}
\newcommand{\ZZZ}{\mathbb{Z}}

\newcommand{\CCC}{\mathbb{C}}

\newcommand{\A}{{\cal A}}
\newcommand{\B}{{\cal B}}
\newcommand{\C}{{\cal C}}
\newcommand{\D}{{\cal D}}

\renewcommand{\H}{{\cal H}}

\renewcommand{\L}{{\cal L}}
\newcommand{\M}{{\cal M}}

\renewcommand{\O}{{\cal O}}

\newcommand{\R}{{\cal R}}
\newcommand{\T}{{\cal T}}

\newcommand{\W}{{\cal W}}

\begin{document}

\title{\vspace{-2cm}Effective Hamiltonians\\ for Constrained Quantum Systems} 

\author{Jakob Wachsmuth$^*$, Stefan Teufel\thanks {Supported by the DFG within the SFB/Transregio 71. University of T\"ubingen, Institute of Mathematics, Auf der Morgenstelle 10, 72076 T\"ubingen, Germany.\hspace{4cm}
Email: {\it jakob.wachsmuth@student.uni-tuebingen.de \& stefan.teufel@uni-tuebingen.de. }
}}

\maketitle


\vspace{-5mm}
\begin{abstract}
We consider the time-dependent Schr\"odinger equation on a Riemannian
manifold $\mathcal{A}$ with a  potential that localizes a certain subspace of states close
to a fixed submanifold $\mathcal{C}$.  When we scale the
potential in the directions normal to $\mathcal{C}$ by a parameter $\varepsilon\ll 1$, the
solutions concentrate  in an $\veps$-neighborhood of $\mathcal{C}$. This situation
occurs for example in quantum wave guides and for the motion of nuclei
in electronic potential surfaces in quantum molecular dynamics. We derive an
effective Schr\"odinger equation on the submanifold  $\mathcal{C}$ and show that its
solutions, suitably lifted to $\mathcal{A}$, approximate the solutions of the original
equation on $\mathcal{A}$ up to errors of order $\varepsilon^3|t|$ at time~$t$. Furthermore,
we prove that the eigenvalues of the corresponding effective Hamiltonian
below a certain energy coincide up to errors of order $\varepsilon^3$ with
those of the full Hamiltonian under reasonable conditions.

Our results hold   in the situation  where tangential and normal energies are
of the same order, and where exchange between these energies occurs. In earlier results tangential energies were assumed to be
small compared to normal energies, and rather restrictive assumptions were
needed, to ensure that the separation of energies is maintained during the
time evolution. Most importantly,  we   can allow for   constraining
potentials that change their shape along the submanifold, which is the
typical  situation  in the applications mentioned above.

Since we consider a very general
situation, our effective Hamiltonian contains many non-trivial terms of different origin. In particular,
the geometry of the normal bundle of $\mathcal{C}$ and a generalized  Berry connection on
an eigenspace bundle over $\mathcal{C}$ 
play a crucial role.  In order to explain the meaning and the relevance of
some of the terms in the effective Hamiltonian, we analyze in some detail
the application to quantum wave guides, where $\mathcal{C}$ is a curve
in $\mathcal{A}=\mathbb{R}^3$. 
This allows us to generalize two recent  results
on spectra of such wave guides.

\medskip

\begin{footnotesize}

\end{footnotesize}
\end{abstract}
\newpage 
 
\tableofcontents

 
 
\section{Introduction}
 
Although the mathematical structure of the linear Schr\"odinger equation
\begin{equation}\label{SE1}
\I \partial_t \psi = - \Delta  \psi + V\psi =: H\psi \,,\qquad \psi|_{t=0}\in L^2(\mathcal{A}, d\tau)
\end{equation}
is quite simple, in many cases the high dimension of the underlying configuration space   $\mathcal{A}$
makes even a numerical solution  impossible. Therefore it is important to identify situations where the dimension can be reduced by approximating the solutions of the original equation~(\ref{SE1}) on the high dimensional configuration space  $\mathcal{A}$ by solutions of an \textit{effective equation} 
\begin{equation}\label{SE2}
\I \partial_t \phi = H_{\rm eff}\phi \,,\qquad \phi|_{t=0}\in L^2(\mathcal{C}, d\mu)
\end{equation}
on a lower dimensional configuration space   $\mathcal{C}$.

\pagebreak

The physically most straight forward situation where such a dimensional reduction is possible are constrained mechanical systems. In these systems   strong forces   effectively constrain the system to remain in the vicinity of a submanifold $\mathcal{C}$ of the configuration space $\mathcal{A}$. 

For classical Hamiltonian systems on a Riemannian manifold $(\A,G)$ there is a straight forward mathematical reduction procedure. 
One just restricts the Hamilton function  to $T^*\C$ by embedding $T^*\C$ into $T^*\A$ via the metric~$G$ and then studies the induced dynamics on $T^*\C$. 
For  quantum systems Dirac \cite{D} proposed to quantize the restricted classical Hamiltonian system on the submanifold following an 'intrinsic' quantization procedure. However, for curved submani\-folds $\mathcal{C}$ there is no unique quantization procedure. One natural guess would be
 an effective Hamiltonian   $H_{\rm eff}$ in (\ref{SE2}) of the form
\begin{equation}\label{Heff0}
H_{\rm eff} = -\Delta_{\mathcal{C}} + V|_\mathcal{C}\,,
\end{equation}
where $\Delta_\mathcal{C}$ is the Laplace-Beltrami operator on $\mathcal{C}$ with respect to the induced metric and $V|_\mathcal{C}$ is the restriction of the potential $V:\mathcal{A}\to \RRR$ to $\mathcal{C}$. 

However, to justify or invalidate the above procedures from first principles, one needs to model  the constraining forces within the   dynamics (\ref{SE1}) on the full space $\mathcal{A}$. This is done by adding a  localizing part to the potential $V$.  Then one analyzes the behavior of solutions of (\ref{SE1}) in the asymptotic limit where the constraining forces become very strong and tries to extract a limiting equation on $\mathcal{C}$. This limit of strong confining forces has been studied in  classical mechanics and in quantum mechanics many times in the literature. 
The classical case was first investigated by Rubin and Ungar \cite{RU}, who found that in the limiting dynamics  an extra potential appears that accounts for the energy contained in the normal oscillations. Today there is a wide literature on the subject. We mention the monograph by Bornemann \cite{B} for a result based on weak convergence and a survey of older results, as well as the book of Hairer, Lubich and Wanner \cite{HLW},  Section XIV.3, for an approach based on classical adiabatic invariants. 

For the quantum mechanical case Marcus \cite{Mar} and later on Jensen and Koppe~\cite{JK} pointed out that  the limiting dynamics depends, in addition, 
also on the embedding of the submanifold $\mathcal{C}$ into the ambient space $\mathcal{A}$. In the sequel Da Costa \cite{DC} deduced a geometrical condition (often called the no-twist condition) ensuring that the effective dynamics does not depend on the localizing potential. This condition is equivalent to the flatness of the normal bundle of $\mathcal{C}$. It fails to hold for a generic submanifold of dimension and codimension both strictly greater than one, which is a typical situation   when applying these ideas to    molecular dynamics.

Thus the hope to obtain  a generic  'intrinsic' effective dynamics as in~(\ref{Heff0}), i.e.\ a Hamiltonian that depends only on the intrinsic geometry of $\mathcal{C}$ and the restriciton of the potential $V$ to $\mathcal{C}$, is unfounded. In both, classical and quantum mechanics, the limiting dynamics on the constraint manifold depends, in general,  on the detailed nature of the constraining forces, on the embedding of $\mathcal{C}$ into $\mathcal{A}$ and on the initial data of (\ref{SE1}). In this work we present and prove a  general result concerning the precise form of the limiting dynamics~(\ref{SE2}) on $\mathcal{C}$ starting from (\ref{SE1}) on the ambient space $\mathcal{A}$ with a strongly confining potential $V$. However, as we explain next, our result generalizes existing results in the mathematical and physical literature not only on a  technical level, but improves the range of applicability in a deeper sense.

Da Costa's statement (like the more refined results by Froese-Herbst \cite{FrH}, Maraner \cite{Ma1} and Mitchell \cite{Mit}, which we discuss in Subsection \ref{models}) requires that the constraining potential is the same at each point on the submanifold. The reason behind this assumption is that the energy stored in the normal modes diverges in the limit of strong confinement. As in the classical result by Rubin and Ungar, variations in the constraining potential lead to exchange of energy between normal and tangential modes, and thus also the energy in the tangential direction grows in the limit of strong confinement. However, the problem can be treated with the methods used in \cite{DC,Ma1,FrH,Mit} only for solutions with bounded kinetic energies in the tangential directions. Therefore the transfer of energy between normal and tangential modes was excluded in those articles by the assumption that the confining potential has the same shape in the normal direction at any point of the submanifold.
In many important applications this assumption is violated, for example for the reaction paths of molecular reactions. The reaction valleys vary in shape depending on the configuration of the nuclei. In the same applications also the typical normal and tangential energies are of the same order. 

Therefore the most important new aspect of our result is that we allow for confining potentials that vary in shape and for solutions with normal and tangential energies of the same order. As a consequence, our limiting dynamics on the constraint manifold has a richer structure than earlier results and resembles, at leading order, the results from classical mechanics. In the limit of small tangential energies we recover the limiting dynamics by Mitchell \cite{Mit}.

The key observation for our analysis is that the problem is an adiabatic limit and has, at least locally, a structure similar to the Born-Oppenheimer approximation in molecular dynamics. In particular, we transfer ideas from adiabatic perturbation theory, which were developed by Nenciu-Martinez-Sordoni and Panati-Spohn-Teufel in \cite{MS,MS2,NS,PST,S,Te}, to a non-flat geometry. We note that the adiabatic nature of the problem  was observed many times before in the physics literature, e.g.\ in the context of adiabatic quantum wave guides \cite{BDNT}, but we are not aware of any work considering constraint manifolds with general geometries in quantum mechanics from this point of view. In particular, we believe that our effective equations have not been derived or guessed before and are new not only as a mathematical but also as a physics result.
In the mathematics literature we are aware of two predecessor works:  in \cite{Te} the problem was solved for constraint manifolds $\mathcal{C}$ which are $d$-dimensional subspaces of $\RRR^{d+k}$, while  
Dell'Antonio and Tenuta~\cite{DT} considered  the leading order behavior of semiclassical  Gaussian wave packets for general geometries. 

Another result about submanifolds of any dimension is due to Wittich \cite{W}, who considers the heat equation on thin tubes of manifolds. Finally, there are related results in the wide literature on thin tubes of quantum graphs. A good starting point for it is \cite{Gr} by Grieser, where mathematical techniques used in this context are reviewed. Both works and the papers cited there, properly translated, deal with the case of small tangential energies. 

\medskip

We now give a non-technical sketch of the structure of our result.  The detailed statements given in Section~\ref{results} require some preparation.

 We implement the limit of strong confinement by mapping the problem to the normal bundle $N\mathcal{C}$ of $\mathcal{C}$ and then scaling one part of the potential in the normal direction by $\veps^{-1}$.   With decreasing  $\veps$ the  normal derivatives of the potential  and thus the constraining forces increase. 
 In order to obtain a non-trivial scaling behavior of the equation, the Laplacian is multiplied with a  prefactor~$\veps^2$. The   reasoning behind this scaling, which is the same as in  \cite{FrH,Mit}, is explained in Section~\ref{models}.
 With $q$ denoting coordinates on $\mathcal{C}$ and $\nu$ denoting normal coordinates our starting equation on $N\mathcal{C}$ has, still somewhat formally, the form
 \begin{eqnarray}\label{SE3}
 \I\partial_t\psi^\veps &=& -\veps^2\Delta_{N\mathcal{C}} \psi^\veps\,+\, V_{\rm c}(q,\veps^{-1}\nu)\psi^\veps\,+\,W(q,\nu)\psi^\veps\ \,=:\ \,H^\veps \psi^\veps\,
 \end{eqnarray}
 for $\psi^\veps|_{t=0}\in L^2(N\mathcal{C})$. Here $\Delta_{N\mathcal{C}} $ is the Laplace-Beltrami operator on $N\mathcal{C}$, where the metric  on $N\mathcal{C}$ is obtained by  pulling back   the metric on    a tubular neighborhood of $\mathcal{C}$ in $\mathcal{A}$ to a tubular neighborhood of the zero section in $N\mathcal{C}$ and then suitably extending it to all of $N\mathcal{C}$.
  We study the asymptotic behavior of (\ref{SE3}) as $\veps$ goes to zero uniformly for initial data with energies of order one. This means that initial data are allowed to oscillate on a scale of order $\veps$ not only in the normal direction, but also in the tangential direction, i.e.\ that tangential kinetic energies are of the same order as the normal energies. More precisely, we assume  that $\|\veps \nabla^{\rm h} \psi_0^\veps\|^2 = \langle \psi_0^\veps\,|\,- \veps^2 \Delta_{\rm h} \psi_0^\veps\rangle$ is of order one, in contrast to the earlier works \cite{FrH,Mit}, where it was assumed to be of order $\veps^2$. Here $\nabla^{\rm h}$ is a suitable horizontal derivative to be introduced in Definition~\ref{deriv}.
  
Our final result is basically an effective equation of the form (\ref{SE2}). It is presented in two steps. In 
 Section \ref{results1} it is stated that on certain subspaces of $L^2(N\mathcal{C})$ the unitary   group $\exp(-\I H^\veps t)$   generating solutions of (\ref{SE3}) is unitarily equivalent to an 'effective' unitary group $\exp(-\I H_{\rm eff}^\veps t)$ associated with (\ref{SE2}) up to errors     of 
 order $\veps^3|t|$ uniformly for bounded initial energies.
In 
Section \ref{results2} we provide the asymptotic expansion of $H^\veps_{\rm eff}$ up to  terms of order $\veps^2$, i.e.\ we compute $H_{{\rm eff},0}$, $H_{{\rm eff},1}$ and $H_{{\rm eff},2}$ in 
 $H_{\rm eff} = H_{{\rm eff},0} + \veps H_{{\rm eff},1} +\veps^2 H_{{\rm eff},2} + \mathcal{O}(\veps^3)$.
 
Furthermore, in Section \ref{eigenvalues} and \ref{waveguides} we explain how to obtain quasimodes of $H^\veps$ from the eigenfunctions of $H_{{\rm eff},0} + \veps H_{{\rm eff},1} +\veps^2 H_{{\rm eff},2}$ and quasimodes of $H_{{\rm eff},0} + \veps H_{{\rm eff},1} +\veps^2 H_{{\rm eff},2}$ from the eigenfunctions of $H^\veps$ and 
apply our formulas to quantum wave guides, i.e. the special case of curves in $\RRR^3$. As corollaries we obtain results generalizing in some respects those by Friedlander and Solomyak obtained in \cite{FS} and   by Bouchitt\'e et al. in \cite{BMT}. In addition, we discuss how   twisted closed wave guides display phase shifts somewhat similar to the Aharanov-Bohm effect but without magnetic fields! 

\medskip

The crucial step in the proof is the construction of closed infinite dimensional subspaces of $L^2(N\mathcal{C})$ which are invariant under the dynamics (\ref{SE3}) up to small errors and which can be mapped unitarily to $L^2(\mathcal{C})$, where the effective dynamics takes place. To construct these 'almost invariant subspaces', we define at each point $q\in\mathcal{C}$ a Hamiltonian operator $H_{\rm f}(q)$ acting on the fibre $N_q\mathcal{C}$. If it has a simple eigenvalue band $E_{\rm f}(q)$ that depends smoothly on $q$ and is isolated from the rest of the spectrum for all $q$, then the corresponding eigenspaces define a smooth line bundle over $\mathcal{C}$. Its $L^2$-sections define a closed subspace of $L^2(N\mathcal{C})$, which after a modification of order $\veps$ becomes the almost invariant subspace associated with the eigenvalue band $E_{\rm f}(q)$. In the end,  to each isolated eigenvalue band $E_{\rm f}(q)$ there is an associated line bundle over $\mathcal{C}$, an associated  almost invariant subspace and an associated effective Hamiltonian $H_{\rm eff}^\veps$.

We now come to the form of the effective Hamiltonian associated with a band $E_{\rm f}(q)$.
For $H_{{\rm eff},0}$  we obtain, as expected, the Laplace-Beltrami operator of the submanifold as kinetic energy term and the eigenvalue band $E_{\rm f}(q)$ as an effective potential,
\[
H_{{\rm eff},0} = -\veps^2 \Delta_\mathcal{C} + E_{\rm f}.
\]
We note that $(V_{\rm c}+W)|_\mathcal{C}$ is contained in $E_{\rm f}$. This is the quantum version of the result of Rubin and Ungar \cite{RU} for classical mechanics.
However,   the   time scale for which the solutions of (\ref{SE3}) propagate along finite distances are times $t$ of order $\veps^{-1}$. On this longer time scale the first order correction $\veps H_{{\rm eff},1}$ to the effective Hamiltonian has effects of order one and must be included in the effective dynamics. We do not give the details of $H_{{\rm eff},1}$ here and just mention that at next to leading order the kinetic energy term, i.e.\ the Laplace-Beltrami operator, must be modified in two ways. First, the metric on $\mathcal{C}$ needs to be changed by terms of order $\veps$ depending on exterior curvature, whenever 
the center of mass of the normal eigenfunctions does not lie exactly on the submanifold $\mathcal{C}$.
Furthermore, the connection on the trivial line bundle over $\mathcal{C}$ (where the wave function $\phi$ takes its values) must be changed from the trivial one to a non-trivial one, the so-called generalized Berry connection. For the variation of the eigenfunctions associated with the eigenvalue band $E_{\rm f}(q)$  along the submanifold induces a non-trivial connection on the associated eigenspace bundle.
This was already discussed by Mitchell in the case that the potential (and thus the eigenfunctions) only twists. 

When $E_{\rm f}$ is constant as in the earlier works, there is no non-trivial potential term up to first order and so the second order corrections in $H_{{\rm eff},2}$ become relevant. They are quite numerous. In addition to terms similar to those at first order, we find generalizations of the Born-Huang potential and the off-band coupling both known from the Born-Oppenheimer setting, and an extra potential depending on the inner and the exterior curvature of $\C$, whose occurence had originally lead to Marcus' reply to Dirac's proposal. Finally, when the ambient space is not flat, there is another extra potential already obtained by Mitchell.  

We note that in the earlier works it was assumed that $-\veps^2\Delta_{\mathcal{C}}$
is of order $\veps^2$ and thus of the same size as the terms in $H_{{\rm eff},2}$. That is why  the extra potential depending on curvature appeared at leading order in these works, while it appears only in $H_{{\rm eff},2}$ for us.
And this is also the reason that assumptions were necessary, assuring that all other terms appearing in our $H_{{\rm eff},0}$ and $H_{{\rm eff},1}$ are of higher order or  trivial, including  that $E_{\rm f}(q)\equiv E_{\rm f}$ is constant.

\medskip

We end this section with some more technical comments concerning our result and the difficulties encountered in its proof. 

In this work we do not assume the potential to become large away from the submanifold. That means we achieve the confinement solely through large potential gradients, not through high potential barriers.
This leads to several additional technical difficulties, not encountered in other rigorous results on the topic that mostly consider harmonic constraints. One aspect of this is the fact that the normal Hamiltonian $H_{\rm f}(q)$ has also continuous spectrum. While its eigenfunctions defining the adiabatic subspaces decay exponentially, the superadiabatic subspaces, which are relevant for our analysis, are slightly tilted  spectral subspaces with small components in the continuous spectral subspace. 

Let us finally mention two technical lemmas, which may both be of independent interest.  After extending the pull back metric from a tubular neighborhood of $\mathcal{C}$ in $\mathcal{A}$ to the whole normal bundle, $N\mathcal{C}$ with this metric has curvature increasing linearly with the distance to $\mathcal{C}$. As a consequence we have to prove weighted elliptic estimates for a manifold of unbounded curvature (Lemmas \ref{opestimates} \& \ref{keyest}).
Moreover, since we aim at uniform results, we need to introduce energy cutoffs. A result of possibly wider applicability is that the smoothing by energy cutoffs preserves polynomial decay (Lemma \ref{notspoiled}).
 

\subsection{The model}\label{model}
Let $(\A,G)$ be a Riemannian manifold of dimension $d+k$ ($d,k\in\NNN$) with associated volume measure $d\tau$. Let furthermore $\C\subset\A$ be a smooth   submanifold without boundary and of dimension $d$/codimension $k$, which is equipped with the induced metric $g=G|_\C$ and the associated volume measure $d\mu$. We will call $\A$ the \textit{ambient manifold} and $\C$ the \textit{constraint manifold}. 

On $\C$ there is a natural decomposition $T\A|_\C=T\C\times N\C$ of $\A$'s tangent bundle into the tangent and the normal bundle of $\C$. We assume that there exists a tubular neighborhood $\B\subset\A$ of $\C$ with globally fixed diameter, that is there is $\delta>0$ such that normal geodesics $\gamma$ (i.e. $\gamma(0)\in\C,\dot{\gamma}(0)\in N\C$) of length $\delta$ do not intersect. We will call a tubular neighborhood of radius $r$ an \emph{$r$-tube}.
Furthermore, we assume that 
\begin{equation}\label{bndcurv1}
\A\ and\ \C\ are\ of\ bounded\ geometry
\end{equation}
(see the appendix for the definition) and that the embedding
\begin{equation}\label{bndcurv2}
\C\hookrightarrow\A\ has\ globally\ bounded\ derivatives\ of\ any\ order,
\end{equation} 
where boundedness is measured by the metric $G$! In particular, these assumptions are satisfied for  $\A=\RRR^{d+k}$ and a smoothly embedded $\C$ that is (a covering of) a compact manifold or asymptotically flatly embedded, which are the cases arising mostly in the applications we are interested in (molecular dynamics and quantum wave\-guides).

\medskip

Let $\Delta_\A$ be the Laplace-Beltrami operator on $\A$. We want to study the Schr\"odinger equation
\begin{equation}\label{preequation}
\I \partial_t \psi \;=\; -\veps^2\Delta_\A\psi \,+\,V_\A^\veps\psi\,, \qquad\psi|_{t=0}\in L^2(\A,d\tau)\,,
\end{equation}
under the assumption that the potential $V_\A^\veps$ localizes at least a certain subspace of states in an $\veps$-tube of $\C$ with $\veps\ll\delta$. The localization will be realized by simply imposing that the potential is squeezed by $\veps^{-1}$ in the directions normal to the submanifold and not by assuming $V_\A^\veps$ to become large away from $\C$, which makes the proof of localization more difficult. To ensure proper scaling behavior, we have multiplied the Laplacian in (\ref{preequation})  by $\veps^{2}$. 
The physical meaning of this is explained at the end of the next subsection. Here we only emphasize that an analogous scaling was used implicitly or explicitly in all other previous works on the problem of constraints in quantum mechanics. The crucial difference in our work is, as explained before, that we allow for $\veps$-dependent initial data $\psi_0^\veps$ with tangential kinetic energy 
of order one instead of   order~$\veps^2$. 

In order to actually implement  the scaling in the normal directions,  we will now construct a related problem on the normal bundle of $\C$ by mapping $N\C$ diffeomorphically to the tubular neighborhood $\B$ of $\C$ in a specific way and then choosing a suitable metric $\overline{g}$ on $N\C$ (considered as a manifold). On the normal bundle the scaling of the potential in the normal directions is straight forward. The theorem we prove for the normal bundle will later be translated back to the original setting. On a first reading it may be convenient to skip the technical construction of $\overline{g}$ and of the horizontal and vertical derivatives $\nabla^{\rm h}$ and $\nabla^{\rm v}$ and to  immediately jump to the end of Definiton \ref{deriv}.

\medskip

The mapping to the normal bundle is performed in the following way. There is a natural diffeomorphism from the $\delta$-tube $\B$ to the $\delta$-neighborhood $\B_\delta$ of the zero section of the normal bundle $N\C$. This diffeomorphism corresponds to choosing coordinates on $\B$ that are geodesic in the directions normal to~$\C$. These coordinates are called (generalized) Fermi coordinates. They will be examined in detail in Section \ref{expansion}. In the following, we will always identify $\C$ with the zero section of the normal bundle.
Next we choose any diffeomorphism $\tilde\Phi\in C^\infty\big(\RRR,(-\delta,\delta)\big)$ which is the identity on $(-\delta/2,\delta/2)$ and satisfies
\begin{equation}\label{blowup}
\forall\;j\in\NNN\quad\exists\;C_j<\infty\quad\forall\;r\in\RRR:\quad|\tilde\Phi^{(j)}(r)| \;\leq\; C_j\,(1+r^2)^{-(j+1)/2}
\end{equation}
(see Figure \ref{fig4}). 
Now a diffeomorphism $\Phi\in C^\infty(N\C,\B)$ is obtained by first applying $\tilde\Phi$ to the radial coordinate on each fibre $N_q\C$ (which are all isomorphic to $\RRR^k$) and then using Fermi charts in the normal directions. 
\Bild{8cm}{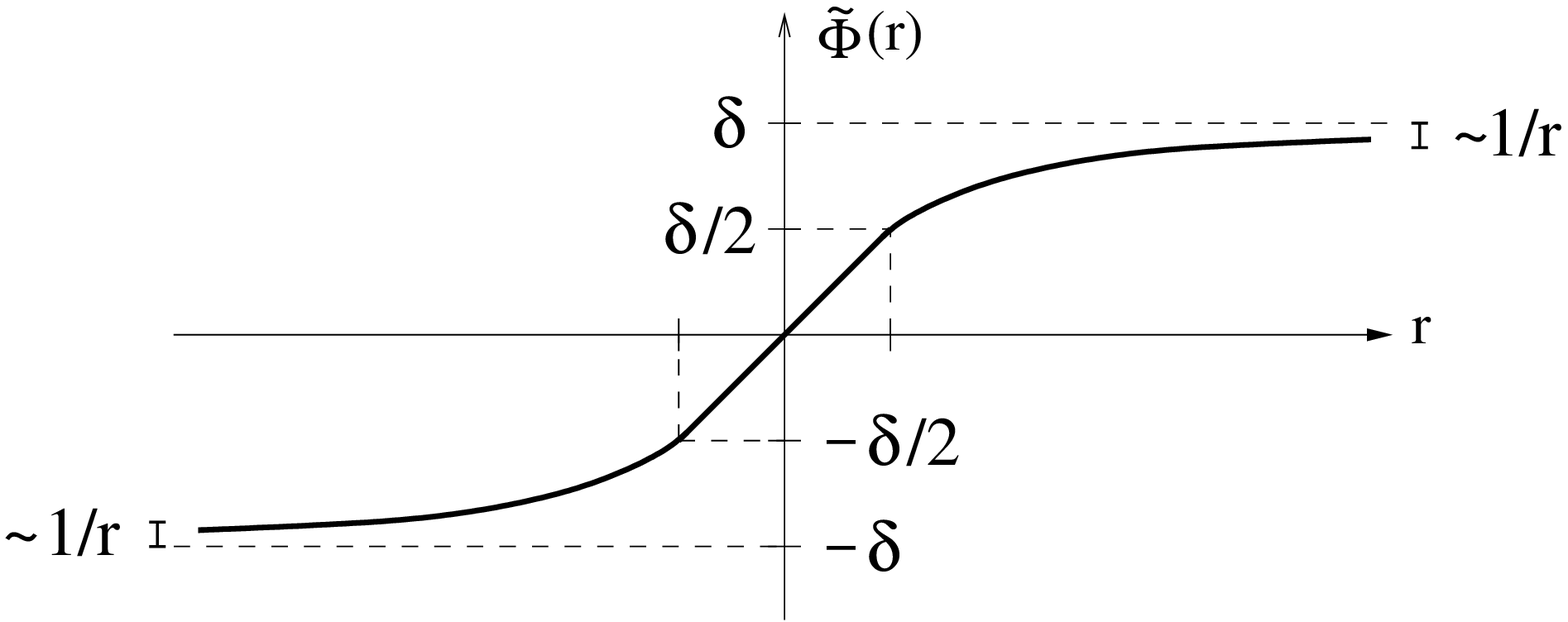}{$\tilde\Phi$ converges to $\pm\delta$ like $1/r$.}{fig4}

The important step now is to choose a suitable metric and corresponding measure on $N\C$. On the one hand we want it to be the pullback $\Phi^*G$ of $G$ on~$\B_{\delta/2}$. On the other hand,  we require that the distance to $\C$ asymptotically behaves like the radius in each fibre and that the associated volume measure on $N\C\setminus\B_{\delta}$ is $d\mu\otimes d\nu$, where $d\nu$ is the Lebesgue measure on the fibers of $N\C$ and $d\mu\otimes d\nu$ is the product measure (the Lebesgue measure and the product measure are defined after locally choosing an orthonormal trivializing frame of $N\C$; they do not depend on the choice of the trivialization because the Lebesgue measure is isotropic). The latter two requirements will help to obtain the decay that is needed to translate the result back to $\A$.

A metric satisfying the latter two properties globally is the so-called Sasaki metric which is defined in the following way (see e.g.\ Ch. 9.3 of \cite{Bl}):  The Levi-Civita connection on $\A$ induces a connection $\nabla$ on $T\C$, which coincides with the Levi-Civita connection on $(\C,g)$, and a connection $\nabla^\perp$ on $N\C$, which is called the \emph{normal connection} (see the appendix). The normal connection itself induces the connection map $K:TN\C\to N\C$ which identifies the vertical subspace of $T_{(q,\nu)}N\C$ with $N_q\C$. Let $\pi:N\C\to\C$ be the bundle projection. 
The Sasaki metric is then given by
\begin{equation}\label{Sasaki}
g^{\rm S}_{(q,\nu)}(v,w) \;:=\; g_q({\rm D}\pi\,v,{\rm D}\pi\,w) \,+\, G_{(q,0)}(K v,K w).
\end{equation}
It was studied by Wittich in \cite{W} in a similar context. The completeness of  $(N\C,g^{\rm S})$  follows from the completeness of $\C$ (see the proof for $T\C$ by Liu in~\cite{Li}). $\mathcal{C}$ is complete because it is of bounded geometry. But $(N\C,g^{\rm S})$ is, in general, not of bounded geometry, as it has curvatures growing polynomially in the fibers. However, $(\B_r\subset N\C,g^{\rm S})$ is a subset of bounded geometry for any $r<\infty$. Both can be seen directly  from the formulas for the curvature in \cite{Bl}. Now we simply fade the pullback metric into the Sasaki metric by defining
\begin{equation}\label{pullback}
\overline{g}_{(q,\nu)}(v,w)\;:=\;\Theta(|\nu|)\,G_{\Phi(q,\nu)}({\rm D}\Phi\,v,{\rm D}\Phi\,w) \,+\, \big(1-\Theta(|\nu|)\big)\,g^{\rm S}_{(q,\nu)}(v,w)
\end{equation}
with $|\nu|:=\sqrt{G_{\Phi(q,0)}({\rm D}\Phi\nu,{\rm D}\Phi\nu)}$ and a cutoff function $\Theta\in C^\infty([0,\infty),[0,1])$ satisfying $\Theta\equiv 1$ on $[0,\delta/2]$ and $\Theta\equiv 0$ on $[\delta,\infty)$. Then we have 
\begin{equation}\label{distance}
|\nu|=\sqrt{\overline{g}_{(q,0)}(\nu,\nu)}.
\end{equation}
The Levi-Civita connection on $(N\C,\overline{g})$ will be denoted by $\overline{\nabla}$ and the volume measure associated with $\overline{g}$ by $d\overline{\mu}$. We note that $\C$ is still isometrically imbedded and that $\overline{g}$ induces the same bundle connections $\nabla$ and $\nabla^\perp$ on $T\C$ and $N\C$ as $G$.  Since $\A$ is of bounded geometry and $(\B_\delta,g^{\rm S})$ is a subset of bounded geometry, $(\B_\delta,\overline{g})$ is a subset of bounded geometry. Furthermore, $(N\C,\overline{g})$ is complete due to the metric completeness of $(\overline{\B_\delta},\Phi^*G)$ (implied by the bounded geometry of $\A$) and the completeness of $(N\C,g^{\rm S})$.

The volume measure associated with $g^{\rm S}$ is, indeed, $d\mu\otimes d\nu$ and its density with respect to the measure associated with $G$ equals $1$ on $\C$ (see Section 6.1 of~\cite{W}). 
Together with the bounded geometry of $(\B_\delta,\overline{g})$ and $(\B_\delta,g^{\rm S})$, which implies that all small enough balls with the same radius have comparable volume (see~\cite{Sh}), we obtain that
\begin{equation}\label{density}
\frac{d\overline{\mu}}{d\mu\otimes d\nu}\Big|_{(N\C\setminus\B_{\delta/2})\cup\,\C}\,\equiv\,1,\ \,
\frac{d\overline{\mu}}{d\mu\otimes d\nu}\,\in\,C^\infty_{\rm b}(N\C),\ \, \frac{d\overline{\mu}}{d\mu\otimes d\nu}\,\geq\,c>0,
\end{equation}
where $C^\infty_{\rm b}(N\C)$ is the space of smooth functions on $N\C$ with all its derivatives globally bounded with respect to $\overline{g}$.

\medskip

Since we will think of the functions on $N\C$ as mappings from $\C$ to the functions on the fibers, the following derivative operators will play a crucial role.


\begin{definition}\label{deriv} 
Denote by $\Gamma(\mathcal{E})$ the set of all smooth sections of a hermitian bundle $\mathcal{E}$ and by $\Gamma_{\rm b}(\mathcal{E})$ the ones with globally bounded derivatives up to any order. 

\smallskip

i) Fix $q\in\C$. 
The fiber $(N_q\C,\overline{g}_{(q,0)})$ is isometric to the euclidean $\RRR^k$. Therefore there is a canonical identification $\iota$ of normal vectors at $q\in\C$ with tangent vectors at $(q,\nu)\in N_q\C$. 

Let $\varphi\in C^1(N_q\C)$. The \emph{vertical derivative} $\nabla^{\rm v}\varphi\in N^*_q\C$ at $\nu\in N_q\C$ is the pullback via $\iota$ of the exterior derivative of $\varphi\in C^1(N_q\C)$ to $N^*_q\C$.
i.e. 
\[(\nabla^{\rm v}_\zeta\varphi)(\nu)\;=\;\big({\rm d}\varphi\big)_\nu\big(\iota(\zeta)\big)\]  
for $\zeta\in N_q\C$. 
The Laplacian associated with $-\int_{N_q\C}\overline{g}_{(q,0)}(\nabla^{\rm v}\varphi,\nabla^{\rm v}\varphi)d\nu$ is denoted by $\Delta_{\rm v}$ and the set of bounded functions with bounded derivatives of arbitrary order by $C^\infty_{\rm b}(N_q\C)$.

\smallskip

ii) Let $\mathcal{E}_{\rm f}:=\{(q,\varphi)\,|\,q\in\C,\,\varphi\in C^\infty_{\rm b}(N_q\C)\}$ be the bundle over $\C$ which is obtained by replacing the fibers $N_q\C$ of the normal bundle with $C^\infty_{\rm b}(N_q\C)$ and canonically lifting the action of ${\rm SO}(k)$ and thus the bundle structure of $N\C$.

The \emph{horizontal connection} $\nabla^{\rm h}$
on $\mathcal{E}_{\rm f}$ is defined by 
\begin{equation}
(\nabla^{\rm h}_\tau\varphi)(q,\nu)\;:=\;\frac{d}{ds}\Big|_{s=0}\varphi(w(s),v(s)),
\end{equation}  
where $\tau\in \Gamma(T\C)$ and $(w,v)\in C^1([-1,1],N\C)$ with
\[w(0)\;=\;q,\ \dot{w}(0)\;=\;\tau(q), \quad\&\quad v(0)\;=\;\nu,\ \nabla^\perp_{\dot{w}}v\;=\;0.\]
Furthermore, $\Delta_{\rm h}$ is the Bochner Laplacian associated with $\nabla^{\rm h}$:
\begin{eqnarray*}
\int_{N\C}\psi^*\,\Delta_{\rm h}\psi\,d\mu\otimes d\nu 
&=& -\int_{N\C}g(\nabla^{\rm h}\psi^*,\nabla^{\rm h}\psi)\,d\mu\otimes d\nu,
\end{eqnarray*}
where we have used the same letter $g$ for the canonical shift of $g$ from the tangent bundle to the cotangent bundle of $\C$.

\pagebreak

Higher order horizontal derivatives are inductively defined by
\[\nabla^{\rm h}_{\tau_1,\dots,\tau_m}\varphi\;:=\;\nabla^{\rm h}_{\tau_1}\nabla^{\rm h}_{\tau_2,\dots,\tau_m}\varphi\,-\,\sum_{j=2}^m\nabla^{\rm h}_{\tau_2,\dots,\nabla_{\tau_1}\tau_j,\dots,\tau_m}\varphi\]
for arbitrary $\tau_1,\dots,\tau_m\in\Gamma(T\C)$.
The set of bounded sections $\varphi$ of $\mathcal{E}_{\rm f}$ such that $\nabla^{\rm h}_{\tau_1,\dots,\tau_m}\varphi$ is also a bounded section for all $\tau_1,\dots,\tau_m\in\Gamma_{\rm b}(T\C)$ is denoted by 
$C^m_{\rm b}(\C,C^\infty_{\rm b}(N_q\C))$.
\end{definition}

Coordinate expressions for $\nabla^{\rm v}$ and $\nabla^{\rm h}$ are given at the beginning of Section~\ref{wholestory}.

\medskip

In the following, we consider the Hilbert space $\overline{\H}:=L^2\big((N\C,\overline{g}),d\overline{\mu}\big)$ of complex-valued square-integrable functions. We emphasize that the elements of $\overline{\H}$ take values in the trivial complex line bundle over $N\C$. This will be the case for all functions throughout the whole text and we will omit this in the definition of Hilbert spaces. However, there will come up non-trivial connections on such line bundles! In addition, we notice that the Riemannian metrics on $N\C$ and $\C$ have canonical continuations on the associated trivial complex line bundles. 

The scalar product of a Hilbert space $\H$ will be denoted by $\langle \,.\,|\,.\,\rangle_\H$ and the induced norm by $\|\,.\,\|_\H$. The upper index $*$ will be used for both the adjoint of an operator and the complex conjugation of a function. 

\medskip

Instead of (\ref{preequation}) we now consider a Schr\"odinger equation on the normal bundle, thought of as a Riemannian manifold $(N\C,\overline{g})$. There we can immediately implement the idea of squeezing the potential in the normal directions: Let 
\[
V^\veps(q,\nu)\,=\,V_{\rm c}(q,\veps^{-1}\nu)\,+\,W(q, \nu)
\]
 for fixed real-valued potentials $V_{\rm c},W\in C^\infty_{\rm b}(\C,C^\infty_{\rm b}(N_q\C))$.   Here we have split up any $Q\in N\C$ as $(q,\nu)$ where $q\in\C$ is the base point and $\nu$ is a vector in the fiber $N_q\C$  at $q$. We allow for an 'external potential' $W$ which does not contribute to the confinement and is not scaled. 
Then  $\veps\ll 1$ corresponds to the regime  of strong confining forces. The setting is sketched in Figure~\ref{fig1}.
\Bild{7.5cm}{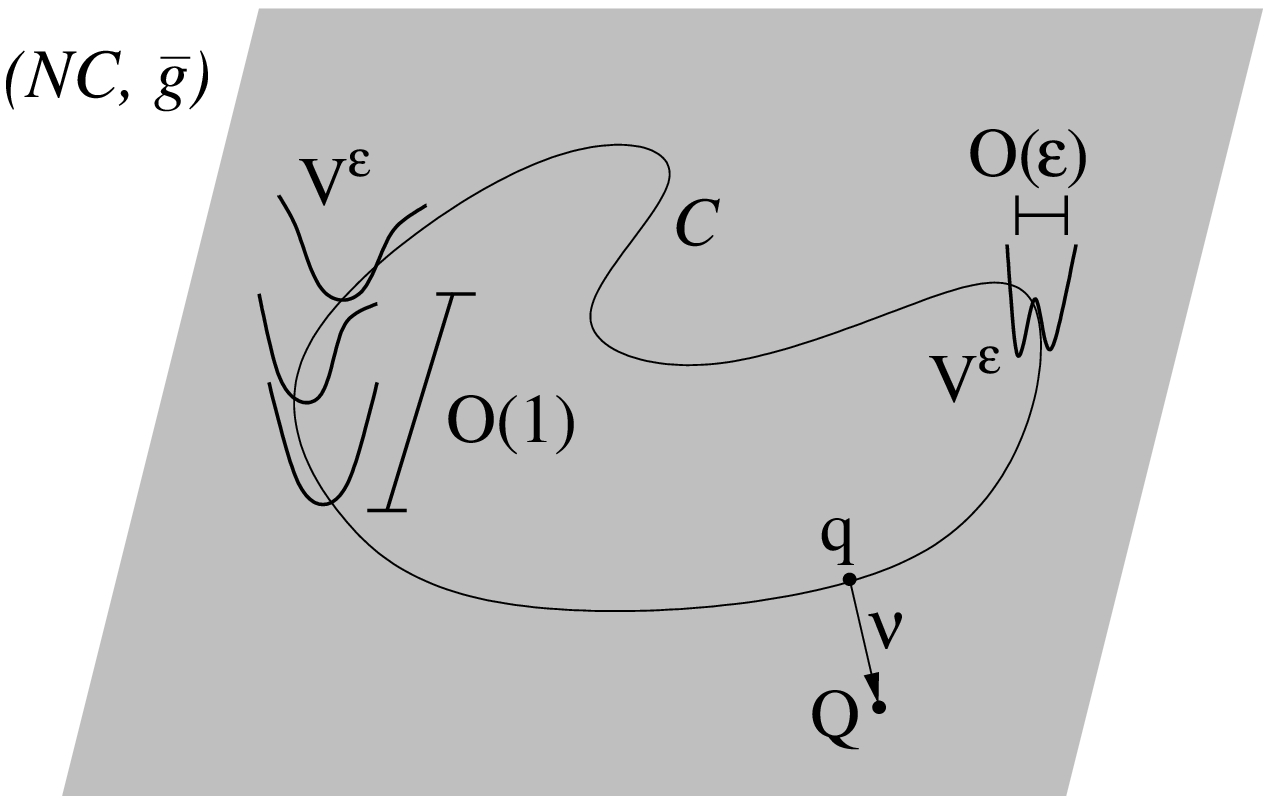}{The width of $V_\veps$ is $\veps$ but it varies on a scale of order one along $\C$.}{fig1}

\medskip

So we will investigate the Schr\"odinger equation 
\begin{equation}\label{equation}
\I \partial_t \psi \;=\;H^\veps\psi\;:=\; -\veps^2\Delta_{N\C} \psi \,+\,V^\veps\psi\,, \qquad\psi|_{t=0} = \psi^\veps_0 \in \overline{\H}\,,
\end{equation}
where  $\Delta_{N\C}$ is the Laplace-Beltrami operator on $(N\C,\overline{g})$, i.e.\ the operator associated with $-\int_{N\C}\overline{g}(d\psi,d\psi)d\overline{\mu}$. The operator $H^\veps$ will be called the Hamiltonian. We note that $H^\veps$ is real, i.e.\ it maps real-valued functions to real-valued functions. Furthermore, it is bounded from below because we assumed $V_{\rm c}$ and $W$ to be bounded. In Section 1.3 of \cite{Sh} $H^\veps$ is shown to be selfadjoint on its maximal domain $\D(H^\veps)$
for any complete Riemannian manifold $\M$, thus in particular for $(N\C,\overline{g})$. Let $W^{2,2}(N\C,\overline{g})$ be the second Sobolev space, i.e.\ the set of all $L^2$-functions with square-integrable covariant derivatives up to second order. We emphasize that, in general, $W^{2,2}(N\C,\overline{g})\subset\D(H^\veps)$ but $W^{2,2}(N\C,\overline{g})\neq\D(H^\veps)$ for a manifold of unbounded geometry.

\medskip

We only need one additional assumption on the potential, that ensures localization in normal direction. Before we state it, we clarify the structure of separation between vertical and horizontal dynamics: 

After a unitary transformation $H^\veps$ can at leading order be split up into an operator which acts on the fibers only and a horizontal operator. That unitary transformation $M_{\rho}$ is given by multiplication with the square root of the relative density $\rho:=\frac{d\overline{\mu}}{d\mu\otimes d\nu}$ of our starting measure and the product measure on $N\C$ that was introduced above. We recall from (\ref{density}) that this density is bounded and strictly positive. After the transformation it is helpful to rescale the normal directions. 

\begin{definition}\label{dilation}
Set $\H:=L^2(N\C,d\mu\otimes d\nu)$ and $\rho:=\frac{d\overline{\mu}}{d\mu\otimes d\nu}$. 

\smallskip

i) The unitary transform $M_\rho$ is defined by
$M_\rho:\H\to\overline{\H},\,\psi\mapsto\rho^{-\frac{1}{2}}\psi.$

\smallskip

ii) The \emph{dilation operator} $D_\veps$ is defined by 
$(D_\veps\psi)(q,\nu):=\veps^{-k/2}\,\psi(q,\nu/\veps).$

\smallskip

iii) The \emph{dilated} Hamiltonian $H_\veps$ and potential $V_\veps$ are defined by
\begin{eqnarray*}
H_\veps \;:=\; D_\veps^*M_{\rho}^*H^\veps M_{\rho}D_\veps,\quad
V_\veps \;:=\; D_\veps^*M_{\rho}^*V^\veps M_{\rho}D_\veps \;=\;  V_{\rm c} + D_\veps^*WD_\veps\,.
\end{eqnarray*}
The index $\veps$ will consistently be placed down to denote dilated objects, while it will placed up to denote objects in the original scale.
\end{definition}

The leading order of $H_\veps$ will turn out to be the sum of $-\Delta_{\rm v}+V_{\rm c}(q,\cdot)+W(q,0)$ and $-\veps^2\Delta_{\rm h}$ (for details on $M_\rho$ and the expansion of $H_\veps$ see Lemmas \ref{transform} \& \ref{expH} below). 
When $-\veps^2\Delta_{\rm h}$ acts on functions that are constant on each fibre, it is simply the Laplace-Beltrami operator on $\C$ carrying an $\veps^2$. Hereby the analogy with the Born-Oppenheimer setting is revealed where the kinetic energy of the nuclei carries the small parameter given by the ratio of the electron mass and the nucleon mass (see e.g.\ \cite{PST}).  

\medskip

We need that the family of $q$-dependent operators  $-\Delta_{\rm v}\,+\,V_{\rm c}(q,\cdot) \,+ \,W(q,0)$ has a family of exponentially decaying bound states in order to construct a subspace of states that are localized close to the constraint manifold.
The following definition makes this precise. We note that the conditions are simpler to verify than one might have thought in the manifold setting, since the space and the operators involved are euclidean! 

\begin{definition}\label{gapcondition}
Let $\H_{\rm f}(q):=L^2(N_q\C,d\nu)$ and $V_0(q,\nu) := V_{\rm c} (q,\nu) \,+\, W(q,0)$. 

\smallskip

i) The selfadjoint operator $(H_{\rm f}(q),H^2(N_q\C,d\nu))$ defined by
\begin{equation}\label{fastH}
H_{\rm f}(q)\;:=\;-\Delta_{\rm v}\,+\,V_0(q,.)
\end{equation}
is called the \emph{fiber Hamiltonian}. Its spectrum is denoted by $\sigma\big(H_{\rm f}(q)\big)$. 

\smallskip

ii) A function $E_{\rm f}:\C\to\RRR$ is called an \emph{energy band}, if $E_{\rm f}(q)\in\sigma\big(H_{\rm f}(q)\big)$ for all $q\in\C$.
$E_{\rm f}$~is called \emph{simple}, if $E_{\rm f}(q)$ is a simple eigenvalue for all $q\in\C$. 

\smallskip

iii) An energy band $E_{\rm f}:\C\to\RRR$ is called \emph{separated}, if there are a constant $c_{\rm gap}>0$ and two bounded continuous functions $f_\pm:\C\to\RRR$ defining an interval $I(q):=[f_-(q),f_+(q)]$ such that
\begin{equation}\label{gap}
E_{\rm f}(q)=   I(q) \cap \sigma(H_{\rm f}(q))  \,,\qquad \inf_{q\in\C}{\rm dist}\big(\sigma\big(H_{\rm f}(q)\big)\setminus E_{\rm f}(q),\,E_{\rm f}(q)\big)\,=\,c_{\rm gap}.
\end{equation}

iv) Set $\langle\nu\rangle:=\sqrt{1+|\nu|^2}=\sqrt{1+\overline{g}_{(q,0)}(\nu,\nu)}$. A separated energy band $E_{\rm f}$ is called a \emph{constraint energy band}, if there is $\Lambda_0>0$ such that the family of spectral projections $P_0:\C\to\L\big(\H_{\rm f}(q)\big)$ corresponding to $E_{\rm f}$ satisfies 
\[\textstyle{\sup}_{q\in\C}\,\|{\rm e}^{\Lambda_0\langle\nu\rangle}P_0(q){\rm e}^{\Lambda_0\langle\nu\rangle}\|_{\L(\H_{\rm f}(q))}<\infty.\]
\end{definition}

\smallskip

\begin{remark}
Condition iii) is known to imply condition iv) in lots of cases (see \cite{H} for a review of known results), in particular for eigenvalues below the continuous spectrum, which is the most important case in the applications. Besides, condition iii) is a uniform but local condition (see Figure \ref{fig2}).
\Bild{7.5cm}{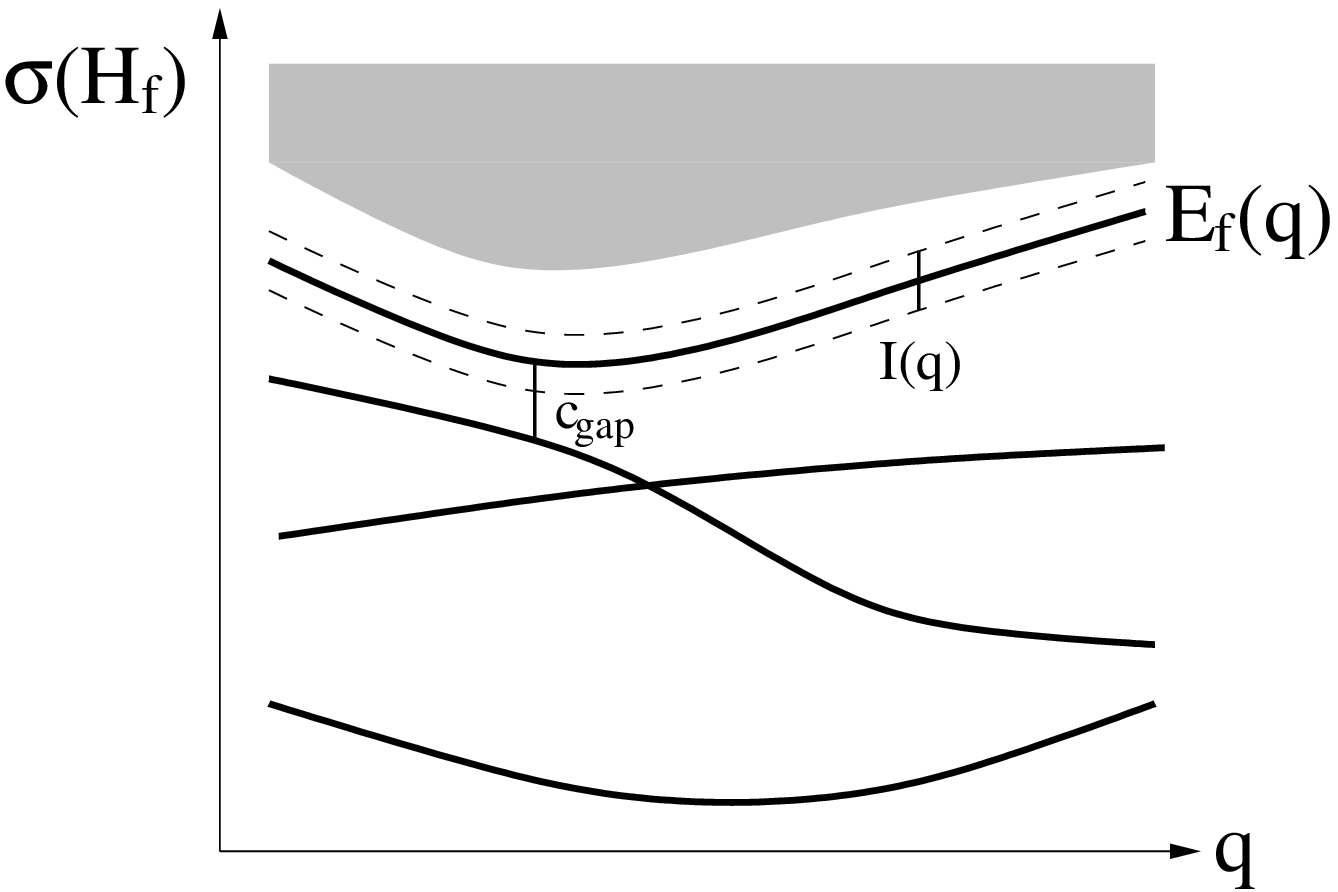}{$E_{\rm f}(q)$ has to be separated by a local gap that is uniform in $q$.}{fig2}
\end{remark}

The family of spectral projections $P_0:\C\to\L\big(\H_{\rm f}(q)\big)$ associated with a simple energy band $E_{\rm f}$ corresponds to a line bundle over $\C$. If this bundle has a global section $\varphi_{\rm f}:\C\to\H_{\rm f}(q)$ of normalized eigenfunctions, it holds for all $q\in\C$ that
$(P_0\psi)(q,\nu)=\langle\varphi_{\rm f}(q,\cdot)|\psi(q,\cdot)\rangle_{\H_{\rm f}(q)}\,\varphi_{\rm f}(q,\nu)$. Furthermore, $\varphi_{\rm f}$ can be used to define a unitary mapping between the corresponding subspace $P_0\H$ and $L^2(\C,d\mu)$: 

\begin{definition}\label{U0}
Let the eigenspace bundle corresponding to a simple constraint energy band $E_{\rm f}$ admit a smooth global section $\varphi_{\rm f}:\C\to\H_{\rm f}(q)$ of normalized eigenfunctions. 
The partial unitary operator $U_0:\H\to L^2(\C,d\mu)$ is defined by $(U_0\psi)(q):=\langle\varphi_{\rm f}(q,\cdot)|\psi(q,\cdot)\rangle_{\H_{\rm f}(q)}$. Then 
$U_0^*U_0=P_0$ and $U_0U_0^*=1$
with $U_0^*$ given by $(U_0^*\psi)(q,\nu)=\varphi_{\rm f}(q,\nu)\psi(q)$. 
\end{definition}

So any $\psi\in P_0\H$ has the product structure $\psi=(U_0\psi)\varphi_{\rm f}$.
Since $V_0$ and therefore $\varphi_{\rm f}$ depends on $q$, such a product will, in general, not be invariant under the time evolution. However, it will turn out to be at least approximately invariant. For short times this follows from the fact that  the commutator 
$[H_\veps,P_0] = [ -\veps^2\Delta_{\rm h} , P_0] +\mathcal{O}(\veps)$
 is of order~$\veps$. For long times this is a consequence of adiabatic decoupling. 

On the macroscopic scale the corresponding  eigenfunction $D_\veps\varphi_{\rm f}$ is more and more localized close to the submanifold: most of its mass is contained in the $\veps$-tube around $\C$ and it decays like ${\rm e}^{-\Lambda_0|\zeta|/\veps}$. This is visualized in Figure \ref{fig3}. 
\Bild{13cm}{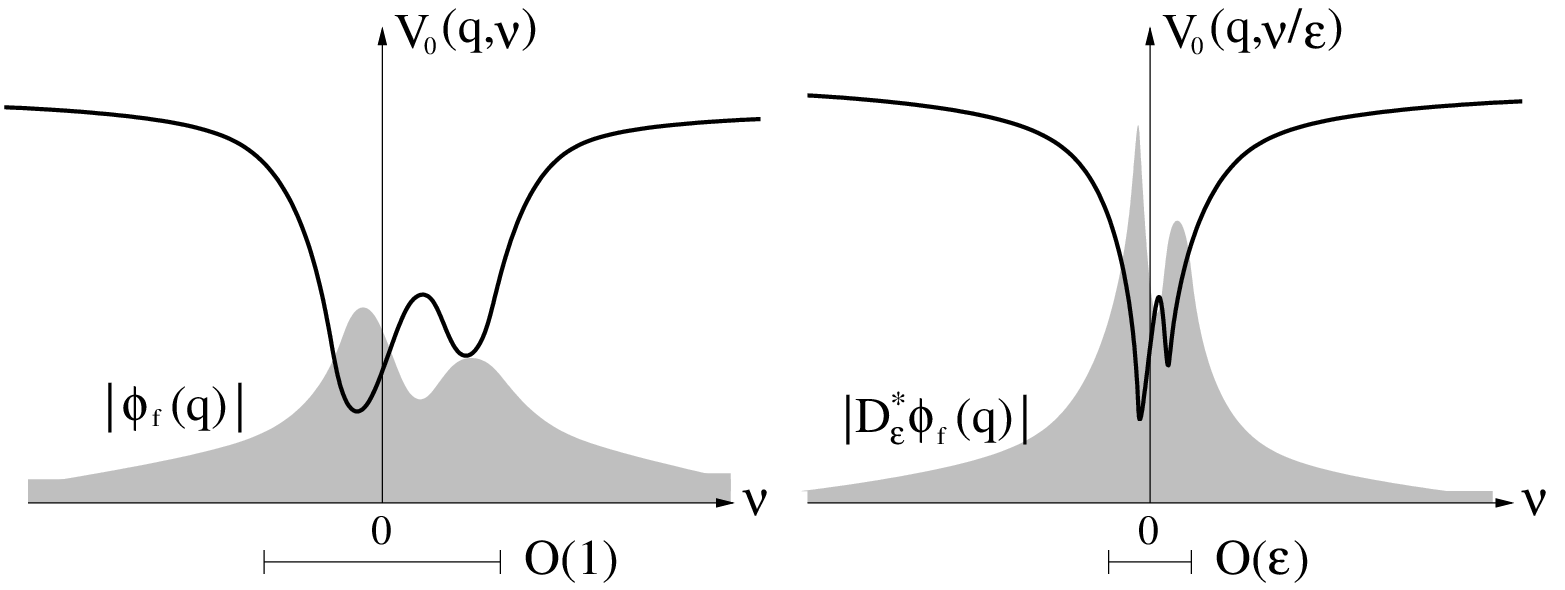}{On the macroscopic level $\varphi_{\rm f}$ is localized on a scale of oder $\veps$.}{fig3}

\medskip

Our goal is to obtain an effective equation of motion on the submanifold for states that are localized close to the submanifold in that sense. More precisely, for each subspace $P_0\H$ corresponding to a constraint energy band~$E_{\rm f}$ we will derive an effective equation using the map $U_0$. However, in order to control errors with higher accuracy we will have to add corrections of order~$\veps$ to $P_0\H$ and $U_0$. 

\subsection{Comparison with existing results}\label{models}

Since similar settings have been considered several times in the past, we want to point out the similarities and the differences with respect to our result. 
We mostly focus on the papers by Mitchell \cite{Mit} and Froese-Herbst \cite{FrH}, since \cite{Mit} is the most general one on a theoretical physics level and \cite{FrH} is the only mathematical paper concerned with deriving effective dynamics on the constraint manifold. Both works deal with a Hamiltonian that is of the form
\begin{equation}\label{Othermodels}
\tilde H^\veps\;=\;-\Delta_{N\C}\,+\,\veps^{-2}V^\veps_{\rm c}\,+\,W\,.
\end{equation}  
The confining potential $V^\veps_{\rm c}$ is chosen to be the same everywhere on $\C$ up to rotations, i.e.\ in any local bundle chart $(q,\nu)$ there exists a smooth family of rotations $R(q)\in {\rm SO}(k)$ such that 
\[
V^\veps_{\rm c}(q,\nu)\;=\;V_{\rm c}(q,\veps^{-1}\nu) \;=\; V_{\rm c}(q_0,\veps^{-1}R(q)\nu)
\]
for some fixed point $q_0$ on $\C$. As a consequence,  the eigenvalues of the resulting fiber Hamiltonian $H_{\rm f}(q) = -\Delta_{\rm v} + V_{\rm c}(q,\cdot)$ are constant, $E_{\rm f}(q)\equiv E_{\rm f}$. 
As our Theorems~\ref{effdyn} and \ref{calcHeff},   the final result  in \cite{Mit} and somewhat disguised also in \cite{FrH} is about effective Hamiltonians acting on~$L^2(\C)$ which approximate the full dynamics on corresponding subspaces of $L^2(N\C)$. In the following we explain how the results in \cite{FrH,Mit} about~(\ref{Othermodels}) are related to  our results on the seemingly different problem (\ref{equation}). It turns out that they indeed follow from our general results under the special assumptions on the confining potential and in a low energy limit.

To see this and to better understand the meaning of the scaling, note that
when we multiply $\tilde H^\veps$ by $\veps^2$, the resulting Hamiltonian 
\begin{equation*} 
  \veps^2\tilde H^\veps\;=\;-\veps^2\Delta_{N\C}\,+\, V_{\rm c}^\veps\,+\,\veps^2W\,,
\end{equation*}  
is the same as $H^\veps$ in (\ref{equation}), however, with very restrictive assumptions on the confining part $V_{\rm c}$ and with a non-confining part of  order $\veps^2$. As one also has to multiply the left hand side of the Schr\"odinger equation (\ref{equation}) by $\veps^2$, this should be interpreted in the following way. Results valid for times of order one for the group generated by $\tilde H^\veps$ would be valid for times of order $\veps^{-2}$ for the group generated by $\veps^2 \tilde H^\veps$. On this time scale our result still yields an approximation with small errors (of order $\veps$). Thus the results in \cite{FrH,Mit} are valid  on the same physical time scale as ours. 

We look at (\ref{equation}) for initial data with  horizontal kinetic energies $\langle \psi^\veps_0 | -\veps^2\Delta_{\rm h}\psi^\veps_0\rangle$   of order one.
This   corresponds to horizontal kinetic energies $\langle \psi^\veps_0 | - \Delta_{\rm h}\psi^\veps_0\rangle$ of order $\veps^{-2}$ in (\ref{Othermodels}), i.e.\ to the situation where potential and kinetic energies are of the same order.
However, in \cite{FrH,Ma1,Mit} it is assumed that horizontal  kinetic energies are of order one, i.e.\ smaller by a factor $\veps^2$ than the potential energies.  
And to ensure that the horizontal kinetic energies remain bounded during the time evolution, the huge effective potential $\veps^{-2}E_{\rm f}(q)$ given by the normal 
eigenvalue must be constant.
 This is achieved in \cite{FrH,Ma1,Mit} by assuming that, up to rotations, the confining potential is  the same  everywhere on 
$\mathcal{C}$. 

Technically, the assumption that (in our units) $\langle \psi^\veps_0 | -\veps^2\Delta_{\rm h}\psi^\veps_0\rangle$  is of order $\veps^2$ simplifies the analysis significantly. This is because the first step in proving effective dynamics for states in a subspace $P_0\mathcal{H}$ for times of order $\veps^{-2}$ is to prove that it is approximately invariant under the time evolution for such times. 
Now the above assumption   implies that
the commutator  $[H_\veps,P_0] $ is of order $\veps^2$, and, as a direct consequence,  that the subspace $P_0\mathcal{H}$ is approximately invariant up to times of order $\veps^{-1}$,
\[
\left\| \left[ \e^{-\I H_\veps t}, P_0\right] \right\|= \mathcal{O}(\veps^2 |t|)\,.
\]
To get approximate invariance for times of order $\veps^{-2}$ one needs an additional adiabatic argument, which is missing in \cite{Mit}. Still, the effective Hamiltonian in \cite{Mit} is correct  for the same reason that the textbook derivation of the Born-Oppenheimer approximation is incomplete but yields the correct result including the first order Berry connection term. In \cite{FrH} it is observed that one either has to assume spherical symmetry of the confining potential, which implies that $[H_\veps,P_0] $ is of order $\veps^3$, or that one has to do an additional averaging argument in order to determine an effective Hamiltonian valid for times of order $\veps^{-2}$. For our case of large kinetic energies the simple argument just gives 
 \[
\left\| \left[ \e^{-\I H_\veps t}, P_0\right] \right\|= \mathcal{O}(\veps  |t|)\,.
\]
Therefore we need to replace the adiabatic subspaces $P_0\mathcal{H}$ by 
 so called super-adiabatic subspaces $P_\veps\mathcal{H}$, for which $\left\| \left[ \e^{-\I H_\veps t}, P_\veps\right] \right\|= \mathcal{O}(\veps^3  |t|)$, in order to pass to the relevant time scale. 

\medskip

We end the introduction with a short discussion on the physical meaning of the scaling. While it is natural to model strong confining forces by dilating the confining potential in the normal direction, the question remains, why in~(\ref{Othermodels}) there appears the factor $\veps^{-2}$ in front of the confining potential, or, in our units, why there appears the factor $\veps^2$ in front of the Laplacian in (\ref{equation}). The short answer is that without this factor no solutions of the corresponding Schr\"odinger equation would exist that remain $\veps$-close to  
$\mathcal{C}$. 
Any solution initially localized in a $\veps$-tube around $\mathcal{C}$ would immediately spread out because its normal kinetic energy would be of order $\veps^{-2}$, allowing it to overcome any confining potential of order one. 
Thus by the prefactor $\veps^{-2}$ in (\ref{Othermodels}) the confining potential is scaled to the level of normal kinetic energies for $\veps$-localized solutions, while in (\ref{equation}) we instead bring down the normal kinetic energy of $\veps$-localized solutions to the level of the finite potential energies. 

The longer answer forces us to look at the physical situation for which we want to derive asymptotically correct effective equations. The prime examples where our results are relevant are molecular dynamics, which was the   motivation for  \cite{Ma1,Ma,Mit}, and nanotubes and -films (see e.g.\ \cite{BDNT}). 
In both cases one is not   interested in the situation of infinite confining forces and perfect constraints. One rather has a regime where the confining potential is given and fixed by the physics, but where the variation of all other potentials and of the geometry is small on the scale defined by the confining potential. This is exactly the regime described by the asymptotics $\veps\ll 1$ in~(\ref{equation}).


\section{Main results}\label{results}

\subsection{Effective dynamics on the constraint manifold}\label{results1}
Since the constraining potential $V_{\rm c}$ is varying along the submanifold, the vertical and the horizontal dynamics do not decouple completely at leading order and, as explained above, the product structure of states in $P_0\H$ is not invariant under the time evolution. In order to get a higher order approximation valid also for   times of order $\veps^{-2}$, we need to construct so-called superadiabatic subspaces $P_\veps\H$.
  These are close to the adiabatic subspaces $P_0\H$ in the sense that the corresponding projections $P_\veps$ have an expansion in $\veps$ starting with the projection $P_0$.
 
\medskip 

Furthermore, when there is a global orthonormal frame of the eigenspace bundle defined by $P_0(q)$, the dynamics inside the superadiabatic subspaces can be mapped unitarily to dynamics on a space over the submanifold only. 

\begin{remark}\label{global}
Let $E_{\rm f}$ be a simple constraint energy band (see Definition \ref{gapcondition}). If $\C$ is contractible or if $E_{\rm f}$ is the ground state energy of $H_{\rm f}$, then the associated eigenspace bundle has a smooth global section $\varphi_{\rm f}:\C\to\H_{\rm f}(q)$ of normalized eigenfunctions. 
\end{remark}

To see this we notice that, on the one hand, all bundles over a contractible manifold are trivializable. On the other hand, $E_{\rm f}$ has to be an eigenvalue for all $q$ due to the gap condition and the eigenfunctions of $H_{\rm f}(q)$ can be chosen real-valued because $H_{\rm f}(q)$ is a real operator for all $q\in\C$. Furthermore, the groundstate of a Schr\"odinger operator with a bounded potential can always be chosen strictly positive (see \cite{RS4}). This defines an orientation on the real eigenspace bundle and thus a trivialization on the real line bundle. 

\smallskip

In the following, we restrict ourselves here to a simple energy band, i.e.\ with one-dimensional eigenspaces because  we do not want to overburden the result about the effective Hamiltonian (Theorem \ref{calcHeff}). To circumvent the possible non-triviality of the eigenspace bundle, we simply assume the existence of a trivializing frame for our main results. Generalizations to non-trivial bundles are in preparation.

\begin{theorem}\label{effdyn}
Let $V_{\rm c}, W \in C^\infty_{\rm b}(\C,C^\infty_{\rm b}(N_q\C))$ and let $E_{\rm f}$ be any associated simple constraint energy band whose eigenspace bundle   has a smooth section of normalized eigenfunctions. 

\smallskip

Fix $E<\infty$. Then there are $C<\infty$ and $\veps_0>0$ which satisfy that for all~$\veps < \veps_0$ there are
\begin{itemize}
\item a closed subspace $ P^\veps\overline{\H}\subset\overline{\H}$ with orthogonal projection $ P^\veps$, 
\item a Riemannian metric $g^\veps_{\rm eff}$ on $\C$ with associated measure $d\mu^\veps_{\rm eff}$,
\item $ U^\veps: \overline{\H}\to \H_{\rm eff}:=L^2(\C,d\mu^\veps_{{\rm eff}})$ with $ U^{\veps*} U^{\veps}= P^\veps$ and $ U^{\veps} U^{\veps*}=1$,
\end{itemize}
such that $\big(H^\veps_{\rm eff}:= U^\veps H^\veps U^{\veps*},\, U^\veps\D(H^\veps)\big)$ is self-adjoint on $\H_{\rm eff}$ and
\begin{equation}\label{difference}
\left\|\left(\e^{-\I H^\veps t}- U^{\veps*}\e^{-\I H^\veps_{\rm eff}t}\, U^\veps\right) P^\veps\chi(H^\veps)\,\right\|_{\L(\overline{\H})}\;\leq\;C\,\veps^3\,|t|
\end{equation}
for all $t\in\RRR$ and each   Borel function $\chi:\RRR\to[-1,1]$ with ${\rm supp}\,\chi\subset (-\infty,E]$. Here $\chi(H^\veps)$ is defined via the spectral theorem. 
\end{theorem}
The proof of this result can be found in Section \ref{proofT1}. The estimate (\ref{difference}) means that, after cutting off large energies, the superadiabatic subspace $ P^\veps \overline{\H}$ is invariant up to errors of order $\veps^3|t|$ and that on this subspace of $\overline{\H}$ the unitary group $\e^{-\I H^\veps t}$ is unitarily equivalent to the effective unitary group $\e^{-\I H^\veps_{{\rm eff}}t}$ on $L^2(\C,d\mu^\veps_{{\rm eff}})$ with the same error. In particular, there is adiabatic decoupling of the horizontal and vertical dynamics. 

\smallskip

The energy cutoff $\chi(H^\veps)$ is necessary in order to obtain a uniform error estimate, since the adiabatic decoupling breaks down for large energies because of the quadratic dispersion relation. It should be pointed out here that, while $ P^\veps\chi(H^\veps)$ is not a projection, it is really the difference of the unitary groups that is small and not $P^\veps\chi(H^\veps)$. More precisely, $P^\veps\chi(H^\veps)$ can be replaced by $U^{\veps*} \chi(H_{\rm eff}^\veps)U^\veps$, which is a projector for any characteristic function $\chi$, as is done in the following corollary.
 
\medskip

Before we come to the form of the effective Hamiltonian, we state our result about effective dynamics for $\A$, which follows from the one above.

\begin{definition}\label{liftop}
Set $A\psi:=\big(\frac{d\overline{\mu}}{\Phi^*d\tau}\big)^{-1/2}
\,(\psi\circ\Phi)$ with $\Phi:N\C\to\B_\delta$ as constructed in Section \ref{model} and $\Phi^*d\tau$ the pullback of $d\tau$ via $\Phi$. This defines an operator $A\in\L\big(L^2(\A,d\tau),\overline{\H}\big)$ with $AA^*=1$.
\end{definition}

The stated properties of $A$ are easily verified by using the substitution rule. Of course, the choice of our metric (\ref{pullback}) changes the metric in a singular way because it blows up a region of finite volume to an infinite one. However, it will turn out that the range of $ P^\veps$ consists of functions that decay faster than any negative power of $|\zeta|/\veps$ away from the zero section of the normal bundle. Then leaving the metric invariant on $\B_{\delta/2}$ is sufficient; 
due to the fast decay the error in the blown up region will be smaller than any power of $\veps$ for $\veps\ll\delta$. 

\begin{corollary}\label{effdyn2}
Fix $\delta>0$ and $E<\infty$. Let $H_\A^\veps:=-\veps^2\Delta_\A+V_\A^\veps$ be self-adjoint on $L^2(\A,d\tau)$. 
Assume that $D_\veps^*A V_\A^\veps A^*D_\veps= V_{\rm c}+D_\veps^*WD_\veps$ for some $ V_{\rm c},W$ satisfying the assumptions of Theorem \ref{effdyn}. Then there are $C<\infty$ and $\veps_0>0$ such that  
\begin{equation*}
\left\|\left(\e^{-\I H_\A^\veps t}-A^*\, U^{\veps*}\e^{-\I H^\veps_{{\rm eff}}t} U^\veps A\right)A^*\, U^{\veps*} \chi(H_{\rm eff}^\veps)U^\veps A\,\right\|_{\L(L^2(\A,d\tau))}\;\leq\;C\,\veps^3\,|t|
\end{equation*}
for all $0<\veps\leq\veps_0$, $t\in\RRR$, and each Borel function $\chi:\RRR\to[-1,1]$ with ${\rm supp}\,\chi\subset (-\infty,E]$. 
\end{corollary}

The proof of this result can be found in Section \ref{proofT3}. We note that for any charateristic function $\chi$ the operator $A^*\, U^{\veps*} \chi(H_{\rm eff}^\veps)U^\veps A$ is a projector whose image is canonically identified with $\chi(H_{\rm eff}^\veps)L^2(\C,d\mu_{\rm eff}^\veps)$. The assumption $D_\veps^*A V_\A^\veps A^*D_\veps= V_{\rm c}+D_\veps^*WD_\veps$ means that $V_\A^\veps$ should be of the same form as $V^\veps$ in the $\delta$-tube of $\C$.
Moreover, the assumptions on $ V_{\rm c}$ and $W$ in Theorem~\ref{effdyn} translate into assumptions on $V^\veps_\A$ on the $\delta$-tube only. 


\subsection{The effective Hamiltonian}\label{results2}

Here we write down the expansion of the effective Hamiltonian $H_{\rm eff}$. We do this only for states with high energies cut off. Then the terms in the expansion do not depend on any cutoff, which is a non-trivial fact, since we will need cutoffs to construct $H_{\rm eff}$!

\begin{theorem}\label{calcHeff}
In addition to the assumptions of Theorem~\ref{effdyn}, assume that the global family of eigenfunctions $\varphi_{\rm f}$ associated with $E_{\rm f}$ is in $C^\infty_{\rm b}\big(\C,\H_{\rm f}(q)\big)$.  

For all $\veps$ small enough there is a self-adjoint operator $H^{(2)}_{{\rm eff}}$ on $\H_{\rm eff}$ such that for 
 each Borel function $\chi:\RRR\to[-1,1]$ with ${\rm supp}\,\chi\subset (-\infty,E\,]$, for every  $\xi\in\{ U^\veps\chi(H^\veps) U^{\veps*},\chi(H_{\rm eff}^\veps),\chi(H^{(2)}_{{\rm eff}})\}$, and for all $\psi,\phi\in \H_{\rm eff}$ satisfying $\psi=\chi(-\veps^2\Delta_\C+E_{\rm f})\psi$ it holds that $\|\,(H^\veps_{{\rm eff}}-H^{(2)}_{{\rm eff}})\,\xi\,\|_{\L(\H_{\rm eff})}=\O(\veps^3)$ and
\begin{eqnarray*}
\lefteqn{\langle\,\phi\,|\,H^{(2)}_{{\rm eff}}\,\psi\,\rangle_{\H_{\rm eff}}}\\ 
&& \;=\; \int_\C \Big( g_{{\rm eff}}^\veps\big((p^\veps_{{\rm eff}}\phi)^*,p^\veps_{{\rm eff}}\psi\big)\,+\,\phi^*\big(E_{\rm f}+\veps\, \langle\varphi_{\rm f}|(\nabla^{\rm v}_\cdot W)\varphi_{\rm f}\rangle_{\H_{\rm f}}+\veps^2\,W^{(2)}\big)\,\psi\\
&& \qquad\qquad\qquad\,-\,\veps^2\,\M\big(\Psi^*(\veps\nabla p^\veps_{{\rm eff}}\phi,p^\veps_{{\rm eff}}\phi,\phi),\Psi(\veps\nabla p^\veps_{{\rm eff}}\psi,p^\veps_{{\rm eff}}\psi,\psi)\big)\Big)\,d\mu^\veps_{{\rm eff}},
\end{eqnarray*}
where for $\tau_1,\tau_2\in\Gamma(T^*\C)$
\begin{eqnarray*}
g^\veps_{{\rm eff}}(\tau_1,\tau_2) &=& 
g(\tau_1,\tau_2) \ +\ \veps\ \langle\,\varphi_{\rm f}\,|\,2{\rm II}(\,.\,)(\tau_1,\tau_2)\,\varphi_{\rm f}\,\rangle_{\H_{\rm f}}\\
&& \ +\ \veps^2\ \Big\langle\,\varphi_{\rm f}\,\Big|\,3g\big(\W(\,.\,)\tau_1,\W(\,.\,)\tau_2\big)\,\varphi_{\rm f}\,+\,\overline{\R}\big(\tau_1,\,.\,,\tau_2,\,.\,\big)\varphi_{\rm f}\Big\rangle_{\H_{\rm f}},\\
\vspace{0.3cm}
p^\veps_{{\rm eff}}\psi &=& -\,\I\veps{\rm d}\psi \,-\,{\rm Im}\,\Big(\veps\,\langle\varphi_{\rm f}|\nabla^{\rm h}\varphi_{\rm f}\rangle_{\H_{\rm f}}\,
-\,\veps^2 \int_{N_q\C}{\textstyle \frac{2}{3}}\,\varphi_{\rm f}^*\,\overline{{\rm R}}\big(\nabla^{\rm v}\varphi_{\rm f},\nu\big)\nu\,d\nu\\
&& \qquad\,+\ \veps^2\,\big\langle\,\varphi_{\rm f}\,\big|\,2\,\big(\W(\,.\,)\,
-\,\langle\,\varphi_{\rm f}\,|\,\W(\,.\,)\varphi_{\rm f}\,\rangle_{\H_{\rm f}}\,\big)\,
\nabla^{\rm h}\varphi_{\rm f}\,\big\rangle_{\H_{\rm f}}\Big)\,\psi,
\end{eqnarray*}
with $\W$ the Weingarten mapping, ${\rm II}$ the second fundamental form, $\overline{{\rm R}}$ the curvature mapping, $\overline{\R}$ the Riemann tensor,
and $T^{(*)}_q\C$ and $N^{(*)}_q\C$ canonically included into $T^{(*)}_{(q,0)}N\C$. The arguments $'\,.\,'$ are integrated over the fibers.

\smallskip

Furthermore, $W^{(2)}=\langle\varphi_{\rm f}|{\textstyle\frac{1}{2}}(\nabla^{\rm v}_{\cdot,\cdot}W)\varphi_{\rm f}\rangle_{\H_{\rm f}}+V_{{\rm BH}}+V_{{\rm geom}}+V_{{\rm amb}}$ and
\begin{eqnarray*}
V_{{\rm BH}} &=& \int_{N_q\C}g_{{\rm eff}}^\veps(\nabla^{\rm h}\varphi_{\rm f}^*\,,\,(1-P_0)\nabla^{\rm h}\varphi_{\rm f})\,d\nu,\\
V_{{\rm geom}} &=& -\,{\textstyle \frac{1}{4}}\,\overline{g}(\eta,\eta)\,+\,{\textstyle \frac{1}{2}}\,\kappa\,-\,{\textstyle \frac{1}{6}}\,\big(\overline{\kappa}+{\rm tr}_\C\,\overline{{\rm Ric}}+{\rm tr}_\C\,\overline{\R}\big),\\
V_{{\rm amb}} &=& \int_{N_q\C}{\textstyle \frac{1}{3}}\,\overline{\R}\big(\nabla^{\rm v}\varphi_{\rm f}^*,\nu,\nabla^{\rm v}\varphi_{\rm f},\nu\big)\,d\nu,\\
\M(\varphi_1^*,\varphi_2) &=& \big\langle\,\varphi_1\,\big|\,(1-P_0)\big(H_{\rm f}-E_{\rm f}\big)^{-1} (1-P_0)\,\varphi_2,\big\rangle_{\H_{\rm f}}\\
\Psi(A,p,\phi) &=& -\,\varphi_{\rm f}\,{\rm tr}_\C\big(\W(\nu)A\big)\,-\,2g_{{\rm eff}}^\veps\big(\nabla^{\rm h} \varphi_{\rm f}^*,p\big)\,+\,\varphi_{\rm f} (\nabla^{\rm v}_\nu W)\phi
\end{eqnarray*}
with $\eta$ the mean curvature vector, $\kappa,\overline{\kappa}$ the scalar curvatures of $\C$ and $\A$, and ${\rm tr}_\C\,\overline{{\rm Ric}},{\rm tr}_\C\,\overline{\R}$ the partial traces with respect to $\C$ of the Ricci and the Riemann tensor of $\A$ (see the appendix for definitions of all the geometric objects).
\end{theorem}

This result will be derived in Section~\ref{proofT2}.  One might wonder whether the complicated form of the effective Hamiltonian renders the result useless for practical purposes. However, as explained in the introduction, the possibly much lower dimension of  $\C$ compared to that of $\A$
outweighs the more complicated form of the Hamiltonian. Moreover, the effective Hamiltonian is of a form that allows the use of semiclassical techniques  for a  further analysis. Finally, in practical applications typically only some of the terms appearing in the effective Hamiltonian are relevant. As an example we discuss the case of a quantum wave guide in Section~\ref{waveguides}.
At this point we only add some general remarks concerning  the numerous terms in $H_{\rm eff}$ and their consequences.

\begin{remark}
\begin{enumerate}
\item If $\C$ is compact or contractible or if $E_{\rm f}$ is the ground state energy of $H_{\rm f}$, the assumption $V_0\in C^\infty_{\rm b}\big(\C,C^\infty_{\rm b}(N_q\C)\big)$ implies the extra assumption that $\varphi_{\rm f}\in C^\infty_{\rm b}(\C,\H_{\rm f})$ (see Lemma \ref{expdecay} in Section \ref{subspace}). We do not know if this implication holds true in general, but expect this for all relevant applications.
\item 
$\nabla^{\rm eff}_\tau\psi:=(\I\,p_{\rm eff}^\veps\psi)(\tau)$ is a metric connection on the trivial complex line bundle over $\C$ where $\psi$ takes its values, a so-called Berry connection.
It is flat because $\varphi_{\rm f}$ can be chosen real-valued locally which follows from $H_{\rm f}$ being real.
The first order correction in $p_{{\rm eff}}^\veps$ is a geometric generalization of the Berry term appearing in the Born-Oppenheimer setting. 
When the constraining potential is constant up to rotations,
the first-order correction reduces to the Berry term discussed by Mitchell in \cite{Mit}. 
\item The correction of the metric tensor by exterior curvature is a feature not realized before because tangential kinetic energies were taken to be small as a whole. Its origin is that the dynamics does not take place exactly on the submanifold. Therefore the mass distribution of $\varphi_{\rm f}$ has to be accounted for when measuring distances. 
\item The off-band coupling $\M$ and $V_{{\rm BH}}$, an analogue of the so-called Born-Huang potential, also appear when adiabatic perturbation theory is applied to the Born-Oppenheimer setting (see \cite{PST}). However, $\M$ contains a new fourth order differential operator which comes from the exterior curvature. Both $\M$ and $V_{{\rm BH}}$ can easily be checked to be gauge-invariant, i.e.\ not depending on the choice of $\varphi_{\rm f}$ but only on $P_0$.
\item The existence of the geometric extra potential $V_{{\rm geom}}$ has been stressed in the literature, in particular as the origin of curvature-induced bound states in quantum wave guides (reviewed by 
Duclos and Exner in \cite{DE}). 
In our setting, these are relevant for sending signals over long distances only (see Remark \ref{signals} below). A simple example where the inner curvature of the ambient manifold plays a role was given by Freitas and Krej$\check{{\rm c}}$i$\check{\rm r}$\'ik in Section 5 of \cite{FK}.
The potential $V_{{\rm amb}}$ was also found in \cite{Mit}. 
\item If $H^{(2)}_{{\rm eff}}$ was defined by the expression in the theorem, the statement would be wrong for $\xi=\chi(H^{(2)}_{{\rm eff}})$ because the fourth order term in $\M$ would be dominant. Therefore $\M$ is modified in the definition of $H^{(2)}_{{\rm eff}}$ so that the associated operator is bounded (see (\ref{truncated}) below). However, when energies of $H^{(2)}_{{\rm eff}}$ are approximated by perturbation theory or the WKB method, that modification is of lower order as the leading order of a quasimode $\psi$ satisfies $\psi=\chi(-\veps^2\Delta_\C+E_{\rm f})\psi+\O(\veps)$ for some $\chi$.
\end{enumerate} 
\end{remark}

Using Theorem \ref{calcHeff} we may exchange $H^\veps_{{\rm eff}}$ with $H^{(2)}_{{\rm eff}}$ in Theorem~\ref{effdyn}. After replacing $P^\veps$ and $ U^\veps$ by their leading order expressions, which adds a time-independent error of order~$\veps$, it is not difficult to derive the following result.

\begin{corollary}\label{effdyn3}
Fix $E<\infty$ and set $ U_0^\veps:=U_0 D_\veps^*$. Under the assumptions of Theorem \ref{calcHeff} there are $C<\infty$ and $\veps_0>0$ such that for all $\veps\leq\veps_0$, $t\in\RRR$, and each Borel function $\chi:\RRR\to[-1,1]$ with ${\rm supp}\,\chi\subset(-\infty,E]$ it holds
\begin{equation}\label{difference2}
\left\|\left(\e^{-\I H^\veps t}- U_0^{\veps*}\e^{-\I H^{(2)}_{{\rm eff}}t}\, U_0^\veps\right) U_0^{\veps*}\chi(H^{(2)}_{{\rm eff}}) U_0^\veps\,\right\|_{\L(\overline{\H})}\;\leq\;C\,\veps\,(\veps^2|t|+1).
\end{equation}

\end{corollary}
 
Corollary \ref{effdyn3} will also be proved in Section \ref{proofT2}. While (\ref{difference2}) is somewhat weaker than (\ref{difference}), it is much better suited for applications, since $ U_0^\veps$ is given in terms of the eigenfunction $\varphi_{\rm f}$ and depends on $\veps$ only via the dilation $D_\veps$. So, in view of Theorem~\ref{calcHeff}, all relevant expressions in (\ref{difference2}) can be computed explicitly.


\subsection{Approximation of eigenvalues}\label{eigenvalues}

In this section we discuss in which way our effective Hamiltonian allows us to approximate certain parts of the discrete spectrum and the associated eigenfunctions of the original Hamiltonian. 
The following result shows how to obtain quasimodes of $H^\veps$ from the eigenfunctions of $H^{(2)}_{{\rm eff}}$ and vice versa.
\begin{theorem}\label{quasimodes}
Let $E_{\rm f}$ be a simple constraint energy band whose associated eigenvalue bundle is smoothly trivializable, and let $ U^\veps,H^{(2)}_{{\rm eff}}$ be the operators associated with $E_{\rm f}$ via Theorems \ref{effdyn} \& \ref{calcHeff}.

a) Let $E\in\RRR$. Then there are $\veps_0>0$ and $C<\infty$ such that for any family $(E_\veps)$ with $\limsup_{\veps\to0} E_\veps\,<\,E$ and all $\veps\leq\veps_0$ the following implications hold:
\begin{enumerate}
\item $H^{(2)}_{{\rm eff}}\psi_\veps\;=\;E_\veps\psi_\veps\quad\Longrightarrow\quad \|\,( H^\veps\,  -\,E_\veps)\, U^{\veps*}\psi_\veps\,\|_{\overline{\H}} \;\leq\;  C\,\veps^3\,\|U^{\veps*}\psi_\veps\|_{\overline{\H}}$, 
\item $\,H^\veps\,\psi^\veps\;=\;E_\veps\psi^\veps \,\quad\Longrightarrow\quad \| \,(H^{(2)}_{{\rm eff}}\,  -\,E_\veps)\, U^\veps\psi^\veps\,\|_{\H_{\rm eff}} \;\leq\; C\,\veps^3 \,\|\psi^\veps\|_{\overline{\H}}$.
\end{enumerate}

\pagebreak

b) Let $E_{\rm f}(q)=\inf\sigma\big(H_{\rm f}(q)\big)$ for some (and thus for all) $q\in\C$ and define $E_1(q):=\inf\,\sigma\big(H_{\rm f}(q)\big)\setminus E_{\rm f}(q)$.
Let $(\psi^\veps)$ be a family with
\begin{equation}\label{threshold}
\limsup_{\veps\to0}\,\big\langle\psi^\veps\big|\big(-\veps^2 M_\rho\Delta_{\rm v}M_\rho^*+V_0(q,\nu/\veps)\big)\psi^\veps\big\rangle\;<\;\inf_{q\in\C}E_1.
\end{equation}
Then there are $\veps_0>0$, $c>0$ such that $\|U^\veps\psi^\veps\|_{\H_{\rm eff}}\geq c\,\|\psi^\veps\|_{\overline{\H}}$ for all $\veps\leq\veps_0$.
\end{theorem}

We recall that for any self-adjoint operator $H$ the bound $\|(H-\lambda)\psi\|<\delta\|\psi\|$ for $\lambda\in\RRR$ implies that $H$ has spectrum in the interval $[\lambda-\delta,\lambda+\delta]$. So a) i) entails that $H^\veps$ has an eigenvalue in an interval of length $2C\veps^3$ around $E_\veps$, if one knows a priori that the spectrum of $H^\veps$ is discrete below the energy~$E$.
The statement b) ensures that a) ii) really yields a quasimode for normal energies below $\inf_{q\in\C}E_1$, i.e.\ that 
\[
\,H^\veps\,\psi^\veps\;=\;E_\veps\psi^\veps \quad\Longrightarrow\quad \| \,(H^{(2)}_{{\rm eff}}\,  -\,E_\veps)\, U^\veps\psi^\veps\,\|_{\H_{\rm eff}} \;\leq\; {\textstyle\frac{C}{c}}\,\veps^3 \,\|U^\veps\psi^\veps\|_{\H_{\rm eff}}\,.
\]

\begin{remark}\label{flatcase}
If the ambient manifold $\A$ is flat, $-\veps^2M_\rho\Delta_{\rm v}M_\rho^*$ is 
form-bounded by $-\veps^2\Delta_{N\C}+C\veps^2$ for some $C<\infty$ independent of $\veps$ (this follows from Lemma~\ref{transform} below and the expression (5.5) for $-\veps^2\Delta_{N\C}$ in \cite{FrH}). Then, since $H^\veps=-\veps^2\Delta_{N\C}+V_0(q,\nu/\veps)+W(q,\nu)-W(q,0)$,  (\ref{threshold}) follows from
\[\limsup_{\veps\to0}\,\langle\psi^\veps|H^\veps\psi^\veps\rangle\;<\;\inf_{q\in\C}E_1-\sup_{(q,\nu)}\big(W(q,0)-W(q,\nu)\big)\;=:\;E_*.\]
Therefore Theorem \ref{quasimodes}, in particular, implies that at least for flat $\A$ there is a one-to-one correspondence between the spectra of $H^\veps$ and $H^{(2)}_{{\rm eff}}$ below $E_*$.
\end{remark}

One may ask whether a family $(E_\veps)$ of energies of $H^\veps$ with $\limsup E_\veps<E_*$ exists at all. A sufficient condition is that $\sup_{(q,\nu)}\big(W(q,0)-W(q,\nu)\big)$ is strictly smaller than $\inf_{q\in\C}E_1-\inf_{q\in\C}E_{\rm f}$. For this implies $\inf_{q\in\C}E_{\rm f}<E_*$,
and the spectrum of $H^{(2)}_{{\rm eff}}$ in the interval $[\inf_{q\in\C}E_{\rm f},E_*]$ has either a continuous part or the number of eigenvalues is at least of order $\veps^{-1}$ because $H^{(2)}_{{\rm eff}}$'s leading order term $-\veps^2\Delta_\C+E_{\rm f}$ is a semiclassical operator. Then by a) i) this is also true for $H^\veps$. 

\medskip

The eigenvalues of $H^{(2)}_{{\rm eff}}$ can be approximated by the WKB construction, which is quite standard (see e.g.\  \cite{HS}). In the simplest case one obtains:

\begin{corollary}
Assume that $\A$ is flat and let $E_{\rm f}$ is a constraint energy band with $\inf E_{\rm f}<E_*$ and
$E_{\rm f}(q)=\inf\sigma\big(H_{\rm f}(q)\big)$ for all $q\in\C$. 
Let there be $q_0$ such that $E_{\rm f}(q_0)<E_{\rm f}(q)$ for all $q\neq q_0$ and $\big(\nabla^2_{\partial_{x^i},\partial_{x^j}}E_{\rm f}\big)(q_0)$ is positive definite. 

Denote by $E_\ell(A)$ the $\ell$-th eigenvalue of a semi-bounded operator $A$, counted from the bottom of the spectrum. Then for any $\ell\in\NNN$
\begin{equation*}
E_\ell(H^\veps)\;=\;E_{\rm f}(q_0)\,+\,\veps E_\ell(H_{\rm HO})\,+\,\O(\veps^2),
\end{equation*}
where $H_{\rm HO}:=-\Delta_{\RRR^d}+\tfrac{1}{2}(\nabla^2_{\partial_{x^i},\partial_{x^j}}E_{\rm f})(q_0)x^ix^j$ is a harmonic oscillator on $\RRR^d$.
\end{corollary}

We want to relate this to results by Friedlander and Solomyak in \cite{FS} and Borisov and Freitas in \cite{BF}. In both works the spectrum of the Dirichlet Laplacian $-\Delta_{\rm D}$ on the thin domain $\{(x,y):\,x\in \Omega,\,-\veps h_-(x)<y<\veps h_+(x)\}$ for positive functions $h_\pm:\Omega\to\RRR$ is considered, where $\Omega$ is any interval in $\RRR$ in \cite{FS} and a bounded domain in $\RRR^d$ in \cite{BF}.
It is shown that, if $h:=h_+-h_-$ has a global maxmimum at $x_0\in\Omega$ which is non-degenerate, then 
\[E_\ell(-\veps^2\Delta_{\rm D})\;=\;\tilde E_{\rm f}(x_0)\,+\,\veps E_\ell\big(-\partial^2_{x^ix^i}+\tfrac{1}{2}(\nabla^2_{\partial_{x^i},\partial_{x^j}}E_{\rm f})(x_0)x^ix^j\big)\,+\,o(\veps),\] 
where $\tilde E_{\rm f}(x):=\pi^2/h^2(x)$ is the lowest eigenvalue of the Dirichlet Laplacian on the interval $[0,h(x)]$. 
So our result generalizes this to an $\Omega$ that is curved and of arbitrary codimension, but with the Dirichlet Laplacian replaced by a constraining potential. 
For a set with smooth boundary, however, we do not see a problem in using the Dirichlet Laplacian instead of the constraining potential. The strict localization to an $\veps$-tube around $\mathcal{C}$ would even simplify many steps in our proof considerably.


\subsection{Application to quantum wave guides}\label{waveguides}

In this section we look at the special case of a curve $\C$ in $\A=\RRR^3$ equipped with the euclidean metric. Such curves may model quantum wave guides which have been discussed theoretically for long times but are nowadays also investigated experimentally. We will provide the expression for our effective Hamiltonian in this case and discuss which terms remain, if we add restrictions on the constraining potential or the geometry. Furthermore, we will apply Theorem \ref{quasimodes} to obtain a statement about the spectrum of a twisted wave guide.
For the sake of brevity, we assume that $W$, the non-constraining part of the potential, vanishes. Its contributions  could be trivially added in all formulas and as long as $\sup_{(q,\nu)}\big(W(q,0)-W(q,\nu)\big)$ is small enough also in the statements (see the preceding subsection).

\medskip

We first look at an infinite quantum wave guide. So let the curve $\C$ be given as a smooth injective $c:\RRR\to\RRR^3,x\mapsto c(x)$ that has bounded derivatives of any order and is parametrized by arc length ($|\dot c|=1$). The mean curvature vector of $c$ is $\eta=\ddot c$ and its (exterior) curvature is $|\eta|$. Denoting by $\cdot$ the usual scalar product in $\RRR^3$ we define $n(\nu):=\nu\cdot\eta/|\eta|$ where $\eta\neq0$ and $n(\nu):=0$ elsewhere. By the Frenet formulas the Weingarten mapping satisfies $\W(\eta)=|\eta|^2$ (see e.g.\ \cite{DoC}) and $\W\equiv0$ on the orthogonal complement of $\eta$ (which is meant to be $N_q\C$ if $\eta(q)=0$). 

A normalized section of $T\C$ is given by $\tau:=\dot c$. We extend this to an orthonormal frame of $T\C\times N\C$ in the following way: We fix $q\in\C$, choose an arbitrary orthonormal basis of $N_q\C$, and take $\nu_1,\nu_2$ to be the parallel transport of this basis with respect to the normal connection $\nabla^\perp$ along the whole curve. This yields an orthonormal frame of $N\C$. Together with $\tau$ we obtain an orthonormal frame of $T\C\times N\C$, which is sometimes called the Tang frame. We denote the coordinates with respect to $\tau$, $\nu_1$, and $\nu_2$ by $x$, $n_1$, and $n_2$ respectively. In these coordinates it holds $\nabla^{\rm h}=\partial_x$ (as can for example be seen from the general coordinate formula (\ref{horder}) below).

\smallskip

Now let $E_{\rm f}$ be a simple constraint energy band and $\varphi_{\rm f}$ a global family of eigenfunctions associated with it. We start by spelling out the formula for $H_{\rm eff}^{(2)}$ from Theorem \ref{calcHeff}. Of course, all terms containing the inner curvature of $\C$ and $\A=\RRR^3$ vanish due to the flatness of $\C$ and $\A$ with the euclidean metric. Since $\C$ is one-dimensional and contractible, $\varphi_{\rm f}$ can be chosen such that 
$p_{\rm eff}^\veps\equiv-\I\veps\partial_x$ globally. Then the effective Hamiltonian is 
\begin{eqnarray}
H_{\rm qwg}^\veps &=& -\,\veps\partial_x\big(1+\veps|\eta|\langle\varphi_{\rm f}|n\varphi_{\rm f}\rangle+3\veps^2|\eta|^2\langle\varphi_{\rm f}|n^2\varphi_{\rm f}\rangle\big)\veps\partial_x\,+\,E_{\rm f}\nonumber\\
&&\,-\,\veps^2\,|\eta|^2/4\,+\,\veps^2\,\big(\langle\partial_x\varphi_{\rm f}|\partial_x\varphi_{\rm f}\rangle-|\langle\varphi_{\rm f}|\partial_x\varphi_{\rm f}\rangle|^2\big)\nonumber\\
&& \,+\,\veps^2\,\Big(4\,\veps\partial_x\,\langle\partial_x\varphi_{\rm f}|R_{H_{\rm f}}(E_{\rm f})\partial_x\varphi_{\rm f}\rangle\,\veps\partial_x\nonumber\\
&& \qquad\quad\,+\,4\,|\eta|\,{\rm Re}\,\veps\partial_{x}\,\langle \partial_x\varphi_{\rm f}|R_{H_{\rm f}}(E_{\rm f})n\varphi_{\rm f}\rangle\,\veps^2\partial^2_{xx}\label{mixed}\nonumber\\
&& \qquad\qquad\,+\,|\eta|^2\,\veps^2\partial^2_{xx}\,\langle n\varphi_{\rm f}|R_{H_{\rm f}}(E_{\rm f})n\varphi_{\rm f}\rangle\,\veps^2\partial^2_{xx}\Big) \quad\label{fullwaveguide}
\end{eqnarray} 
with $R_{H_{\rm f}}(E_{\rm f}):=(1-P_0)(H_{\rm f}-E_{\rm f})^{-1}(1-P_0)$ and $\langle\,\phi\,|\,\psi\,\rangle:=\int_{\RRR^2}\phi^*\,\psi\,dn_1dn_2$. 

\smallskip

We emphasize that  formula (\ref{fullwaveguide}) is only valid when applied to states with high energies cut off because this is required for the application of Theorem~\ref{calcHeff}. In particular, this explains why the differential operator of fourth order is not to be thought of as the dominant term but only as of order $\veps^2$. But still $\|\partial_x\psi\|\sim\veps^{-1}$ is possible for a $\psi$ of finite energy! Before we consider some special cases, we want to make the following crucial remark about sending signals through wave guides with constant 'cross section'.
\begin{remark}\label{signals}
For highly oscillating states $\psi$, i.e.\ with $\langle\psi|-\veps^2\partial^2_{xx}\psi\rangle\sim1$, the only term of order one besides $-\veps^2\partial^2_{xx}$ is $E_{\rm f}$. In particular, if $E_{\rm f}$ is constant, the dynamics is free at leading order and, even more, the potential terms are of order $\veps^2$. So they only become relevant for times of order $\veps^{-2}$. However, a semiclassical wave packet $\psi$ covers distances of order $\veps^{-1}$ on this time scale. Hence, for such $\psi$ note-worthy trapping occurs only for very long wave guides!  
\end{remark}

\smallskip

If we consider a straight wave guide, i.e.\ $\eta\equiv0$, the formula we end up with is a complete analogue of the one derived by Panati, Spohn, and Teufel in~\cite{PST} in the case of the Born-Oppenheimer approximation:
\begin{eqnarray}
H_{\rm qwg}^\veps|_{\eta\equiv0} &=& -\,\veps^2\partial^2_{xx}\,+\,E_{\rm f} \,+\,\veps^2\,\big(\langle\partial_x\varphi_{\rm f}|\partial_x\varphi_{\rm f}\rangle-|\langle\varphi_{\rm f}|\partial_x\varphi_{\rm f}\rangle|^2\big)\nonumber\\
&& \quad\,+\,\veps^2\,4\veps\partial_x\,\langle\partial_x\varphi_{\rm f}|R_{H_{\rm f}}(E_{\rm f})\partial_x\varphi_{\rm f}\rangle\,\veps\partial_x.\label{straightwaveguide}
\end{eqnarray} 
We note that, although $\eta\equiv0$, the $x$-dependence of the constraining potential still allows us to model interesting situations, e.g.\ a beam splitter \cite{JS}.

\smallskip

Now we drop the assumption $\eta\equiv0$ and assume that the constraining potential $V_{\rm c}$ is parallel with respect to $\nabla^{\rm h}$ instead. This means $\nabla^{\rm h}V_{\rm c}\equiv0$. Then we obtain a global family of eigenfunctions $\varphi_{\rm f}$ with $\nabla^{\rm h}\varphi_{\rm f}\equiv0$ by taking it to be the parallel transport with respect to $\nabla^{\rm h}$ of $\varphi_{\rm f}(q_0)$ for any $q_0\in\C$. 
In addition, since $V_{\rm c}$ does not change its shape, $E_{\rm f}$ is constant and thus may be removed by redefining zero energy. Therefore we have
\begin{eqnarray}
H_{\rm qwg}^\veps|_{\nabla^{\rm h}V_{\rm c}\equiv0} &=& -\,\veps\partial_x\big(1+\veps|\eta|\langle\varphi_{\rm f}|n\varphi_{\rm f}\rangle+3\veps^2|\eta|^2\langle\varphi_{\rm f}|n^2\varphi_{\rm f}\rangle\big)\veps\partial_x\nonumber\\
&& \,-\,\veps^2\,|\eta|^2/4\,+\,\veps^2\,|\eta|^2\,\veps^2\partial^2_{xx}\,\langle n\varphi_{\rm f}|R_{H_{\rm f}}(E_{\rm f})n\varphi_{\rm f}\rangle\,\veps^2\partial^2_{xx}.\quad
\end{eqnarray} 

There is a wide literature on quantum wave guides where the effects of bending and twisting on the spectrum of the Dirichlet Laplacian on an $\veps$-tube with a fixed cross section is investigated (see the review \cite{K} by Krej$\check{{\rm c}}$i$\check{\rm r}$\'ik). If we consider the corresponding situation that $V_{\rm c}$ does not change its shape but is allowed to twist, $E_{\rm f}$ is the only term in (\ref{fullwaveguide}) that may be neglected. 
Since the remaining potential terms are, however, of order $\veps^2$, the kinetic energy operator $-\veps^2\partial^2_{xx}$ will also be of order $\veps^2$ for low eigenvalues. So $H_{\rm qwg}^\veps$ may be devided by $\veps^2$. Keeping only the leading order terms we arrive at
\begin{eqnarray}\label{twist}
H_{\rm twist}^\veps &:=& -\partial^2_{xx}\,-\,|\eta|^2/4\,+\,\langle\partial_x\varphi_{\rm f}|\partial_x\varphi_{\rm f}\rangle-|\langle\varphi_{\rm f}|\partial_x\varphi_{\rm f}\rangle|^2
\end{eqnarray} 
The twisting assumption means that there is $\tilde V_{\rm c}\in C^\infty_{\rm b}(\RRR^2)$ and $\alpha\in C^\infty_{\rm b}(\RRR)$ such that the constraining potential has the form:
\begin{equation*}
\big(V_{\rm c}^\alpha(x)\big)(n_1,n_2)\,:=\, 
\tilde V_{\rm c}\big(n_1\cos\alpha(x)-n_2\sin\alpha(x),n_1\sin\alpha(x)+n_2\cos\alpha(x)\big).
\end{equation*} 
Then the family of eigenfunctions $\varphi_{\rm f}$ may be chosen as
\begin{equation*}
\big(\varphi_{\rm f}(x)\big)(n_1,n_2)\,:=\, 
\Phi_{\rm f}\big(n_1\cos\alpha(x)-n_2\sin\alpha(x),n_1\sin\alpha(x)+n_2\cos\alpha(x)\big)
\end{equation*} 
for an eigenfunction $\Phi_{\rm f}$ of $-\Delta_{\RRR^2}+\tilde V_{\rm c}(x)$ with eigenvalue $E_{\rm f}$. 
A simple calculation yields 
\begin{equation*}
\langle\partial_x\varphi_{\rm f}|\partial_x\varphi_{\rm f}\rangle-|\langle\varphi_{\rm f}|\partial_x\varphi_{\rm f}\rangle|^2 \;=\; \dot\alpha^2\int_{\RRR^2}|n_1\partial_{n_2}\Phi_{\rm f}-n_2\partial_{n_1}\Phi_{\rm f}|^2dn_1dn_2.
\end{equation*}
We note that the integral is the expectation value of the squared angular momentum of $\Phi_{\rm f}$ and thus vanishes for a rotation-invariant $\Phi_{\rm f}$. So (\ref{twist}) shows that bending is attractive, while twisting is repulsive.
Now Theorem~\ref{quasimodes} together with Remark~\ref{global} and Remark~\ref{flatcase} implies the following result.

\begin{corollary}\label{twist2}
Let $\C\subset\RRR^3$ be an infinite curve, $W\equiv0$, and $E_{\rm f}$ be a constraint energy band with $E_{\rm f}(q)=\inf\sigma\big(H_{\rm f}(q)\big)$ for all $q\in\C$. Let $L\in\NNN_0\cup\{\infty\}$ be the number of eigenvalues of $H_{\rm twist}$ below the continuous spectrum, where $H_{\rm twist}$ is the following operator on $\RRR$:
\begin{equation*}
H_{\rm twist}\;:=\;-\partial^2_{xx}\,-\,|\eta|^2/4\,+\,C(\Phi_{\rm f})\,\dot\alpha^2
\end{equation*}
with $C(\Phi_{\rm f}):=\int_{\RRR^2}|n_1\partial_{n_2}\Phi_{\rm f}-n_2\partial_{n_1}\Phi_{\rm f}|^2dn_1dn_2$.

\smallskip

Denote by $E_\ell(A)$ the $\ell$-th eigenvalue of a semi-bounded operator $A$, counted from the bottom of the spectrum. 
If $V_{\rm c}$ only twists, i.e.\ $V_{\rm c}=V_{\rm c}^\alpha$ for some $\alpha$ as above, then for any $\ell<L$
\begin{equation*}
E_\ell(H^\veps)\;=\;E_{\rm f}\,+\,\veps^2 E_\ell(H_{\rm twist})\,+\,\O(\veps^3).
\end{equation*}
\end{corollary}

This is an analogue of the result by Bouchitt\'e, Mascarenhas and Trabucho in~\cite{BMT} for $\veps$-tubes twisted by $\alpha$, extended by Borisov and Cardone in \cite{BC}. In~\cite{K} it was posed as an open problem to generalize this result to an infinite tube. Corollary \ref{twist2} achieves this for a constraining potential that twists instead of the Dirchlet tube.

\medskip

Up to now we have considered an infinite wave guide which is topological trivial. The only possible non-trivial topology for a one-dimensional manifold is that of a circle. So let $\C$ now be diffeomorphic to a circle. We refer to such a $\C$ as a \emph{quantum wave circuit}. Because of the non-trivial topology our choices of the family $\varphi_{\rm f}$ made above are only possible locally but in general not globally. Therefore we rewrite (\ref{straightwaveguide}) without those choices and ignoring the terms of order $\veps^2$ for the moment:  
\begin{equation}\label{wavecircuit}
H_{\rm qwc}^\veps \;=\; p_\veps^*\big(1+\veps|\eta|\langle\varphi_{\rm f}|n\varphi_{\rm f}\rangle\big)p_{\veps}\,+\,E_{\rm f}\,+\,\O(\veps^2)
\end{equation} 
with $p_\veps=-\I\veps\partial_x+\veps\,\big\langle\varphi_{\rm f}\big|\I\partial_x\varphi_{\rm f}\big\rangle$. Although the curvature of the connection $\I p_\veps$ always vanishes, it may have a non-trivial holonomy over the circle, which we will discuss next.

\medskip

For the sake of simplicity we consider a round circle, i.e.\ with constant $|\eta|$. Let $x$ be a $2\pi$-periodic coordinate for it. The eigenfunction $\varphi_{\rm f}(x)$ can be chosen real-valued for each fixed $x$ because $H_{\rm f}$ is real. This associates a real line bundle to $E_{\rm f}$. From the topological point of view, there are two real line bundles over the sphere: the trivial one and the non-trivializable M\"obius band. In the former case the global section $\varphi_{\rm f}$ can be chosen real everywhere. This implies $\langle\varphi_{\rm f}|\partial_x\varphi_{\rm f}\rangle\equiv0$ which results in $\I p_\veps=\veps\partial_x$. Thus the holonomy group of $\I p_\veps$ is trivial in this case. We will now provide an example for the realization of the M\"obius band by a suitable constraining potential and show that the holonomy group of $\I p_\veps$ becomes $\ZZZ/2\ZZZ$! 

\smallskip

Let $\tilde V_{\rm c}\in C^\infty_{\rm b}(\RRR^2)$ have two orthogonal axes of reflection symmetry, i.e. in suitable coordinates 
\begin{equation}\label{symmetryV}
\tilde V_{\rm c}(-N_1,N_2)\;=\;\tilde V_{\rm c}(N_1,N_2)\;=\;\tilde V_{\rm c}(N_1,-N_2).
\end{equation} 
Then the real ground state $\Phi_0$  of $-\Delta_{\RRR^2}+\tilde V_{\rm c}$ with energy $E_0$ is symmetric with respect to both reflections,
\[
\Phi_0(N_1,N_2)\;=\;\Phi_0(-N_1,N_2)\;=\;\Phi_0(N_1,-N_2)\,,
\]
while the first excited state $\Phi_1$, also taken real-valued, with energy $E_1$ is typically only symmetric with respect to one reflection and anti-symmetric with respect to the other one, e.g.\
\begin{equation}\label{symmetryPhi}
\Phi_1(N_1,N_2)\;=\;-\,\Phi_1(-N_1,N_2)\;=\;\Phi_1(N_1,-N_2)\,.
\end{equation} 
This is true in particular for a harmonic oscillator with different frequencies. 
As the potential constraining to the round circle we choose the twisting potential $V_{\rm c}^\alpha$ introduced above with $\alpha(x)=x/2$, i.e.
\begin{equation*}
\big(V_{\rm c}^{x/2}(x)\big)(n_1,n_2)\,:=\, 
\tilde V_{\rm c}\big(\cos(x/2)n_1-\sin(x/2)n_2,\sin(x/2)n_1+\cos(x/2)n_2\big).
\end{equation*} 
We note that due to (\ref{symmetryV}) this defines a $V_{\rm c}^{x/2}\in C^\infty_{\rm b}\big(\C,C^\infty_{\rm b}(N\C)\big)$. Then
\begin{equation*}
\big(\tilde\varphi_j(x)\big)(n_1,n_2)\,:=\, 
\Phi_j\big(\cos(x/2)n_1-\sin(x/2)n_2,\sin(x/2)n_1+\cos(x/2)n_2\big)
\end{equation*} 
is an eigenfunction of $H_{\rm f}(x):=-\Delta_{\rm v}+V_{\rm c}(x)$ with eigenvalue $E_j$ for every $x$ and $j\in\{0,1\}$. However, while $\tilde \varphi_0$ is a smooth section of the corresponding eigenspace bundle,  $\tilde\varphi_1$ is not. For by (\ref{symmetryPhi}) it holds $\tilde\varphi_1(x)=-\tilde\varphi_1(x+2\pi)$. 
Still the complex eigenspace bundle admits a smooth non-vanishing section. A possible choice is $\varphi_1(x):=\e^{\I x/2}\tilde\varphi_1(x)$. Using (\ref{symmetryPhi})  we obtain that for the first excited band the effective Hamiltonian (\ref{wavecircuit}) reduces to
\begin{eqnarray*}
H_{\rm qwc,1}^\veps&=& E_1\,+\, (-\I\veps\partial_x+\veps/2)^2 \,+\,\O(\veps^2),
\end{eqnarray*} 
while for the ground state band it is
\begin{eqnarray*}
H_{\rm qwc,0}^\veps&=& E_0\,-\, \veps^2\partial^2_{xx} \,+\,\O(\veps^2)
\end{eqnarray*} 
This shows that depending on the symmetry of the normal eigenfunction
the twist by an angle of $\pi$ has different effects on the effective momentum operator in the effective Hamiltonian. 
With respect to the connection appearing in  $H_{\rm qwc,1}^\veps$ 
the holonomy of a closed loop $\gamma$ winding around the circle once is given by $h(\gamma)=\e^{\I\int_0^{2\pi}1/2\,dx}=-1$. Hence, the holonomy group of $\I p_\veps$ is indeed $\ZZZ/2\ZZZ$ and the $1/2$ cannot be gauged away. Furthermore, a wave packet which travels around the circuit once accumulates a $\pi$-phase. This can be seen as an analogue of the Aharanov-Bohm effect, though with the only possible phase~$\pi$.  

The effect of this phase can also be seen in the level spacing of $H_j^\veps$ and thus, with Theorem~\ref{quasimodes}, also in the spectrum of $H^\veps$. The arguments that led to $(\ref{twist})$ for an infinite wave guide may be applied here, too, except that, of course,  $-\partial^2_{xx}$ has to be replaced by $(-\I\partial_x+1/2)^2$ for $H_{\rm qwc,1}^\veps$.
Since $|\eta|$ and $\dot\alpha$ are constant, the eigenvalues of $H_{\rm qwc,1}^\veps$ are 
\[
E_\ell(H_{\rm qwc,1}^\veps) = E_1 + \veps^2\left[(\ell+{\textstyle\frac{1}{2}})^2 + {\textstyle\frac{C(\Phi_1)-|\eta|}{4}}    \right]\,+\,\O(\veps^3)    \,,\qquad
\ell\in \NNN_0\,,
\]
while for $H_{\rm qwc,0}^\veps$ we find
\[
E_\ell(H_{\rm qwc,0}^\veps) = E_0 + \veps^2\left[\ell^2 + {\textstyle\frac{ C(\Phi_0)-|\eta|}{4}}  \right]\,+\,\O(\veps^3)    \,,\qquad
\ell\in \NNN_0\,.
\]

Although a constraining potential that twists along a circle was investigated by Maraner in detail in \cite{Ma1} and by Mitchell in \cite{Mit}, the effect discussed above was not found in both treatments. The reason for this is that they allowed only for whole rotations and not for half ones to avoid the non-smoothness of $\tilde\varphi_1$. Finally, we note that it easy to generalize the statements above to a circuit whose curvature and potential twist are non-constant.


\section{Proof of the main results}\label{proofs}

In the following, $\L(X,Y)$ is the Banach space of bounded operators between two Banach spaces $X$ and $Y$. $\D(A)$ will \emph{always denote the maximal domain} of an operator $A$, equipped with the graph norm. For convenience we set $\D(H^0):=\H$. $A^*$ will always be used for the adjoint of $A$ on $\H$ if not stated differently. We recall that we have set $\langle\nu\rangle:=\sqrt{1+|\nu|^2}$. $A=\langle\nu\rangle^l$ is meant to be the multiplication with $\langle\nu\rangle^l$. 
Finally, we write $a\lesssim b$, if $a$ is bounded by $b$ times a constant independent of $\veps$, and $a=\O(\veps^l)$, if $\|a\|\lesssim\veps^l$. 

\smallskip

Throughout this section we assume that $V_{\rm c},W\in C^\infty_{\rm b}(\C,C^\infty_{\rm b}(N_\cdot\C))$ and that $E_{\rm f}$ is a constraint energy band as defined in Definition \ref{gapcondition}. 

\subsection{Proof of adiabatic decoupling}\label{proofT1}

As explained in the introduction the first step in proving Theorem \ref{effdyn} is the unitary transformation of $H^\veps$ by multiplication with the square root of the relative density $\rho:=\frac{d\overline{\mu}}{d\mu\otimes d\nu}$ of the volume measure associated with $\overline{g}$ and the product measure on $N\C$. This transformation factorizes the measure, which will allow us to easily split the integral over $N\C$ later on, but it also yields an additional potential term.
The abstract statement reads as follows:
\begin{lemma}\label{transform}
Let $(\M,g)$ be a Riemannian manifold. Let $d\sigma_1,d\sigma_2$ be two measures on $\M$ with smooth and positive relative density $\rho:=\frac{d\sigma_1}{d\sigma_2}$. Define
\[M_{\rho}:L^2(\M,d\sigma_2)\to L^2(\M,d\sigma_1),\,\psi\mapsto\rho^{-\frac{1}{2}}\psi.\]
Then $M_{\rho}$ is unitary and it holds
\begin{eqnarray*}
M_{\rho}^*(-\Delta_{d\sigma_1})M_{\rho}\psi &=& -\,\Delta_{d\sigma_2}\psi\,-\,\Big({\textstyle \frac{1}{4}}g\big({\rm d} (\ln\rho),{\rm d} (\ln\rho)\big)-{\textstyle \frac{1}{2}}\Delta_{d\sigma_1}(\ln\rho)\Big)\psi\\
&=:& -\,\Delta_{d\sigma_2}\psi\,+\,V_\rho\psi,
\end{eqnarray*}
with $\Delta_{d\sigma_i}:={\rm div}_{d\sigma_i}\,{\rm grad}\,\psi$, where ${\rm grad}\,\psi$ is the vector field associated with ${\rm d}\psi$ via $g$ and ${\rm div}_{d\sigma_i}$ is the adjoint of ${\rm grad}$ on $L^2(\M,d\sigma_i)$.
\end{lemma}
The proof is a simple calculation, which can be found in the sequel to the proof of Theorem \ref{effdyn}. We recall from (\ref{density}) that $\rho=\frac{d\overline{\mu}}{d\mu\otimes d\nu}$ is in $C^\infty_{\rm b}(N\C)$ and strictly positive. Therefore $V_\rho$ is in $C^\infty_{\rm b}(N\C)$ for our choice of $\rho$. Since $\rho$ is equal to $1$ outside of $\B_{\delta}$, $V_\rho$ is even in $C^\infty_{\rm b}(\C,C^\infty_{\rm b}(N_q\C))$ which coincides with $C^\infty_{\rm b}(N\C)$ inside $\B_r$ for any $r<\infty$.

\medskip

The heart of Theorem \ref{effdyn} is the existence of a subspace $P_\veps\H\subset\H$ that can be mapped unitarily to $L^2(\C,d\mu)$ and approximately commutes with $H_\veps$:

\begin{lemma}\label{projector}
Under the assumptions of Theorem \ref{effdyn} there is $\veps_0>0$ such that for all $\veps<\veps_0$ there are an orthogonal projection $P_\veps\in\L(\H)$ and a unitary $\tilde U_\veps\in\L(\H)$ with $P_\veps=\tilde U_\veps^* P_0\tilde U_\veps$ and 
\begin{eqnarray}\label{invariance}
&\bullet& \|\tilde U_\veps-1\|_{\L(\H)}\;=\;\O(\veps)\,,\quad
\|P_\veps\|_{\L(\D(H_\veps^m))}\;\lesssim\;1,\nonumber\\
&\bullet& 
\|\langle\nu\rangle^l P_\veps\langle\nu\rangle^j\|_{\L(\H)}\;\lesssim\;1\,,\quad
\|\langle\nu\rangle^l P_\veps\langle\nu\rangle^j\|_{\L(\D(H_\veps))}\;\lesssim\;1,\nonumber\\
&\bullet& \|[H_\veps,P_\veps]\|_{\L(\D(H_\veps^{m}),\D(H_\veps^{m-1}))} = \O(\veps),\nonumber\\
&\bullet& 
\|[H_\veps,P_\veps] \,\chi(H_\veps)\|_{\L(\H,\D(H_\veps^m))} = \O(\veps^3)
\end{eqnarray}
for all $j,l,m\in\NNN_0$ and each Borel function $\chi:\RRR\to[-1,1]$ satisfying ${\rm supp}\,\chi\subset(-\infty,E]$. 
\end{lemma}

The construction of $P_\veps$ and $\tilde U_\veps$ is carried out in Section \ref{subspace}. There is a heuristic discussion at the beginning of that section that the reader may find instructive to get an idea why $P_\veps$ and $\tilde U_\veps$ exist. When we take its existence for granted, it is not difficult to prove that the effectice dynamics on the submanifold is a good approximation.

\pagebreak

{\sc Proof of Theorem \ref{effdyn} (Section \ref{results1}):}
\newline
Let $d\mu^\veps_{\rm eff}$ be the volume measure associated with $g^\veps_{\rm eff}$ which we define by the expression in Theorem \ref{calcHeff}. For any fixed $E<\infty$, Lemma \ref{projector} yields some unitary~$\tilde U_\veps$ for all $\veps$ below a certain $\veps_0$. Since we assumed that the eigenspace bundle associated with $E_{\rm f}$ is trivializable, there is $U_0:\H\to L^2(\C,d\mu)$ as in Definition~\ref{U0}. We define $U_\veps:=U_0\tilde U_\veps$. Using Definition~\ref{U0} and Lemma \ref{projector} we have $U_\veps^* U_\veps\;=\;\tilde U_\veps^*U_0^* U_0 \tilde U_\veps\;=\;\tilde U_\veps^*P_0\tilde U_\veps\;=\;P_\veps$ and
\begin{equation}\label{unitarity}
U_\veps U_\veps^*\;=\;U_0 \tilde U_\veps\tilde U_\veps^*U_0^* \;=\;U_0U_0^*\;=\;1.
\end{equation}
In view of Lemma \ref{transform}, we next set $ U^\veps:=M_{\tilde{\rho}}^*\,U_\veps D_\veps^* M_\rho^*$ with $\rho:=\frac{d\overline{\mu}}{d\mu\otimes d\nu}$ and $\tilde{\rho}:=\frac{d\mu}{d\mu^\veps_{\rm eff}}$. In view of (\ref{unitarity}), the unitarity of $M_{\tilde{\rho}}, M_\rho$, and $D_\veps$ implies $ U^\veps U^{\veps*}=1$.
Furthermore, we simply define  $ P^\veps$ by $P^\veps:= U^{\veps*} U^\veps$. Then $U^\veps$ is unitary from $ P^\veps\H$ to $L^2(\C,d\mu_{\rm eff}^\veps)$. Finally, we set
\begin{equation}\label{Heff}
H^\veps_{{\rm eff}}\;:=\; U^\veps H^\veps U^{\veps*}\;=\;M_{\tilde{\rho}}\,U_\veps H_\veps U_\veps^*M_{\tilde{\rho}}^*.
\end{equation}
We notice that $H_{{\rm eff}}^\veps$ is symmetric by definition. Since $M_{\tilde{\rho}}$ is unitary and $U_\veps$ is unitary when restricted to $P_\veps \H$ due to Lemma \ref{projector}, the self-adjointness of  $\big(H_{{\rm eff}}^\veps, U^\veps\D(H^\veps)\big)$ on $\H_{\rm eff}:=L^2(\C,d\mu_{\rm eff}^\veps)$ is implied by the self-adjointness of $\big(P_\veps H_\veps P_\veps,P_\veps\D(H_\veps)\big)$ on $P_\veps\H$, which is in turn a consequence of the self-adjointness of $\big(P_\veps H_\veps P_\veps+(1-P_\veps)H_\veps (1-P_\veps),\D(H_\veps)\big)$ on $\H$. For $\veps$ small enough this last self-adjointness can be verified using Lemma \ref{projector} and the Kato-Rellich theorem (see e.g.\ \cite{RS2}):   
\begin{eqnarray*}
\lefteqn{H_\veps\,-\,\big(P_\veps H_\veps P_\veps+(1-P_\veps)H_\veps (1-P_\veps)\big)}\\
&& \ =\ \;(1-P_\veps) H_\veps P_\veps\,+\,P_\veps H_\veps (1-P_\veps)\\
&& \ =\ \;(1-P_\veps) [H_\veps, P_\veps]\,-\,P_\veps [H_\veps, P_\veps]\\
&& \ =\ \;(1-2P_\veps)\,[H_\veps, P_\veps].
\end{eqnarray*}
Lemma \ref{projector} entails that $[H_\veps, P_\veps]$ is operator-bounded by $\veps H_\veps$. Hence, for $\veps$ small enough (we adjust $\veps_0$ if nescessary) the difference above is operator-bounded by $H_\veps$ with relative bound smaller than one. Now the Kato-Rellich theorem yields the claim, because $\big(H_\veps,\D(H_\veps)\big)$ is self-adjoint (as it is unitarily equivalent to the self-adjoint $H^\veps$).

\smallskip

We now turn to the derivation of the estimate (\ref{difference}). To do so we first pull out the unitaries $M_{\tilde{\rho}},M_\rho$ and $D_\veps$. Using that $D_\veps^* M_\rho^*\,\chi(H^\veps)\,M_\rho D_\veps=\chi(D_\veps^* M_\rho^*H^\veps  M_\rho D_\veps)=\chi(H_\veps)$ due to the spectral theorem we obtain by a straight forward calculation that
\begin{eqnarray*}
\lefteqn{\left(\e^{-\I H^\veps t}- U^{\veps*}\e^{-\I H_{{\rm eff}}t} U^\veps\right) P^\veps\,\chi(H^\veps)}\\
&=& M_\rho D_\veps\left(\e^{-\I H_\veps t}-U_\veps^*\e^{-\I U_\veps H_\veps U_\veps^* t}U_\veps\right)\,U_\veps^*U_\veps\,\chi(H_\veps)\,D_\veps^* M_\rho^*.
\end{eqnarray*}
Since $M_\rho$ and $D_\veps$ are unitary, we can ignore them for the estimate and continue with the term in the middle.
Next we use Duhamel's principle to express the difference of the unitary groups as a difference of its generators. Because of $U_\veps U_\veps^*=1$ and $P_\veps=U_\veps^*U_\veps$ we have that
\begin{eqnarray}\label{cook}
\lefteqn{\left(\e^{-\I H_\veps t}-U_\veps^*\e^{-\I U_\veps H_\veps U_\veps^*t}U_\veps\right)U_\veps^*U_\veps\,\chi(H_\veps )}\nonumber\\
&=& \left(P_\veps-U_\veps^*\e^{-\I U_\veps H_\veps U_\veps^*t}U_\veps \e^{\I H_\veps t} \right)\e^{-\I H_\veps t}\,\chi(H_\veps )\,+\,[\e^{-\I H_\veps t},P_\veps]\,\chi(H_\veps )\nonumber\\
&=& \I\int_0^tU_\veps^*\e^{-\I U_\veps H_\veps U_\veps^*s}\left(U_\veps H_\veps U_\veps^*U_\veps-U_\veps H_\veps \right)\e^{\I H_\veps s}\,ds\,\e^{-\I H_\veps t}\,\chi(H_\veps )\nonumber\\
&& \qquad\qquad\qquad\qquad\qquad\qquad\qquad\qquad\ \ \,+\,[\e^{-\I H_\veps t},P_\veps]\,\chi(H_\veps )\nonumber\\
&=& \I\int_0^tU_\veps^*\e^{-\I U_\veps H_\veps U_\veps^*s}U_\veps\left(H_\veps P_\veps-P_\veps H_\veps \right)\chi(H_\veps )\,\e^{\I H_\veps s}\,ds\,\e^{-\I H_\veps t}\nonumber\\
&& \qquad\qquad\qquad\qquad\qquad\qquad\qquad\qquad\ \ \,+\,[\e^{-\I H_\veps t},P_\veps]\,\chi(H_\veps ),
\end{eqnarray}
where we used that $[\e^{-\I H_\veps s},\chi(H_\veps )]=0$ for any $s$ due to the spectral theorem.
Now we observe that (\ref{invariance}) 
implies that 
\begin{equation}\label{groupinv}
\left\|\,[\e^{-\I H_\veps t},P_\veps]\,\chi(H_\veps )\,\right\|_{\L(\H)}\;=\;\O(\veps^3|t|),
\end{equation}
as it holds
\begin{eqnarray*}
[\e^{-\I H_\veps t},P_\veps]\,\chi(H_\veps ) &=& \e^{-\I H_\veps t}\left(P_\veps-\e^{\I H_\veps t}P_\veps\e^{-\I H_\veps t}\right)\chi(H_\veps )\\
&=& -\e^{-\I H_\veps t}\,\I\int_0^t\e^{\I H_\veps s}\left(H_\veps P_\veps-P_\veps H_\veps \right)\e^{-\I H_\veps s}\,ds\,\chi(H_\veps )\\
&=& -\e^{-\I H_\veps t}\,\I\int_0^t\e^{\I H_\veps s}[H_\veps ,P_\veps]\,\chi(H_\veps )\e^{-\I H_\veps s}\,ds
\ \,\stackrel{(\ref{invariance})}{=}\;\O(\veps^3|t|)
\end{eqnarray*}
because of Lemma \ref{projector} and $\|\e^{-\I H_\veps s}\|_{\L(\H)}=1$ for any $s$. So, in view of (\ref{cook}), 
\begin{eqnarray*}
\lefteqn{\left\|\left(\e^{-\I H_\veps t}-U_\veps^*\e^{-\I U_\veps H_\veps U_\veps^*t}U_\veps\right)P_\veps\,\chi(H_\veps )\,\right\|_{\L(\H)}}\\
&& \ \;\stackrel{(\ref{groupinv})}{\leq}\ \; \left\|\int_0^t U_\veps^*\e^{-\I U_\veps H_\veps U_\veps^*s}U_\veps\,[H_\veps ,P_\veps]\,\chi(H_\veps )\,\e^{\I H_\veps s}\,ds\,\right\|_{\L(\H)}\,+\,\O(\veps^3|t|)\\
&& \ \;\leq\ \; |t|\,\underbrace{\left\|\,U_\veps^*\e^{-\I U_\veps H_\veps U_\veps^*s}U_\veps\,\right\|_{\L(\H)}}_{\leq\,1}\,\left\|\,[H_\veps ,P_\veps]\,\chi(H_\veps )\,\right\|_{\L(\H)}\ \,+\,\O(\veps^3|t|)\\
&& \ \;\stackrel{(\ref{invariance})}{=}\ \; \O(\veps^3|t|).
\end{eqnarray*}
This proves the error estimate (\ref{difference}). \qed

\pagebreak

{\sc Proof of Lemma \ref{transform}:}
\newline
$M_\rho$ is an isometry because for all $\psi,\varphi\in L^2(\M,d\sigma_2)$
\[\int_\M M_\rho\psi^*\,M_\rho\varphi\,d\sigma_1\;=\;\int_\M \psi^*\,\varphi\,\rho^{-1}\,d\sigma_1\;=\;\int_\M \overline{\psi}\,\varphi\,d\sigma_2.\]
Therefore it is clear that
\[M_\rho^*\psi\;=\;\rho^{\frac{1}{2}}\psi\]
which is well-defined because $\rho$ is positive. One immediately concludes
\[M_\rho M_\rho^*\;=\;1\;=\;M_\rho^* M_\rho\]
and thus $M_\rho$ is unitary.
Now we note that $[{\rm grad},\rho^{-\frac{1}{2}}]=-{\textstyle \frac{1}{2}}\,\rho^{-\frac{1}{2}}\,{\rm grad}\ln\rho\,$. So we have
\begin{eqnarray*}
M_\rho^*(-\Delta_{d\sigma_1})M_\rho\psi
&=& -\,\rho^{\frac{1}{2}}{\rm div}_{d\sigma_1}\,{\rm grad}(\rho^{-{\frac{1}{2}}}\psi)\\
&=& -\,\rho^{\frac{1}{2}}{\rm div}_{d\sigma_1}\,\rho^{-\frac{1}{2}}\big({\rm grad}\psi-{\textstyle \frac{1}{2}}({\rm grad}\ln\rho)\psi\big)\\
&=& -\,\rho^{\frac{1}{2}}{\rm div}_{d\sigma_1}\,\rho^{-\frac{1}{2}}\,{\rm grad}\psi
\,+\,\rho^{\frac{1}{2}}{\rm div}_{d\sigma_1}\,\Big(\rho^{-\frac{1}{2}}{\textstyle \frac{1}{2}}({\rm grad}\ln\rho)\psi\Big)
\end{eqnarray*}
On the one hand,
\begin{eqnarray*}
\rho^{\frac{1}{2}}{\rm div}_{d\sigma_1}\,\rho^{-\frac{1}{2}}\,{\rm grad}\psi
&=& \rho\,{\rm div}_{d\sigma_1}\,\rho^{-1}\,{\rm grad}\psi \,+\,{\textstyle \frac{1}{2}}\,g({\rm grad}\ln\rho,{\rm grad}\psi)
\end{eqnarray*}
and on the other hand,
\begin{eqnarray*}
\rho^{\frac{1}{2}}{\rm div}_{d\sigma_1}\,\Big(\rho^{-\frac{1}{2}}{\textstyle \frac{1}{2}}({\rm grad}\ln\rho)\psi\Big)
&=&
-\,{\textstyle \frac{1}{4}}\,g({\rm grad}\ln\rho,{\rm grad}\ln\rho)\psi\\
&&\,+\,{\textstyle \frac{1}{2}}\,({\rm div}_{d\sigma_1}\,{\rm grad}\ln\rho)\psi\\
&&\,+\,{\textstyle \frac{1}{2}}\,g({\rm grad}\ln\rho,{\rm grad}\,\psi).
\end{eqnarray*}
Together we obtain
\begin{eqnarray*}
M_\rho^*(-\Delta_{d\sigma_1})M_\rho\psi
&=& -\,\rho\,{\rm div}_{d\sigma_1}\,\rho^{-1}\,{\rm grad}\psi\\ 
&&\ \ \ \,-\,\Big({\textstyle \frac{1}{4}}\,g({\rm grad}\ln\rho,{\rm grad}\ln\rho)\,-\,{\textstyle \frac{1}{2}}\,{\rm div}_{d\sigma_1}\,{\rm grad}\ln\rho\Big)\psi\\
&=&-\Delta_{d\sigma_2}\psi\,-\,\Big({\textstyle \frac{1}{4}}\,g({\rm grad}\ln\rho,{\rm grad}\ln\rho)\,-\,{\textstyle \frac{1}{2}}\,\Delta_{d\sigma_1}\,\ln\rho\Big)\psi,
\end{eqnarray*}
which is the claim.\qed


\subsection{Pullback of the results to the ambient space}\label{proofT3}

In this section we show how to derive Corollary \ref{effdyn2} about effective dynamics on the ambient manifold $\A$ from Theorem \ref{effdyn}. To do so we first state some immediate consequences of Lemma \ref{projector} for $P^\veps$ and $U^\veps$ from Theorem \ref{effdyn}.

\pagebreak 

\begin{corollary}\label{projector3}
For $\veps$ small enough $ P^\veps$ and $ U^\veps$ from Theorem \ref{effdyn} satisfy
\begin{eqnarray}
&\bullet& \| P^\veps\|_{\L(\D(H^{\veps^m}))}\;\lesssim\;1,\nonumber\\
&\bullet& 
\|\langle\nu/\veps\rangle^l  P^\veps\langle\nu/\veps\rangle^j\|_{\L(\overline{\H})}\;\lesssim\;1\,,\quad
\|\langle\nu/\veps\rangle^l  P^\veps\langle\nu/\veps\rangle^j\|_{\L(\D(H^\veps))}\;\lesssim\;1,\nonumber\\
&\bullet& \|[H^\veps, P^\veps]\|_{\L(\D(H^{\veps^{m+1}},\D(H^{\veps^m}))}\;=\; \O(\veps),\label{invariance2}\\
&\bullet& 
\|[H^\veps, P^\veps] \,\chi(H^\veps)\|_{\L(\overline{\H},\D(H^{\veps^m})} \;=\; \O(\veps^3)\label{invariance3},\\
&\bullet& \| U^{\veps}\|_{\L(\D(H^{\veps^m}),\D(H_{\rm eff}^{\veps^m}))}\lesssim1\,,\quad\| U^{\veps*}\|_{\L(\D(H_{\rm eff}^{\veps^m}),\D(H^{\veps^m}))}\lesssim1\label{surprise}
\end{eqnarray}
for all $j,l,m\in\NNN_0$ and each Borel functions $\chi:\RRR\to[-1,1]$ satisfying ${\rm supp}\,\chi\subset(-\infty,E]$.
\end{corollary}

The proof can be found at the end of this subsection. Now we gather some facts about the operator $A$ defined in (\ref{liftop}) and its adjoint.

\begin{lemma}\label{lift}
Let $A$ be defined by $A\psi\,:=\,(\frac{d\overline{\mu}}{\Phi^*d\tau})^{-\frac{1}{2}}
\,(\psi\circ\Phi)$ with $\Phi:N\C\to\B$ as constructed in Section \ref{model}. 

\smallskip

i) It holds $A\in\L\big(L^2(\A,d\tau),\overline{\H}\big)$ with
\[\|A\psi\|_{L^2(N\C,d\overline{\mu})}\;\leq\;\|\psi\|_{L^2(\A,d\tau)}\qquad\forall\;\psi\in L^2(\A,d\tau).\] 

\smallskip

ii) For $\varphi\in \overline{\H}$ the adjoint $A^*\in\L\big(\overline{\H},L^2(\A,d\tau)\big)$ of $A$ is given by
\begin{equation*}
A^*\varphi\;=\;
\begin{cases}
\big((\frac{d\overline{\mu}}{\Phi^*d\tau})^{\frac{1}{2}}\,\varphi\big)\circ\Phi^{-1} & \text{on }\B,\\
0 & \text{on }\A\setminus\B.
\end{cases}
\end{equation*}
It satisfies $\|A^*\varphi\|_{L^2(\A,d\tau)}=\|\varphi\|_{L^2(N\C,d\overline{\mu})}$, $A^*A=\chi_{\B}$, and $AA^*=1$. 

\smallskip
 
iii) It holds $A^*P^\veps\in\L\big(\D(H^\veps),\D(H_\A^\veps)\big)$ and
\begin{equation}\label{lifterror}
\|(H_\A^\veps A^*-A^* H^\veps) P^\veps\|_{\L(\D(H^\veps),L^2(\A,d\tau))}\;\lesssim\;\veps^3.
\end{equation}
\end{lemma}

The last estimate is crucial for the proof of Corollary \ref{effdyn2}. It results from the two facts that $H_\A^\veps A^*=A^* H^\veps$ on $\B_{\delta/2}$ by construction and that $ P^\veps$ is 'small' on the complement. Lemma \ref{lift} will be proved at the end of Section \ref{preliminaries}. The following estimates for energy cutoffs will be useful not only for the proof of Corollary \ref{effdyn2} but also in the following sections.

\begin{lemma}\label{microlocal}
Assume that $\big(H,\D(H)\big)$ is self-adjoint on some Hilbert space $\H$. Let $\chi_1\in C^\infty_0(\RRR)$ and $\chi_2:\RRR\to\RRR$ be a bounded Borel function. 

\smallskip
 
a) Let $A\in\L(\H)$. If 
$\big\|[H,A]\,\chi_2(H)\big\|_{\L(\D(H^l),\D(H^{m-1}))}\leq\delta$ for some $l,m\in\NNN$,
then there is $C<\infty$ depending only on $\chi_1$ such that
\[\|[\chi_1(H),A]\,\chi_2(H)\|_{\L(\D(H^{l-1}),\D(H^{m}))}\;\leq\;C\,\delta.\]
b) Let $\big(\tilde H,\D(\tilde H)\big)$ be also self-adjoint on $\H$. If there are $l,m\in\NNN$ with 
$\big\|(H-\tilde H)\,\chi_2(\tilde H)\big\|_{\L(\D(\tilde H^l),\D(H^{m-1}))}\leq\delta$,
then there is $C<\infty$ depending only on $\chi_1$ such that
\[\|(\chi_1(H)-\chi_1(\tilde H))\,\chi_2(\tilde H)\|_{\L(\D(\tilde H^{l-1}),\D(H^{m}))}\;\leq\;C\,\delta.\]
c) Let $\tilde\H$ be another Hilbert space and $B\in\L(\H,\tilde\H)$ such that $BB^*=1$ and $\big(\tilde H:=BHB^*,\D(\tilde H)\big)$ is self-adjoint on $\tilde\H$. Assume that there is $m\in\NNN$ such that $B\in\L\big(\D(H^{l}),\D(\tilde H^l)\big)$ and $B^*\in\L\big(\D(\tilde H^{l}),\D(H^l)\big)$ for all $l\leq m$.
\begin{enumerate}
\item If $\chi_2\in C^\infty_0(\RRR)$ and 
$\big\|[H,B^*B]\,\chi_2(H)\big\|_{\L(\H,\D(H^m))}\leq\delta$, then there is $C<\infty$ depending only on $\chi_1$, $\chi_2$, $\|B\|_{\L(\D(H^{l}),\D(\tilde H^l))}$, $\|B^*\|_{\L(\D(\tilde H^{l}),\D(H^l))}$ for $l\leq m$ such that
\[\big\|\big(\chi_1(\tilde H)-B\chi_1(H)B^*\big)B\,\chi_2^2(H)\big\|_{\L(\H,\D(\tilde H^m))}\;\leq\;C\,\delta.\] 
 
\item 
If 
$\|[H,B^*B]\|_{\L(\D(H^m),\D(H^{m-1}))}\leq\delta$,
then there is $C<\infty$ depending only on $\chi_1$, $\|B\|_{\L(\D(H^{l}),\D(\tilde H^l))}$, and $\|B^*\|_{\L(\D(\tilde H^{l}),\D(H^l))}$ for $l\leq m$ such that
\[\big\|\chi_1(\tilde H)-B\chi_1(H)B^*\big\|_{\L(\D(\tilde H^{m-1}),\D(\tilde H^m))}\;\leq\;C\,\delta^2.\]
\end{enumerate}
\end{lemma}

These statements can be generalized in many ways. Here we have given versions which are sufficient for the situations that we encounter in the following. We emphasize that the support of $\chi_2$ in a) and b) need not be compact, in particular $\chi_2\equiv 1$ is allowed there. 
We now turn to the short derivation of Corollary \ref{effdyn2}. Lemma \ref{microlocal} will be proved afterwards.

\medskip

{\sc Proof of Corollary \ref{effdyn2} (Section \ref{results1}):}
\newline
Let $\chi:\RRR\to[-1,1]$ be a Borel function with ${\rm supp}\,\chi\subset(-\infty,E]$. 
Let $E_-:=\min\{\inf\sigma(H^\veps),\inf\sigma(H^{\veps}_{{\rm eff}})\}$ and $\tilde\chi,\tilde{\tilde\chi}\in C^\infty_0(\RRR)$ with $\tilde\chi|_{[E_-,E]}\equiv1$ and $\tilde{\tilde\chi}|_{{\rm supp}\,\tilde\chi}\equiv1$. The spectral calculus implies $\tilde\chi(H^\veps)=\tilde{\tilde\chi}^2(H^\veps)\tilde\chi(H^\veps)$. 
We recall from Theorem \ref{effdyn} that $H^\veps_{{\rm eff}}= U^\veps H^\veps U^{\veps*}$ and $ P^\veps= U^{\veps*} U^\veps$. 
In view of Corollary~\ref{projector3}, $ U^\veps$ satisfies the assumptions on $B$ in Lemma~\ref{microlocal} c) i) with $\delta=\veps^3$ and c) ii) with $\delta=\veps$. Thus in the norm of $\L\big(L^2(\C,d\mu_{\rm eff}^\veps),\D(H^{\veps}_{{\rm eff}})\big)$ it holds
\begin{eqnarray}\label{funny}
\tilde\chi^2(H^{\veps}_{{\rm eff}})
&=&  U^{\veps}\tilde\chi(H^{\veps}) U^{\veps*}\,\tilde\chi(H^{\veps}_{{\rm eff}}) \,+\,\big(\tilde\chi(H^{\veps}_{{\rm eff}}) - U^{\veps}\tilde\chi(H^{\veps}) U^{\veps*}\big)^2\nonumber\\
&&\,+\,\big(\tilde\chi(H^{\veps}_{{\rm eff}})- U^{\veps}\tilde\chi(H^{\veps}) U^{\veps*}\big)\, U^{\veps}\tilde\chi(H^{\veps}) U^{\veps*}\nonumber\\
&=&  U^{\veps}\tilde\chi(H^{\veps}) U^{\veps*}\,\tilde\chi(H^{\veps}_{{\rm eff}}) \,+\,\O(\veps^4)\nonumber\\
&&\,+\,\big(\tilde\chi(H^{\veps}_{{\rm eff}})- U^{\veps}\tilde\chi(H^{\veps}) U^{\veps*}\big)\, U^{\veps}\tilde{\tilde\chi}^2(H^{\veps})\tilde\chi(H^{\veps}) U^{\veps*}\nonumber\\
&=&  U^{\veps}\tilde\chi(H^{\veps}) U^{\veps*}\,\tilde\chi(H^{\veps}_{{\rm eff}})\,+\,\O(\veps^3).
\end{eqnarray}
Then Corollary~\ref{projector3} yields that in the norm of $\L\big(\H,\D(H^\veps)\big)$ it holds
\begin{eqnarray}\label{cutoffexchange}
U^{\veps*}\chi(H_{{\rm eff}}^\veps)U^\veps
&=&  P^\veps\tilde\chi(H^\veps) U^{\veps*}\chi(H_{{\rm eff}}^\veps)U^\veps\,+\,\O(\veps^3)
\end{eqnarray}
because the spectral calculus also implies $\chi(H^\veps_{\rm eff})=\tilde\chi^2(H^\veps_{\rm eff})\chi(H^\veps_{\rm eff})$. 
By Lemma \ref{lift} we have $AA^*=1$. Therefore
\begin{eqnarray*}
\lefteqn{(\e^{-\I H_\A^\veps t}-A^*\, U^{\veps*}\e^{-\I H^\veps_{{\rm eff}}t} U^\veps A)\,A^*\, P^\veps \tilde\chi(H^\veps)}\\
&&=\; \Big((\e^{-\I H_\A^\veps t}A^*-A^*\,\e^{-\I H^\veps t})\;+\; A^*(\e^{-\I H^\veps t}- U^{\veps*}\e^{-\I H^\veps_{{\rm eff}}t} U^\veps )\Big) P^\veps \tilde\chi(H^\veps)
\end{eqnarray*}
Since $A$ and $A^*$ are bounded independent of $\veps$ by Lemma \ref{lift}, Theorem \ref{effdyn} implies that the second difference is of order $\veps^3 |t|$. So it suffices to estimate the first difference. 
The estimate (\ref{invariance3}) implies $[\e^{-\I H^\veps t}, P^\veps]\,\tilde\chi(H^\veps)=\O(\veps^3|t|)$ analogously with the proof of (\ref{groupinv}). So
\begin{eqnarray*}\label{cook2}
\lefteqn{(\e^{-\I H_\A^\veps t}A^*-A^*\,\e^{-\I H^\veps t}) P^\veps \tilde\chi(H^\veps)}\nonumber\\
&=& \e^{-\I H_\A^\veps t}\big(A^* P^\veps-\e^{\I H_\A^\veps t}A^*\, P^\veps\e^{-\I H^\veps t} \big)\tilde\chi(H^\veps)A \,+\,A^*[\e^{-\I H^\veps t}, P^\veps]\,\chi(H^\veps)\nonumber\\
&=& \I\e^{-\I H_\A^\veps t}\int_0^t \e^{\I H_\A^\veps s}\left(A^* P^\veps H^\veps - H_\A^\veps A^* P^\veps\,\tilde\chi(H^\veps)\right)\e^{-\I H_\veps s}\,ds\,+\,\O(\veps^3|t|)\nonumber\\
&\stackrel{(\ref{invariance3})}{=}& \I\e^{-\I H_\A^\veps t}\int_0^t \e^{\I H_\A^\veps s}\left(A^*H^\veps - H_\A^\veps A^*\right) P^\veps\,\tilde\chi(H^\veps)\e^{-\I H_\veps s}\,ds\,+\,\O(\veps^3|t|)\\
&=& \O(\veps^3|t|)
\end{eqnarray*}
due to (\ref{lifterror}) and $\|\tilde\chi(H^\veps)\|_{\L(\overline{\H},\D(H^\veps))}\lesssim1$. The latter holds because $H^\veps$ is bounded from below and the support of $\tilde\chi$ is bounded from above, both independent of $\veps$. Because of (\ref{cutoffexchange}) and $\|A\|\lesssim1$ we have shown that
\[\left\|\left(\e^{-\I H_\A^\veps t}-A^*\, U^{\veps*}\e^{-\I H^\veps_{{\rm eff}}t} U^\veps A\right)A^*\, U^{\veps*} \chi(H_{\rm eff}^\veps)U^\veps A\,\right\|_{\L(L^2(\A,d\tau))}\;\leq\;C\,\veps^3\,|t|,\]
which was the claim. \qed

\bigskip

{\sc Proof of Lemma \ref{microlocal}:}

We want to apply the so called Helffer-Sj\"ostrand formula (see e.g.\ chapter~2 in \cite{Dav}) to $\chi_1$. It states that for any $\chi\in C^\infty_0(\RRR)$
\begin{equation*}
\chi(H)\;=\;\frac{1}{\pi}\,\int_\CCC \partial_{\overline{z}}\tilde\chi(z)\,R_H(z)\,dz,
\end{equation*}
where $R_H(z):=(H-z)^{-1}$ denotes the resolvent and $\tilde\chi:\CCC\to\CCC$ is a so-called almost analytic extension of $\chi$. We emphasize that by $dz$ we mean the usual volume measure on $\CCC$. With $z=x+\I y$ a possible choice for $\tilde\chi$ is
\[\tilde\chi(x+\I y)\;:=\; \tau(y)\,\sum_{j=0}^l \chi^{(j)}(x)\frac{(\I y)^j}{j!}\]
with arbitrary $\tau\in C^\infty(\RRR,[0,1])$ satisfying $\tau|_{[-1,1]}\equiv 1$ and ${\rm supp}\,\tau\subset[-2,2]$ and $l\geq2$. Then obviously $\tilde\chi=\chi$ for $y=0$ and there is $C_{\chi}<\infty$ depending only on $\chi$ such that
\begin{equation}\label{almostanal}
\partial_{\overline{z}}\tilde\chi(z)\;:=\;\partial_x\tilde\chi\,+\,\I\partial_y\tilde\chi\;\leq\;C_{\chi}\,|{\rm Im}z|^{l},
\end{equation}
which is the reason why it is called an almost analytic extension.
We choose such an extension $\tilde\chi_1\in C^\infty_0(\CCC)$ of $\chi_1$ with $l=2$.
Next we observe that for all $j\in\NNN_0$
\begin{equation}\label{resestimate}
\big\|R_H(z)\|_{\L(\D(H^j),\D(H^{j+1}))}\;\leq\;\frac{\sqrt{1+2|{\rm Im}z|^2+2|z|^2}}{|{\rm Im}z|}
\end{equation}
because for all $\psi\in\H$
\begin{eqnarray*}
\big\|H^{j+1}R_H(z)\psi\big\|^2 \,+\,\big\|R_H(z)\psi\big\| ^2
&=& \big\|HR_H(z)H^j\psi\big\|^2 \,+\,\big\|R_H(z)\psi\big\|^2 \\
&\leq& \|(1+zR_H(z))H^j\psi\|^2\,+\,\big\|R_H(z)\psi\big\|^2 \\
&\leq& \Big(2+\frac{2|z|^2}{|{\rm Im}z|^2}\Big)\|H^j\psi\|^2+\frac{1}{|{\rm Im}z|^2}\|\psi\|^2 \\
&\leq& \frac{1+2|{\rm Im}z|^2+2|z|^2}{|{\rm Im}z|^2}\big(\|\psi\|^2+\|H^j\psi\|^2\big).
\end{eqnarray*}
Now by the Helffer-Sj\"ostrand formula
\begin{eqnarray*}
[\chi_1(H),A]\,\chi_2(H) 
&=& \frac{1}{\pi}\int_\CCC\partial_{\overline{z}}\tilde\chi_1(z)\,[R_H(z),A]\,dz\,\chi_2(H)\\
&=& \frac{1}{\pi}\int_\CCC\partial_{\overline{z}}\tilde\chi_1(z)\,R_H(z)[A,H]\,R_H(z)\,dz\,\chi_2(H)\\
&=& \frac{1}{\pi}\int_\CCC\partial_{\overline{z}}\tilde\chi_1(z)\,R_H(z)[A,H]\,\chi_2(H)\,R_H(z)\,dz,
\end{eqnarray*}
where in the last step we used that $[R_H(z),\chi_2(H)]=0$ due to the spectral theorem. 
Using the assumption $\big\|[A,H]\,\chi_2(H)\big\|_{\L(\D(H^l),\D(H^m))}\leq\delta$ we obtain 
\begin{eqnarray*}
\lefteqn{\|[\chi_1(H),A]\,\chi_2(H)\|_{\L(\D(H^{l-1}),\D(H^{m+1}))}}\\
&\qquad\leq& \frac{1}{\pi}
\int_\CCC|\partial_{\overline{z}}\tilde\chi_1(z)|\,\|R_H(z)\|_{\L(\D(H^m),\D(H^{m+1}))}\\
&& \qquad \times \,\big\|[H,A]\,\chi_2(H)\big\|_{\L(\D(H^l),\D(H^m))}\|R_H(z)\|_{\L(\D(H^{l-1}),\D(H^l))}\,dz\\ 
&\qquad\stackrel{(\ref{almostanal}),(\ref{resestimate})}{\leq}& C_{\chi_1}\,\delta\,\int_{{\rm supp}\tilde\chi_1}\,|{\rm Im}z|^{2}\,\frac{1+2|{\rm Im}z|^2+2|z|^2}{|{\rm Im}z|^2}\,dz\\
&\qquad\leq& C\,\delta,
\end{eqnarray*}
with a $C$ depending only on $C_{\chi_1}$  and the support of $\tilde\chi_1$. This shows a). The proof of b) can be carried out analogously because 
\begin{eqnarray*}
\big(\chi_1(H)-\chi_1(\tilde H)\big)\,\chi_2(\tilde H) 
&=& \frac{1}{\pi}\int_\CCC\partial_{\overline{z}}\tilde\chi_1(z)\,\big(R_H(z)-R_{\tilde H}(z)\big)\,dz\,\chi_2(\tilde H)\\
&=& \frac{1}{\pi}\int_\CCC\partial_{\overline{z}}\tilde\chi_1(z)\,R_H(z)(\tilde H-H)\,R_{\tilde H}(z)\,dz\,\chi_2(\tilde H)\\
&=& \frac{1}{\pi}\int_\CCC\partial_{\overline{z}}\tilde\chi_1(z)\,R_H(z)(\tilde H-H)\,\chi_2(\tilde H)\,R_{\tilde H}(z)\,dz.
\end{eqnarray*} 
due to the Helffer-Sj\"ostrand formula. For c) the formula yields:
\begin{eqnarray}\label{observation}
\chi_1(\tilde H)-B\chi_1(H)B^*
\;=\; \frac{1}{\pi}\int_\CCC\partial_{\overline{z}}\tilde\chi_1(z)\big(R_{\tilde H}(z)-BR_H(z)B^*\big)\,dz.&&
\end{eqnarray}
So we have to estimate $R_{\tilde H}(z)-BR_H(z)B^*$. We set $A:=B^*B$ and note that $BB^*=1$ implies that $BA=B$, $AB^*=B^*$ and $A^2=A$. By definition $\tilde H=BHB^*$. Therefore
\begin{eqnarray}\label{resdiff}
R_{\tilde H}(z)-BR_H(z)B^* &=& R_{\tilde H}(z)\,\big(1-(BHB^*-z)BR_H(z)B^*\big)\nonumber\\
&=& R_{\tilde H}(z)\,\big(1-B(H-z)AR_H(z)B^*\big)\nonumber\\
&=& R_{\tilde H}(z)\,\big(1-BAB^*-B[H,A]R_H(z)B^*\big)\nonumber\\
&=& -R_{\tilde H}(z)\,B[H,A]R_H(z)B^*.
\end{eqnarray}

For the first part of c) we compute
\begin{eqnarray}\label{observation2}
\lefteqn{B^*\big(R_{\tilde H}(z)-BR_H(z)B^*\big)B\,\chi_2^2(H)}\nonumber\\ 
&&\quad=\ \, -B^*R_{\tilde H}(z)\,B[H,A]R_H(z)A\chi_2(H)\chi_2(H)\nonumber\\
&&\quad=\ \, -B^*R_{\tilde H}(z)\,B[H,A]R_H(z)\big(\chi_2(H)A+[A,\chi_2(H)]\big)\chi_2(H).\quad
\end{eqnarray}
We will write $C_B$ for a constant depending only on $\|B\|_{\L(\D(H^l),\D(\tilde H^l))}$ and  $\|B^*\|_{\L(\D(\tilde H^l),\D(H^l))}$ for $l\leq m$. We note that the estimate (\ref{resestimate}) holds true with $H$ replaced by $\tilde H$ because $\tilde H$ is assumed to be self-adjoint. 
Then, on the one hand, $B\in\L\big(\D(H^{m-1}),\D(\tilde H^{m-1})\big)$ implies  
\begin{eqnarray*}
\lefteqn{\|R_{\tilde H}(z)B\,[H,A]R_H(z)\,\chi_2(H)A\chi_2(H)\|_{\L(\H,\D(\tilde H^{m}))}}\\ 
&&\quad=\ \,\|R_{\tilde H}(z)B\,[H,A]\chi_2(H)\,R_H(z)A\chi_2(H)\|_{\L(\H,\D(\tilde H^{m}))}\\
&&\quad\leq\ \,C_B\,\frac{\sqrt{1+2|{\rm Im}z|^2+2|z|^2}}{|{\rm Im}z|}\,\|[H,A]\chi_2(H)\|_{\L(\H,\D(H^{m-1}))}\,|{\rm Im}z|^{-1}\\
&&\quad\leq\ \,C_B\,\delta\,\frac{\sqrt{1+2|{\rm Im}z|^2+2|z|^2}}{|{\rm Im}z|^2}
\end{eqnarray*}
by the assumption on the commutator term. On the other hand, the assumptions on $B$ and $B^*$ imply that 
\begin{eqnarray*}
\lefteqn{\|R_{\tilde H}(z)\,B[H,A]R_H(z)\,[A,\chi_2(H)]\chi_2(H)\|_{\L(\H,\D(\tilde H^{m}))}}\\ 
&&\quad=\ \,\|R_{\tilde H}(z)\,B(HB^*B-B^*BH)R_H(z)\,[A,\chi_2(H)]\chi_2(H)\|_{\L(\H,\D(\tilde H^{m}))}\\
&&\quad\leq\ \,C_B\,\frac{2(1+2|{\rm Im}z|^2+2|z|^2)}{|{\rm Im}z|^2}\,\|[A,\chi_2(H)]\chi_2(H)\|_{\L(\H,\D(H^{m-1}))}\\
&&\quad\leq\ \,C_{B,\chi_2}\,\delta\,\frac{2(1+2|{\rm Im}z|^2+2|z|^2)}{|{\rm Im}z|^2},
\end{eqnarray*}
where $C_{B,\chi_2}$ depends also on $\chi_2$ because in the last step we used that $H,A$, and $\chi_2$ satisfy the assumptions of a). After plugging (\ref{observation2}) into (\ref{observation}) the latter two estimates allow us to deduce the first part of c) analogously with a).

\smallskip

For the second part of c)
we observe that $A^2=A$ entails $A[H,A]A=0$. Then we may derive from (\ref{resdiff}) that
\begin{eqnarray*}
R_{\tilde H}(z)-BR_H(z)B^* 
&=& -R_{\tilde H}(z)\,BA[H,A](1-A)R_H(z)AB^*\\
&=& -R_{\tilde H}(z)\,BA[H,A](1-A)[R_H(z),A]B^*\\
&=& R_{\tilde H}(z)\,BA[H,A]R_H(z)[H,A]R_H(z)B^*\\
&=& R_{\tilde H}(z)\,B[H,A]R_H(z)[H,A]R_H(z)B^*.
\end{eqnarray*}
Hence, with $B\in\L\big(\D(H^{m-1}),\D(\tilde H^{m-1})\big)$ and $B^*\in\L\big(\D(\tilde H^{m-1}),\D(H^{m-1})\big)$ we obtain
\begin{eqnarray*}
\lefteqn{\big\|R_{\tilde H}(z)-BR_H(z)B^*\big\|_{\L(\D(\tilde H^{m-1}),\D(\tilde H^{m}))}}\\ 
&&\quad=\ \,\big\|R_{\tilde H}(z)B\,[H,A]R_H(z)\,[H,A]R_H(z)\,B^*\big\|_{\L(\D(\tilde H^{m-1}),\D(\tilde H^{m}))}\\
&&\quad\leq\ \,C_B\,\frac{(1+2|{\rm Im}z|^2+2|z|^2)^{3/2}}{|{\rm Im}z|^3}\,\|[H,A]\|^2_{\L(\D(H^{m}),\D(H^{m-1}))}\\
&&\quad\leq\ \,C_B\,\delta^2\,\frac{(1+2|{\rm Im}z|^2+2|z|^2)^{3/2}}{|{\rm Im}z|^3}
\end{eqnarray*}
by assumption. Together with (\ref{observation}) this yields the claim as in a) when we put $l=3$ in the choice of the almost analytic extension.
\qed

\bigskip

{\sc Proof of Corollary \ref{projector3}:}
\newline
We will only prove that (\ref{surprise}) is a consequence of the other statements. These follow directly from Lemma \ref{projector} by making use of the unitarity of $M_\rho$ and $D_\veps$ as well as of $D_\veps\langle\nu\rangle D_\veps^*=\langle\nu/\veps\rangle$, when we recall that $P^\veps=M_\rho D_\veps P_\veps D_\veps^*M_\rho^*$ from the proof of Theorem \ref{effdyn}.

\smallskip

We prove (\ref{surprise}) by induction. For $m=0$ both statements are clear. Now we assume that it is true for some fixed $m\in\NNN_0$. Theorem \ref{effdyn} yields that $P^\veps= U^{\veps*} U^{\veps}$ and $H_{\rm eff}^\veps= U^\veps H^\veps U^{\veps*}$. On the one hand, this implies
\[H_{\rm eff}^{\veps^{m+1}} U^\veps\;=\;H_{\rm eff}^{\veps^{m}} U^\veps H^\veps P^\veps.\]
Then $\| P^\veps\|_{\L(\D(H^{\veps^{m+1}}))}\lesssim1$ and the induction assumption immediately imply $\| U^\veps\|_{\L(\D(H^{\veps^{m+1}}),D(H_{\rm eff}^{\veps^{m+1}}))}\lesssim1$. On the other hand, we have  
\begin{eqnarray*}
H^{\veps^{m+1}} U^{\veps*} &=& H^{\veps^{m}} P^\veps H^\veps U^{\veps*}\,+\,H^{\veps^{m}}[H^\veps, P^\veps]  U^{\veps*}\\
&=& H^{\veps^{m}} U^{\veps*}H_{\rm eff}^{\veps}\,+\,H^{\veps^{m}}[H^\veps, P^\veps] U^{\veps*}.
\end{eqnarray*}
By the induction assumption and (\ref{invariance2}) it holds for all $\psi$ that
\begin{eqnarray*}
\|H^{\veps^{m+1}} U^{\veps*}\psi\| 
&\leq& \|H^{\veps^{m}} U^{\veps*}H_{\rm eff}^{\veps}\psi\|\,+\,\|H^{\veps^{m}}[H^\veps, P^\veps] U^{\veps*}\psi\|\\
&\lesssim& \|H_{\rm eff}^{\veps^{m+1}}\psi\|+\|H_{\rm eff}^{\veps}\psi\|\,+\,\veps\,\big(\|H^{\veps^{m+1}} U^{\veps*}\psi\|+\|H^{\veps^{m}} U^{\veps*}\psi\|\big)\\
&\lesssim& \|H_{\rm eff}^{\veps^{m+1}}\psi\|\,+\,\veps\,\|H^{\veps^{m+1}} U^{\veps*}\psi\|\,+\,\|\psi\|,
\end{eqnarray*}
where we used that lower powers of a self-adjoint operator are operator-bounded by higher powers. For $\veps$ small enough, we can absorb the term with the $\veps$ on the left-hand side, which yields  $\| U^{\veps*}\|_{\L(D(H_{\rm eff}^{\veps^{m+1}}),\D(H^{\veps^{m+1}}))}\lesssim1$.
\qed


\subsection{Derivation of the effective Hamiltonian}\label{proofT2}
 
The goal of this section is to prove Theorem \ref{calcHeff}. We first take a closer look at the horizontal connection $\nabla^{\rm h}$ (see Definition \ref{deriv}):

\begin{lemma}\label{berryphase2}
It holds $\langle\nabla^{\rm h}_\tau\phi|\psi\rangle_{\H_{\rm f}} +\langle\phi|\nabla^{\rm h}_\tau\psi\rangle_{\H_{\rm f}}=\big({\rm d}\langle\phi|\psi\rangle_{\H_{\rm f}} \big)(\tau)$ and
\begin{equation}\label{curvature2}
{\rm R}^{\rm h}(\tau_1,\tau_2)\psi \,:=\, \Big(\nabla^{\rm h}_{\tau_1} \nabla^{\rm h}_{\tau_2} -\nabla^{\rm h}_{\tau_2} \nabla^{\rm h}_{\tau_1} -\nabla^{\rm h}_{[\tau_1,\tau_2]}\Big)\psi
\,=\, -\nabla^{\rm v}_{{\rm R}^\perp(\tau_1,\tau_2)\nu}\psi,
\end{equation}
where ${\rm R}^\perp$ is the normal curvature mapping (defined in the appendix). 
\end{lemma}

The proof of this result can be found at the beginning of Section \ref{wholestory}. 
%
%
In order to deduce the formula for the effective Hamiltonian we need that $H_\veps$ can be expanded with respect to the normal directions when operating on functions that decay fast enough. For this purpose we split up the integral over $N\C$ into an integral over the fibers $N_q\C$, isomorphic to $\RRR^k$, followed by an integration over $\C$, which is always possible for a measure of the form $d\mu\otimes d\nu$ (see e.g.\ chapter XVI, §4 of \cite{L}).

\begin{lemma}\label{expH}
Let $m\in\NNN_0$. If a densely defined operator $A$  satisfies 
\[\|A \langle\nu\rangle^l\|_{\L(\D(H_\veps^m),\H)}\;\lesssim\;1,\quad\|\langle\nu\rangle^l A \|_{\L(\D(H_\veps^{m+1}),\D(H_\veps))}\;\lesssim\;1\] 
for every $l\in\NNN$, then the operators $H_\veps A,A H_\veps\in\L\big(\D(H_\veps^{m+1}),\H\big)$ can be expanded in powers of $\veps$:
\begin{eqnarray*}
H_\veps\,A &=& \big(H_0\,+\,\veps H_1\,+\,\veps^2H_2\big)\,A\,+\,\O(\veps^3),\\
A\,H_\veps &=& A\,\big(H_0\,+\,\veps H_1\,+\,\veps^2H_2\big)\,+\,\O(\veps^3),
\end{eqnarray*}
where $H_0,H_1,H_2$ are the operators associated with 
\begin{eqnarray}\label{expcompl}
\langle\phi|H_0\psi\rangle_{\H} &=& \int_\C\int_{N_q\C}g(\veps\nabla^{\rm h}\phi^*,\veps\nabla^{\rm h}\psi)\,d\nu\,d\mu\;+\;\langle\phi|H_{\rm f}\psi\rangle_{\H},\\
\langle\phi|H_1\psi\rangle_{\H} &=& \int_\C\int_{N_q\C} 2\,{\rm II}_\nu\big(\veps\nabla^{\rm h}\phi^*,\veps\nabla^{\rm h}\psi\big)\;+\;\phi^*\,(\nabla^{\rm v}_{\nu}W)\psi\;d\nu\,d\mu,\nonumber\\
\langle\phi|H_2\psi\rangle_{\H} &=& \int_\C\int_{N_q\C} 3\,g\big(\W_\nu\,\veps\nabla^{\rm h}\phi^*,\W_\nu\,\veps\nabla^{\rm h}\psi\big)+\R\big(\veps\nabla^{\rm h}\phi^*,\nu,\veps\nabla^{\rm h}\psi,\nu\big)\nonumber\\
&& \qquad\,+\,{\textstyle\frac{2}{3}}\,\R\big(\veps\nabla^{\rm h}\phi^*,\nu,\nabla^{\rm v}\psi,\nu\big)\,+\,{\textstyle\frac{2}{3}}\,\R\big(\nabla^{\rm v}\phi^*,\nu,\veps\nabla^{\rm h}\psi,\nu\big)\nonumber\\
&& \qquad\,+\,{\textstyle\frac{1}{3}}\,\R\big(\nabla^{\rm v}\phi^*,\nu,\nabla^{\rm v}\psi,\nu\big)
\,+\phi^*({\textstyle\frac{1}{2}}\nabla^{\rm v}_{\nu,\nu}W+V_{{\rm geom}})\psi
\;d\nu\,d\mu,\nonumber
\end{eqnarray}
where ${\rm II}$ is the second fundamental form, $\W$ is the Weingarten mapping, and $\R$ is the Riemann tensor (see the appendix for the definitions).
%
%
Furthermore, for $l\in\{0,1,2\}$
\begin{equation}\label{Hlbound}
\|H_lA\|_{\L(\D(H_\veps^{m+1}),\H)}\,\lesssim\,1\,,\quad \|A H_l\|_{\L(\D(H_\veps^{m+1}),\H)}\,\lesssim\,1. 
\end{equation}
\end{lemma}

This will be proved in Section \ref{expansion}. Definition \ref{gapcondition}, Lemma \ref{projector}, and the following lemma imply that Lemma \ref{expH} can be applied to the projectors $P_0$ and $P_\veps$ with $m=0$. In the next lemma we gather some useful properties of the spectral projector $P_0$, the unitary $\tilde U_\veps$ satisfying $P_\veps=\tilde U_\veps^* P_0\tilde U_\veps$ from Lemma \ref{projector}, and the global family of eigenfunctions $\varphi_{\rm f}$ associated with $P_0$ (see Definition \ref{U0}):

\begin{lemma}\label{props}
It holds $E_{\rm f}\in C^\infty_{\rm b}(\C)$, as well as:
\vspace{-3pt}
\begin{enumerate}
\item $\forall\;l,j\in\NNN_0:\ 
\|\langle\nu\rangle^l P_0\langle\nu\rangle^j\|_{\L(\D(H_\veps))}\,\lesssim\,1\,,\,\|[-\veps^2\Delta_{\rm h},P_0]\|_{\L(\D(H_\veps),\H)} \,\lesssim\,\veps.$

\item There are $U_1^\veps,U_2^\veps\in\L(\H)\,\cap\,\L(\D(H_\veps))$ with norms bounded independently of $\veps$ satisfying $P_0U_1^\veps P_0=0$ and $U_2^\veps P_0=P_0U_2^\veps P_0=P_0U_2^\veps$ such that $\tilde U_\veps= 1+\veps U_1^\veps+\veps^2 U_2^\veps$. 

\item $\|P_0U_1^\veps \langle\nu\rangle^l\|_{\L(\D(H_\veps^m))}\lesssim1$ for all $l\in\NNN_0$ and $m\in\{0,1\}$.

\item For $B_\veps:=P_0\tilde U_\veps\chi(H_\veps)$ and all $u\in\{1,(U_1^\veps)^*,(U_2^\veps)^*\}$ it holds
\begin{equation}\label{alleklein}
\big\|\,[-\veps^2\Delta_{\rm h}+E_{\rm f},u P_0]\,B_\veps\,\big\|_{\L(\H)}\;=\;\O(\veps).
\end{equation} 

\item For $R_{H_{\rm f}}(E_{\rm f}):=(1-P_0)\big(H_{\rm f}-E_{\rm f}\big)^{-1}(1-P_0)$ it holds
\begin{eqnarray}\label{U1}
\big\|U_1^{\veps\,*}B_\veps \,+\,R_{H_{\rm f}}(E_{\rm f})\,([-\veps\Delta_{\rm h},P_0]+H_1)P_0B_\veps\big\|_{\L(\H,\D(H_\veps))} =\, \O(\veps)
\end{eqnarray}

\item If $\varphi_{\rm f}\in C^\infty_{\rm b}(\C,\H_{\rm f})$, it holds 
\[\|U_0\|_{\L(\D(H_\veps),\D(-\veps^2\Delta_\C+E_{\rm f}))}\,\lesssim\,1,\quad \|U_0^*\|_{\L(\D(-\veps^2\Delta_\C+E_{\rm f}),\D(H_\veps))}\,\lesssim\,1,\] 
and there is $\lambda_0\gtrsim1$ with $\sup_{q}\|\e^{\lambda_0\langle\nu\rangle}\varphi_{\rm f}(q)\|_{\H_{\rm f}(q)}\lesssim1$ and
\[\textstyle{\sup}_{q\in\C}\,\|\e^{\lambda_0\langle\nu\rangle}\nabla^{\rm v}_{\nu_1,\dots,\nu_l}\nabla^{\rm h}_{\tau_1,\dots,\tau_m}\varphi_{\rm f}(q)\|_{\H_{\rm f}(q)}\;\lesssim\;1\]
for all $\nu_1,\dots,\nu_l\in\Gamma_{\rm b}(N\C)$ and $\tau_1,\dots,\tau_m\in\Gamma_{\rm b}(T\C)$.
\end{enumerate}
\end{lemma}

The proof of this lemma can be found in Section \ref{subspace}. Since $U_2^\veps$ does only effect $P_\veps\H$ but not the effective Hamiltonian, we have not stated its particular form here, as we did for $U_1^\veps$ in v). Now we are ready to derive the theorem about the form of the effective Hamiltonian. We deduce its corollary concerning the unitary groups before. 

\bigskip

{\sc Proof of Corollary \ref{effdyn3} (Section \ref{results2}):}
\newline 
In order to check that 
\begin{eqnarray}\label{appldiff}
\left\|\left(\e^{-\I H^\veps t}- U_0^{\veps*}\e^{-\I H^{(2)}_{{\rm eff}}t} U_0^\veps\right) U_0^{\veps*}\chi(H^{(2)}_{{\rm eff}}) U_0^\veps\,\right\|_{\L(\overline{\H})} &\lesssim& \veps\,(1+\veps^2|t|),
\end{eqnarray}
with $U_0^\veps=U_0D_\veps^*$, indeed, follows from Theorem \ref{effdyn} and Theorem \ref{calcHeff} we start by verifying that $\| U^\veps- U_0^\veps\|_{\L(\overline{\H},\H_{\rm eff})}=\O(\veps)$. 

We recall that we defined $\tilde{\rho}:=\frac{d\mu}{d\mu^\veps_{\rm eff}}$ as well as $ U^\veps:=M_{\tilde\rho}^*U_0\tilde U_\veps D_\veps^* M_{\rho}^*$ in the proof of Theorem \ref{effdyn}. Since $d\mu^\veps_{\rm eff}$ is the volume measure associated with $g^\veps_{\rm eff}$, which is given by the expression in Theorem \ref{calcHeff}, we have $\|\tilde\rho-1\|_\infty=\O(\veps)$ and thus $\|M_{\tilde\rho}-1\|_{\L(L^2(\C,d\mu))}=\O(\veps)$. Using in addition that $\|\tilde U_\veps-1\|_{\L(\H)}=\O(\veps)$ by Lemma \ref{props} 
and $M_{\tilde\rho}^*M_{\tilde\rho}=1$ we obtain that 
\begin{eqnarray*}
\| U^\veps- U_0^\veps\|_{\L(\overline{\H},\H_{\rm eff})}
&=& \|M_{\tilde\rho}^*(U_0\tilde U_\veps D_\veps^* M_{\rho}^*-M_{\tilde\rho}U_0D_\veps^+)\|_{\L(\overline{\H},\H_{\rm eff})}\\
&=& \|U_0\tilde U_\veps D_\veps^* M_{\rho}^*-M_{\tilde\rho}U_0D_\veps^*\|_{\L(\overline{\H},L^2(\C,d\mu))}\\
&=& \|U_0D_\veps^* (M_{\rho}^*-1)\|_{\L(\overline{\H},L^2(\C,d\mu))}\,+\,\O(\veps)\\
&=& \|U_0P_0D_\veps^* (M_{\rho}^*-1)\|_{\L(\overline{\H},L^2(\C,d\mu))}\,+\,\O(\veps)\\
&\lesssim& \|\langle\nu\rangle^{-1}D_\veps^* (M_{\rho}-1)\|_{\L(\overline{\H},\H)}\,+\,\O(\veps)
\end{eqnarray*}
because $U_0=U_0P_0$ and the projector $P_0$ associated with the constraint energy band $E_{\rm f}$ satisfies $\|P_0\langle\nu\rangle\|_{\L(\H)}\lesssim1$ by assumption (see Definition \ref{gapcondition}). In view of (\ref{density}), a first order Taylor expansion of $\rho$ in normal directions yields that $D_\veps^* (M_{\rho}^*-1)$ is globally bounded by a constant times  $\veps\langle\nu\rangle$. Hence, we end up with $\| U^\veps- U_0^\veps\|_{\L(\overline{\H},\H_{\rm eff})}=\O(\veps)$ and may thus replace $ U_0^\veps$ by $U^\veps$ in (\ref{appldiff}). 

\smallskip

Now let $\chi:\RRR\to[-1,1]$ be a Borel function with ${\rm supp}\,\chi\subset(-\infty,E]$. Using the triangle inequality and $ U^{\veps} U^{\veps*}=1$ we see that 
\begin{eqnarray}\label{triangle}
\lefteqn{\left\|\left(\e^{-\I H^\veps t}- U^{\veps*}\e^{-\I H^{(2)}_{{\rm eff}}t} U^\veps\right) U^{\veps*}\chi(H^{(2)}_{{\rm eff}}) U^\veps\,\right\|_{\L(\overline{\H})}}\nonumber\\ 
&& \;\leq\; \left\|\left(\e^{-\I H^\veps t}- U^{\veps*}\e^{-\I H^\veps_{{\rm eff}}t} U^\veps\right) U^{\veps*}\chi(H^{(2)}_{{\rm eff}}) U^\veps\,\right\|_{\L(\overline{\H})}\nonumber\\
&& \qquad\,+\,\left\| U^{\veps*}\left(\e^{-\I H^\veps_{{\rm eff}} t}-\e^{-\I H^{(2)}_{{\rm eff}}t}\right)\chi(H^{(2)}_{{\rm eff}}) U^\veps\,\right\|_{\L(\overline{\H})}.
\end{eqnarray}
The second term is of order $\veps^3|t|$ because
\begin{eqnarray*}\label{cook3}
\lefteqn{\left(\e^{-\I H^\veps_{{\rm eff}} t}-\e^{-\I H^{(2)}_{{\rm eff}}t}\right)\chi(H^{(2)}_{{\rm eff}})}\nonumber\\
&& \;=\; \I\e^{-\I H^\veps_{{\rm eff}} t}\int_0^t \e^{\I H^\veps_{{\rm eff}} s}\left(H^{(2)}_{{\rm eff}} - H^\veps_{{\rm eff}}\right)\e^{\I H^{(2)}_{{\rm eff}} s}\,\chi(H^{(2)}_{{\rm eff}})\,ds\nonumber\\
&& \;=\;\I\e^{-\I H^\veps_{{\rm eff}} t}\int_0^t \e^{\I H^\veps_{{\rm eff}} s}\left(H^{(2)}_{{\rm eff}} - H^\veps_{{\rm eff}}\right)\,\chi(H^{(2)}_{{\rm eff}})\e^{\I H^{(2)}_{{\rm eff}} s}\,ds \quad=\; \O(\veps^3|t|)
\end{eqnarray*}
by Theorem \ref{calcHeff}. Let $\tilde\chi\in C^\infty_0(\RRR)$ with ${\rm supp}\,\tilde\chi|_{[\inf\sigma(H^{(2)}_{{\rm eff}}),E]}\equiv1$.
By Theorem \ref{calcHeff} and Lemma \ref{microlocal} b) we have
\begin{eqnarray*}
 U^{\veps*}\chi(H^{(2)}_{{\rm eff}}) &=&  U^{\veps*}\tilde\chi(H^{(2)}_{{\rm eff}})\chi(H^{(2)}_{{\rm eff}})\\
&=&  U^{\veps*}\tilde\chi(H^\veps_{{\rm eff}})\chi(H^{(2)}_{{\rm eff}})\,+\,\O(\veps^3).
\end{eqnarray*} 
We recall from Theorem \ref{effdyn} that $H^\veps_{{\rm eff}}= U^\veps H^\veps U^{\veps*}$ and $ P^\veps= U^{\veps*} U^\veps$. In view of Corollary~\ref{projector3}, $U^\veps$ satisfies the assumptions on $B$ in Lemma~\ref{microlocal} c) ii) with $\delta=\veps$. Therefore
\begin{eqnarray*}
 U^{\veps*}\chi(H^{(2)}_{{\rm eff}})
&=&  U^{\veps*}\tilde\chi( H^\veps_{{\rm eff}})\chi(H^{(2)}_{{\rm eff}})\,+\,\O(\veps^3)\\ 
&=&  U^{\veps*} U^{\veps}\tilde\chi(H^\veps) U^{\veps*}\chi(H^{(2)}_{{\rm eff}})\,+\,\O(\veps^2)\\
&=&  P^\veps\tilde\chi(H^\veps) U^{\veps*}\chi(H^{(2)}_{{\rm eff}})\,+\,\O(\veps^2).
\end{eqnarray*}
After plugging this into the first term in (\ref{triangle}) we may apply Theorem \ref{effdyn} to it. This yields the claim. \qed

\pagebreak

{\sc Proof of Theorem \ref{calcHeff} (Section \ref{results2}):}
\newline 
Let $\chi:\RRR\to[-1,1]$ be a Borel function with ${\rm supp}\,\chi\subset(-\infty,E]$. 
We recall that $\D(A)$ always denotes the maximal domain of an operator $A$ (i.e.\ all $\psi$ with $\|A\psi\|+\|\psi\|<\infty$) equipped with the graph norm. A differential operator $A$ of order $m$ will be called elliptic on $(\C,g)$, if it satisfies $\big[\dots[A,\underbrace{f]\dots,f\big]}_{m-times}\geq c|{\rm d} f|_g^m$ for some $c>0$ and any $f$.

\medskip

We set $H^{(0)}_{{\rm eff}}:=-\veps^2\Delta_\C+E_{\rm f}$ with $\Delta_\C$ the Laplace-Beltrami operator on $(\C,g)$. Since $E_{\rm f}\in C^\infty_{\rm b}(\C)$ due to Lemma \ref{props}, all powers of $H^{(0)}_{{\rm eff}}$ are obviously elliptic operators of class $C^\infty_{\rm b}(\C)$ on $\H_{\rm eff}$. This implies that $\big(H^{(0)}_{\rm eff},\,\D(H^{(0)}_{\rm eff})\big)$ is self-adjoint on $\H_{\rm eff}$ because $\C$ is of bounded geometry (see Section 1.4. of \cite{Sh}; in particular, this entails that $\D(H^{(0)}_{\rm eff})$ is the Sobolev space $W^{2,2}(\C)$, but equipped with an $\veps$-dependent norm). Let $E_-:=\min\{\inf\sigma(H^\veps),\inf\sigma(H^{(0)}_{{\rm eff}})\}$ and $\tilde\chi,\tilde{\tilde\chi}\in C^\infty_0(\RRR)$ with $\tilde\chi|_{[E_-,E]}\equiv1$ and $\tilde{\tilde\chi}|_{{\rm supp}\,\tilde\chi}\equiv1$. Then we define $H^{(2)}_{{\rm eff}}$ for $\phi,\psi\in\D(H^{(0)}_{\rm eff})$ by 
\begin{eqnarray}\label{truncated}
\langle\,\phi\,|\,H^{(2)}_{{\rm eff}}\,\psi\,\rangle
&:=& \int_\C \Big( g_{{\rm eff}}^\veps\big((p^\veps_{{\rm eff}}\phi)^*,p^\veps_{{\rm eff}}\psi\big)\,+\,\phi^*\big(E_{\rm f}+\veps\, \langle\varphi_{\rm f}|(\nabla^{\rm v}_\cdot W)\varphi_{\rm f}\rangle_{\H_{\rm f}}\big)\,\psi\nonumber\\
&& \qquad\,+\,\phi^*\veps^2\,W^{(2)}\,\psi\,-\,\veps^2\,\M\big(\Phi^*(\phi),\Phi(\tilde{\tilde\chi}(H^{(0)}_{{\rm eff}})\psi)\big)\\
&& \quad \qquad\,-\,\veps^2\,\M\big(\Phi^*(\tilde{\tilde\chi}(H^{(0)}_{{\rm eff}})\phi),\Phi(\psi-\tilde{\tilde\chi}(H^{(0)}_{{\rm eff}})\psi)\big)\Big)\,d\mu^\veps_{{\rm eff}}\nonumber
\end{eqnarray}
where $\Phi(\psi):=\Psi(\veps\nabla p^\veps_{{\rm eff}}\psi,p^\veps_{{\rm eff}}\psi,\psi)$ and all the other objects are defined by the expressions in Theorem \ref{calcHeff}. Because of $\tilde{\tilde\chi}(H^{(0)}_{{\rm eff}})\chi(H^{(0)}_{{\rm eff}})=\chi(H^{(0)}_{{\rm eff}})$ this definition immediately implies that $H^{(2)}_{{\rm eff}}$ operates on $\psi$ with $\psi=\chi(H^{(0)}_{{\rm eff}})\psi$ as stated in the theorem.

\smallskip

The rest of the proof will be devided into several steps.

\bigskip

\begin{step}\label{Step1}
$\big(H^{(2)}_{\rm eff},\,\D(H^{(0)}_{\rm eff})\big)$ is self-adjoint on $\H_{\rm eff}$ and
\[\|H^{(2)}_{\rm eff}-H^{(0)}_{\rm eff}\|_{\L(\D(H^{(0)}_{\rm eff}),\H_{\rm eff})}\,=\,\O(\veps).\]
\end{step} 

\vspace{-0.6cm}
It easy to verify that $H^{(2)}_{{\rm eff}}$ is symmetric. Then it suffices to prove the stated estimate because by the Kato-Rellich theorem (see e.g.\ \cite{RS2}) the estimate implies that $\big((H^{(2)}_{\rm eff},\,\D(H^{(0)}_{\rm eff})\big)$ is self-adjoint on $\H_{\rm eff}$ for $\veps$ small enough.

\smallskip

Since $\C$ is of bounded geometry, maximal regularity estimates hold true there (see Appendix 1 of \cite{Sh}), in particular, differential operators of order $m\in\NNN$ with coefficients in $C^\infty_{\rm b}(\C)$ are bounded by elliptic operators of same order and class. 

\smallskip

The operator $M$ associated with $\int_\C\M\big(\Phi(\phi),\Phi(\psi)\big)d\mu_{\rm eff}^\veps$ is a fourth order differential operator which, in view of Lemma \ref{props} vi), has coefficients in $C^\infty_{\rm b}(\C)$. Hence, it is bounded by $(H^{(0)}_{{\rm eff}})^2$ with a constant independent of $\veps$ because all derivatives carry an $\veps$. We notice that  $\|\tilde{\tilde\chi}(H_\veps)\|_{\L(\H,\D({H^{(0)}_{\rm eff}}^m))}\lesssim1$ for all $m\in\NNN_0$ because the support of $\tilde{\tilde\chi}$ is bounded independently of $\veps$. Thus we obtain that $M\tilde{\tilde\chi}(H^{(0)}_{{\rm eff}})$ is bounded. The same is true for $\tilde{\tilde\chi}(H^{(0)}_{{\rm eff}})M\big(1-\tilde{\tilde\chi}(H^{(0)}_{{\rm eff}})\big)$ because it is operator-bounded by the adjoint of $M\tilde\chi(H^{(0)}_{{\rm eff}})$. Therefore the $\M$-terms in (\ref{truncated}) correspond to bounded operators! All the other terms are associated with differential operators of second order whose coeffcients are in $C^\infty_{\rm b}(\C)$ by Lemma \ref{props} vi) and whose derivatives carry at least one $\veps$ each. Therefore they are bounded by the elliptic $H^{(0)}_{{\rm eff}}$. 

\smallskip

So we obtain that $\|H^{(2)}_{\rm eff}-H^{(0)}_{\rm eff}\|_{\L(\D(H^{(0)}_{\rm eff}),\H_{\rm eff})}\,=\,\O(\veps)$ by observing that the leading order of $H^{(2)}_{\rm eff}$ is indeed $H^{(0)}_{\rm eff}$. 

\bigskip

\begin{step}\label{Step2}
$\D(H^\veps_{\rm eff})=\D(H^{(0)}_{\rm eff})$ and
$\|H^\veps_{\rm eff}-H^{(0)}_{\rm eff}\|_{\L(\D(H^\veps_{\rm eff}),\H_{\rm eff})}\,=\,\O(\veps)$.
\end{step} 

We recall that we defined $U_\veps:=U_0\tilde U_\veps$, $ U^\veps:=M_{\tilde{\rho}}^*\,U_\veps D_\veps^*M_\rho^*$, $ P^\veps:= U^{\veps*} U^\veps$, and $
H^\veps_{{\rm eff}}:= U^\veps H^\veps U^{\veps*}$ in the proof of Theorem \ref{effdyn}, which implied $P_\veps=U_\veps^*U_\veps$. 

\smallskip

Since $\|\tilde U_\veps\|_{\L(\D(H_\veps))}\lesssim1$ and $\|U_0\|_{\L(\D(H_\veps),\D(H^{(0)}_{\rm eff}))}\lesssim1$ by Lemma \ref{props}, it also holds $\| U^\veps\|_{\L(\D(H^\veps),\D(H^{(0)}_{\rm eff}))}\lesssim1$. Using, in addition, that $\| U^{\veps*}\|_{\L(\D(H^\veps),\D(H^{\veps}_{\rm eff}))}\lesssim1$ due to Corollary \ref{projector3} and $ U^\veps  U^{\veps*}=1$ we conclude that for all $\psi\in\D(H^{\veps}_{\rm eff})$
\[\|\psi\|_{\D(H^{(0)}_{\rm eff})}\;=\;\| U^\veps  U^{\veps*}\psi\|_{\D(H^{(0)}_{\rm eff})}\;\lesssim\;\| U^{\veps*}\psi\|_{\D(H^\veps)}\;\lesssim\;\|\psi\|_{\D(H^{\veps}_{\rm eff})}.\]
On the other hand, Lemma \ref{props} and Corollary \ref{projector3} imply via the analogous arguments that for all $\psi\in\D(H^{(0)}_{\rm eff})$

\[\|\psi\|_{\D(H^{\veps}_{\rm eff})}\;=\;\| U^\veps  U^{\veps*}\psi\|_{\D(H^{\veps}_{\rm eff})}\;\lesssim\;\| U^{\veps*}\psi\|_{\D(H^\veps)}\;\lesssim\;\|\psi\|_{\D(H^{(0)}_{\rm eff})}.\]
Hence, $\D(H^{\veps}_{\rm eff})=\D(H^{(0)}_{\rm eff})$.

\smallskip

Using $H^\veps_{\rm eff}= U^{\veps} H^\veps U^{\veps*}= U^{\veps} P^\veps H^\veps P^\veps U^{\veps*}$ and again Corollary \ref{projector3} we get
\begin{eqnarray*}
\lefteqn{\|H^\veps_{\rm eff}-H^{(0)}_{\rm eff}\|_{\L(\D(H^\veps_{\rm eff}),\H_{\rm eff})}}\\
&& \;=\; \| U^{\veps}( P^\veps H^\veps P^\veps- U^{\veps*}H^{(0)}_{\rm eff} U^\veps) U^{\veps*}\|_{\L(\D(H^\veps_{\rm eff}),\H_{\rm eff})}\\
&& \;\lesssim\; \| P^\veps H^\veps P^\veps- U^{\veps*}H^{(0)}_{\rm eff} U^\veps\|_{\L(\D(H^\veps),\overline{\H})}\\
&& \;=\; \|P_\veps H_\veps P_\veps-U_\veps^*M_{\tilde{\rho}}H^{(0)}_{\rm eff}M_{\tilde{\rho}}^*U_\veps\|_{\L(\D(H_\veps),\H)}\\
&& \;=\; \|P_0 H_\veps P_0-U_0^*M_{\tilde{\rho}}H^{(0)}_{\rm eff}M_{\tilde{\rho}}^*U_0\|_{\L(\D(H_\veps),\H)}\,+\,\O(\veps)
\end{eqnarray*}
because $P_\veps=U_\veps^*U_\veps$, $U_\veps=U_0\tilde U_\veps$, and by Lemma \ref{props} ii) it holds $\tilde U_\veps-1=\O(\veps)$ both in $\L(\H)$ and in $\L(\D(H_\veps))$. Lemma \ref{expH} implies that $P_0(H_\veps-H_0)P_0=\O(\veps)$ in $\L(\D(H_\veps),\H)$. Hence, \begin{eqnarray}\label{dies}
\lefteqn{\|H^\veps_{\rm eff}-H^{(0)}_{\rm eff}\|_{\L(\D(H^\veps_{\rm eff}),\H_{\rm eff})}}\nonumber\\
&& \;=\; \|P_0 H_0 P_0-U_0^*M_{\tilde{\rho}}H^{(0)}_{\rm eff}M_{\tilde{\rho}}^*U_0\|_{\L(\D(H_\veps),\H)}\,+\,\O(\veps)\nonumber\\
&& \;\lesssim\; \|U_0 H_0 U_0^*-M_{\tilde{\rho}}H^{(0)}_{\rm eff}M_{\tilde{\rho}}^*\|_{\L(\D(H^{(0)}_{\rm eff}),L^2(\C,d\mu))}\,+\,\O(\veps),
\end{eqnarray}
where in the last step we used $P_0=U_0^*U_0$ and  $\|U_0\|_{\L(\D(H_\veps),\D(H^{(0)}_{\rm eff}))}\lesssim1$ due to Lemma \ref{props} vi). It holds $U_0\psi=\varphi_{\rm f}\psi$ by definiton of $U_0$ and $H_0=-\Delta_{\rm h}+H_{\rm f}$ by Lemma \ref{expH}.
In view of Definition \ref{deriv}, we have
\begin{eqnarray}\label{leibniz}
\veps\nabla^{\rm h}\,\psi\,\varphi_{\rm f} &=& \varphi_{\rm f}\,\veps{\rm d}\psi\;+\;\psi\,\veps\nabla^{\rm h}\varphi_{\rm f},\\
\veps^2\Delta_{\rm h}\,\psi\,\varphi_{\rm f} &=& \varphi_{\rm f}\,\veps^2\Delta_\C\psi\;+\;2g(\veps{\rm d}\psi,\veps\nabla^{\rm h}\varphi_{\rm f})\;+\;\psi\,\veps^2\Delta_{\rm h}\varphi_{\rm f},\nonumber
\end{eqnarray} 
where ${\rm d}$ is the exterior derivative on $\C$. We note that $\sup_q\|\veps\nabla^{\rm h}\varphi_{\rm f}\|_{\H_{\rm f}(q)}$ and $\sup_q\|\veps^2\Delta_{\rm h}\varphi_{\rm f}\|_{\H_{\rm f}(q)}$ are of order~$\veps$ and $\veps^2$ respectively by Lemma \ref{props}. Therefore 
\begin{eqnarray}\label{jenes}
U_0H_0U_0^*\psi\;=\;U_0(-\veps^2\Delta_{\rm h}+H_{\rm f})U_0^*\psi &=& \langle\varphi_{\rm f}|(-\veps^2\Delta_{\rm h}+E_{\rm f})\varphi_{\rm f}\psi\rangle_{\H_{\rm f}}\nonumber\\
&=&
H^{(0)}_{\rm eff}\psi\,+\,\O(\veps).
\end{eqnarray}
We recall that $\tilde\rho=d\mu^\veps_{\rm eff}/d\mu$ with $d\mu^\veps_{\rm eff}$ the measure associated with~$g^\veps_{\rm eff}$. Since  $d\mu$ and $d\mu_{\rm eff}^\veps$ coincide at leading order and $\tilde\rho\in C^\infty_{\rm b}(\C,g)$ due to Lemma~\ref{props}~vi), we have 
$\|M_{\tilde\rho}^*-1\|_{\L(\D(H^{(0)}_{\rm eff}))}=\O(\veps)$ and $\|M_{\tilde\rho}-1\|_{\L(L^2(\C,d\mu))}=\O(\veps)$. So we obtain that $\|M_{\tilde{\rho}}H^{(0)}_{\rm eff}M_{\tilde{\rho}}^*-H^{(0)}_{\rm eff}\|_{\L(\D(H^{(0)}_{\rm eff}),L^2(\C,d\mu))}=\O(\veps)$. Together with (\ref{dies}) and (\ref{jenes}) this yields $\|H^\veps_{\rm eff}-H^{(0)}_{\rm eff}\|_{\L(\D(H^\veps_{\rm eff}),\H_{\rm eff})}\,=\,\O(\veps)$.

\bigskip

\begin{step}\label{Step3}
It holds
$\|(H^\veps_{\rm eff}-H^{(2)}_{\rm eff})\, U^{\veps}\chi(H^{\veps}) U^{\veps*}\|_{\L(\H_{\rm eff})}\,=\,\O(\veps^3)$.
\end{step} 

This step contains the central order-by-order calculation of $H^\veps_{\rm eff}$ and is therefore by far the longest one. For any $\psi$ we set $\tilde\psi:=M_{\tilde\rho}\psi$, $\psi^\chi:= U^\veps\chi(H^\veps) U^{\veps*}\psi$, and $\tilde\psi_\chi:=U_\veps\chi(H_\veps)U_\veps^*\tilde\psi$. Of course, we have $\widetilde{\psi^\chi}=\tilde\psi_\chi$, $\|\tilde\psi\|_{L^2(\C,d\mu)}=\|\psi\|_{\H_{\rm eff}}$, and $\|\tilde\psi_\chi\|_{L^2(\C,d\mu)}\leq\|\tilde\psi\|_{L^2(\C,d\mu)}$ for all $\psi\in\H_{\rm eff}$.
 
\smallskip
 
We first explain why the cut off in the definition of $H^{(2)}_{\rm eff}$ does not matter here.  
We note that $P^\veps$ and $U^\veps$ satisfy the assumption on $A$ and $B$ in Lemma~\ref{microlocal}~a) and c)~ii) with $\delta=\veps$ by Corollary~\ref{projector3}. In addition, $H_{\rm eff}^{(0)}$ and $H_{\rm eff}^{\veps}$ satisfy the assumption of Lemma~\ref{microlocal}~b) with the same $\delta$ by Step~\ref{Step2}. Therefore
\begin{eqnarray*}
\tilde{\tilde\chi}(H_{\rm eff}^{(0)})U^{\veps}\chi(H^{\veps}) U^{\veps*} &=& \tilde {\tilde\chi}(H_{\rm eff}^{\veps})U^{\veps}\chi(H^{\veps}) U^{\veps*} \,+\,\O(\veps)\\
&=& U^{\veps}\tilde{\tilde\chi}(H^{\veps}) P^{\veps}\chi(H^{\veps}) U^{\veps*} \,+\,\O(\veps)\\
&=& U^{\veps}P^{\veps}\tilde{\tilde\chi}(H^{\veps})\chi(H^{\veps}) U^{\veps*}\,+\,\O(\veps)\\
&=& U^{\veps}\chi(H^{\veps}) U^{\veps*}\,+\,\O(\veps),
\end{eqnarray*}
which shows that 
\begin{eqnarray}\label{goal}
\langle\,\phi\,|\,H^{(2)}_{{\rm eff}}\,\psi^\chi\,\rangle
&=& \int_\C \Big( g_{{\rm eff}}^\veps\big((p^\veps_{{\rm eff}}\phi)^*,p^\veps_{{\rm eff}}\psi^\chi\big)\,+\,\phi^*\big(E_{\rm f}+\veps\, \langle\varphi_{\rm f}|(\nabla^{\rm v}_\cdot W)\varphi_{\rm f}\rangle_{\H_{\rm f}}\big)\psi^\chi\nonumber\\
&& \qquad\qquad\qquad \,+\,\phi^*\veps^2\,W^{(2)}\,\psi^\chi\,-\,\veps^2\,\M\big(\Phi^*(\phi),\Phi(\psi^\chi)\big)\,d\mu_{\rm eff}^\veps \nonumber\\
&&\;+\;\O(\veps^3\|\phi\|_{\H_{\rm eff}}\|\psi\|_{\H_{\rm eff}}).
\end{eqnarray}
So now we aim at showing that the same is true for $\langle\,\phi\,|\,H^{\veps}_{{\rm eff}}\,\psi^\chi\,\rangle$.
In the following, we omit the $\veps$-scripts of $H_{{\rm eff}}^\veps,U_1^\veps$, $U_2^\veps$, and $\tilde U_\veps$ and set $\H_{\rm b}:=L^2(\C,d\mu)$. Next we will show that  
\begin{eqnarray}\label{naiveexp}
\langle\,\phi\,|\,H_{{\rm eff}}\,\psi^\chi\,\rangle_{\H_{\rm eff}} 
&=& \langle\,\tilde\phi\,|\,U_0\,(H_0\,+\,\veps H_1\,+\,\veps^2H_2)\,U^*_0\,\tilde\psi_\chi\,\rangle_{\H_{\rm b}} \nonumber\\
&& \,+\,\veps\, \langle\,\tilde\phi\,|\,U_0\big(U_1\, (H_0+\veps H_1)\,+\,(H_0 +\veps H_1) \,U_1^{*}\big)\,U^*_0\,\tilde\psi_\chi\,\rangle_{\H_{\rm b}} \nonumber\\
&& \,+\,\veps^2\,\langle\,\tilde\phi\,|\,U_0\,\big(U_1\,H_0\,U^*_1\,+\,U_2\,H_0 \,+\,\,H_0\,U^*_2\big)\,U^*_0\,\tilde\psi_\chi\,\rangle_{\H_{\rm b}}\nonumber\\
&& 
\,+\,\O(\veps^3\|\phi\|_{\H_{\rm eff}}\|\psi\|_{\H_{\rm eff}}).
\end{eqnarray}
By definition of $H_{{\rm eff}}$ it holds
\begin{eqnarray*}
\langle\,\phi\,|\,H_{{\rm eff}}\,\psi^\chi\,\rangle_{\H_{\rm eff}} 
\ \;=\ \; \langle\,\tilde\phi\,|\,M_{\tilde{\rho}}H_{{\rm eff}}M_{\tilde{\rho}}^*\,\tilde\psi_\chi\,\rangle_{\H_{\rm b}}
&=& \langle\,\tilde\phi\,|\,U_\veps\,H_\veps\,U_\veps^*\,\tilde\psi_\chi\,\rangle_{\H_{\rm b}}\\
&=& \langle\,\tilde\phi\,|\,U_0\tilde U\,H_\veps\,\tilde U^*U_0^*\,\tilde\psi_\chi\,\rangle_{\H_{\rm b}}. 
\end{eqnarray*}
If we could just count the number of $\veps$'s after plugging in the expansion of $H_\veps$ from Lemma \ref{expH} and the one of $\tilde U$ from Lemma \ref{props}, the claim (\ref{naiveexp}) would be clear. But the expansion of $H_\veps$ yields polynomially growing coefficients. So we have to use carefully the estimate (\ref{Hlbound}). 


By Lemma \ref{props} it holds $\|uP_0\tilde U\|_{\L(\D(H_\veps))}\lesssim1$ for each $u\in\{\tilde U^*,1,U_1^*,U_2^*\}$. Since $\tilde U^*P_0=P_\veps\tilde U^*P_0$ and $U_2^*P_0=P_0U_2^*P_0$ by Lemma \ref{props}, $u\,P_0\tilde U$ satisfies the assumptions on $A$ in Lemma \ref{expH} with $m=0$ for all those $u$ due to the decay properties of  $P_\veps$, $P_0$, and $U_1^*P_0$ from Lemma \ref{projector} and Lemma \ref{props}. We notice that $\|\chi(H_\veps)\|_{\L(\H,\D(H_\veps))}\lesssim1$ because $H_\veps$ is bounded from below and the support of $\chi$ is bounded from above, both independently of $\veps$. Hence, using $U_0^*U_\veps=P_0\tilde U$ we may conclude from (\ref{Hlbound}) that
\begin{equation}\label{tildebound}
\|h \,u\,U_0^*\tilde\psi_\chi\|_{\H}\;=\;\|h\,u\,P_0\,\tilde U\,\chi(H_\veps)\,U_\veps^*\,\tilde\psi\|_{\H}\;\lesssim\;\|\psi\|_{\H_{\rm eff}} 
\end{equation}
for each $h\in\{H_\veps,H_0,H_1,H_2\}$. Furthermore, Lemma \ref{expH} implies in the same way that
\[\big\|\big(H_\veps-(H_0\,+\,\veps H_1\,+\,\veps^2H_2)\big)\tilde U^*U_0^*\,\tilde\psi_\chi\big\|_{\H_{\rm eff}} \;=\;\O(\veps^3).\]

So we have
\begin{eqnarray*}
H_\veps\,\tilde U^*U_0^*\,\tilde\psi_\chi
&=& (H_0+\veps H_1+\veps^2H_2)\,\tilde U^*U_0^*\,\tilde\psi_\chi \;+\;\O(\veps^3\|\psi\|)\\
&=& (H_0+\veps H_1+\veps^2H_2)\,(1+\veps U_1^*+\veps^2 U_2^*)U_0^*\,\tilde\psi_\chi \,+\,\O(\veps^3\|\psi\|)\\
&=& \Big((H_0+\veps H_1+\veps^2H_2)
\\
&& \qquad\,+\,\veps \,(H_0+\veps H_1)U_1^*\,+\,\veps^2 H_0U_2^*\Big)\,U_0^*\,\tilde\psi_\chi \,+\,\O(\veps^3\|\psi\|).
\end{eqnarray*}
For the rest of the proof we write $\O(\veps^l)$ for bounded by $\veps^l\|\phi\|_{\H_{\rm eff}}\|\psi\|_{\H_{\rm eff}}$ times a constant independent of $\veps$. The above yields
\begin{eqnarray*}
\langle\,\phi\,|\,H_{{\rm eff}}\,\psi\,\rangle 
&=& \langle\,\tilde\phi\,|\,U_0\tilde U\,H_\veps\,\tilde U^*U_0^*\,\tilde\psi_\chi\,\rangle \nonumber\\
&=& \langle\,\,\tilde\phi\,|\,U_0\tilde U(H_0\,+\,\veps H_1\,+\,\veps^2H_2)\,U_0^*\,\tilde\psi_\chi\,\rangle \nonumber\\
&& \;+\;\veps\, \langle\,\tilde\phi\,|\,U_0\tilde U\,(H_0+\veps H_1)\,U_1^*\,U_0^*\tilde\psi_\chi\,\rangle  
\nonumber\\
&& \;+\;\veps^2\,\langle\,U_0^*\,\tilde\phi\,|\,\tilde U\,H_0 U_2\,U_0^*\tilde\psi_\chi\,\rangle
\;+\;\O(\veps^3),
\end{eqnarray*}
After plugging $\tilde U=1+\veps U_1+\veps^2 U_2$ we may  drop the terms with three or more $\veps$'s in it because of (\ref{tildebound}). Gathering all the remaining terms we, indeed, end up with (\ref{naiveexp}).

\medskip

Now we calculate all the terms in (\ref{naiveexp}) separately.  By Definition \ref{U0} 
\begin{equation}\label{null}
\langle\tilde\phi\,|\,U_0\,A\,U^*_0\,\tilde\psi_\chi\rangle_{\H_{\rm b}}
\;=\; \langle\varphi_{\rm f}\tilde\phi\,|\,A\,\varphi_{\rm f}\tilde\psi_\chi\rangle_{\H}.
\end{equation}
for any operator $A$. Furthermore, the exponential decay of $\varphi_{\rm f}$ and its derivatives due to the Lemma \ref{props} guarantees that, in the following, all the fiber integrals are bounded in spite of the terms growing polynomially in $\nu$. 

We observe that $\tilde\psi_\chi=U_0\tilde U\chi(H_\veps)U_\veps^*\tilde\psi$ implies that
\[\|H_{\rm eff}^{(0)}\tilde\psi_\chi\|_{\H_{\rm b}}\;\lesssim\;\|H_\veps\chi(H_\veps)U_\veps^*\tilde\psi\|_{\H}\,+\,\|\chi(H_\veps)U_\veps^*\tilde\psi\|_{\H}\;\lesssim\;1\] 
because $\|U_0\|_{\L(\D(H_\veps),\D(H_{\rm eff}^{(0)})}\lesssim1$ and $\|\tilde U\|_{\L(\D(H_\veps))}\lesssim1$ by Lemma \ref{props}. As explained in Step \ref{Step1} every differential operator of second order with coefficients in~$C^\infty_{\rm b}(\C)$ on $\H_{\rm eff}$ is operator-bounded by $H_{\rm eff}^{(0)}$. Therefore derivatives that hit $\tilde\psi_\chi$ do not pose any problem, either. These facts will be used throughout the computations below. We write down the calculations via quadratic forms for the sake of readability. However, one should think of all the operators applied to $\phi$ as the adjoint applied to the corresponding term containing $\psi$. Since $\|\varphi_{\rm f}\|_{\H_{\rm f}(q)}=1$ for all $q\in\C$, Lemma \ref{berryphase2} implies
\begin{equation*}
2\,{\rm Re}\langle\varphi_{\rm f}|\nabla^{\rm h}\varphi_{\rm f}\rangle_{\H_{\rm f}} 
\,=\, \langle\nabla^{\rm h}_\tau\varphi_{\rm f}|\varphi_{\rm f}\rangle_{\H_{\rm f}} +\langle\varphi_{\rm f}|\nabla^{\rm h}_\tau\varphi_{\rm f}\rangle_{\H_{\rm f}} 
\,=\, \big({\rm d}\langle\varphi_{\rm f}|\varphi_{\rm f}\rangle_{\H_{\rm f}} \big)(\tau)\\
\,=\, 0.
\end{equation*}
Thus $\langle\varphi_{\rm f}|\nabla^{\rm h}\varphi_{\rm f}\rangle_{\H_{\rm f}}={\rm Im}\langle\varphi_{\rm f}|\nabla^{\rm h}\varphi_{\rm f}\rangle_{\H_{\rm f}}$.
Therefore the product rule (\ref{leibniz}) implies
\begin{eqnarray}\label{eins} 
\lefteqn{\langle\varphi_{\rm f}\tilde\phi\,|\,H_0\,\varphi_{\rm f}\tilde\psi_\chi\rangle_{\H}} \nonumber\\ 
&& \stackrel{(\ref{expcompl})}{=}\; \int_\C \tilde\phi^*\langle \varphi_{\rm f}|H_{\rm f}\varphi_{\rm f}\rangle_{\H_{\rm f}}\,\tilde\psi_\chi\,d\mu\,+\,\int_\C\int_{N_q\C}g(\veps\nabla^{\rm h}\varphi_{\rm f}^*\tilde\phi^*,\veps\nabla^{\rm h}\varphi_{\rm f}\tilde\psi_\chi)\,d\nu\,d\mu\nonumber\\
&& =\; \int_\C \tilde\phi^* E_{\rm f}\,\tilde\psi_\chi\,d\mu \,+\,\int_\C\int_{N_q\C}|\varphi_{\rm f}|^2\,g\big(\veps{\rm d}\tilde\phi^*,\veps{\rm d}\tilde\psi_\chi\big)\,+\veps\,g\big(\varphi_{\rm f}^*\,\veps{\rm d}\tilde\phi^*,\tilde\psi_\chi\,\nabla^{\rm h}\varphi_{\rm f}\big)\nonumber\\
&& \qquad\qquad\qquad\;+\;\veps\,g\big(\tilde\phi^*\,\nabla^{\rm h}\varphi_{\rm f}^*,\varphi_{\rm f}\,\veps{\rm d}\tilde\psi_\chi\big)\,+\,\veps^2\,g\big(\tilde\phi^*\,\nabla^{\rm h}\varphi_{\rm f}^*,\tilde\psi_\chi\,\nabla^{\rm h}\varphi_{\rm f}\big)\;d\nu\,d\mu\nonumber\\
&& =\; \int_\C g\big((p_{\rm eff}\tilde\phi)^*,p_{\rm eff}\tilde\psi_\chi\big)\,+\, \tilde\phi^* E_{\rm f}\,\tilde\psi_\chi\,+\,\veps^2\,\tilde\phi^* V_{{\rm BH}}\,\tilde\psi_\chi\;d\mu\nonumber\\
&&\quad\,-\,\veps^2\int_\C g\big((-\I\veps{\rm d}\tilde\phi)^*,\tilde\psi_\chi (r_1+r_2)\big)+g\big(\tilde\phi^*(r_1+r_2)^*,-\I\veps{\rm d}\tilde\psi_\chi\big)\,d\mu
\end{eqnarray}
with 
\begin{eqnarray*}
V_{{\rm BH}} &=& \int_{N_q\C}g_{{\rm eff}}^\veps(\nabla^{\rm h}\varphi_{\rm f}^*\,,\,(1-P_0)\nabla^{\rm h}\varphi_{\rm f})\,d\nu,\\
p^\veps_{{\rm eff}}\psi &=& -\,\I\veps{\rm d}\psi \,-\,{\rm Im}\,\Big(\veps\,\langle\varphi_{\rm f}|\nabla^{\rm h}\varphi_{\rm f}\rangle_{\H_{\rm f}}\,
-\,\veps^2 \int_{N_q\C}{\textstyle \frac{2}{3}}\,\varphi_{\rm f}^*\,\overline{{\rm R}}\big(\nabla^{\rm v}\varphi_{\rm f},\nu\big)\nu\,d\nu\\
&& \qquad\,+\ \veps^2\,\big\langle\,\varphi_{\rm f}\,\big|\,2\,\big(\W(\,.\,)\,
-\,\langle\,\varphi_{\rm f}\,|\,\W(\,.\,)\varphi_{\rm f}\,\rangle_{\H_{\rm f}}\,\big)\,
\nabla^{\rm h}\varphi_{\rm f}\,\big\rangle_{\H_{\rm f}}\Big)\,\psi,
\end{eqnarray*}
as well as $r_1:={\rm Im}\,R_1$ for  $R_1:=\big\langle\,\varphi_{\rm f}\,\big|\,2\big(\W(\,.\,)-\langle\,\varphi_{\rm f}\,|\,\W(\,.\,)\varphi_{\rm f}\,\rangle_{\H_{\rm f}}\big)\,\nabla^{\rm h}\varphi_{\rm f}\,\big\rangle_{\H_{\rm f}}$ and $r_2:={\rm Im}\,R_2$ for $R_2:=\int_{N_q\C}\frac{2}{3}\,\varphi_{\rm f}^*\,{\rm R}\big(\nabla^{\rm v}\varphi_{\rm f},\nu\big)\nu\,d\nu$. When we split up $R_i$ into real and imaginary part for $i\in\{1,2\}$, an integration by parts shows 
\begin{eqnarray*}
\lefteqn{\int_\C g\big((-\I\veps{\rm d}\tilde\phi)^*,\tilde\psi_\chi R_i\big)+g\big(\tilde\phi^*R_i^*,-\I\veps{\rm d}\tilde\psi_\chi\big)\,d\mu}\\
&&\ \;=\ \;\int_\C g\big((-\I\veps{\rm d}\tilde\phi)^*,\tilde\psi_\chi r_i\big)+g\big(\tilde\phi^*r_i^*,-\I\veps{\rm d}\tilde\psi_\chi\big)\,d\mu\,+\,\O(\veps).
\end{eqnarray*}
Therefore the $r_1$-terms are cancelled by terms coming from $H_1$:
\begin{eqnarray}\label{zwei} 
\lefteqn{\langle\varphi_{\rm f}\tilde\phi\,|\,H_1\,\varphi_{\rm f}\tilde\psi_\chi\rangle_{\H}}\nonumber\\ 
&& \stackrel{(\ref{expcompl})}{=}\;\int_\C\int_{N_q\C}2{\rm II}(\nu)\big(\veps\nabla^{\rm h}\varphi_{\rm f}^*\tilde\phi^*,\veps\nabla^{\rm h}\varphi_{\rm f}\tilde\psi_\chi\big)\,+\,\tilde\phi^*\,(\nabla^{\rm v}_{\nu}W)|\varphi_{\rm f}|^2\,\tilde\psi_\chi\;d\nu\,d\mu\nonumber\\
&& =\;\int_\C\int_{N_q\C}|\varphi_{\rm f}|^2\,2{\rm II}(\nu)\big(\veps{\rm d}\tilde\phi^*,\veps{\rm d}\tilde\psi_\chi\big)\,+\,\veps\,2{\rm II}(\nu)\big(\varphi_{\rm f}^*\,\veps{\rm d}\tilde\phi^*,\tilde\psi_\chi\,\nabla^{\rm h}\varphi_{\rm f}\big)\nonumber\\
&& \qquad\quad+\,\veps\,2{\rm II}(\nu)\big(\tilde\phi^*\,\nabla^{\rm h}\varphi_{\rm f}^*,\varphi_{\rm f}\,\veps{\rm d}\tilde\psi_\chi\big)\,+\,\tilde\phi^*\,(\nabla^{\rm v}_{\nu}W)|\varphi_{\rm f}|^2\,\tilde\psi_\chi\;d\nu\,d\mu\,+\,\O(\veps^2)\nonumber\\
&& =\;\int_\C\langle\varphi_{\rm f}|2{\rm II}(\,.\,)\big((p_{\rm eff}\tilde\phi)^*,p_{\rm eff}\tilde\psi_\chi\big)\varphi_{\rm f}\rangle_{\H_{\rm f}}\,d\mu \,+\,\int_\C\tilde\phi^*\langle\varphi_{\rm f}|(\nabla^{\rm v}_{\cdot}W)\varphi_{\rm f}\rangle_{\H_{\rm f}}\,\tilde\psi_\chi\,d\mu\nonumber\\
&&\qquad\,+\,\veps\int_\C g\big((-\I\veps{\rm d}\tilde\phi)^*,\tilde\psi_\chi R_1\big)\,+\,g\big(\tilde\phi^*R_1^*,-\I\veps{\rm d}\tilde\psi_\chi\big)\,d\mu\,+\,\O(\veps^2),
\end{eqnarray}
where we used that $g(\tau_1,\W(\nu)\tau_2)={\rm II}(\nu)(\tau_1,\tau_2)=g(\W(\nu)\tau_1,\tau_2)$ (see the second appendix).
At second order we first omit all the terms involving the Riemann tensor:
\begin{eqnarray}\label{drei}
\lefteqn{\langle\varphi_{\rm f}\tilde\phi\,|\,H_2\,\varphi_{\rm f}\tilde\psi_\chi\rangle_{\H}\ -\ \text{'Riemann-terms'}}\nonumber\\
&& \;\stackrel{(\ref{expcompl})}{=}\ \; \int_\C\int_{N_q\C}3g\big(\W(\nu)\veps\nabla^{\rm h}\varphi_{\rm f}^*\tilde\phi^*,\W(\nu)\veps\nabla^{\rm h}\varphi_{\rm f}\tilde\psi_\chi\big)\nonumber\\
&&\qquad\quad\,+\,\tilde\phi^*\,({\textstyle\frac{1}{2}}\nabla^{\rm v}_{\nu,\nu}W+V_{\rm geom})|\varphi_{\rm f}|^2\,\tilde\psi_\chi\;d\nu\,d\mu\nonumber\\
&& \ \;=\ \; \int_\C\big\langle\varphi_{\rm f}\big|3g\big(\W(\,.\,)\veps{\rm d}\tilde\phi^*,\W(\,.\,)\veps{\rm d}\tilde\psi_\chi\big)\varphi_{\rm f}\big\rangle_{\H_{\rm f}}\,d\mu\,+\,\O(\veps)\nonumber\\
&&\qquad\quad\,+\,\int_\C \tilde\phi^*\,\big(\langle\varphi_{\rm f}|({\textstyle\frac{1}{2}}\nabla^{\rm v}_{\cdot,\cdot}W)\varphi_{\rm f}\rangle_{\H_{\rm f}}\,+\,V_{\rm geom}\big)\tilde\psi_\chi\;d\mu\nonumber\\
&& \ \;=\ \; \int_\C\big\langle\varphi_{\rm f}\big|3g\big(\W(\,.\,)(p_{\rm eff}\tilde\psi_\chi)^*,\W(\,.\,)p_{\rm eff}\tilde\psi_\chi\big)\,\varphi_{\rm f}\big\rangle_{\H_{\rm f}}\,d\mu\nonumber\\
&& \qquad\quad\,+\,\int_\C\tilde\phi^*\,\big(\langle\varphi_{\rm f}|({\textstyle\frac{1}{2}}\nabla^{\rm v}_{\cdot,\cdot}W)\varphi_{\rm f}\rangle_{\H_{\rm f}}\,+\,V_{\rm geom}\big)\tilde\psi_\chi\;d\mu\ \,+\,\O(\veps),
\end{eqnarray}
where we used that $-\I\veps{\rm d}\tilde\psi_\chi=p_{\rm eff}\tilde\psi_\chi+\O(\veps)$ in the last step.
Now we take care of the omitted second order terms. Noticing that $\nabla^{\rm v}\,\tilde\psi_\chi\varphi_{\rm f} = \tilde\psi_\chi\,\nabla^{\rm v}\varphi_{\rm f}$ we have
\begin{eqnarray}\label{vier}
\lefteqn{\text{'Riemann-terms'}}\nonumber\\
&\stackrel{(\ref{expcompl})}{=}& \int_\C\int_{N_q\C}\overline{\R}\big(\veps\nabla^{\rm h}\varphi_{\rm f}^*\tilde\phi^*,\nu,\veps\nabla^{\rm h}\varphi_{\rm f}\tilde\psi_\chi,\nu\big)
\,+\,\textstyle{\frac{2}{3}}\,\overline{\R}\big(\veps\nabla^{\rm h}\varphi_{\rm f}^*\tilde\phi^*,\nu,\nabla^{\rm v}\varphi_{\rm f}\tilde\psi_\chi,\nu\big)\nonumber\\
&&\quad+\,\textstyle{\frac{2}{3}}\,\overline{\R}\big(\nabla^{\rm v}\varphi_{\rm f}^*\tilde\phi^*,\nu,\veps\nabla^{\rm h}\varphi_{\rm f}\tilde\psi_\chi,\nu\big)\,+\,\textstyle{\frac{1}{3}}\,\overline{\R}\big(\nabla^{\rm v}\varphi_{\rm f}^*\tilde\phi^*,\nu,\nabla^{\rm v}\varphi_{\rm f}\tilde\psi_\chi,\nu\big)\,d\nu\,d\mu\nonumber\\
&=& \int_\C\int_{N_q\C}|\varphi_{\rm f}|^2\,\overline{\R}\big(\veps{\rm d}\tilde\phi^*,\nu,\veps{\rm d}\tilde\psi_\chi,\nu\big)
\,+\,\textstyle{\frac{2}{3}}\,\overline{\R}\big(\varphi_{\rm f}^*\veps{\rm d}\tilde\phi^*,\nu,\tilde\psi_\chi\,\nabla^{\rm v}\varphi_{\rm f},\nu\big)\nonumber\\
&& \hspace{-0.5cm}+\textstyle{\frac{2}{3}}\,\overline{\R}\big(\tilde\phi^*\nabla^{\rm v}\varphi_{\rm f}^*,\nu,\varphi_{\rm f}\,\veps{\rm d}\tilde\psi_\chi,\nu\big)+\textstyle{\frac{1}{3}}\,\tilde\phi^*\overline{\R}\big(\nabla^{\rm v}\varphi_{\rm f}^*,\nu,\nabla^{\rm v}\varphi_{\rm f},\nu\big)\tilde\psi_\chi\,d\nu\,d\mu+\O(\veps)\nonumber\\
&=&  \int_\C\big\langle\varphi_{\rm f}\,\big|\,\overline{\R}\big(\veps{\rm d}\tilde\phi^*,\,.\,,\veps{\rm d}\tilde\psi_\chi,\,.\,\big)\varphi_{\rm f}\big\rangle_{\H_{\rm f}}\,d\mu
\,+\,\int_\C \tilde\phi^*V_{\rm amb}\,\tilde\psi_\chi\,d\mu\nonumber\\
&& \,+\,\int_\C g\big((-\I\veps{\rm d}\tilde\phi)^*,\tilde\psi_\chi R_2\big)\,+\,g\big(\tilde\phi^*R_2^*,-\I\veps{\rm d}\tilde\psi_\chi\big)\,d\mu\,+\,\O(\veps)
\end{eqnarray}
with $V_{\rm amb}=\int_{N_q\C}{\textstyle \frac{1}{3}}\,\overline{\R}\big(\nabla^{\rm v}\varphi_{\rm f}^*,\nu,\nabla^{\rm v}\varphi_{\rm f},\nu\big)\,d\nu$.
Again replacing $-\I\veps{\rm d}$ with $p_{\rm eff}$ and $g$ with $g_{\rm eff}$ yields errors of order $\veps$ only. In view of (\ref{null})-(\ref{vier}), we have 
\begin{eqnarray}\label{basic}
\lefteqn{\langle\tilde\phi\,|\,U_0\,(H_0\,+\,\veps H_1\,+\,\veps^2H_2)\,U^*_0\,\tilde\psi_\chi\rangle_{\H_{\rm b}}}\nonumber\\ 
&& \quad=\ \, \int_\C g_{{\rm eff}}^\veps\big((p_{{\rm eff}}\tilde\phi)^*,p_{{\rm eff}}\tilde\psi_\chi\big) \,+\,\tilde\phi^*\,E_{\rm f}\,\tilde\psi_\chi\nonumber\\
&&\qquad\qquad \,+\,\tilde\phi^*\,\big(\veps\langle\varphi_{\rm f}|\nabla^{\rm v}_{\cdot}W\varphi_{\rm f}\rangle_{\H_{\rm f}}+\veps^2 W^{(2)}\big)\,\tilde\psi_\chi\;d\mu 
\,+\O(\veps^3)
\end{eqnarray}
with 
\begin{eqnarray*}
g_{\rm eff}^\veps(\tau_1,\tau_2) &=& g(\tau_1,\tau_2) \ +\ \veps\ \langle\,\varphi_{\rm f}\,|\,2{\rm II}(\,.\,)(\tau_1,\tau_2)\,\varphi_{\rm f}\,\rangle_{\H_{\rm f}}\\
&& \ +\ \veps^2\ \Big\langle\,\varphi_{\rm f}\,\Big|\,3g\big(\W(\,.\,)\tau_1,\W(\,.\,)\tau_2\big)\,\varphi_{\rm f}\,+\,\overline{\R}\big(\tau_1,\,.\,,\tau_2,\,.\,\big)\varphi_{\rm f}\Big\rangle_{\H_{\rm f}}.
\end{eqnarray*}
We define $P_0^\perp:=(1-P_0)$. Before we deal with the corrections by $U_1$ and $U_2$ in (\ref{naiveexp}), we notice that due to $P_0=U_0^*U_0^*$ and $P_0^\perp U_0^*=0$  
\begin{eqnarray}\label{Psi}
\lefteqn{P_0^\perp\big([-\veps\Delta_{\rm h},P_0]\,+\,H_1\big)\,U_0^*\tilde\psi_\chi}\nonumber\\
&\stackrel{(\ref{expcompl})}{=}& P_0^\perp\Big([-\veps\Delta_{\rm h},U_0^*U_0]\,-\,{\rm tr}_\C\,\veps\nabla^{\rm h}\W(\nu)\,\veps\nabla^{\rm h}\,+\,(\nabla^{\rm v}_\nu W)\Big)\,U_0^*\tilde\psi_\chi\nonumber\\
&=&  P_0^\perp\Big((\nabla^{\rm v}_\nu W)\,-\,{\rm tr}_\C\big(2(\nabla^{\rm h}\varphi_{\rm f})U_0\,+\,\veps\nabla^{\rm h}\W(\nu)\big)\,\veps\nabla^{\rm h}\Big)\,U_0^*\tilde\psi_\chi\,+\,\O(\veps)\nonumber\\
&=&  P_0^\perp\Big(\varphi_{\rm f} (\nabla^{\rm v}_\nu W)\tilde\psi_\chi\,-\,2g(\nabla^{\rm h} \varphi_{\rm f}^*,\veps{\rm d} \tilde\psi_\chi)\,-\,\varphi_{\rm f}\,{\rm tr}_\C\,\W(\nu)\veps^2\nabla {\rm d} \tilde\psi_\chi\Big)\,+\O(\veps)\nonumber\\
&=& P_0^\perp\,\Psi(\veps\nabla {\rm d}\tilde\psi_\chi,{\rm d}\tilde\psi_\chi,\tilde\psi_\chi)\,+\,\O(\veps). 
\end{eqnarray}
with $\Psi(A,p,\phi) = -\,\varphi_{\rm f}\,{\rm tr}_\C\big(\W(\nu)A\big)\,-\,2g_{{\rm eff}}^\veps\big(\nabla^{\rm h} \varphi_{\rm f}^*,p\big)\,+\,\varphi_{\rm f} (\nabla^{\rm v}_\nu W)\phi$.

We note that $U_0^*\tilde\psi_\chi=B^\veps U^*\tilde\psi$ with $B^\veps=P_0\tilde U\chi(H_\veps)$. So we may apply (\ref{alleklein}) und (\ref{U1}) in the following.
Since $U_0=U_0P_0$ by definition and we know from Lemma~\ref{props} that $P_0 U_1P_0=0$, the first corrections by $U_1$ are an order of $\veps$ higher than expected:
\begin{eqnarray}\label{correction}
\lefteqn{\Big\langle\tilde\phi\,\Big|\,U_0\,\Big((H_0\,+\,\veps H_1)\,U^*_1\,+\,\,U_1\,(H_0\,+\,\veps H_1)\Big)\,U^*_0  \,\tilde\psi_\chi\Big\rangle_{\H_{\rm b}}}\nonumber\\
&=& \Big\langle\tilde\phi\,\Big|\,U_0\Big(\big([P_0,H_0]+\veps H_1\big)\,U_1^* \,+\,U_1\,\big([H_0,P_0]+\veps H_1\big)\Big)U^*_0\,\tilde\psi_\chi\Big\rangle_{\H_{\rm b}}\nonumber\\
&=& \veps\,\Big\langle\tilde\phi\,\Big|\,U_0\Big(\big([\veps\Delta_{\rm h},P_0]+H_1\big)\,U_1^* \,+\,U_1\,\big([-\veps\Delta_{\rm h},P_0]+H_1\big)P_0\Big)U^*_0\,\tilde\psi_\chi\Big\rangle_{\H_{\rm b}}\nonumber\\
&\stackrel{(\ref{U1})}{=}& -\veps\,\Big\langle\tilde\phi\,\Big|\,U_0\,\big([\veps\Delta_{\rm h},P_0]+H_1\big)\,R_{H_{\rm f}}(E_{\rm f})\,\big([-\veps\Delta_{\rm h},P_0]+H_1\big)\,U^*_0\,\tilde\psi_\chi\Big\rangle_{\H_{\rm b}}\nonumber\\
&& \quad\,-\,\veps\,\langle \tilde\phi\,|\,U_0U_1\,(H_{\rm f}-E_{\rm f})\,U^*_1U_0^*\,\tilde\psi_\chi\rangle_{\H_{\rm b}}\nonumber\\
&\stackrel{(\ref{Psi})}{=}& -\veps\,\Big\langle\Psi(\veps^2\nabla {\rm d}  \tilde\phi,\veps {\rm d}  \tilde\phi,\tilde\phi)\,\Big|\,R_{H_{\rm f}}(E_{\rm f})\,\Psi(\veps^2\nabla {\rm d}  \tilde\psi_\chi,\veps {\rm d}  \tilde\psi_\chi,\tilde\psi_\chi)\Big\rangle_{\H_{\rm b}}\nonumber\\
&& \quad\,-\,\veps\,\langle \tilde\phi\,|\,U_0U_1\,(H_{\rm f}-E_{\rm f})\,U^*_1U_0^*\,\tilde\psi_\chi\rangle_{\H_{\rm b}}\nonumber\\
&=& -\veps\,\int_\C
\M\big(\Psi^*(\veps^2\nabla {\rm d}  \tilde\phi,\veps {\rm d}  \tilde\phi,\tilde\phi),\Psi(\veps^2\nabla {\rm d}  \tilde\psi_\chi,\veps {\rm d}  \tilde\psi_\chi,\tilde\psi_\chi)\big)\Big)\,d\mu\nonumber\\
&& \quad\,-\,\veps\,\langle \tilde\phi\,|\,U_0U_1\,(H_{\rm f}-E_{\rm f})\,U^*_1U_0^*\,\tilde\psi_\chi\rangle_{\H_{\rm b}}.
\end{eqnarray}
with $\M(\varphi_1^*,\varphi_2) = \big\langle\,\varphi_1\,\big|\,(1-P_0)\big(H_{\rm f}-E_{\rm f}\big)^{-1} (1-P_0)\,\varphi_2\,\big\rangle_{\H_{\rm f}}$.
Furthermore,
\begin{eqnarray}\label{cancel}
\lefteqn{\langle\tilde\phi\,|\,U_0\,\big(U_2\,H_0\,+\,H_0\,U^*_2 \big)\,U^*_0\,\tilde\psi_\chi\rangle_{\H_{\rm b}}}\nonumber\\
&=& \langle U^*_0\tilde\phi\,|\,P_0\big(U_2\,(-\veps^2\Delta_{\rm h}+H_{\rm f})\,+\,(-\veps^2\Delta_{\rm h}+H_{\rm f})\,U^*_2\big)\,P_0U^*_0\tilde\psi_\chi\rangle_{\H_{\rm b}}\nonumber\\
&=& \langle U^*_0\tilde\phi\,|\,\big(P_0U_2\,(-\veps^2\Delta_{\rm h}+E_{\rm f})P_0\,+\,P_0(-\veps^2\Delta_{\rm h}+E_{\rm f})\,U^*_2P_0\big)\,U^*_0\tilde\psi_\chi\rangle_{\H_{\rm b}}\nonumber\\
&\stackrel{(\ref{alleklein})}{=}& \langle U^*_0\tilde\phi\,|\,P_0\,(U_2+U^*_2)P_0(-\veps^2\Delta_{\rm h}+E_{\rm f})\,U^*_0\tilde\psi_\chi\rangle_{\H_{\rm b}}\ \,+\,\O(\veps)\nonumber\\
&=& -\,\langle U^*_0\tilde\phi\,|\,P_0\,U_1 U^*_1P_0(-\veps^2\Delta_{\rm h}+E_{\rm f})\,U^*_0\tilde\psi_\chi\rangle_{\H_{\rm b}}\ \,+\,\O(\veps),
\end{eqnarray}
because $\tilde U= 1+\veps U_1+\veps^2 U_2$ implies via $P_0\tilde U\tilde U^*P_0=P_0$ and $P_0U_1P_0=0$ that $P_0(U_2+U^*_2)P_0=-\,P_0U_1U^*_1P_0\,+\,\O(\veps)$. Finally, the remaining second order term cancels the term from (\ref{cancel}) and the second term from (\ref{correction}): 
\begin{eqnarray}\label{cancelled}
\lefteqn{\big\langle\tilde\phi\,\big|\,U_0\,U_1\,H_0\,U^*_1\, U_0^*\,\tilde\psi_\chi\big\rangle_{\H_{\rm b}}}\nonumber\\
&=& \langle \tilde\phi\,|\,U_0\,U_1\,(-\veps^2\Delta_{\rm h}+H_{\rm f})\,U^*_1 U_0^*\,\tilde\psi_\chi\rangle_{\H_{\rm b}}\nonumber\\
&=& \langle \tilde\phi\,|\,U_0U_1\,(H_{\rm f}-E_{\rm f})\,U^*_1U_0^*\,\tilde\psi_\chi\,+\,U_0U_1(-\veps^2\Delta_{\rm h}+E_{\rm f})\,U^*_1 P_0 U_0^*\,\tilde\psi_\chi\rangle_{\H_{\rm b}}\nonumber\\
&\stackrel{(\ref{alleklein})}{=}& \langle \tilde\phi\,|\,U_0U_1\,(H_{\rm f}-E_{\rm f})\,U^*_1U_0^*\,\tilde\psi_\chi\rangle_{\H_{\rm b}}\nonumber\\
&& \quad\,+\,\langle \tilde\phi\,|\,U_0U_1U^*_1 P_0\,(-\veps^2\Delta_{\rm h}+E_{\rm f})\, U_0^*\,\tilde\psi_\chi\rangle_{\H_{\rm b}}\,+\,\O(\veps).
\end{eqnarray}
We gather the terms from (\ref{basic}) to (\ref{cancelled}) and replace ${\rm d}\tilde\psi_\chi$ by $p^\veps_{\rm eff}\tilde\psi_\chi$ in the argument of $\Psi$, which only yields an error of order~$\veps^3$. Then we obtain that $\langle\,\tilde\phi\,|\,H^{\veps}_{{\rm eff}}\,\tilde\psi_\chi\,\rangle$ equals the right-hand side of (\ref{goal}) up to errors of order~$\veps$,
only with $d\mu$ instead of $d\mu_{\rm eff}$. 
Here $\tilde\psi=M_{\tilde\rho}\psi$ enters. By Lemma \ref{transform} $M_{\tilde\rho}$ interchanges the former with the latter but may add extra terms. However, $g$ and $g_{\rm eff}$ coincide at leading order and so do their associated volume measures. Therefore ${\rm d}(\ln\tilde\rho)$ and $\Delta_\C\ln\tilde\rho$ are of order $\veps$. This shows that the extra potential from Lemma \ref{transform}, given by
$-\frac{\veps^2}{4}\,g\big({\rm d} (\ln\tilde\rho),{\rm d} (\ln\tilde\rho)\big)+\frac{\veps^2}{2}\Delta_\C(\ln\tilde\rho)$, is of order $\veps^3$. Exploiting ${\rm d}(\ln\tilde\rho)=\O(\veps)$ we easily obtain that all the other extra terms are also only of order $\veps^3$, which finishes the proof of Step \ref{Step3}.

\bigskip

\begin{step}\label{Step4}
It holds
$\|(H^\veps_{\rm eff}-H^{(2)}_{\rm eff})\chi(H^{\veps}_{\rm eff})\|_{\L(\H_{\rm eff})}\,=\,\O(\veps^3)$.
\end{step} 

The spectral calculus implies $\chi(H^\veps_{\rm eff})=\tilde\chi^2(H^\veps_{\rm eff})\chi(H^\veps_{\rm eff})$. 
As in (\ref{funny}) this implies that
\[\|\,\chi(H^{\veps}_{\rm eff})- U^{\veps}\tilde\chi(H^{\veps}) U^{\veps*}\,\chi(H^{\veps}_{\rm eff})\|_{\L(\H_{\rm eff},\D(H^{\veps}_{{\rm eff}}))}\,=\,\O(\veps^3).\]
Now Step \ref{Step4} follows from Step \ref{Step3} and $\D(H^{(2)}_{{\rm eff}})=\D(H^{\veps}_{{\rm eff}})$ due to Step \ref{Step2}.

\pagebreak

\begin{step}\label{Step5}
It holds
$\|(H^\veps_{\rm eff}-H^{(2)}_{\rm eff})\chi(H^{(2)}_{\rm eff})\|_{\L(\H_{\rm eff})}\,=\,\O(\veps^3)$.
\end{step} 

We note that Step \ref{Step1} \& \ref{Step2} imply that
$\|H^\veps_{\rm eff}-H^{(2)}_{\rm eff}\|_{\L(\D(H^\veps_{\rm eff}),\H_{\rm eff})}\;=\;\O(\veps)$. So in the norm of $\L\big(\H_{\rm eff},\D(H^{\veps}_{{\rm eff}})\big)$ it holds that
\begin{eqnarray}\label{funny2}
\tilde\chi^3(H^{(2)}_{{\rm eff}})
&=& \tilde\chi(H^{\veps}_{{\rm eff}})\,\tilde\chi^2(H^{(2)}_{{\rm eff}}) \,+\,\big(\tilde\chi(H^{(2)}_{{\rm eff}})-\tilde\chi(H^{\veps}_{{\rm eff}})\big)^3\nonumber\\
&&\,+\,\big(\tilde\chi(H^{(2)}_{{\rm eff}})-\tilde\chi(H^{\veps}_{{\rm eff}})\big)^2\,\tilde\chi(H^{\veps}_{{\rm eff}})\nonumber\\
&&\,+\,\big(\tilde\chi(H^{(2)}_{{\rm eff}})-\tilde\chi(H^{\veps}_{{\rm eff}})\big)\,\tilde\chi(H^{\veps}_{{\rm eff}})\,\tilde\chi(H^{(2)}_{{\rm eff}})\nonumber\\ 
&=& \tilde\chi(H^{\veps}_{{\rm eff}})\,\tilde\chi^2(H^{(2)}_{{\rm eff}}) \,+\,\O(\veps^3)\nonumber
\end{eqnarray}
by Lemma \ref{microlocal} b) and Step \ref{Step2} \& \ref{Step4}. Hence, Step \ref{Step5} can be reduced to Step \ref{Step4} in the same way as we reduced Step \ref{Step4} to Step \ref{Step3}. 

\medskip

Theorem \ref{calcHeff} is entailed by Step \ref{Step3} to \ref{Step5} and the remark preceding Step \ref{Step1}.\qed


\subsection{Proof of the approximation of eigenvalues}\label{proofquasimodes}

With Theorem \ref{calcHeff}, Corollary \ref{projector3}, and Lemma \ref{microlocal} we have already everything at hand we need to prove Theorem \ref{quasimodes}, which relates the spectra of $H^\veps$ and $H^{(2)}_{{\rm eff}}$.

\medskip

{\sc Proof of Theorem \ref{quasimodes} (Section \ref{eigenvalues}):}
\newline
We fix $E<\infty$ and set $E_-:=\min\{\inf\sigma(H^\veps),\inf\sigma(H^{(2)}_{{\rm eff}})\}-1$. Let $\chi$ be the characteristic function of $[E_-,E]$ and $\tilde\chi\in C^\infty_0(\RRR)$ with $\tilde\chi|_{[E_-,E]}\equiv1$. 

To show a) i) we assume we are given a family of eigenvalues $(E_\veps)$ of $H^{(2)}_{{\rm eff}}$ with $\limsup E_\veps\,<\,E$ and a corresponding family of eigenfunctions $(\psi_\veps)$. Since $\psi_\veps$ is an eigenfunction of $H^{(2)}_{{\rm eff}}$, we have that $\psi_\veps=\chi(H^{(2)}_{{\rm eff}})\psi_\veps$ for $\veps$ small enough. 
By Theorem \ref{calcHeff} and Lemma \ref{microlocal} b) it holds in the norm of~$\L\big(L^2(\C,d\mu_{\rm eff}),\D(H^{\veps}_{{\rm eff}})\big)$
\begin{eqnarray}\label{cutoffexchange2}
\chi(H^{(2)}_{{\rm eff}}) &=& \tilde\chi^2(H^{(2)}_{{\rm eff}})\,\chi(H^{(2)}_{{\rm eff}})\nonumber\\
&=& \tilde\chi^2(H^{\veps}_{{\rm eff}})\,\chi(H^{(2)}_{{\rm eff}})\,+\,\O(\veps^3)\nonumber\\ 
&\stackrel{(\ref{funny})}{=}&  U^{\veps}\tilde\chi(H^{\veps}) U^{\veps*}\,\tilde\chi(H^{\veps}_{{\rm eff}})\,\chi(H^{(2)}_{{\rm eff}})\,+\,\O(\veps^3)\nonumber\\
&=&  U^{\veps}\tilde\chi(H^{\veps}) U^{\veps*}\,\chi(H^{(2)}_{{\rm eff}})\,+\,\O(\veps^3).
\end{eqnarray}
Therefore with $ U^{\veps*}= P^\veps U^{\veps*}$, $ U^{\veps*} U^{\veps}= P^\veps$, and $H^{\veps}_{{\rm eff}}= U^{\veps}H^\veps U^{\veps*}$
\begin{eqnarray*}
H^\veps\, U^{\veps*}\psi_\veps &=& \big(P^\veps+(1- P^\veps)\big)H^\veps U^{\veps*}\chi(H^{(2)}_{{\rm eff}})\psi_\veps\\
&=&  U^{\veps*}H^\veps_{{\rm eff}}\chi(H^{(2)}_{{\rm eff}})\psi_\veps \,+\, (1- P^\veps)[H^\veps, P^\veps]\, U^{\veps*}\chi(H^{(2)}_{{\rm eff}})\psi_\veps\\
&\stackrel{(\ref{cutoffexchange2})}{=}&  U^{\veps*}H^{(2)}_{{\rm eff}}\psi_\veps \,+\, (1- P^\veps)[H^\veps, P^\veps]\,\tilde\chi(H^\veps)\, U^{\veps*}\chi(H^{(2)}_{{\rm eff}})\psi_\veps\\
&& \qquad\qquad\qquad\qquad\qquad\qquad\qquad\qquad\qquad\,+\,\O(\veps^3\|\psi_\veps\|_{\H_{\rm eff}})\\
&=& E_\veps\, U^{\veps*}\psi_\veps\,+\,\O(\veps^3\|\psi_\veps\|_{\H_{\rm eff}}),
\end{eqnarray*}
where we made use of the assumption and Corollary \ref{projector3} in the last step. This proves a) i) because $U^\veps U^{\veps*}=1$ and thus $\|\psi_\veps\|_{\H_{\rm eff}}=\|U^{\veps*}\psi_\veps\|_{\overline{\H}}$. 

\smallskip

To show a)~ii) we now assume that we are given a family of eigenvalues $(E_\veps)$ of $H^\veps$ with $\limsup E_\veps<E$ and a corresponding family of eigenfunctions~$(\psi^\veps)$. Here this implies $\psi^\veps=\chi(H^\veps)\psi^\veps$ for $\veps$ small enough. With $ U^{\veps}= U^{\veps} P^\veps$ and $ U^{\veps*} U^{\veps}= P^\veps$ we obtain
\begin{eqnarray*}
H^{(2)}_{{\rm eff}}\, U^{\veps}\psi^\veps &=& H^{(2)}_{{\rm eff}}\, U^{\veps} P^\veps\,\tilde\chi(H^\veps)\chi(H^\veps)\psi^\veps\\
&=& H^{(2)}_{{\rm eff}}\, U^{\veps}\tilde\chi(H^\veps)\, P^\veps\chi(H^\veps)\psi^\veps\,+\,\O(\veps^3)\\
&=& H^{(2)}_{{\rm eff}}\,\tilde\chi(H^{\veps}_{{\rm eff}})\, U^{\veps}\chi(H^\veps)\psi^\veps\,+\,\O(\veps^3),
\end{eqnarray*}
where we used Lemma \ref{microlocal} a) \& c) in the two last steps. In view of Theorem~\ref{calcHeff}, we get
\begin{eqnarray*}
H^{(2)}_{{\rm eff}}\, U^{\veps}\psi^\veps &=& H^{\veps}_{{\rm eff}}\tilde\chi(H^{\veps}_{{\rm eff}})\, U^{\veps}\chi(H^\veps)\psi^\veps\,+\,\O(\veps^3)\\
&=&  U^{\veps}H^\veps U^{\veps*}\, U^{\veps}\tilde\chi(H^\veps) P^\veps\chi(H^\veps)\psi^\veps\,+\,\O(\veps^3).
\end{eqnarray*}
Using again Lemma \ref{microlocal} a) \& c) and the assumption we end up with
\begin{eqnarray*}
H^{(2)}_{{\rm eff}}\, U^{\veps}\psi^\veps \ \;=\ \;  U^{\veps}\,H^\veps\,P^\veps\,\chi(H^\veps)\psi^\veps\,+\,\O(\veps^3)&=&  U^{\veps}\,H^\veps\chi(H^\veps)\psi^\veps\,+\,\O(\veps^3)\\
&=& E_\veps\, U^{\veps}\psi^\veps\,+\,\O(\veps^3).
\end{eqnarray*}
This finishes the proof of a) ii). 

\smallskip

For b) we set $\psi_\veps:=D_\veps^* M_\rho^*\psi^\veps$ and observe that  $-\veps^2\Delta_{\rm v}=D_\veps \Delta_{\rm v}D_\veps^*$ by Definition~\ref{deriv} and thus $-\veps^2M_\rho\Delta_{\rm v}M_\rho^*+V_0(q,\nu/\veps)=M_\rho D_\veps  H_{\rm f}D_\veps^*M_\rho^*$. Therefore the statement is equivalent to 
\[\limsup\,\langle\psi_\veps|H_{\rm f}\psi_\veps\rangle\,<\,\inf_{q\in\C}E_1\|\psi_\veps\|^2\quad\Longrightarrow\quad \|U_\veps\psi_\veps\|\,\gtrsim\,\|\psi_\veps\|\]
because  $U^\veps:=M_{\tilde \rho}^*U_\veps D_\veps^* M_\rho^*$ by definition in the proof of Theorem \ref{effdyn}.
We have
\begin{eqnarray*}
\langle \psi_\veps|H_{\rm f} \psi_\veps\rangle
&=& \langle P_0\psi_\veps|H_{\rm f} P_0\psi_\veps\rangle \,+\, \langle(1-P_0)\psi_\veps|H_{\rm f}(1-P_0)\psi_\veps\rangle\\
&\geq& \inf_{q\in\C}E_{\rm f}\,\|P_0\psi_\veps\|^2 \,+\, \inf_{q\in\C}E_1\,\|(1-P_0)\psi_\veps\|^2 \\
&=& \inf_{q\in\C}E_{\rm f}\,\|\psi_\veps\|^2 +(\inf_{q\in\C}E_1-\inf_{q\in\C}E_{\rm f})\,\|(1-P_0)\psi_\veps\|^2\\
&=& \inf_{q\in\C}E_{\rm f}\,\|\psi_\veps\|^2 +(\inf_{q\in\C}E_1-\inf_{q\in\C}E_{\rm f})\,\|(1-P_\veps)\psi_\veps\|^2 \,+\,\O(\veps),
\end{eqnarray*}
where we used that $P_\veps-P_0=\O(\veps)$ by Lemma \ref{projector} in the last step.
Since $E_{\rm f}$ is a constraint energy band, hence, separated by a gap from $E_1$, and $\limsup \langle \psi_\veps|H_{\rm f} \psi_\veps\rangle<\inf_{q\in\C} E_1\|\psi_\veps\|^2$ by assumption, we may conclude that
\[\limsup\|(1-P_\veps)\psi_\veps\|^2<\limsup\|\psi_\veps\|^2.\] 
Because of $P_\veps=U_\veps^*U_\veps$ this implies $\|U_\veps\psi_\veps\|\gtrsim\|\psi_\veps\|$ for all $\veps$ small enough.
\qed


\section {The whole story}\label{wholestory}

In Section \ref{proofs} we proved our main theorems with the help of Lemmas \ref{transform} to~\ref{microlocal}. We still have to derive Lemmas \ref{projector} to \ref{props}, which is the task of this section. Before we can start with it, we have to carry out some technical preliminaries. 

\medskip

\begin{remark}\label{coordinates}
Since $\C$ is of bounded geometry, it has a countable covering $(\Omega_j)_j$ of finite multiplicity (i.e.\ there is $l_0\in\NNN$ such that each $\Omega_j$ overlaps with not more than $l_0$ of the others) by contractable geodesic balls of fixed diameter, and there is a corresponding partition of unity $(\xi_j\in C_0^\infty(\Omega_j))$ whose  derivatives of any order are bounded uniformly in $j$ (see e.g.\ App.\ 1 of~\cite{Sh}).

\smallskip

We fix $j\in\NNN$. By \emph{geodesic coordinates} with respect to the center $q\in\Omega_j$ we mean to choose an orthonormal basis $(v_i)_i$ of $T_q\C$ and to use the exponential mapping as a chart on $\Omega_j$. Let $(x^i)_{i=1,\ldots,d}$ be geodesic coordinates on $\Omega_j$. The bounded geometry of $\C$ that we assumed in (\ref{bndcurv1}) yields bounds uniform in $j$ on the metric tensor $g_{il}$ and its partial derivatives, thus, in particular, on all the inner curvatures of $\C$ and their partial derivatives. For the same reason the inverse of the metric tensor $g_{il}$ is positive definite with a constant greater than zero uniform in~$j$. 

\smallskip

We choose an orthonormal basis of the normal space at the center of $\Omega_j$ and extend it radially to $N\C|_{\Omega_j}=N\Omega_j$ via the parallel transport by the normal connection $\nabla^\perp$ (defined in the appendix). In this way we obtain an orthonormal trivializing frame $(\nu_\alpha)_\alpha$ over $\Omega_j$. Let $(n^\alpha)_{\alpha=1,\ldots,k}$ be bundle coordinates with respect to this frame. The connection coefficients $\Gamma^\gamma_{i\alpha}$ of the normal connection are given by $\nabla^\perp_{\partial_{x_i}}\nu_\alpha=\sum_{\gamma=1}^k\Gamma^\gamma_{i\alpha}\nu_\gamma$. Due to the smooth embedding of~$\C$ assumed in (\ref{bndcurv2}) the exterior curvatures of~$\C$,  the curvature of~$N\C$, as well as all their derivatives are globally bounded. This implies that all the partial derivatives of~$\Gamma^\gamma_{i\alpha}$ and of the exterior curvatures of $\C$ are bounded uniformly in $j$ in the coordinates $(x^i)_{i=1,\ldots,d}$ and $(n^\alpha)_{\alpha=1,\ldots,k}$.
\end{remark}

\smallskip

From now on we implicitly sum over repeated indices. 
The vertical derivative in local coordinates is given by
\begin{equation}\label{verder}
(\nabla^{\rm v}_{\nu_\alpha}\psi)(x,n)\;=\; \partial_{n_\alpha}\psi(x,n).
\end{equation} 
and the horizontal connection is given by 
\begin{equation}\label{horder}
(\nabla^{\rm h}_{\partial_{x^i}}\psi)(x,n) \;=\; \partial_{x^i}\psi(x,n)\,-\,\Gamma^\gamma_{i\alpha}\,n^\alpha\,\partial_{n^\gamma}\psi(x,n).
\end{equation}
The former directly follows from the definition of $\nabla^{\rm v}$ (see Definition \ref{deriv}). To obtain the latter equation we note first that for a normal vector field $v=n^\alpha\nu_\alpha$ over~$\C$ it holds \begin{equation}\label{Christoffel}
(\nabla^\perp_{\partial_{x_i}}v)^\gamma\;=\;\partial_{x_i}n^\gamma\,+\,\Gamma^\gamma_{i\alpha}n^\alpha.
\end{equation}
Now let $(w,v)\in C^1([-1,1],N\Omega_j)$ with
\[w(0)\;=\;x,\ \dot{w}(0)\;=\;\partial_{x_i}, \quad\&\quad v(0)\;=\;n,\ \nabla^\perp_{\dot{w}}v\;=\;0.\]
Then by definition of $\nabla^{\rm h}$ we have
\begin{eqnarray*}
(\nabla^{\rm h}_{\partial_{x_i}}\psi)(x,n) &=& \textstyle{\frac{d}{ds}}\big|_{s=0}\psi(w(s),v(s))\\
&=& \textstyle{\frac{d}{ds}}\big|_{s=0}\psi(w(s),n) \,+\, \textstyle{\frac{d}{ds}}\big|_{s=0}\psi(x,v(s))\\
&=& \partial_{x^i}\psi(x,n) \,+\, (\partial_{x^i}n^\gamma)\partial_{n^\gamma}\psi(x,n)\\
&=& \partial_{x^i}\psi(x,n) \,-\, \Gamma^\gamma_{i\alpha}\,n^\alpha\,\partial_{n^\gamma}\psi(x,n),
\end{eqnarray*}
where we used (\ref{Christoffel}) and the choice of the curve $v$ in the last step.

\medskip

With the formulas (\ref{verder}) and (\ref{horder}) it is easy to derive the properties of $\nabla^{\rm h}$ that were stated in Lemma \ref{berryphase2}.

\medskip

{\sc Proof of Lemma \ref{berryphase2} (Section \ref{proofT2}):}
\newline
Let $\tau,\tau_1,\tau_2\in\Gamma(T\C)$ and $\psi,\psi_1,\psi_2\in C^2\big(\C,\H_{\rm f}(q)\big)$. We fix a geodesic ball $\Omega\in\C$ and choose $(x^i)_{i=1,\ldots,d}$ and $(n^\alpha)_{\alpha=1,\ldots,k}$ as above. 
We first verify that $\nabla^{\rm h}$ is metric, i.e. $\big({\rm d}\,\langle\psi_1|\psi_2\rangle_{\H_{\rm f}}\big)(\tau)=\langle\nabla^{\rm h}_{\tau}\psi_1|\psi_2\rangle_{\H_{\rm f}}+\langle\psi_1|\nabla^{\rm h}_{\tau}\psi_2\rangle_{\H_{\rm f}}$. Since $\nabla^\perp$ is a metric connection, $\Gamma^\gamma_{i\alpha}$ is anti-symmetric in $\alpha$ and $\gamma$, in particular $\Gamma^\alpha_{i\alpha}=0$ for all $\alpha$. Therefore an integration by parts yields that
\begin{equation*}
\big\langle\, \Gamma^\gamma_{i\alpha}n^\alpha\partial_{n^\gamma}\psi_1\,\big|\,\psi_2\,\big\rangle_{\H_{\rm f}}\,+\,\big\langle\,\psi_1\,\big|\,\Gamma^\gamma_{i\alpha}n^\alpha\partial_{n^\gamma}\psi_2\,\big\rangle_{\H_{\rm f}}\;=\;0.
\end{equation*}
Therefore we have
\begin{eqnarray*}
\big({\rm d}\langle\psi_1|\psi_2\rangle\big)(\tau) 
&=& \tau^i\langle\partial_{x_i}\psi_1|\psi_2\rangle\,+\,\tau^i\langle\psi_1|\partial_{x_i}\psi_2\rangle\\
&=& \tau^i\big\langle(\partial_{x_i}-\Gamma^\gamma_{i\alpha}n^\alpha\partial_{n^\gamma})\psi_1\big|\psi_2\big\rangle \,+\,\tau^i\big\langle\psi_1\big|(\partial_{x_i}-\Gamma^\gamma_{i\alpha}n^\alpha\partial_{n^\gamma})\psi_2\big\rangle\\
&=& \langle\nabla^{\rm h}_{\tau}\psi_1|\psi_2\rangle+\langle\psi_1|\nabla^{\rm h}_{\tau}\psi_2\rangle.
\end{eqnarray*}
To compute the curvature we notice that 
\begin{eqnarray*}
{\rm R}^{\rm h}(\tau_1,\tau_2)\psi
&=& \big(\nabla^{\rm h}_{\tau_1} \nabla^{\rm h}_{\tau_2}  \,-\, \nabla^{\rm h}_{\tau_2} \nabla^{\rm h}_{\tau_1} -\nabla^{\rm h}_{[\tau_1,\tau_2]}\big)\psi\\
&=& \tau_1^i\tau_2^j\,\big(\nabla^{\rm h}_{\partial_{x_i}} \nabla^{\rm h}_{\partial_{x_j}} \,-\,\nabla^{\rm h}_{\partial_{x_j}} \nabla^{\rm h}_{\partial_{x_i}}\big)\,\psi \\
&=& \tau_1^i\tau_2^j\Big(\big(\partial_{x_i}\Gamma^\gamma_{j\alpha}-\partial_{x_j}\Gamma^\gamma_{i\alpha}\big)n^\alpha\partial_{n^\gamma}\psi
+ \big[\Gamma^\delta_{i\alpha}n^\alpha\partial_{n^\delta},\Gamma^\gamma_{j\beta}n^\beta\partial_{n^\gamma}\big]\psi\Big).
\end{eqnarray*}

Using the commutator identity
\begin{eqnarray*}
\big[\Gamma^\delta_{i\alpha}n^\alpha\partial_{n^\delta},\Gamma^\gamma_{j\beta}n^\beta\partial_{n^\gamma}\big]\psi
&=& \big(\Gamma^\beta_{i\alpha}\Gamma^\gamma_{j\beta}\,-\,\Gamma^\beta_{j\alpha}\Gamma^\gamma_{i\beta}\big)n^\alpha\partial_{n^\gamma}\psi
\end{eqnarray*}
we obtain that
\begin{eqnarray*}
{\rm R}^{\rm h}(\tau_1,\tau_2)\psi
&=& \tau_1^i\tau_2^j\,\big(\partial_{x_i}\Gamma^\gamma_{j\alpha}-\partial_{x_j}\Gamma^\gamma_{i\alpha}+\Gamma^\beta_{i\alpha}\Gamma^\gamma_{j\beta}\,-\,\Gamma^\beta_{j\alpha}\Gamma^\gamma_{i\beta}\big)n^\alpha\partial_{n^\gamma}\psi\\
&=& \tau_1^i\tau_2^j\, \overline{R}^\gamma_{\;\alpha ij}n^\alpha\partial_{n^\gamma}\psi\\
&=& -\nabla^{\rm v}_{{\rm R}^\perp(\tau_1,\tau_2)\nu}\psi,
\end{eqnarray*}
which was the claim. \qed

\subsection{Elliptic estimates for the Sasaki metric}\label{preliminaries}

In the following, we deduce important properties of differential operators related to the Sasaki metric defined in the introduction (see (\ref{Sasaki})), in particular we will provide a-priori estimates for the associated Laplacian. 

\medskip

In bundle coordinates the Sasaki metric has a simple form. Here we keep the convention that it is summed over repeated indices and write $a^{ij}$ for the inverse of $a_{ij}$.  

\begin{proposition}\label{dualGSasaki}
Let $g^{\rm S}$ be the Sasaki metric on $N\C$ defined in (\ref{Sasaki}). Choose $\Omega\subset\C$ where the normal bundle $N\C$ is trivializable and an orthonormal frame $(\nu_{\alpha})_\alpha$ of $N\C|_{\Omega}$. Define $\Gamma^\gamma_{i\alpha}$ by $\nabla^\perp_{\partial_{x_i}}\nu_\alpha=\Gamma^\gamma_{i\alpha}\nu_\gamma$. In the corresponding bundle coordinates the dual metric tensor $g_{\rm S}\in\T^2_{\,\,0}(TN\C)$ for all $q\in\Omega$ is given by:
\begin{equation*}
g_{\rm S}\;=\;\begin{pmatrix}
			1 & 0\\
			C^T & 1
                 \end{pmatrix}
		 \begin{pmatrix}
			A & 0\\
			0 & B
                 \end{pmatrix}
		 \begin{pmatrix}
			1 & C\\
			0 & 1
                 \end{pmatrix},
\end{equation*} 
where for $i,j=1,...,d$ and $\alpha,\gamma,\delta=1,..,k$
\begin{eqnarray*}
A^{ij}(q,n) &=& g^{ij}(q),\quad B^{\gamma\delta}(q,n) \ \;=\ \;\delta^{\gamma\delta},\\
C_{\,i}^{\gamma}(q,n) &=& -\,n^\alpha\,\Gamma^\gamma_{i\alpha}(q).
\end{eqnarray*}
In particular, $(\det (g_{\rm S})_{ab})(q,n)=(\det g_{ij})(q)$ for $a,b=1,...,d+k$.
\end{proposition}

The proof was carried out by Wittich in \cite{W}. From this expression we deduce the form of the associated Laplacian.

\begin{corollary}\label{SasakiLaplace}
The Laplace-Beltrami operator associated with $g_{\rm S}$ is
\begin{equation*}
\Delta_{\rm S} \;=\; \Delta_{\rm h} \,+\, \Delta_{\rm v}.
\end{equation*}
\end{corollary}

{\sc Proof of Corollary \ref{SasakiLaplace}:}
\newline
We set $\mu:=\det g_{ij}$ and $\mu_{\rm S}:=\det (g_{\rm S})_{ab}$. Since $(\nu_\alpha)_{\alpha=1}^{k}$ is an orthonormal frame, we have that $g_{(q,0)}(\nu\alpha,\nu_\beta)=\delta^{\alpha\beta}$. So (\ref{verder}) and  (\ref{horder}) imply that 
\begin{equation}\label{Laplace}
\Delta_{\rm v}=\partial_{n^\alpha}\delta^{\alpha\beta}\partial_{n^\beta}\ \,\&\ \, \Delta_{\rm h}=\mu^{-1}\big(\partial_{x^i}-\Gamma^\gamma_{i\alpha}\,n^\alpha\,\partial_{n^\gamma}\big)\mu g^{ij}\big(\partial_{x^j}-\Gamma^\gamma_{i\alpha}\,n^\alpha\,\partial_{n^\gamma}\big).
\end{equation}
Now plugging the expression for $g_{\rm S}^{ab}$ and $\det g_{\rm S}^{ab}$ from Proposition \ref{dualGSasaki} into the general formula $\Delta_{\rm S}=\sum_{a,b=1}^{d+k}(\mu_{\rm S})^{-1}\partial_a\, \mu_{\rm S}\, g_{\rm S}^{ab}\partial_b$ yields the claim. \qed

\bigskip

Next we gather some useful properties of $\Delta_{\rm v},\,\Delta_{\rm h}$, and $\nabla^{\rm h}$. We recall that in Definition \ref{dilation} we introduced the unitary operator $D_\veps$ for the isotropic dilation of the fibers with $\veps$.

\begin{lemma}\label{opequations}
Let $f\in C^2(\RRR)$ and $\tau\in\Gamma(T\C)$ be arbitrary. Fix $\lambda\in\RRR$.
It holds
\begin{enumerate}
\item $D_\veps \Delta_{\rm v} D_\veps^* \;=\; \veps^2\Delta_{\rm v},\quad D_\veps \Delta_{\rm h} D_\veps ^*\;=\; \Delta_{\rm h},\quad D_\veps V_\veps D_\veps ^*\;=\; V^\veps$,
\item $[\nabla^{\rm h}_\tau,\Delta_{\rm v}] \;=\; 0,\quad [\Delta_{\rm h},\Delta_{\rm v}] \;=\; 0,\quad [\nabla^{\rm h}_\tau,f(\langle\lambda\nu\rangle)] \;=\; 0\,$, 
\item $[\Delta_{\rm v},f(\langle\lambda\nu\rangle)] \,=\, \lambda f'(\langle\lambda\nu\rangle)\big(\lambda\frac{k\langle\lambda\nu\rangle^2-|\lambda\nu|^2}{\langle\lambda\nu\rangle^3}+\frac{2}{\langle\lambda\nu\rangle}\nabla^{\rm v}_{\lambda\nu}\big)\,+\,\lambda^2f''(\langle\lambda\nu\rangle)\frac{|\lambda\nu|^2}{\langle\lambda\nu\rangle^2}\,$. 
\end{enumerate}
\end{lemma}

In the following, we write $A\prec B$ when $A$ is operator-bounded by $B$ with a constant independent of $\veps$, i.e.\ if $\D(B)\subset\D(A)$ and $\|A\psi\|\lesssim\|B\psi\|+\|\psi\|$ for all $\psi\in\D(B)$. We will have to estimate multiple applications of $\nabla^{\rm v}$ and $\nabla^{\rm h}$ by powers of $H_\veps$, which was defined as $H_\veps:=D_\veps^*M_\rho^*H^\veps M_\rho D_\veps$ with $H^\veps:=-\veps^2\Delta_{N\C}+V^\veps$. Essential for our analysis, especially for the proofs of Lemmas \ref{projector} \& \ref{props}, are the following statements: 

\begin{lemma}\label{opestimates}
Fix $m\in\NNN_0$ and $M\in\{0,1,2\}$.
For all $l\in\ZZZ$, $\lambda\in[0,1]$ and $m_1+m_2\leq 2m$ the following operator estimates hold true on $\H$:
\begin{enumerate}
\item $H_\veps^{m}  \;\prec\; \big(-\veps^2\Delta_{\rm h}-\Delta_{\rm v}+V_\veps\big)^{m} \;\prec\; H_\veps^{m} \,$, 
\item $\big(-\Delta_{\rm v}\big)^{m}\, \big(-\veps^2\Delta_{\rm h}\big)^M  \;\prec\; H_\veps^{M+m}\,$, 
\item $\lambda^{-1}\langle\lambda\nu\rangle^l\,[H_\veps^{M+1},\langle\lambda\nu\rangle^{-l}]
 \;\prec\; H_\veps^{M+1}\,$ with a constant independent of $\lambda$, 
\item $\langle\nu\rangle^{-4m_1-5m_2}
(\nabla^{\rm v})^{m_1}(\veps\nabla^{\rm h})^{m_2} \;\prec\; H_\veps^{m} \,$. 
\end{enumerate}
\end{lemma}

The last three estimates rely on the following estimates in local coordinates. Here we a use covering $(\Omega_j)_j$ of $\C$ and coordinates $(x^i)_{i=1,\dots,d}$ and $(n^\alpha)_{\alpha=1,\dots,k}$ as in Remark \ref{coordinates} in the introduction to Section \ref{wholestory}.

\begin{lemma}\label{keyest}
Let $\alpha,\beta,\gamma$ be multi-indices with $|\alpha|\leq 2l$, $|\alpha|+|\beta|\leq 2m$ and $|\gamma|=2$. Set $\mu:=\det g_{ij}$. For all smooth and compactly supported $\psi$ it holds
\begin{enumerate}
\item $\Big(\sum_j
\int_{\Omega_j}\int_{\RRR^k}
|\,\partial_{n}^{\alpha}\psi|^2\,dn\,\mu\,dx\Big)^{1/2} \ \lesssim\ 
\|(-\Delta_{\rm v})^l\psi\| \,+\, \|\psi\|$,
\item $\Big(\sum_j\int_{\Omega_j}\int_{\RRR^k}
|\,\partial_{n}^{\gamma}\psi|^2\,dn\,\mu\,dx\Big)^{1/2} \ \lesssim\ 
\|(-\veps^2\Delta_{\rm h}-\Delta_{\rm v})\psi\| \,+\, \|\psi\|$,
\item $\Big(\sum_j
\int_{\Omega_j}\int_{\RRR^k}
\langle \nu\rangle^{-8(|\alpha|+|\beta|)}|\partial_{n}^{\alpha}(\veps^{|\beta|}\partial_{x}^{\beta})\psi|^2\,dn\,\mu\,dx\Big)^{1/2}\vspace{0.2cm}\\
 \hspace*{5.5cm}\;\lesssim\,
\big\|\big(-\veps^2\Delta_{\rm h}-\Delta_{\rm v}+V_\veps\big)^{m}\psi\big\| \,+\, \|\psi\|$,
\item $\Big(\sum_j
\int_{\Omega_j}\int_{\RRR^k}
\langle \nu/\veps\rangle^{-8(|\alpha|+|\beta|)}|\veps^{|\alpha|}\partial_{N}^{\alpha}(\veps^{|\beta|}\partial_{x}^{\beta})\psi|^2\,dN\,\mu\,dx\Big)^{1/2}\vspace{0.2cm}\\
 \hspace*{5cm}\;\lesssim\,
\big\|\big(-\veps^2\Delta_{\rm h}-\veps^2\Delta_{\rm v}+V^\veps\big)^{m}\psi\big\| \,+\, \|\psi\|$.
\end{enumerate}
\end{lemma}

We now provide the proofs of these three technical lemmas.

\medskip

{\sc Proof of Lemma \ref{opequations}:}
\newline
We fix a geodesic ball $\Omega\subset\C$. Let $(\nu_\alpha)_{\alpha=1,\dots,k}$ be an orthonormal trivializing frame of $N\Omega$ with associated coordinates $(n^\alpha)_{\alpha=1,\dots,k}$ and $(x^i)_{i=1,\dots,d}$ be any coordinates on $\Omega$. Observing that $D_\veps\psi(x,n)=\veps^{-k/2}\psi(x, n/\veps)$ and $D_\veps^*\psi(x,n)=\veps^{k/2}\psi(x, \veps n)$ we immediately obtain i) due to (\ref{Laplace}).

\smallskip

Since $\nabla^\perp$ is a metric connection, $\Gamma^\gamma_{i\alpha}$ is anti-symmetric in $\alpha$ and $\gamma$ and so (\ref{horder}) implies 
\begin{eqnarray*}
\nabla^{\rm h}_{\partial_{x^i}}\psi(q,\nu) &=& \partial_{x^i}\psi(x,n)\,-\,{\textstyle \frac{1}{2}} \Gamma^\gamma_{i\alpha}\,\big(n^\alpha\partial_{n^\gamma}-n^\gamma\partial_{n^\alpha}\big)\psi(x,n).
\end{eqnarray*}
Using that $\Delta_{\rm v}=\delta^{\alpha\beta}\partial_{n^\alpha}\partial_{n^\beta}$ by (\ref{Laplace}) we obtain that for any $\tau=\tau^i\partial_{x^i}$
\[[\nabla^{\rm h}_\tau,\Delta_{\rm v}] \;=\; \tau^i\Gamma^{\gamma\alpha}_{i}\,\big(\partial_{n^\alpha}\partial_{n^\gamma}-\partial_{n^\gamma}\partial_{n^\alpha}\big) \;=\; 0.\]
We recall that $\langle\nu\rangle=\sqrt{1+g_{(q,0)}(\nu,\nu)}$. Since $(\nu_\alpha)_{\alpha=1}^{k}$ is an orthonormal frame, we have that $g_{(q,0)}(\nu\alpha,\nu_\beta)=\delta^{\alpha\beta}$. This entails that $\langle\nu\rangle=\sqrt{1+\delta_{\alpha\beta}n^\alpha n^\beta}$. With this the remaining statements follow by direct computation. \qed

\bigskip

{\sc Proof of Lemma \ref{opestimates}:}
\newline
We recall from Definition \ref{dilation} that $V_\veps=V_{\rm c}+D_\veps^*WD_\veps$ and that we assumed that $V_{\rm c}$ and $W$ are in $C^\infty_{\rm b}\big(\C,C^\infty_{\rm b}(N_q\C)\big)$. These facts together imply that $V_\veps\in C^\infty_{\rm b}\big(\C,C^\infty_{\rm b}(N_q\C)\big)$.

\smallskip

Since $D_\veps$ and $M_\rho$ are unitary, Lemma \ref{opequations} i) yields that Lemma \ref{opestimates} i) is equivalent to 
\begin{equation}\label{macrobound}
(H^\veps)^{m}   \;\prec\; M_\rho \big(-\veps^2\Delta_{\rm h}-\veps^2\Delta_{\rm v}+V^\veps\big)^{m}  M_\rho^* \;\prec\; (H^\veps)^{m} 
\end{equation} 
for all $m\in\NNN$. By choice of $\overline{g}$ it coincides with the Sasaki metric $g^{\rm S}$ outside of $\B_{\delta}$ and, hence, so do $\Delta_{N\C}$ and $\Delta_{\rm S}$. In addition, this means $\rho\equiv 1$ outside of $\B_{\delta}$ and so $M_\rho$ is multiplication by $1$ there. Then Corollary~\ref{SasakiLaplace} implies $H^\veps = M_\rho \big(-\veps^2\Delta_{\rm h}-\veps^2\Delta_{\rm v}+V^\veps) M_\rho^*$ on $N\C\setminus\B_{\delta}$. Hence, by introducing suitable cutoff functions it suffices to prove (\ref{macrobound}) for functions with support in $\B_{2\delta}\cap N\Omega_j$. The set $\B_{2\delta}\cap N\Omega_j$ is easily seen to be bounded with respect to both $\overline{g}$ and $g^{\rm S}$ and thus relatively compact because $N\C$ is complete with both $\overline{g}$ and $g^{\rm S}$ as explained in the sequel to the definition of $g^{\rm S}$ in (\ref{Sasaki}). Furthermore, on $\B_{2\delta}\cap N\Omega_j$ both $(H^\veps)^m$ and $M_\rho \big(-\veps^2\Delta_{\rm h}-\veps^2\Delta_{\rm v}+V^\veps\big)^m M_\rho^*$ are elliptic operators with bounded coefficients of order $2m$. Therefore (\ref{macrobound}) follows from the usual elliptic estimates. These are uniform in $j$ because $\B_{2\delta}$ is a subset of bounded geometry of $N\C$ with respect to both $\overline{g}$ and $g^{\rm S}$, which was also explained in the sequel to (\ref{Sasaki}).

\smallskip

In the following, we prove the estimates only on smooth and compactly supported functions, where  we may apply Lemma \ref{keyest}. Then it is just a matter of standard approximation arguments to extend them to the maximal domains of the operators on the right hand side of each estimate. In this context one should note that the mamixal domains $\D(H_\veps^m)$ and $\D((-\veps^2\Delta_{\rm h}-\Delta_{\rm v}+V_\veps)^m)$ coincide for all $m\in\NNN$ by i).

\smallskip

We recall that $V_\veps\in C^\infty_{\rm b}\big(\C,C^\infty_{\rm b}(N_q\C)\big)$ and turn to ii). By i) we may replace $H_\veps$ by $-\veps^2\Delta_{\rm h}-\Delta_{\rm v}+V_\veps$. We first prove the statement for $M=0$ inductively in $m$. In view of (\ref{Laplace}), Lemma \ref{keyest} ii) implies that $-\Delta_{\rm v}\prec-\veps^2\Delta_{\rm h}-\Delta_{\rm v}$ and thus also $-\veps^2\Delta_{\rm h}\prec-\veps^2\Delta_{\rm h}-\Delta_{\rm v}$. So due to the boundedness of $V_\veps$ the triangle inequality yields the statement for $m=0$ as well as
\begin{equation}\label{basicest}
-\veps^2\Delta_{\rm h}\;\prec\;-\,\veps^2\Delta_{\rm h}-\Delta_{\rm v}+V_\veps.
\end{equation}
In the following, we will write $A\prec B\dotplus C$, if $\|A\psi\|\lesssim\|B\psi\|+\|C\psi\|+\|\psi\|$. We note that with this notation $A\prec B$ implies $AC\prec BC\dotplus C$. 

Now we assume that the statement is true for some $m\in\NNN_0$. Since $V^\veps\in C^\infty_{\rm b}$ and $N\C$ with the Sasaki metric $g^{\rm S}$ is complete, the operator $-\veps^2\Delta_{S}+V_\veps$ is self-adjoint on $\H$ and so is $-\veps^2\Delta_{\rm h}-\Delta_{\rm v}+V_\veps$, as it is unitary equivalent to $-\veps^2\Delta_{S}+V^\veps$ via $D_\veps$. Therefore by the spectral calculus lower powers of $-\veps^2\Delta_{\rm h}-\Delta_{\rm v}+V_\veps$ are operator-bounded by higher powers. In addition, $\Delta_{\rm v}$ and $\Delta_{\rm h}$ commute by Lemma \ref{opequations}. Then we obtain the statement for $m+1$ via
\begin{eqnarray*}
(-\Delta_{\rm v})^{m+1} &\prec& (-\veps^2\Delta_{\rm h}-\Delta_{\rm v}+V_\veps)\,(-\Delta_{\rm v})^{m}\,\dotplus\,(-\Delta_{\rm v})^{m}\\
&=& (-\Delta_{\rm v})^{m}(-\veps^2\Delta_{\rm h}-\Delta_{\rm v}+V_\veps)\,+\,\big[V_\veps,(-\Delta_{\rm v})^{m}\big]\,\dotplus\,(-\Delta_{\rm v})^{m}\\
&\prec& (-\veps^2\Delta_{\rm h}-\Delta_{\rm v}+V_\veps)^{m+1}\,\dotplus\,(-\Delta_{\rm v})^{m}\\
&\prec& (-\veps^2\Delta_{\rm h}-\Delta_{\rm v}+V_\veps)^{m+1}.
\end{eqnarray*}
Here we used $V_\veps\in C^\infty_{\rm b}(\C,C^\infty_{\rm b}(N_q\C))$, $\Delta_{\rm v}=\delta^{\alpha\beta}\partial_{n^\alpha}\partial_{n^\beta}$ locally, and i) of Lemma \ref{keyest} to bound $\big[V_\veps,(-\Delta_{\rm v})^{m}\big]$ by $(-\Delta_{\rm v})^{m}$.
Using $[\Delta_{\rm v},\Delta_{\rm h}]=0$ and (\ref{basicest}) we have
\begin{eqnarray*}
(-\Delta_{\rm v})^{m}\,(-\veps^2\Delta_{\rm h}) &=& (-\veps^2\Delta_{\rm h})\,(-\Delta_{\rm v})^{m}\\
&\prec& (-\veps^2\Delta_{\rm h}-\Delta_{\rm v}+V)\,(-\Delta_{\rm v})^{m}\,\dotplus\,(-\Delta_{\rm v})^{m}.
 \end{eqnarray*}
Continuing as before we obtain the claim for $M=1$. Furthermore,
\begin{eqnarray*}
(-\Delta_{\rm v})^{m}(-\veps^2\Delta_{\rm h})^2 &=& (-\veps^2\Delta_{\rm h})\,(-\Delta_{\rm v})^{m}\,(-\veps^2\Delta_{\rm h})\\
&\prec& (-\veps^2\Delta_{\rm h}-\Delta_{\rm v}+V_\veps)(-\Delta_{\rm v})^{m}(-\veps^2\Delta_{\rm h})\\
&& \qquad\,\dotplus\,(-\Delta_{\rm v})^{m}(-\veps^2\Delta_{\rm h})\\
&\prec& (-\Delta_{\rm v})^{m}(-\veps^2\Delta_{\rm h})(-\veps^2\Delta_{\rm h}-\Delta_{\rm v}+V_\veps)\\
&& \ \,+\,\big[V_\veps,(-\Delta_{\rm v})^{m}(-\veps^2\Delta_{\rm h})\big]\,\dotplus\,(-\veps^2\Delta_{\rm h}-\Delta_{\rm v}+V_\veps)^{m+1}\\
&\prec& (-\veps^2\Delta_{\rm h}-\Delta_{\rm v}+V_\veps)^{m+2}\,\dotplus\,\big[V_\veps,(-\Delta_{\rm v})^{m}(-\veps^2\Delta_{\rm h})\big],
\end{eqnarray*}
where in the last step we used the statement for $M=1$ and again that lower powers of $(-\veps^2\Delta_{\rm h}-\Delta_{\rm v}+V_\veps)$ are operator-bounded by higher powers.
To handle the remaining term on the right hand side we choose a partition of unity~$(\xi_j)_j$ corresponding to the covering $(\Omega_j)_j$ as in Remark~\ref{coordinates} and orthonormal sections $(\tau^j_i)_{i=1,\dots,d}$ of $T\Omega_j$ for all $j$. Then it holds
\begin{equation}\label{Laplace2}
\Delta_{\rm h}\;=\;\sum_{j,i} \xi_j\nabla^{\rm h}_{\tau^j_i,\tau^j_i}\;=\;\sum_{j,i} \xi_j(\nabla^{\rm h}_{\tau^j_i}\nabla^{\rm h}_{\tau^j_i}-\nabla^{\rm h}_{\nabla_{\tau^j_i}{\tau^j_i}}).
\end{equation}
The finite multiplicity of our coverings implies
\begin{eqnarray*}
\sum_{i,j}\int_{\Omega_j\times\RRR^k}\xi_j^2\,\veps\nabla^{\rm h}_{\tau_i^j}\psi^*\,\veps\nabla^{\rm h}_{\tau_i^j}\psi\,d\mu\otimes d\nu 
&\lesssim&
\int_{N\C}g(\veps\nabla^{\rm h}\psi^*,\veps\nabla^{\rm h}\psi)d\mu\otimes d\nu\\ 
&=& \langle\psi|-\veps^2\Delta_{\rm h}\psi\rangle\\ 
&\leq& \|-\veps^2\Delta_{\rm h}\psi\|+\|\psi\|.
\end{eqnarray*}
Therefore
\begin{eqnarray*}
\big[V_\veps,(-\Delta_{\rm v})^{m}(-\veps^2\Delta_{\rm h})\big]\hspace{-2pt}
&=& \big[V_\veps,(-\Delta_{\rm v})^{m}\big]\,(-\veps^2\Delta_{\rm h}) \,+\, (-\Delta_{\rm v})^{m}\,\big[V_\veps,(-\veps^2\Delta_{\rm h})\big]\\
&\prec& (-\Delta_{\rm v})^{m}(-\veps^2\Delta_{\rm h})\dotplus\sum_{j,i}\xi_j(-\Delta_{\rm v})^{m}\veps\nabla^{\rm h}_{\tau_i^j}\dotplus(-\Delta_{\rm v})^{m}\\
&=& (-\veps^2\Delta_{\rm h})(-\Delta_{\rm v})^{m}\dotplus\sum_{j,i}\xi_j\veps\nabla^{\rm h}_{\tau_i^j}(-\Delta_{\rm v})^{m} \dotplus(-\Delta_{\rm v})^{m}\\
&\prec& (-\veps^2\Delta_{\rm h})(-\Delta_{\rm v})^{m}\,\dotplus\,(-\Delta_{\rm v})^{m}\\
&\prec& (-\veps^2\Delta_{\rm h}-\Delta_{\rm v}+V_\veps)^{m+2}.
\end{eqnarray*}

\medskip

We prove iii) only for $M=2$ which is the hardest case. We notice that
\begin{eqnarray}\label{commsplit}
\langle\lambda\nu\rangle^m\,[H_\veps^{3},\langle\lambda\nu\rangle^{-m}] &=&
\langle\lambda\nu\rangle^m\,[H_\veps,\langle\lambda\nu\rangle^{-m}]\,H_\veps^2
+\langle\lambda\nu\rangle^m\,H_\veps\,[H_\veps,\langle\lambda\nu\rangle^{-m}]\,H_\veps \nonumber\\
&& \qquad\qquad\qquad\qquad
\,+\, \langle\lambda\nu\rangle^m\,H_\veps^2\,[H_\veps,\langle\lambda\nu\rangle^{-m}].\nonumber
\end{eqnarray}
We also only treat the hardest of these summands which is the last one. The arguments below also work for the other summands and for $M\in\{0,1\}$.
Inside of $\B_{2\delta}$ the estimate iii) can be reduced to standard elliptic estimates as in i). Therefore we may replace $H_\veps$ by $-\veps^2\Delta_{\rm h}-\Delta_{\rm v}+V_\veps$ because both operators coincide outside $\B_\delta$. 
In view of ii) of Lemma \ref{opequations}, we have 
\begin{eqnarray*}
\lefteqn{\lambda^{-1}\,\langle\lambda\nu\rangle^m\,(-\veps^2\Delta_{\rm h}-\Delta_{\rm v}+V_\veps)^2\,[-\veps^2\Delta_{\rm h}-\Delta_{\rm v}+V_\veps,\langle\lambda\nu\rangle^{-m}]}\\
&& =\ \lambda^{-1}\,\langle\lambda\nu\rangle^m\,(-\veps^2\Delta_{\rm h}-\Delta_{\rm v}+V_\veps)^2\,[-\Delta_{\rm v},\langle\lambda\nu\rangle^{-m}]\\
&& =\ \Big(\langle\lambda\nu\rangle^m\,(-\Delta_{\rm v}+V_\veps)^2\,[-\Delta_{\rm v},\langle\lambda\nu\rangle^{-m}]  \,+\, \langle\lambda\nu\rangle^m[-\Delta_{\rm v},\langle\lambda\nu\rangle^{-m}]\,(-\veps^2\Delta_{\rm h})^2\\
&& \qquad\qquad \,+\, 2\,\langle\lambda\nu\rangle^m\,(-\Delta_{\rm v}+V_\veps)\,[-\Delta_{\rm v},\langle\lambda\nu\rangle^{-m}]\,(-\veps^2\Delta_{\rm h}) \\
&& \qquad\qquad \,+\, \langle\lambda\nu\rangle^m\,[-\veps^2\Delta_{\rm h},V_\veps]\,[-\Delta_{\rm v},\langle\lambda\nu\rangle^{-m}]\Big)\lambda^{-1}
\end{eqnarray*}
Because of $\Delta_{\rm v}= \delta^{\alpha\beta}\partial_{n_\alpha}\partial_{n_\beta}$ the operator $\langle\lambda\nu\rangle^{m}(-\Delta_{\rm v}+V_\veps)^l\,[-\Delta_{\rm v},\langle\lambda\nu\rangle^{-m}]\lambda^{-1}$ contains only normal partial derivatives. It has coefficients bounded independently of $\lambda$ for any $l$, as the commutator $[-\Delta_{\rm v},\langle\lambda\nu\rangle^{-m}]$ provides a $\lambda$ due to Lemma~\ref{opequations}~iii). So by i) of Lemma \ref{keyest} it is bounded by $(-\Delta_{\rm v})^{l+1}$. Then ii) of Lemma \ref{opestimates} immediately allows to bound the first three terms by $(-\veps^2\Delta_{\rm h}-\Delta_{\rm v}+V_\veps)^3$.
The last term can be treated as follows. In the proof of ii) we saw that $[-\veps^2\Delta_{\rm h},V_\veps]\prec-\veps^2\Delta_{\rm h}$. Therefore
\begin{eqnarray*}
\lefteqn{\langle\lambda\nu\rangle^m\,[-\veps^2\Delta_{\rm h},V_\veps]\,[-\Delta_{\rm v},\langle\lambda\nu\rangle^{-m}]\lambda^{-1}}\\
&& \ \,=\ \, [-\veps^2\Delta_{\rm h},V_\veps]\,\langle\lambda\nu\rangle^m\,[-\Delta_{\rm v},\langle\lambda\nu\rangle^{-m}]\,\lambda^{-1} \\
&& \ \,\prec\ \, -\veps^2\Delta_{\rm h}\,\langle\lambda\nu\rangle^m\,[-\Delta_{\rm v},\langle\lambda\nu\rangle^{-m}]\,\lambda^{-1} \,\dotplus\,\langle\lambda\nu\rangle^m\,[-\Delta_{\rm v},\langle\lambda\nu\rangle^{-m}]\,\lambda^{-1}\\
&& \ \,=\ \,\langle\lambda\nu\rangle^m\,[-\Delta_{\rm v},\langle\lambda\nu\rangle^{-m}]\,\lambda^{-1}\,(-\veps^2\Delta_{\rm h})\,\dotplus\,\langle\lambda\nu\rangle^m\,[-\Delta_{\rm v},\langle\lambda\nu\rangle^{-m}]\,\lambda^{-1}\\
&& \ \,\prec\ \, (-\Delta_{\rm v})\,(-\veps^2\Delta_{\rm h})\,\dotplus\,(-\veps^2\Delta_{\rm h})\,\dotplus\,(-\Delta_{\rm v})
\end{eqnarray*}
which is bounded independently of $\lambda$ by $(-\veps^2\Delta_{\rm h}-\Delta_{\rm v}+V_\veps)^2$ again due to~ii). Here again $[-\Delta_{\rm v},\langle\lambda\nu\rangle^{-m}]$ has provided the lacking $\lambda$.

\medskip

In view of (\ref{verder}) and (\ref{horder}), the estimate iv) follows directly from i) of this lemma and iii) of Lemma~\ref{keyest}. A polynomial weight is nescessary because here the unbounded geometry of $(N\C,\overline{g})$ really comes into play. In i) we could avoid this using that the operators differ only on a set of bounded geometry, while in ii) and iii) the number of horizontal derivatives was small! \qed

\bigskip

{\sc Proof of Lemma \ref{keyest}:}
\newline
The first estimate is just an elliptic estimate on each fibre and thus a consequence of the usual elliptic estimates on $\RRR^k$. To see this we note that $\Delta_{\rm v}= \delta^{\alpha\beta}\partial_{n_\alpha}\partial_{n_\beta}$ is the Laplace operator on the fibers by (\ref{Laplace}) and that the measure $d\mu\otimes d\nu=dn\,\mu(x)dx$ is independent of $n$.

\smallskip

To deduce the second estimate we aim to show that
\begin{eqnarray}\label{aim}
\lefteqn{\sum_{|\gamma|=2}\int_{\Omega_j}\int_{\RRR^k}
|\,\partial_{n}^{\gamma}\Psi|^2\,dn\,\mu(x)dx}\\
&\lesssim& \sum_{|\gamma|=2}\int_{\Omega_j}\int_{\RRR^k}
|\,\partial_{n}^{\gamma}\Psi|\, \big(|(-\veps^2\Delta_{\rm h}-\Delta_{\rm v})\Psi| +|\veps\nabla^{\rm h}\psi| + |\Psi|\big)\,dn\,\mu(x)dx.\nonumber
\end{eqnarray}
with a constant independent of $j$. Then the claim follows from the Cauchy-Schwarz inequality and $\||\veps\nabla^{\rm h}\psi|\|= \langle\psi|-\veps^2\Delta_{\rm h}\psi\rangle^{\frac{1}{2}}\leq\langle\psi|(-\veps^2\Delta_{\rm h}-\Delta_{\rm v})\psi\rangle^{\frac{1}{2}}$ which is smaller than $\|(-\veps^2\Delta_{\rm h}-\Delta_{\rm v})\Psi\| \,+\,  \|\Psi\|$. We note that here and in the sequel there is no problem to sum up over $j$ because the covering $(\Omega_j)_j$ has finite multiplicity!

On the one hand, there are $\alpha,\beta\in\{1,\dots,k\}$ such that
\begin{eqnarray*}
\int_{\Omega_j}\int_{\RRR^k}
|\,\partial_{n}^{\gamma}\Psi|^2\,dn\,\mu(x)dx &=& \int_{\Omega_j}\int_{\RRR^k}\partial_{n^\alpha}\partial_{n^\beta}\psi^*\, \partial_{n^\alpha}\partial_{n^\beta}\psi\,dn\,\mu(x)dx\\
&=& -\int_{\Omega_j}\int_{\RRR^k}\partial_{n^\beta}\psi^*\, \partial_{n^\alpha}\partial_{n^\alpha}\partial_{n^\beta}\psi\,dn\,\mu(x)dx\\
&=& \int_{\Omega_j}\int_{\RRR^k}\partial_{n^\beta}\partial_{n^\beta}\psi^*\, \partial_{n^\alpha}\partial_{n^\alpha}\psi\;dn\,\mu(x)dx\\
&=& \int_{\Omega_j}\int_{\RRR^k}\partial_{n^\beta}\partial_{n^\beta}\psi^*\, \Delta_{\rm v}\psi\;dn\,\mu(x)dx.
\end{eqnarray*}

On the other hand, 
\begin{eqnarray*}
0 &\leq& \int_{\Omega_j}\int_{\RRR^k}g\big(\veps\nabla^{\rm h}\partial_{n^\beta}\psi^*, \veps\nabla^{\rm h}\partial_{n^\beta}\psi\big)\,dn\,\mu(x)dx\\
&\stackrel{(\ref{horder})}{=}& \int_{\Omega_j}\int_{\RRR^k}g^{il}\veps\big(\partial_{x^i}+\Gamma^{\alpha}_{i\zeta}n^\zeta\partial_{n^\alpha}\big)\partial_{n^\beta}\psi^*\, \veps\big(\partial_{x^l}+\Gamma^{\eta}_{l\delta}n^\delta\partial_{n^\eta}\big)\partial_{n^\beta}\psi\,dn\,\mu(x)dx\\
&=& \int_{\Omega_j}\int_{\RRR^k}-g^{il}\veps\big(\partial_{x^i}+\Gamma^{\alpha}_{i\zeta}n^\zeta\partial_{n^\alpha}\big)\partial_{n^\beta}\partial_{n^\beta}\psi^*\, \veps\big(\partial_{x^l}+\Gamma^{\eta}_{l\delta}n^\delta\partial_{n^\eta}\big)\psi\\
&& \qquad\qquad\qquad\,-\;\veps\, g^{il}\veps\big(\partial_{x^i}+\Gamma^{\alpha}_{i\zeta}n^\zeta\partial_{n^\alpha}\big)\partial_{n^\beta}\psi^*\, \Gamma^{\eta}_{l\beta}\partial_{n^\eta}\psi\\
&& \qquad\qquad\qquad\,-\;\veps\, g^{il}\Gamma^{\alpha}_{i\beta}\partial_{n^\alpha}\partial_{n^\beta}\psi^*\, \veps\big(\partial_{x^l}+\Gamma^{\eta}_{l\delta}n^\delta\partial_{n^\eta}\big)\psi\ dn\,\mu(x)dx\\
&=& \int_{\Omega_j}\int_{\RRR^k}\partial_{n^\beta}\partial_{n^\beta}\psi^*\, \veps^2\Delta_{\rm h}\psi\,+\,\veps^2 g^{ij}\Gamma^{\alpha}_{i\beta}\partial_{n^\alpha}\psi^*\, \Gamma^{\eta}_{l\beta}\partial_{n^\eta}\psi\\
&& \qquad\qquad\,-\;2\veps\,{\rm Im}\Big(g^{il}\Gamma^{\alpha}_{i\beta}\partial_{n^\alpha}\partial_{n^\beta}\psi^*\, \veps\big(\partial_{x^l}+\Gamma^{\eta}_{l\delta}n^\delta\partial_{n^\eta}\big)\psi\Big)\;dn\,\mu(x)dx
\end{eqnarray*}
with ${\rm Im}(a)$ the imaginary part of $a$.
When we add the last two calculations and sum up over all multi-indices $\gamma$ with $|\gamma|=2$, we obtain the desired $(-\veps^2\Delta_{\rm h}-\Delta_{\rm v})$-term. However, we have to take care of the two error terms in the latter estimate:
\begin{eqnarray*}
\lefteqn{\int_{\Omega_j}\int_{\RRR^k}g^{il}\Gamma^{\alpha}_{i\beta}\partial_{n^\alpha}\psi^*\, \Gamma^{\eta}_{l\beta}\partial_{n^\eta}\psi\;dn\,\mu(x)dx}\\
&& \qquad=\quad \int_{\Omega_j}\int_{\RRR^k}-g^{il}\Gamma^{\alpha}_{i\beta}\partial_{n^\eta}\partial_{n^\alpha}\psi^*\, \Gamma^{\eta}_{l\beta}\psi\;dn\,\mu(x)dx \\
&& \qquad\leq\quad \sup |g^{il}\Gamma^{\alpha}_{i\beta}\Gamma^{\eta}_{l\beta}| \sum_{|\gamma|=2}\int_{\Omega_j}\int_{\RRR^k}|\partial_{n^\eta}\partial_{n^\alpha}\psi^*|\, |\psi|\;dn\,\mu(x)dx 
\end{eqnarray*}
and
\begin{eqnarray*}
\lefteqn{\int_{\Omega_j}\int_{\RRR^k}2\,{\rm Im}\Big(g^{il}\Gamma^{\alpha}_{i\beta}\partial_{n^\alpha}\partial_{n^\beta}\psi^*\, \veps\big(\partial_{x^l}+\Gamma^{\eta}_{l\delta}n^\delta\partial_{n^\eta}\big)\psi\Big)\;dn\,\mu(x)dx} \\
&& \quad\leq\quad 2\sup |(g^{il})^{\frac{1}{2}}\Gamma^{\alpha}_{i\beta}| \sum_{|\gamma|=2}\int_{\Omega_j}\int_{\RRR^k}|\partial_{n^\alpha}\partial_{n^\beta}\psi|\, |\veps\nabla^{\rm h}\psi|\;dn\,\mu(x)dx. 
\end{eqnarray*}
This yields (\ref{aim}) because $g^{il}$ and $\Gamma^{\alpha}_{i\beta}$ can be bounded independently of $j$ in our coordinates due to the bounded geometry and the smooth embedding of~$\C$ assumed in (\ref{bndcurv1}) and (\ref{bndcurv2}) as explained in Remark \ref{coordinates}.

\smallskip

To see that iii) is just a reformulation of iv), we replace $n$ by $N=\veps n$ in iii), put in $\psi=D_\veps^*\tilde\psi$, and use that $(-\veps^2\Delta_{\rm h}-\Delta_{\rm v}+V_\veps\big)D_\veps^*= D_\veps^*(-\veps^2\Delta_{\rm h}-\veps^2\Delta_{\rm v}+V^\veps\big)$ by Lemma \ref{opequations}.

\smallskip

So we immediately turn to iv). We notice that the powers of $\veps$ on both sides match because all derivatives carry an $\veps$. Therefore we may drop all the $\veps$'s in our calculations to deduce the last estimate. Since we have stated the estimate with a non-optimal power of $\langle\nu\rangle$, there is also no need to distinguish between normal and tangential derivatives anymore. So the multi-index $\alpha$ will be supposed to allow for both normal and tangential derivatives. We recall that $\Delta_{\rm S}=\Delta_{\rm h}+\Delta_{\rm v}$. We will prove by induction that for all $m\in\NNN_0$
\begin{eqnarray}\label{induction}
\lefteqn{\Big(\sum_{|\alpha|\leq m+2}
\int_{\Omega_j}\int_{\RRR^k}
\langle\nu\rangle^{-8|\alpha|}|\partial^{\alpha}\psi|^2\,dN\,\mu\,dx\Big)^{\frac{1}{2}} }\nonumber\\ 
&& \ \lesssim\ 
\Big(\sum_{|\beta|\leq m}
\int_{\Omega_j}\int_{\RRR^k}
\langle\nu\rangle^{-8|\beta|}|\partial^{\beta}(-\Delta_{\rm S}+V)\psi|^2\,dN\,\mu\,dx\Big)^{\frac{1}{2}}
 \,+\, \|\psi\|
\end{eqnarray}
with a constant independent of $j$. Applying this estimate iteratively we obtain our claim because as explained before $-\Delta_{\rm S}+V$ is self-adjoint and thus $(-\Delta_{\rm S}+V)^l$ is operator-bounded by $(-\Delta_{\rm S}+V)^m$ for $l\leq m$ due to the spectral calculus.

Before we begin with the induction we notice that, in view of Proposition \ref{dualGSasaki}, $g_{\rm S}^{ab}$ is positive definite with a constant that is bounded from below by $\langle\nu\rangle^{-2}$ times a constant depending only on the geometry of $\C$. More precisely, the constant depends on $\sup\Gamma^\beta_{i\gamma}$ and the inverse constant of positive definiteness of $g^{il}$, which are both uniformly bounded in our coordinates again due to (\ref{bndcurv1}) and (\ref{bndcurv2}).

We start the induction with the case $m=0$. For $|\alpha|=0$ there is nothing to prove. Since $\mu=\det g^{\rm S}_{ab}$ by Proposition \ref{dualGSasaki}, it holds $\Delta_{\rm S}=\mu^{-1}\partial_{a}\,\mu\,g_{\rm S}^{ab}\,\partial_{b}$. So for $|\alpha|=1$ we have
\begin{eqnarray}\label{indstart}
\sum_{|\alpha|=1}
\int_{\Omega_j}\int_{\RRR^k}
\langle\nu\rangle^{-8}|\partial^{\alpha}\psi|^2\,dN\mu\,dx &\lesssim&
\int_{\Omega_j}\int_{\RRR^k}
g_{\rm S}^{ab}\partial_{a}\psi^*\,\partial_{b}\psi\,dN\,\mu\,dx\nonumber\\
&=&
-\int_{\Omega_j}\int_{\RRR^k}\psi^*
\,\mu^{-1}\partial_{a}\,\mu\,g_{\rm S}^{ab}\,\partial_{b}\psi\,dN\mu\,dx\nonumber\\
&=&
\int_{\Omega_j}\int_{\RRR^k}
\psi^*\big((-\Delta_{\rm S}+V-V)\psi\big)dN\mu\,dx\nonumber\\
&\leq& \|\psi\| \big(\|(-\Delta_{\rm S}+V)\psi\| + \sup |V| \,\|\psi\|\big)\nonumber\\
&\lesssim& \|(-\Delta_{\rm S}+V)\psi\|^2 \,+\,\|\psi\|^2\nonumber\\
&\leq& \big(\|(-\Delta_{\rm S}+V)\psi\| \,+\,\|\psi\|\big)^2.
\end{eqnarray}
Taking the square root yields the desired estimate in this case. 
For $|\alpha|=2$ we have
\begin{eqnarray*}
\lefteqn{\sum_{|\alpha|=2}
\int_{\Omega_j}\int_{\RRR^k}
\langle\nu\rangle^{-16}|\partial^{\alpha}\psi|^2\,dN\mu\,dx }\\
&\lesssim& 
\sum_{c}\int_{\Omega_j}\int_{\RRR^k}
\langle\nu\rangle^{-14}
g_{\rm S}^{ab}\partial_{a}\partial_{c}\psi^*\,\partial_{b}\partial_{c}\psi\,dN\mu\,dx\\
&=&
\sum_{c}\int_{\Omega_j}\int_{\RRR^k}
-\langle\nu\rangle^{-14}
g_{\rm S}^{ab}\partial_{a}\partial_{c}\partial_{c}\psi^*\,\partial_{b}\psi\\\
&& \qquad\qquad\qquad\qquad\qquad\,-\,\mu^{-1}\big(\partial_{c}\,\mu\langle\nu\rangle^{-14}
g_{\rm S}^{ab}\big)\partial_{a}\partial_{c}\psi^*\,\partial_{b}\psi\,dN\mu\,dx\\
&=&
\sum_{c}\int_{\Omega_j}\int_{\RRR^k} \langle\nu\rangle^{-14}
\partial_{c}\partial_{c}\psi^*\,(\Delta_{\rm S}-V+V)\psi
\\
&& \qquad
-\Big(\mu^{-1}\big(\partial_{c}\,\mu\langle\nu\rangle^{-14}
g_{\rm S}^{ab}\big)\partial_{a}\partial_{c}\psi^* -(\partial_{a}\langle\nu\rangle^{-14})
\,g_{\rm S}^{ab}\partial_{c}\partial_{c}\psi^*\Big)\,\partial_{b}\psi\,dN\mu\,dx\\
&\lesssim& 
\sum_{|\alpha|=2}\int_{\Omega_j}\int_{\RRR^k} \langle\nu\rangle^{-8}
|\partial^\alpha\psi|\Big(|(-\Delta_{\rm S}+V)\psi| +|V| |\psi| +\langle\nu\rangle^{-4}
|\partial_{b}\psi|\Big)dN\mu\,dx
\end{eqnarray*}
which yields (\ref{induction}) via (\ref{indstart}) when we apply the Cauchy-Schwarz inquality and devide both sides by the square root of the left-hand side. Here we used that both $\mu^{-1}\big(\partial_{c}\,\mu\langle\nu\rangle^{-14}
g_{\rm S}^{ab}\big)$ and $(\partial_{a}\langle\nu\rangle^{-14})
\,g_{\rm S}^{ab}$ are bounded by $\langle\nu\rangle^{-12}$. This is due to the facts that the derivatives of $\mu$ are globally bounded due to the bounded geometry of $\C$, that $g_{\rm S}^{ab}$ and its derivatives are bounded by $\langle\nu\rangle^{2}$ due to Proposition \ref{dualGSasaki}, and that any derivative of $\langle\nu\rangle^l=\sqrt{1+\delta_{\alpha\beta}n^\alpha n^\beta}^l$ is bounded by $\langle\nu\rangle^l$. We will use these facts also in the following calculation.

\smallskip

We assume now that (\ref{induction}) is true for some fixed $m\in\NNN_0$. Then it suffices to consider multi-indices $\alpha$ with $|\alpha|=m+3$ to show the statement for $m+1$. We have
\begin{eqnarray*}
\lefteqn{\sum_{|\alpha|=m+3}
\int_{\Omega_j}\int_{\RRR^k}
\langle\nu\rangle^{-8|\alpha|}|\partial^{\alpha}\psi|^2\,dN\mu\,dx }\\
&& \ \;\lesssim\ \;
\sum_{|\tilde\alpha|=m+2}\int_{\Omega_j}\int_{\RRR^k}
\langle\nu\rangle^{-8|\tilde\alpha|-6}
g_{\rm S}^{ab}\partial_{a}\partial^{\tilde\alpha}\psi^*\,\partial_{b}\partial^{\tilde\alpha}\psi\,dN\mu\,dx\\
&& \ \;=\ \;
\sum_{|\tilde\alpha|=m+2}\int_{\Omega_j}\int_{\RRR^k} \langle\nu\rangle^{-8|\tilde\alpha|-6}
\partial^{\tilde\alpha}\psi^*\,(-\Delta_{\rm S})\partial^{\tilde\alpha}\psi\\
&& \qquad\qquad\qquad\qquad\qquad\,-\,\partial^{\tilde\alpha}\psi^*\,(\partial_{a}\langle\nu\rangle^{-8|\tilde\alpha|-6})
\,g_{\rm S}^{ab}\partial_{b}\partial^{\tilde\alpha}\psi\,dN\mu\,dx\\
&& \ \;=\ \;
\sum_{|\tilde\alpha|=m+2}\int_{\Omega_j}\int_{\RRR^k} \langle\nu\rangle^{-8|\tilde\alpha|-6}
\partial^{\tilde\alpha}\psi^*\,\partial^{\tilde\alpha}(-\Delta_{\rm S})\psi\\
&& \qquad\qquad\,-\,\partial^{\tilde\alpha}\psi^*\,\Big((\partial_{a}\langle\nu\rangle^{-8|\tilde\alpha|-6})
\,g_{\rm S}^{ab}\partial_{b}\partial^{\tilde\alpha}\psi+\langle\nu\rangle^{-8|\alpha|-6}[\Delta_{\rm S},\partial^{\tilde\alpha}]\psi\Big)\,dN\mu\,dx\\
&& \ \;\lesssim\ \;
\sum_{|\alpha|=m+3}\sum_{|\beta|=m+1}\int_{\Omega_j}\int_{\RRR^k} \langle\nu\rangle^{-4|\alpha|}
|\partial^{\alpha}\psi|\,\langle\nu\rangle^{-4|\beta|}|\partial^{\beta}(-\Delta_{\rm S})\psi| \,dN\mu\,dx\\
&& \qquad\qquad\,+\,\sum_{|\alpha|=m+3}\sum_{|\tilde\alpha|=m+2}\int_{\Omega_j}\int_{\RRR^k} \langle\nu\rangle^{-4|\tilde\alpha|}
|\partial^{\tilde\alpha}\psi|\,\langle\nu\rangle^{-4|\alpha|}|\partial^{\alpha}\psi|\,dN\mu\,dx,
\end{eqnarray*}
where we used that $[\Delta_{\rm S},\partial^{\tilde\alpha}]$ includes no terms with more than $m+3$ partial derivatives and that its coefficients are bounded by $\langle\nu\rangle^2$ times a constant independent of $\veps$.
Using again the Cauchy-Schwarz inequality, deviding by the square root of the left hand side, and applying the induction assumption  to the $\tilde\alpha$-term we are almost done with the proof of (\ref{induction}) for $m+1$. We only have to introduce $V$ in the Laplace term. 
We recall that it follows from $V_{\rm c},W\in C^\infty_{\rm b}\big(\C,C^\infty_{\rm b}(N_q\C)\big)$ that $V_\veps\in C^\infty_{\rm b}\big(\C,C^\infty_{\rm b}(N_q\C)\big)$.
When we put it in and use the triangle inquality we are left with the following error term:
\begin{eqnarray*}
\lefteqn{\sum_{|\beta|=m+1}\int_{\Omega_j}\int_{\RRR^k}\langle\nu\rangle^{-8|\beta|}|\partial^{\beta}V\psi|^2 \,dN\,\mu\,dx}\\
&& \ \,=\ \,\sum_{|\alpha|+|\beta|=m+1}\int_{\Omega_j}\int_{\RRR^k}\langle\nu\rangle^{-8|\alpha|}|\partial^{\alpha}V|^2\,\langle\nu\rangle^{-8|\beta|}|\partial^{\beta}\psi|^2\,dN\,\mu\,dx.
\end{eqnarray*}
In order to apply the induction assumption to this expression, we have to bound $\sup\langle\nu\rangle^{-8|\alpha|}|\partial^{\alpha}V|^2$. To be able to use $V\in C^\infty_{\rm b}(\C,C^\infty_{\rm b}(N_q\C))$ we first replace the tangential derivatives  in $\partial^{\alpha}$ by $\nabla^{\rm h}$ and afterwards the normal derivatives by $\nabla^{\rm v}$. In view of (\ref{verder}) and (\ref{horder}), this costs at most a factor $\langle\nu\rangle^{-1}$ for each derivative.
\qed

\bigskip

We still have to give the proof of Lemma \ref{lift} from Section \ref{proofT3}. It was postponed because it makes use of Lemma \ref{keyest}.

\medskip

{\sc Proof of Lemma \ref{lift} (Section \ref{proofT3}):}
\newline 
All statements in i) and ii) are easily verified by using the substitution rule.

\smallskip

To show iii) we first verify that $(H_\A^\veps A^*-A^* H^\veps) P^\veps$ is in $\L(\D(H^\veps),L^2(\A,d\tau))$ at all. For $A^* H^\veps P^\veps$ this immediately follows from ii) and Corollary \ref{projector3}. So we have to show that $H_\A^\veps A^* P^\veps\in\L(\D(H^\veps),L^2(\A,d\tau))$. By Corollary \ref{projector3} we have
\[\|H_\A^\veps A^* P^\veps\|_{\L(\D(H^\veps),L^2(\A,d\tau))} \;\lesssim\; \|H_\A^\veps A^*\langle\nu/\veps\rangle^{-l}\|_{\L(\D(H^\veps),L^2(\A,d\tau))}\]
for any $l\in\NNN$. Now we again fix one of the geodesic balls $\Omega_j\subset\C$ of a covering as in Remark \ref{coordinates} and choose geodesic coordinates~$(x^i_j)_{i=1,\ldots,d}$ and bundle coordinates~$(n^\alpha_j)_{\alpha=1,\ldots,k}$ with respect to an orthonormal trivializing frame~$(\nu_\alpha^j)_\alpha$ over~$\Omega_j$. When we write down $A^*$ and $H_\A^\veps$ in these coordinates, we will end up with coefficients that grow polynomially due to our choice of the diffeomorphism~$\Phi$ and the metric $\overline{g}$. However, this is compensated by $\langle\nu/\veps\rangle^{-l}$. Choosing $l$ big enough allows us to apply Lemma~\ref{keyest}~iii) to bound $H_\A^\veps A^*\langle\nu/\veps\rangle^{-l}$ by $-\veps^2\Delta_{\rm h}-\veps^2\Delta_{\rm v}+V^\veps$. The proof of Lemma~\ref{opestimates}~i) also shows that $-\veps^2\Delta_{\rm h}-\veps^2\Delta_{\rm v}+V^\veps\prec H^\veps$. To sum up over $j$ is once more no problem because the covering $(\Omega_j)_j$ has finite multiplicity.
Hence, $H_\A^\veps A^* P^\veps\in\L(\D(H^\veps),L^2(\A,d\tau))$. With the same arguments one also sees that $\|A^*\langle\nu/\veps\rangle^{3}A\,(H_\A^\veps A^*-A^* H^\veps) P^\veps\|_{\L(\D(H^\veps),L^2(\A,d\tau))}\lesssim1$.

\smallskip

Since $\overline{g}$ is by definition the pullback of $G$ on $\B_{\delta/2}$, the operators $H_\A^\veps A^*$ and $A^* H^\veps$ coincide on functions whose support is contained in $\B_{\delta/2}$.
But outside of $\B_{\delta/2}$, i.e.\ for $|\nu|\geq\delta/2$, we have that 
\[\langle\nu/\veps\rangle^{-3}\;=\;\left(\sqrt{\veps^2+|\nu|^2}\,/\veps\right)^{-3} \leq 8\,\veps^3/\delta^3.\]
Hence, denotig by $\chi^{\rm c}_{\B_{\delta/2}}$ the characteristic function of $N\C\setminus\B_{\delta/2}$ we obtain that $\|\chi^{\rm c}_{\B_{\delta/2}}\langle\nu/\veps\rangle^{3}\|_\infty\lesssim\veps^3$.
Using that $A^*\psi\equiv0$ on $\A\setminus\B$ for all $\psi$ and $AA^*=1$ by ii) we may estimate
\begin{eqnarray*}
\lefteqn{\|(H_\A^\veps A^*-A^* H^\veps) P^\veps\|_{\L(\D(H^\veps),L^2(\A,d\tau))}}\\ 
&=& \|A^*\chi^{\rm c}_{\B_{\delta/2}}A\,(H_\A^\veps A^*-A^* H^\veps) P^\veps\|_{\L(\D(H^\veps),L^2(\A,d\tau))}\\
&=& \|A^*\chi^{\rm c}_{\B_{\delta/2}}\langle\nu/\veps\rangle^{-3}AA^*\langle\nu/\veps\rangle^{3}A\,(H_\A^\veps A^*-A^* H^\veps) P^\veps\|_{\L(\D(H^\veps),L^2(\A,d\tau))}\\
&\lesssim& \|A^*\chi^{\rm c}_{\B_{\delta/2}}\langle\nu/\veps\rangle^{-3}A\|_{\L(L^2(\A,d\tau))}\\
&=& \|\chi^{\rm c}_{\B_{\delta/2}}\langle\nu/\veps\rangle^{-3}\|_{\infty}\\
&\lesssim& \veps^3
\end{eqnarray*}
which was the claim. \qed


\subsection{Expansion of the Hamiltonian}\label{expansion}

In order to expand the Hamiltonian $H_\veps$ in powers of $\veps$ it is crucial to expand the metric  $\overline{g}$ around $\C$ because the Laplace-Beltrami operator depends on it. The use of the expansion will be justified by the fast decay of functions from the relevant subspaces $P_0$ and $P_\veps$ in the fibers. 

\begin{proposition}\label{dualG}
Let $\overline{g}$ be the metric on $N\C$ defined in (\ref{pullback}). Choose $\Omega\subset\C$ where the normal bundle $N\C$ is trivializable and an orthonormal frame $(\nu_{\alpha})_\alpha$ of $N\C|_{\Omega}$ as in Remark~\ref{coordinates}. In the corresponding bundle coordinates the inverse metric tensor $\overline{g}\in\T^2_{\,\,0}(N\C)$ has the following expansion for all $q\in\Omega$:
\begin{equation*}
\overline{g}\;=\;\begin{pmatrix}
			1 & 0\\
			C^T & 1
                 \end{pmatrix}
		 \begin{pmatrix}
			A & 0\\
			0 & B
                 \end{pmatrix}
		 \begin{pmatrix}
			1 & C\\
			0 & 1
                 \end{pmatrix}\,+\,r_1,
\end{equation*} 
where for $i,j,l,m=1,...,d$ and $\alpha,\beta,\gamma,\delta=1,..,k$
\begin{eqnarray*}
A^{ij}(q,n) &=& g^{ij}(q)\;+\;n^\alpha\,\big({\rm W}^{\;\;i}_{\alpha l}  g^{lj}\,+\,g^{il}{\rm W}^{\;\;l}_{\alpha j} \big)(q)\\
&&\;+\;n^\alpha n^\beta\,\big(3\,{\rm W}^{\;\;i}_{\alpha m} g^{ml}{\rm W}^{\;\;j}_{\beta l}\,+\,{\rm R}^{i\ \,j}_{\ \alpha\ \beta}\big)(q),\\
B^{\gamma\delta}(q,n) &=& \delta_{\gamma\delta}\;+\;\textstyle{\frac{1}{3}}\,n^\alpha n^\beta\,{\rm R}^{\gamma\ \,\delta}_{\ \alpha\ \beta}(q),\\
C_{\,i}^{\gamma}(q,n) &=& -\,n^\alpha\,\Gamma^\gamma_{i\alpha}(q)\;+\;\textstyle{\frac{2}{3}}\,n^\alpha n^\beta\,{\rm R}^{\gamma}_{\ \alpha i\beta}(q).
\end{eqnarray*}
Here ${\rm R}$ denotes the curvature tensor of $\A$ and ${\rm W}_\alpha$ is the Weingarten mapping  corresponding to $\nu_\alpha$, i.e. $\W(\nu_\alpha)$ (see the appendix for definitions). The remainder term $r_1$ and all its derivatives are bounded by $|n|^3$ times a constant. 
\end{proposition}

For the proof we refer to the recent work of Wittich \cite{W}. He does not calculate the second correction to $C$ but it is easily deducable from his proof. Furthermore, Wittich actually calculates the expansion of the pullback of $G$, which coincides with $\overline{g}$ only on $\B_{\delta/2}$. Then $r_1$ is only locally bounded by~$|n|^3$. To see that the global bound is true for $\overline{g}$ we recall that outside of $\B_{\delta}$ it coincides with $g_{\rm S}$, which was explicitly given in Proposition \ref{dualGSasaki}. Comparing the expressions for $\overline g$ and $g_{\rm S}$ we obtain a bound by $|n|^2$ which is bounded by $|n|^3$ times a constant for $|n|\geq\delta$. 

In addition, we need to know the expansion of the extra potential occuring in Lemma~\ref{transform}, which is also provided in \cite{W}:

\begin{proposition}\label{Vgeom}
For $\rho:=d\overline{\mu}/d\sigma$ with $d\sigma=d\mu\otimes d\nu$ it holds
\begin{eqnarray*}
V_\rho(q,n) &=& -\,{\textstyle \frac{1}{4}}\overline{g}_{(q,0)}(\eta,\eta)\,+\,{\textstyle\frac{1}{2}}\kappa(q)\,-\,{\textstyle\frac{1}{6}}\big(\overline{\kappa}+{\rm tr}_\C\,\overline{{\rm Ric}}+{\rm tr}_\C\,\overline{{\rm R}}\big)(q)\,+\,r_2(q,n)\\ 
&=:& V_{\rm geom}(q)\,+\,r_2(q,n),
\end{eqnarray*} 
where $\eta$ is the mean curvature normal, $\kappa,\overline{\kappa}$ are the scalar curvatures of $\C$ and $\A$, ${\rm tr}_\C\,\overline{{\rm Ric}},{\rm tr}_\C\,\overline{{\rm R}}$ are the partial traces with respect to $T_q\C\subset T_{q}\A$ of the Ricci and the Riemann tensor of $\A$ and $r_2$ is bounded by $|n|$ times a constant.
\end{proposition}

Again there is only a local bound on $r_2$ in \cite{W}. In our setting the global bound follows immediately from the coincidence of  $d\overline{\mu}$ and $d\sigma$ outside of~$\B_{\delta}$, see (\ref{density}).
With these two inputs the proof of Lemma \ref{expH} is not difficult anymore.

\bigskip

{\sc Proof of Lemma \ref{expH} (Section \ref{proofT2}):}
\newline
Let $P$ with $\|\langle\nu\rangle^lP\|_{\L(\D(H_\veps^{m+1}),\D(H_\veps))}\lesssim1$ for all $l\in\NNN_0$ be given. The similar proof for a $P$ with  $\|P\langle\nu\rangle^l\|_{\L(\D(H_\veps^{m}),\H)}\lesssim1$ for all $l\in\NNN_0$ will be omitted. 

\smallskip

We choose a covering of $\C$ of finite multiplicity and local coordinates as at the beginning of Section \ref {wholestory} and start by proving $\|H_jP\|_{\L(\D(H_\veps^{m+1}),\H)}\lesssim1$ for $j\in\{0,1,2\}$.
Exploiting that all the coefficients in $H_j$ are bounded and have bounded derivatives due to the bounded geometry of $\A$ and $\C$ and the bounded derivatives of the embedding of $\C$ assumed in (\ref{bndcurv1}) and (\ref{bndcurv2}) we have
\begin{eqnarray}\label{zuerst}
\|H_jP\|_{\L(\D(H_\veps^{m+1}),\H)} &\lesssim& \|H_j\langle\nu\rangle^{-16}\|_{\L(\D(H_\veps),\H)} \nonumber\\
&\lesssim& \sum_{|\alpha|+|\beta|\leq2}\|\langle\nu\rangle^{-8(|\alpha|+|\beta|)}\partial^\alpha_n\veps^{|\beta|}\partial^\beta_x\|_{\L(\D(H_\veps),\H)} \nonumber\\
&\lesssim& \|H_\veps\|_{\L(\D(H_\veps),\H)} \ \,=\ \,1,
\end{eqnarray}
where we made use of Lemma \ref{keyest} iii) and Lemma \ref{opestimates} for the bound by $H_\veps$. Now we set $\psi_P:=P\psi$. By definition of $H_\veps$ and $V^\veps$ it holds
\begin{eqnarray}\label{tick}
\langle\phi\,|H_\veps\psi_P\rangle &=& \big\langle\phi\,\big|\,D_\veps^* M_\rho^* \big(-\veps^2\Delta_{\overline{g}} +V^\veps\big)M_\rho D_\veps\psi_P\big\rangle\nonumber\\
&=& \big\langle\phi\,\big|\,D_\veps^* M_\rho^* (-\veps^2\Delta_{\overline{g}})M_\rho D_\veps\psi_P\big\rangle  \,+\,\big\langle\phi\,\big|\,(V_{\rm c}+D_\veps^* WD_\veps)\psi_P\big\rangle.\qquad
\end{eqnarray}
Due to $\|\langle\nu\rangle^3P\|\lesssim1$ a Taylor expansion of $D_\veps^* WD_\veps$ in the fiber yields $D_\veps^* WD_\veps (q,\nu)P=\big(W(q,0)+\veps(\nabla^{\rm v}_{\nu}W)(q,0)+\frac{1}{2}\veps^2(\nabla^{\rm v}_{\nu,\nu}W)(q,0)\big)P+\O(\veps^3)$. Recalling that $V_0(q,\nu)=V_{\rm c}(q,\nu)+W(q,0)$ we find that
\begin{eqnarray}\label{tick2}
\lefteqn{\big\langle\phi\,\big|\,(V_{\rm c}+D_\veps^* WD_\veps)\psi_P\big\rangle}\nonumber\\ 
&& \ \,=\ \, \big\langle\phi\,\big|\,\big(V_0+\veps(\nabla^{\rm v}_{\cdot}W)(q,0)+{\textstyle\frac{1}{2}}\veps^2(\nabla^{\rm v}_{\cdot,\cdot}W)(q,0)\big)\psi_P\big\rangle \,+\,\O(\veps^3).\quad
\end{eqnarray}

The error estimate in Proposition \ref{Vgeom} yields that $\|D_\veps^* r_2D_\veps\langle\nu\rangle^{-1}\psi\|\lesssim \veps\|\psi\|$ and thus $\|D_\veps^* r_2D_\veps\psi_P\|\lesssim \veps\|\psi\|$. So Lemma \ref{transform} and Proposition \ref{Vgeom} imply that
\begin{eqnarray}\label{trick}
\lefteqn{
\big\langle\phi\,\big|\,D_\veps^* M_\rho^* (-\veps^2\Delta_{\overline{g}} )M_\rho D_\veps\psi_P\big\rangle}\nonumber\\
&& =\, \int_\C\int_{N_q\C} \veps^2\,\overline{g}\big({\rm d}D_\veps\phi^*,{\rm d}D_\veps\psi_P\big)\,d\nu\,d\mu
\,+\,\veps^2\, \langle\phi|D_\veps^* V_\rho D_\veps\psi_P\rangle\nonumber\\
&& =\, \int_\C\int_{N_q\C} \veps^2\,\overline{g}\big({\rm d}D_\veps\phi^*,{\rm d}D_\veps\psi_P\big)\,d\nu\,d\mu
+\veps^2\, \langle\phi|V_{\rm geom}\psi_P\rangle\,+\,\O(\veps^3),\quad
\end{eqnarray}
where we used that $V_{\rm geom}$ does not depend on $\nu$.

Next we fix one of the geodesic balls $\Omega\subset\C$ of our covering and insert the expansion for $\overline{g}$ from Proposition \ref{dualG} into (\ref{trick}). Noting that $\partial_{x^i}D_\veps=D_\veps\partial_{x^i}$ and $\partial_{n^\alpha}D_\veps=\veps^{-1}D_\veps\partial_{n^\alpha}$ we then obtain that
\begin{eqnarray}\label{track}
\lefteqn{\int_{\Omega}\int_{N_q\C} \veps^2\,\overline{g}\big({\rm d}D_\veps\phi^*,{\rm d}D_\veps\psi_P\big)\,d\nu\,d\mu}\nonumber\\
&=& \int_{\Omega}\int_{\RRR^k} \veps^2\Big( \big(\partial_{x^i}+C_{\,i}^{\alpha}(q,n)\partial_{n^\alpha}\big)D_\veps\phi^*\Big)\, A^{ij}(q,n)\big(\partial_{x^j}+C_{\,j}^{\beta}(q,n)\partial_{n^\beta}\big)D_\veps\psi_P \nonumber\\
&& \qquad\quad \,+\, \veps^2\,\big(\partial_{n^\alpha}D_\veps\phi^*\big)\, B^{\alpha\beta}(q,n)\,\partial_{n^\beta}D_\veps\psi_P\;dn\,d\mu\,+\,\O(\veps^3)\nonumber\\
&=& \int_{\Omega}\int_{\RRR^k} \Big( \big(\veps\partial_{x^i}+C_{\,i}^{\alpha}(q,\veps n)\partial_{n^\alpha}\big)\phi^*\Big)\, A^{ij}(q,\veps n)\,\big(\veps\partial_{x^j}+C_{\,j}^{\beta}(q,\veps n)\partial_{n^\beta}\big)\psi_P \nonumber\\
&& \qquad\quad \,+\, \big(\partial_{n^\alpha}\phi^*\big)\, B^{\alpha\beta}(q,\veps n)\,\partial_{n^\beta}\psi+\phi^*V_\veps(q,n)\psi_P \; dn\,d\mu\,+\,\O(\veps^3)
\end{eqnarray}
because the bound on $r_1$ from Proposition \ref{dualG} allows to conclude that the term containing $D_\veps^* r_1D_\veps$ is of order $\veps^3$. To do so one 
bounds the partial derivatives by $H_\veps$ as in (\ref{zuerst}). After gathering the terms from (\ref{tick}) to (\ref{track}) and plugging in the expressions for $A$, $B$, and $C$ from Proposition \ref{dualG} the rest of the proof is just a matter of identfying  $\nabla^{\rm v}$ and $\nabla^{\rm h}$ via (\ref{verder}) and (\ref{horder}). When we sum up over the whole covering, the error stays of order $\veps^3$ because our covering has finite multiplicity and the bounds are uniform as explained in Remark \ref{coordinates}. \qed


\subsection{Construction of the superadiabatic subspace}\label{subspace}

Let $E_{\rm f}$ be a constraint energy band. We search for 
$P_\veps\in\L(\H)$ with
\begin{enumerate}
\item $P_\veps P_\veps\;=\;P_\veps$,
\item $[H_\veps,P_\veps]\,\chi(H_\veps)\;=\;\O(\veps^3)$
\end{enumerate}
The former simply means that $P_\veps$ is an orthogonal projection, while the latter says that $P_\veps\chi(H_\veps)\H$ is invariant under the Hamiltonian $H_\veps$ up to errors of order $\veps^3$. 

Since the projector $P_0$ associated with $E_{\rm f}$ is a spectral projection of $H_{\rm f}$, we know that $[H_{\rm f},P_0]=0$, $[E_{\rm f},P_0]=0$, and $H_{\rm f}P_0=E_{\rm f}P_0$. Lemma \ref{expH} yields that $H_\veps=H_0+\O(\veps)$ with $H_0=-\veps^2\Delta_{\rm h}+H_{\rm f}$. So $P_0$ satisfies, at least formally, $[H_\veps,P_0]\,\chi(H_\veps)=[-\veps^2\Delta_{\rm h},P_0]\,\chi(H_\veps)+\O(\veps)=\O(\veps)$. Therefore we expect $P_\veps$ to have an expansion in $\veps$ starting with $P_0$: 
\begin{eqnarray*}
P_\veps &=& P_0\,+\,\veps P_1\,+\,\veps^2 P_2\,+\,\O(\veps^3).
\end{eqnarray*}
We first construct $P_\veps$ in a formal way ignoring problems of boundedness. Afterwards we will show how to obtain a well-defined projector and the associated unitary $U_\veps$.
We make the ansatz $P_1:=T_1^*P_0+P_0T_1$ with $T_1:\H\to\H$ to be determined. 
Assuming that $[P_1,-\veps^2\Delta_{\rm h}+E_{\rm f}]=\O(\veps)$ we have
\begin{eqnarray*}
[H_\veps,P_\veps]/\veps
&=& [H_0/\veps+H_1,P_0\,+\,\veps P_1]\,+\,\O(\veps)\\
&=& [H_0/\veps+H_1,P_0]\,+\,[H_0,P_1]\,+\,\O(\veps)\\
&=& [-\veps\Delta_{\rm h}+H_1,P_0]\,+\,[H_{\rm f}-E_{\rm f},P_1]\,+\,\O(\veps)\\
&=& [-\veps\Delta_{\rm h}+H_1,P_0]+(H_{\rm f}-E_{\rm f})T_1^*P_0-P_0T_1(H_{\rm f}-E_{\rm f})\,+\O(\veps)
\end{eqnarray*}
We have to choose $T_1$ such that the first term vanishes. Observing that every term on the right hand side is off-diagonal with respect to $P_0$,
we may multiply with $P_0$ from the right and $1-P_0$ from the left and vice versa to determine $P_1$. This leads to 
\begin{equation}\label{A1}
-\,\big(H_{\rm f}-E_{\rm f}\big)^{-1}\,(1-P_0)\,\big([-\veps\Delta_{\rm h},P_0]+H_1\big)\,P_0 \;=\; (1-P_0)\,T_1^*\,P_0
\end{equation}
and
\begin{equation}\label{A2}
-\,P_0\,\big([P_0,-\veps\Delta_{\rm h}]+H_1\big)\,(1-P_0)\,\big(H_{\rm f}-E_{\rm f}\big)^{-1}\;=\; P_0\,T_1\,(1-P_0),
\end{equation}
where we have used that the operator $H_{\rm f}-E_{\rm f}$ is invertible on $(1-P_0)\H_{\rm f}$. In view of (\ref{A1}) and (\ref{A2}), we define $T_1$ by
\begin{equation}\label{T1}
T_1 \,:=\, -\,P_0\big([P_0,-\veps\Delta_{\rm h}]+H_1\big)\,R_{H_{\rm f}}(E_{\rm f})\,+\,R_{H_{\rm f}}(E_{\rm f})\,\big([-\veps\Delta_{\rm h},P_0]+H_1\big)P_0 
\end{equation}
with $R_{H_{\rm f}}(E_{\rm f})=(1-P_0)\big(H_{\rm f}-E_{\rm f}\big)^{-1}(1-P_0)$. 
$T_1$ is anti-symmetric so that $P^{(1)}:=P_0+\veps P_1=P_0+\veps (T_1^*P_0+ P_0T_1)$ automatically satisfies condition i) for $P_\veps$ up to first order: Due to $P_0^2=P_0$
\begin{eqnarray*}
P^{(1)}P^{(1)} &=& P_0+\veps \big(T_1^*P_0+ P_0T_1+ P_0(T_1^*+T_1)P_0\big)\,+\O(\veps^2)\\
&=& P_0+\veps \big(T_1^*P_0+ P_0T_1\big)\,+\,\O(\veps^2)\\ 
&=& P^{(1)}\,+\O(\veps^2).
\end{eqnarray*}

In order to derive the form of the second order correction, we make the ansatz $P_2=T_1^*P_0T_1+T_2^*P_0+P_0T_2$ with some $T_2:\H\to\H$. The anti-symmetric part of $T_2$ is determined analogously with $T_1$ just by calculating the commutator $[P_\veps,H_\veps]$ up to second order and inverting $H_{\rm f}-E_{\rm f}$. One ends up with
\begin{equation*}
(T_2-T_2^*)/2 \;=\; -\,P_0\,\big([P^{(1)},H^{(2)}]/\veps^2\big)\,R_{H_{\rm f}}(E_{\rm f})\,+\,R_{H_{\rm f}}(E_{\rm f})\,\big([H^{(2)},P^{(1)}]/\veps^2\big)\,P_0
\end{equation*}
with $H^{(2)}:=H_0\,+\,\veps H_1\,+\,\veps^2 H_2$.
The symmetric part is again determined by the first condition for $P_\veps$. Setting $P^{(2)}:=P^{(1)}\,+\,\veps^2 P_2$ we have
\begin{eqnarray*}
P^{(2)}P^{(2)} 
&=& P^{(2)}+\veps^2 \big(P_0T_1T_1^*P_0+ P_0(T_2^*+T_2)P_0\big)\,+\,\O(\veps^3),
\end{eqnarray*}
which forces $T_2^*+T_2=-T_1T_1^*$ in order to satisfy condition i) upto second order. 

\pagebreak

We note that $T_1$ includes a differential operator of second order (and $T_2$ even of fourth order) and will therefore not be bounded on the full Hilbert space and thus neither $P_\veps$. This is related to the well-known fact that for a quadratic dispersion relation adiabatic decoupling breaks down for momenta tending to infinity. The problem can be circumvented by cutting off high energies in the right place, which was carried out by Sordoni for the Born-Oppenheimer setting in \cite{S} and by Tenuta and Teufel for a model of non-relativistic QED in \cite{TT}. 

\smallskip

To do so we fix $E<\infty$. Since $H_\veps$ is bounded from below, $E_-:=\inf\sigma(H_\veps)$ is finite. 
We choose $\chi_{E+1}\in C^\infty_0(\RRR,[0,1])$ with $\chi_{E+1}|_{(E_--1,E+1]}\equiv 1$ and ${\rm supp}\,\chi_{E+1}\subset(E_--2,E+2]$. Then we define
\begin{equation}\label{tildeP}
\tilde P_\veps\ :=\ P^{(2)}-P_0\ =\ \veps(T_1^*P_0+P_0T_1)+\veps^2(T_1^*P_0T_1+T_2^*P_0+P_0T_2)
\end{equation}
and
\begin{eqnarray}\label{preP}
P_\veps^{\chi_{E+1}}&:=& P_0 
\,+\,\tilde P_\veps\chi_{E+1}(H_\veps)\,+\,\chi_{E+1}(H_\veps)\tilde P_\veps\big(1-\chi_{E+1}(H_\veps)\big)
\end{eqnarray}
with $\chi_{E+1}(H_\veps)$ defined via the spectral theorem. We remark that $P_\veps^{\chi_{E+1}}$ is symmetric. 

We will show that $P_\veps^{\chi_{E+1}}-P_0=\O(\veps)$ in the sense of bounded operators. Then for $\veps$ small enough a projector is obtained via the formula 
\begin{equation}\label{P}
P_\veps\;:=\;\frac{\I}{2\pi}\oint_{\Gamma}\big(P_\veps^{\chi_{E+1}}-z\big)^{-1}\,dz,
\end{equation}
where $\Gamma=\{z\in\CCC\,|\,|z-1|=1/2\}$ is the positively oriented circle around~$1$ (see e.g.\ \cite{DS}).
Following here the construction of Nenciu and Sordoni \cite{NS} we define the unitary mapping $\tilde U_\veps:P_\veps\H\to P_0\H$ by the so-called Sz-Nagy formula: 
\begin{equation}\label{Utilde}
\tilde U_\veps \;:=\;\big(P_0P_\veps+(1-P_0)(1-P_\veps)\big)\,\big(1-(P_\veps-P_0)^2\big)^{-1/2}.
\end{equation}
 
We now verify that $P_\veps$ and $\tilde U_\veps$ have indeed all the properties which we stated in Lemmas \ref{projector} \& \ref{props} and state here again for convenience:
\begin{proposition}\label{unitary}
Fix $E<\infty$. Let $E_{\rm f}$ be a simple constraint energy band and $\chi_{E+1}\in C^\infty(\RRR,[0,1])$ with $\chi_{E+1}|_{(-\infty,E+1]}\equiv 1$ and ${\rm supp}\,\chi_{E+1}\subset (-\infty,E+2]$.

\smallskip

For  all $\veps$ small enough $P_\veps$ defined by (\ref{tildeP})-(\ref{P}) is a bounded operator on $\H$ and $\tilde U_\veps$ defined by (\ref{Utilde}) is unitary from $P_\veps\H$ to $P_0\H$. In particular, $P_\veps=\tilde U_\veps^* P_0\tilde U_\veps$. 

\smallskip

For all $m\in\NNN_0$ and Borel function $\chi:\RRR\to[-1,1]$ with ${\rm supp}\,\chi\subset(-\infty,E+1]$ it holds $\|P_\veps\|_{\L(\D(H_\veps^m))}\lesssim1$ and
\begin{equation*}
\|[H_\veps,P_\veps]\|_{\L(\D(H_\veps^{m+1}),\D(H_\veps^m))} \,=\, \O(\veps),\ \;
\|[H_\veps,P_\veps] \,\chi(H_\veps)\|_{\L(\H,\D(H_\veps^m))} \,=\, \O(\veps^3).
\end{equation*}

\pagebreak

Furthermore, it holds $E_{\rm f}\in C^\infty_{\rm b}(\C)$, as well as:
\vspace{-3pt}
\begin{enumerate}
\item $\forall\;j,l\in\NNN_0,\,m\in\{0,1\}:\ 
\|\langle\nu\rangle^l P_\veps\langle\nu\rangle^j\|_{\L(\D(H_\veps^m))}\;\lesssim\;1.$

\item $\forall\;j,l\in\NNN_0:\ 
\|\langle\nu\rangle^l P_0\langle\nu\rangle^j\|_{\L(\D(H_\veps))}\,\lesssim\,1\,,\,\|[-\veps^2\Delta_{\rm h},P_0]\|_{\L(\D(H_\veps),\H)}\,\lesssim\,\veps.$

\item There are $U_1^\veps,U_2^\veps\in\L(\H)\,\cap\,\L(\D(H_\veps))$ with norms bounded independently of $\veps$ satisfying $P_0U_1^\veps P_0=0$ and $U_2^\veps P_0=P_0U_2^\veps P_0=P_0U_2^\veps$ such that $\tilde U_\veps= 1+\veps U_1^\veps+\veps^2 U_2^\veps$. In particular, $\|\tilde U_\veps-1\|_{\L(\H)}=\O(\veps)$.

\item $\|P_0U_1^\veps \langle\nu\rangle^l\|_{\L(\D(H_\veps^m))}\lesssim1$ for all $l\in\NNN_0$ and $m\in\{0,1\}$.

\item For $B_\veps:=P_0\tilde U_\veps\chi(H_\veps)$ and all $u\in\{1,(U_1^\veps)^*,(U_2^\veps)^*\}$ it holds
\begin{equation*}
\big\|\,[-\veps^2\Delta_{\rm h}+E_{\rm f},u P_0]B_\veps\,\big\|_{\L(\H)}\;=\;\O(\veps).
\end{equation*} 

\item For $R_{H_{\rm f}}(E_{\rm f}):=(1-P_0)\big(H_{\rm f}-E_{\rm f}\big)^{-1}(1-P_0)$ it holds 
\begin{eqnarray*}
\big\|U_1^{\veps\,*}B_\veps \,+\,R_{H_{\rm f}}(E_{\rm f})\,([-\veps\Delta_{\rm h},P_0]+H_1)P_0B_\veps\big\|_{\L(\H,\D(H_\veps))} =\, \O(\veps).
\end{eqnarray*}

\item If $\varphi_{\rm f}\in C^\infty_{\rm b}(\C,\H_{\rm f})$, it holds 
\[\|U_0\|_{\L(\D(H_\veps),\D(-\veps^2\Delta_\C+E_{\rm f}))}\,\lesssim\,1,\quad \|U_0^*\|_{\L(\D(-\veps^2\Delta_\C+E_{\rm f}),\D(H_\veps))}\,\lesssim\,1,\] 
 and there is $\lambda_0\gtrsim1$ with $\sup_{q}\|\e^{\lambda_0\langle\nu\rangle}\varphi_{\rm f}(q)\|_{\H_{\rm f}(q)}\lesssim1$ and
\[\sup_{q}\|\e^{\lambda_0\langle\nu\rangle}\nabla^{\rm v}_{\nu_1,\dots,\nu_l}\nabla^{\rm h}_{\tau_1,\dots,\tau_m}\varphi_{\rm f}(q)\|_{\H_{\rm f}(q)}\lesssim1\]
for all $\nu_1,\dots,\nu_l\in\Gamma_{\rm b}(N\C)$ and $\tau_1,\dots,\tau_m\in\Gamma_{\rm b}(T\C)$.
\end{enumerate}
\end{proposition}
The proof relies substantially on the following decay properties of $P_0$ and the associated family of eigenfunctions. 
 
\begin{lemma}\label{expdecay}
Let $V_0\in C^{\infty}_b\big(\C,C^{\infty}_b(N\C)\big)$ and $E_{\rm f}$ be a constraint energy band with family of projections $P_0$ as defined in Definition \ref{gapcondition}. 

Define $\nabla^{\rm h}_{\tau_1}P_0\;:=\;[\nabla^{\rm h}_{\tau_1},P_0]$ and, inductively,
\[\nabla^{\rm h}_{\tau_1,\dots,\tau_m}P_0 \;:=\; [\nabla^{\rm h}_{\tau_1},\nabla^{\rm h}_{\tau_2,\dots,\tau_m}P_0]\,-\,\textstyle{\sum}_{j=2}^m\nabla^{\rm h}_{\tau_2,\dots,\nabla_{\tau_1}\tau_j,\dots,\tau_m}P_0\]
for arbitrary $\tau_1,\dots,\tau_m\in\Gamma(T\C)$. For arbitrary $\nu_1,\dots,\nu_l\in\Gamma(N\C)$ define $\nabla^{\rm v}_{\nu_1,\dots,\nu_l}\nabla^{\rm h}_{\tau_1,\dots,\tau_m}P_0:=\big[\nabla^{\rm v}_{\nu_1},\dots,[\nabla^{\rm v}_{\nu_l},\nabla^{\rm h}_{\tau_1,\dots,\tau_m}P_0]\dots\big]$. 

\pagebreak

i) Then $E_{\rm f}\in C^\infty_{\rm b}(\C)$, $P_0\in C^\infty_{\rm b}(\C,\L(\H_{\rm f}))$, and there is $\lambda_0>0$ independent of $\veps$ such that for all $\lambda\in[-\lambda_0,\lambda_0]$
\[\|\e^{\lambda\langle\nu\rangle}R_{H_{\rm f}}(E_{\rm f})\e^{-\lambda\langle\nu\rangle}\|_{\L(\H)} \;\lesssim\; 1\]
and
\[\big\|\,\e^{\lambda\langle\nu\rangle}\big(\nabla^{\rm v}_{\nu_1,\dots,\nu_l}\nabla^{\rm h}_{\tau_1,\dots,\tau_m}P_0\big)\e^{\lambda\langle\nu\rangle}\,\big\|_{\L(\H)}\lesssim 1\]
for all $\nu_1,\dots,\nu_l\in\Gamma_{\rm b}(N\C)$ and $\tau_1,\dots,\tau_m\in\Gamma_{\rm b}(T\C)$.

\medskip

Let $E_{\rm f}$ be simple and $\varphi_{\rm f}$ be a family of normalized eigenfunctions that define a smooth section of the associated eigenspace bundle . 

\smallskip

ii) If $\varphi_{\rm f}\in C^m_{\rm b}(\C,\H_{\rm f})$, then 
$\varphi_{\rm f}\in C^m_{\rm b}(\C,C^{\infty}_b(N\C))$. Furthermore, 
\begin{equation*}
\sup_{q\in\C}\|\e^{\lambda_0\langle\nu\rangle}\varphi_{\rm f}(q)\|_{\H_{\rm f}(q)}\lesssim1,\quad \sup_{q\in\C}\,\|\e^{\lambda_0\langle\nu\rangle}\nabla^{\rm v}_{\nu_1,\dots,\nu_l}\nabla^{\rm h}_{\tau_1,\dots,\tau_m}\varphi_{\rm f}(q)\|_{\H_{\rm f}(q)}\;\lesssim\;1
\end{equation*} 
for all $\nu_1,\dots,\nu_l\in\Gamma_{\rm b}(N\C)$ and $\tau_1,\dots,\tau_m\in\Gamma_{\rm b}(T\C)$.

\smallskip

iii) If $\C$ is compact or contractable or if $E_{\rm f}(q)=\inf \sigma\big(H_{\rm f}(q)\big)$ for all $q\in\C$, then $\varphi_{\rm f}$ can be chosen such that $\varphi_{\rm f}\in C^\infty_{\rm b}(\C,\H_{\rm f})$. 
\end{lemma}

In addition, we need that the application of $\chi_{E+1}(H_\veps)$ does not completely spoil the exponential decay. This is stated in the following lemma. We notice that we cannot expect it to preserve exponential decay in general, for we do not assume the cutoff energy $E$ to lie below the continuous spectrum of $H_\veps$! 

\begin{lemma}\label{notspoiled}
Let $\chi\in C^\infty_0(\RRR)$ be non-negative and $\big(H,\D(H)\big)$ be self-adjoint on $\H$. 
Assume that there are $l\in\ZZZ,m\in\NNN$ and $C_1<\infty$ such that 
\begin{equation}\label{asslambda}
\|\langle\lambda\nu\rangle^l\,[H^j,\langle\lambda\nu\rangle^{-l}] \|_{\L(\D(H^m),\H)}\;\leq\; C_1\,\lambda
\end{equation}
for all $\lambda\in(0,1]$ and $1\leq j\leq m$. Then there is $C_2<\infty$ independent of $H$ such that
\[\|\langle\nu\rangle^l\,\chi(H)\,\langle\nu\rangle^{-l}\|_{\L(\H,\D(H^m))}\;\leq\;C_1^l\,C_2.\]
\end{lemma}

This lemma can be applied to $H_\veps$ for $m\leq3$ in view of Lemma \ref{opestimates}. Now we give the proof of the proposition. Afterwards we take care of the two technical lemmas.

\bigskip

{\sc Proof of Proposition \ref{unitary}:}
\newline
We recall that  $\D(H_\veps^0):=\H$ and $E_-:=\inf\sigma(H_\veps)$. 
Let $\chi_E\in C^\infty_0(\RRR,[0,1])$ with $\chi_E|_{[E_-,E]}\equiv 1$ and ${\rm supp}\,\chi_E\subset[E_--1,E+1]$. Then by the spectral theorem 
$\chi_E(H_\veps)\chi(H_\veps)=\chi(H_\veps)$ and $\chi_{E+1}(H_\veps)\chi_E(H_\veps)=\chi_E(H_\veps)$ for $\chi$ and $\chi_{E+1}$ as in the proposition. In the sequel, we drop all $\veps$-subscripts except those of $H_\veps$ and write $\chi,\chi_E$, and $\chi_{E+1}$ for $\chi(H_\veps),\chi_E(H_\veps)$, and $\chi_{E+1}(H_\veps)$ respectively.

\medskip

The proof of the proposition will be devided into several steps. 
We will often need that an operator $A\in\L(\H)$ is in $\L(\D(H_\veps^l),\D(H_\veps^m))$ for some $l,m\in\NNN_0$. The strategy to show that will always be to show that there are $l_1,l_2\in\NNN$ with $l_1+l_2\leq 2l$ such that for all $j\in\NNN_0$
\begin{eqnarray}\label{rein}
(-\veps^2\Delta_{\rm h}-\Delta_{\rm v}+V_\veps)^m A \ \prec\ 
\langle\nu\rangle^{-j}(\nabla^{\rm v})^{l_1}(\veps\nabla^{\rm h})^{l_2}.
\end{eqnarray}
Then we can use Lemma \ref{opestimates} to estimate:
\begin{eqnarray}\label{raus}
\|H_\veps^m A\psi\| \,+\, \|A\psi\| &\lesssim& \|(-\veps^2\Delta_{\rm h}-\Delta_{\rm v}+V_\veps)^m A\psi\| \,+\, \|\psi\| \nonumber\\
&\lesssim& \|
\langle\nu\rangle^{-4l_1-5l_2}(\nabla^{\rm v})^{l_1}(\veps\nabla^{\rm h})^{l_2}\psi\| \,+\, \|\psi\| \nonumber\\
&\lesssim& \|H_\veps^l\psi\| \,+\, \|\psi\| ,
\end{eqnarray}
which yields the desired bound. 

\bigskip

\setcounter{stepcounter}{0}
\begin{step}\label{step1}
$\exists\;\lambda_0\gtrsim 1\ 
\forall\;\lambda<\lambda_0,\,m\in\NNN_0:\ \|{\rm e}^{\lambda\langle\nu\rangle}\,P_0\,{\rm e}^{\lambda\langle\nu\rangle}\|_{\L(\D(H_\veps^m))}\;\lesssim\;1$ and
\[\|{\rm e}^{\lambda\langle\nu\rangle}\,[-\veps^2\Delta_{\rm h},P_0]\,{\rm e}^{\lambda\langle\nu\rangle}\|_{\L(\D(H_\veps^{m+1}),\D(H_\veps^m))}\;\lesssim\;\veps.\]
Both statements hold true with ${\rm e}^{\lambda\langle\nu\rangle}$ replaced by $\langle\nu\rangle^l$ for any $l\in\NNN_0$. 
\end{step}

Let $\lambda_0$ be as given by Lemma \ref{expdecay}. When we choose a partition of unity~$(\xi_j)_j$ corresponding to the covering $(\Omega_j)_j$ as in Remark \ref{coordinates} at the beginning of Section~\ref{wholestory} and orthonormal sections $(\nu^j_\alpha)_{\alpha=1,\dots,k}$ of $N\Omega_j$ and $(\tau^j_i)_{i=1,\dots,d}$ of $T\Omega_j$ for all $j$, the coordinate formulas (\ref{Laplace}) imply
\begin{equation}\label{Laplace3}
\Delta_{\rm v}\;=\;\sum_{j,\alpha} \xi_j\,\nabla^{\rm v}_{\nu^j_\alpha}\nabla^{\rm v}_{\nu^j_\alpha}\,, \quad\Delta_{\rm h}\;=\;\sum_{j,i} \xi_j(\nabla^{\rm h}_{\tau^j_i}\nabla^{\rm h}_{\tau^j_i}-\nabla^{\rm h}_{\nabla_{\tau^j_i}{\tau^j_i}}).
\end{equation}
In order to obtain the estimate (\ref{rein}) for $A={\rm e}^{\lambda_0\langle\nu\rangle}\,P_0\,{\rm e}^{\lambda_0\langle\nu\rangle}$ we first commute all horizontal derivatives to the right and then the vertical ones. Using $V_0\in C^{\infty}_b\big(\C,C^{\infty}_b(N\C)\big)$ and Lemma \ref{opequations} we end up with terms of the form $\xi_j\,\e^{\lambda\langle\nu\rangle}\big(\nabla^{\rm v}_{\nu^j_1,\dots,\nu^j_{l_3}}\nabla^{\rm h}_{\tau^j_1,\dots,\tau^j_{l_4}}P_0\big)\e^{\lambda\langle\nu\rangle}(\nabla^{\rm v})^{l_1}(\veps\nabla^{\rm h})^{l_2}$ times a bounded function with $l_1+l_2\leq2m$. By Lemma \ref{expdecay} we have
\[\xi_j\,\e^{\lambda\langle\nu\rangle}\big(\nabla^{\rm v}_{\nu^j_1,\dots,\nu^j_{l_3}}\nabla^{\rm h}_{\tau^j_1,\dots,\tau^j_{l_4}}P_0\big)\e^{\lambda\langle\nu\rangle}(\nabla^{\rm v})^{l_1}(\veps\nabla^{\rm h})^{l_2} \;\prec\; \e^{-(\lambda_0-\lambda)\langle\nu\rangle}(\nabla^{\rm v})^{l_1}(\veps\nabla^{\rm h})^{l_2}\]
which implies (\ref{rein}) due to $\lambda<\lambda_0$. This yields the first claim of Step \ref{step1} via~(\ref{raus}). 

The second claim can easily be proven in the same way. For the last claim it suffices to notice that $\|\langle\nu\rangle^l{\rm e}^{-\lambda_0\langle\nu\rangle}\|_{\L(\D(H_\veps^m))}\lesssim1$ for all $l,m\in\NNN_0$, which is easy to verify.

\bigskip

\begin{step}\label{step2}
It holds $\forall\;\lambda<\lambda_0,\,m\in\NNN_0,\,i\in\{1,2\}:\ $
\begin{equation*}
\|{\rm e}^{\lambda\langle\nu\rangle} T_i^*P_0\,{\rm e}^{\lambda\langle\nu\rangle}\|_{\L(\D(H_\veps^{m+i}),\D(H_\veps^m))}\lesssim\,1,\,\|{\rm e}^{\lambda\langle\nu\rangle}P_0T_i\,{\rm e}^{\lambda\langle\nu\rangle}\|_{\L(\D(H_\veps^{m+i}),\D(H_\veps^m))}\,\lesssim\,1.
\end{equation*}
In particular, $\forall\;\lambda<\lambda_0,m\in\NNN_0:\ \|{\rm e}^{\lambda\langle\nu\rangle}\,\tilde P\,{\rm e}^{\lambda\langle\nu\rangle} \|_{\L(\D(H_\veps^{m+2}),\D(H_\veps^m))}\,\lesssim\, \veps.$
\end{step}

The last statement is an immediate consequence because by definition of $\tilde P$
\begin{equation*}
{\rm e}^{\lambda\langle\nu\rangle}\tilde P{\rm e}^{\lambda\langle\nu\rangle} \;=\;  \veps\,{\rm e}^{\lambda\langle\nu\rangle}\Big((T_1^*P_0+P_0T_1)\,+\,\veps(T_1^*P_0P_0T_1+T_2^*P_0+P_0T_2)\Big){\rm e}^{\lambda\langle\nu\rangle}.
\end{equation*}
We carry out the proof of the first estimate only for $T_1^*P_0$. The same arguments work for the other terms. To obtain (\ref{rein}) for $A={\rm e}^{\lambda\langle\nu\rangle}T_1^* P_0\,{\rm e}^{\lambda\langle\nu\rangle}$ we again commute all derivatives in $(-\veps^2\Delta_{\rm h}-\Delta_{\rm v}+V_\veps)^m$ and $T_1^*P_0$ to the right. 
In view of (\ref{T1}), the definition of $T_1$, we have to compute the commutator of $R_{H_{\rm f}}(E_{\rm f})$ with $\nabla^{\rm h}$ and  $\nabla^{\rm v}$. 
For arbitrary $\tau\in\Gamma_{\rm b}(T\C)$ it holds
\begin{eqnarray*}
\big[\nabla^{\rm h}_\tau,R_{H_{\rm f}}(E_{\rm f})\big] &=& -\,(\nabla^{\rm h}_\tau P_0)R_{H_{\rm f}}(E_{\rm f})\,-\,R_{H_{\rm f}}(E_{\rm f})(\nabla^{\rm h}_\tau P_0)\\
&&\qquad \,-\, R_{H_{\rm f}}(E_{\rm f})\big[\nabla^{\rm h}_\tau,H_{\rm f}-E_{\rm f}\big]R_{H_{\rm f}}(E_{\rm f}).
\end{eqnarray*}
with $\big[\nabla^{\rm h}_\tau,H_{\rm f}-E_{\rm f}\big]=(\nabla^{\rm h}_\tau V_0-\nabla_\tau E_{\rm f})$. 
The latter is bounded because of $V_0\in C^{\infty}_b\big(\C,C^{\infty}_b(N_q\C)\big)$ by assumption and $E_{\rm f}\in C^\infty_{\rm b}(\C)$ by Lemma \ref{expdecay}. An analogous statement  is true for $\nabla^{\rm v}$.
Hence, we end up with all remaining derivatives on the right-hand side after a finite iteration. These are at most $2m+2$. After exploiting that $\|{\rm e}^{\lambda\langle\nu\rangle}\,R_{H_{\rm f}}(E_{\rm f})\,{\rm e}^{-\lambda\langle\nu\rangle}\|_{\L(\H)}\,\lesssim\,1$ by Lemma \ref{expdecay} we may obtain a bound by $H_\veps^{m+1}$ as in Step \ref{step1}.

\bigskip

\begin{step}\label{step4}
$\forall\,m\in\NNN_0:\ 
\| P^{\chi_{E+1}}\|_{\L(\D(H_\veps^m))}\,\lesssim\,1$ 
and
\[\forall\,j,l,\in\NNN_0,m\in\{0,1\}:\ 
\|\langle\nu\rangle^j P^{\chi_{E+1}}\langle\nu\rangle^l\|_{\L(\D(H_\veps^m))}\,\lesssim\,1.\]
\end{step}

\vspace{-0.8cm}
We recall that $P^{\chi_{E+1}}$ was defined as
\[P^{\chi_{E+1}}\;=\;P_0 \,+\,\widetilde{P}\,\chi_{E+1}\,+\,\chi_{E+1}\widetilde{P}(1-\chi_{E+1}).\]
Step \ref{step1} implies that $P_0\in\L(\D(H_\veps^m))$ for all $m\in\NNN_0$. 
So it suffices to bound the second and the third term to show that $P^{\chi_{E+1}}\in\L(\D(H_\veps^m))$. Since $H_\veps$ is bounded from below and the support of $\chi_{E+1}$ is bounded from above, $\|\chi_{E+1}\|_{\L(\H,\D(H_\veps^m))}\lesssim1$ for every $m\in\NNN_0$. So the estimate for $\widetilde{P}$ obtained in 

\pagebreak

Step \ref{step2} implies the boundedness of the second term. By comparing them on the dense subset $\D(H_\veps^2)$ we see that $\chi_{E+1}\widetilde{P}$ is the adjoint of $\widetilde{P}\chi_{E+1}$ and thus also bounded. This finally implies the boundedness of the third term, which establishes $\|P^{\chi_{E+1}}\|_{\L(\D(H_\veps^m))}\lesssim1$ for all $m\in\NNN_0$.

\smallskip

We now address the second claim. We fix $\lambda$ with $0<\lambda<\lambda_0$. Then
\begin{eqnarray*}
\langle\nu\rangle^j P^{\chi_{E+1}} \langle\nu\rangle^l &=& \langle\nu\rangle^j P_0 \langle\nu\rangle^l + \langle\nu\rangle^j\widetilde{P}\,\chi_{E+1}\langle\nu\rangle^l
+ \langle\nu\rangle^j\chi_{E+1}\widetilde{P}(1-\chi_{E+1})\langle\nu\rangle^l\\ 
&=& \langle\nu\rangle^j{\rm e}^{-\lambda\langle\nu\rangle} \;({\rm e}^{\lambda\langle\nu\rangle}P_0{\rm e}^{\lambda\langle\nu\rangle}) \;{\rm e}^{-\lambda\langle\nu\rangle}\langle\nu\rangle^l\\
&& \  \,+\; \langle\nu\rangle^j{\rm e}^{-\lambda_0\langle\nu\rangle} \;({\rm e}^{\lambda\langle\nu\rangle}\widetilde{P}{\rm e}^{\lambda\langle\nu\rangle}) \;({\rm e}^{-\lambda\langle\nu\rangle}\langle\nu\rangle^l)\,\langle\nu\rangle^{-l}\chi_{E+1}\langle\nu\rangle^l\\ 
&& \ \,+\; \langle\nu\rangle^j\chi_{E+1}\langle\nu\rangle^{-l}\,(\langle\nu\rangle^l{\rm e}^{-\lambda\langle\nu\rangle}) \;({\rm e}^{\lambda\langle\nu\rangle}\widetilde{P}{\rm e}^{\lambda\langle\nu\rangle}) \;\\
&& \qquad\qquad\times\,({\rm e}^{-\lambda\langle\nu\rangle}\langle\nu\rangle^l)\,\langle\nu\rangle^{-l}(1-\chi_{E+1})\langle\nu\rangle^l
\end{eqnarray*}

It is straight forward to see that $\|\langle\nu\rangle^j{\rm e}^{-\lambda_0\langle\nu\rangle}\|_{\L(\D(H_\veps^m))}\lesssim1$ for all $j,m\in\NNN_0$. Therefore Step \ref{step1} yields the desired estimate for the first term. In addition, we know from Lemma \ref{notspoiled} that $\|\langle\nu\rangle^{-l}\chi_{E+1}\langle\nu\rangle^l\|_{\L(\H,\D(H_\veps^3))}\lesssim1$ because $H_\veps$ satisfies the assumption of Lemma \ref{notspoiled} due to Lemma \ref{opestimates} iii). So Step \ref{step2} implies the desired estimate for the second term. Then it also follows for the third term again by estimating it by the adjoint of the second one.

\bigskip

\begin{step}\label{step5}
It holds $\forall\;m\in\NNN_0,\,i\in\{1,2\}$
\begin{eqnarray*}
\|[T_i^*P_0,-\veps^2\Delta_{\rm h}+E_{\rm f}]\|_{\L(\D(H_\veps^{m+i+1}),\D(H_\veps^m))} &=& \O(\veps),\\
\|[P_0T_i,-\veps^2\Delta_{\rm h}+E_{\rm f}]\|_{\L(\D(H_\veps^{m+i+1}),\D(H_\veps^m))} &=& \O(\veps).
\end{eqnarray*} 
\end{step}

\vspace{-0.8cm}
We again restrict to $T_1^*P_0$ because the other cases can be treated in quite a similar way. 

We note that $E_{\rm f}$ commutes with all operators contained in $T_1^*P_0$ but $\veps \nabla^{\rm h}$.
Furthermore, $\|[\veps \nabla^{\rm h}_\tau,E_{\rm f}]P_0\|_{\L(\D(H_\veps^m))}=\veps\| (\nabla_\tau E_{\rm f})P_0\|_{\L(\D(H_\veps^m))}=\O(\veps)$ for any $\tau\in\Gamma_{\rm b}(T\C)$ by Lemma \ref{expdecay}. With this $\|[T_1^*P_0,E_{\rm f}]\|_{\L(\D(H_\veps^{m+2}),\D(H_\veps^m))}=\O(\veps)$ is easily verified. 

\smallskip

We will obtain the claim of Step \ref{step5} for $T_1^*P_0$, if we are able to deduce that $\|[T_1^*P_0,-\veps^2\Delta_{\rm h}]\|_{\L(\D(H_\veps^{m+2}),\D(H_\veps^m))}=\O(\veps)$. Again we aim at proving (\ref{rein}) by commuting all derivatives to the right. In Step \ref{step1} and Step \ref{step2} we have already treated the commutators of $-\veps^2\Delta_{\rm h}$ with $P_0$ an $R_{H_{\rm f}}(E_{\rm f})$. So it remains to discuss the commutator of $\veps \nabla^{\rm h}_\tau$ and $-\veps^2\Delta_{\rm h}$, which does not vanish in general! To do so we again fix a covering~$(\Omega_j)_{j\in\NNN}$ of~$\C$ and choose a partition of unity~$(\xi_j)_j$ corresponding to the covering~$(\Omega_j)_j$ as in Remark~\ref{coordinates}, as well as orthonormal sections $(\tau^j_i)_{i=1,\dots,d}$ of $T\Omega_j$ for all $j$. 

Recalling from (\ref{Laplace3}) that~$\Delta_{\rm h}=\sum_{i=1}^d \xi_j(\nabla^{\rm h}_{\tau^j_i}\nabla^{\rm h}_{\tau^j_i}-\nabla^{\rm h}_{\nabla_{\tau^j_i}{\tau^j_i}})$ we have
\begin{eqnarray*}
[\veps \nabla^{\rm h}_\tau,-\veps^2\Delta_{\rm h}] &=& -\sum_{j,i}\xi_j\; [\veps\nabla^{\rm h}_\tau,\veps^2(\nabla^{\rm h}_{\tau^j_i}\nabla^{\rm h}_{\tau^j_i}-\nabla^{\rm h}_{\nabla_{\tau^j_i}{\tau^j_i}})]\\
&=& -\veps^3\sum_{j,i}\xi_j \big([\nabla^{\rm h}_\tau, \nabla^{\rm h}_{\tau^j_i}]\,\nabla^{\rm h}_{\tau^j_i}+ \nabla^{\rm h}_{\tau^j_i}\,[\nabla^{\rm h}_\tau,\nabla^{\rm h}_{\tau^j_i}] -[\nabla^{\rm h}_\tau,\nabla^{\rm h}_{\nabla_{\tau^j_i}{\tau^j_i}})]\big)\\
&=& -\veps^3\sum_{j,i}\xi_j \Big( {\rm R}^{\rm h}(\tau,\tau^j_i)\,\nabla^{\rm h}_{\tau^j_i}\,+\,\nabla^{\rm h}_{[\tau,\tau^j_i]}\nabla^{\rm h}_{\tau^j_i} \\
&& \qquad\qquad\,+\, \nabla^{\rm h}_{\tau^j_i}\,{\rm R}^{\rm h}(\tau,\tau^j_i)\,+\,\nabla^{\rm h}_{\tau^j_i}\nabla^{\rm h}_{[\tau,\tau^j_i]}\,+\,[\nabla^{\rm h}_\tau,\nabla^{\rm h}_{\nabla_{\tau^j_i}{\tau^j_i}})]\Big).
\end{eqnarray*}  
In view of the expression for ${\rm R}^{\rm h}$ in Lemma \ref{berryphase2}, all these terms contain only two derivatives. So we have gained an $\veps$ because, although ${\rm R}^{\rm h}$ and its derivatives grow linearly, we are able to bound the big bracket as required in (\ref{rein}) using the decay provided by $P_0$. The estimate is independent of $\Omega_j$ because ${\rm R}^\perp$ is globally bounded due to our assumption on the embedding of $\C$ in (\ref{bndcurv2}). 

\bigskip

\begin{step}\label{step6}
For all $m\in\NNN_0$
\[\|[H_\veps,P^{\chi_{E+1}}]\|_{\L(\D(H_\veps^{m+1}),\D(H_\veps^m))}=\O(\veps),\ \,\|[H_\veps,P^{\chi_{E+1}}]\,\chi_E\|_{\L(\H,\D(H_\veps^m))}=\O(\veps^3).\] 
\end{step}

\vspace{-0.8cm}
We fix $m\in\NNN_0$. Due to the exponential decay obtained in Steps \ref{step1} \& \ref{step2} for $P_0$ and $\tilde P$ we may plug in the expansion of $H_\veps$ from Lemma \ref{expH} when deriving the stated estimates. The proof of Step \ref{step2} entails that $P^{\chi_{E+1}}-P_0$ is of order~$\veps$ in~$\L(\D(H_\veps^m))$ for any $m\in\NNN_0$. Therefore
\begin{eqnarray*}
\|[H_\veps,P^{\chi_{E+1}}]\|_{\L(\D(H_\veps^{m+1}),\D(H_\veps^m))} &=& \|[H_\veps,P_0]\|_{\L(\D(H_\veps^{m+1}),\D(H_\veps^m))}\,+\,\O(\veps)\\
&=& \|[H_0,P_0]\|_{\L(\D(H_\veps^{m+1}),\D(H_\veps^m))}\,+\,\O(\veps)\\
&=& \|[-\veps^2\Delta_{\rm h},P_0]\|_{\L(\D(H_\veps^{m+1}),\D(H_\veps^m))}\,+\,\O(\veps)\\
&=& \O(\veps),
\end{eqnarray*}  
by Step \ref{step1}. On the other hand we use $[H_\veps,\chi_E]=0$ and $(1-\chi_{E+1})\chi_E=0$ to obtain
\begin{eqnarray*}
\lefteqn{\|[H_\veps,P^{\chi_{E+1}}]\,\chi_E\|_{\L(\H,\D(H_\veps^m))}}\\
&& \quad=\ \, \|[H_\veps,P^{(2)}]\,\chi_E\|_{\L(\H,\D(H_\veps^m))}\\ 
&& \quad=\ \, \|[H_\veps,P_0+\tilde P]\,\chi_E\|_{\L(\H,\D(H_\veps^m))}\\ 
&& \quad=\ \, \|[H_0+\veps H_1+\veps^2 H_2,P_0+\tilde P]\,\chi_E\|_{\L(\H,\D(H_\veps^m))}+\O(\veps^3)
\ \,=\ \, \O(\veps^3),
\end{eqnarray*}  
where the last estimate follows from the construction of $T_1$ and $T_2$ at the beginning of this subsection (which were used to define $\tilde P$). To make precise the formal discussion presented there one uses Step \ref{step5} and once more the decay properties of $P_0$ and $\tilde P$ to bound the error terms by $H_\veps^m$ for some $m\in\NNN$ as in (\ref{rein}) and  (\ref{raus}).

\bigskip

\begin{step}\label{step7}
For $\veps$ small enough $P$ \& $U$ are well-defined, $P^2=P$, and $U|_{P\H}$ is unitary. $\|P\|_{\L(\D(H_\veps^m))}\lesssim1$ and $\|P-P_0\|_{\L(\D(H_\veps^m))}=\O(\veps)$ for all $m\in\NNN_0$.
\end{step}

Since $P_0$ is a projector and $\|P^{\chi_{E+1}}-P_0\|_{\L(\H)}=\O(\veps)$ by the proof of Step \ref{step4}, we have 
\begin{equation}\label{almostPro}
\|(P^{\chi_{E+1}})^2-P^{\chi_{E+1}}\|_{\L(\H)}=\O(\veps).
\end{equation} 
Now the spectral mapping theorem for bounded operators implies that there is a $C<\infty$ such that
\[\sigma(P^{\chi_{E+1}})\;\subset\;[-C\veps,C\veps]\,\cup\,[1-C\veps,1+C\veps].\]
Thus $P:=\frac{\I}{2\pi}\oint_{\Gamma}\big(P_\veps^{\chi_{E+1}}-z\big)^{-1}\,dz$ is an operator on $\H$ bounded independent of $\veps$ for $\veps<1/2C$ and satisfies $P^2=P$ by the spectral calculus (see e.g.\ \cite{DS}). By the spectral theorem $P=\chi_{[1-C\veps,1+C\veps]}(P^{\chi_{E+1}})$ and so $\|P-P^{\chi_{E+1}}\|_{\L(\H)}=\O(\veps)$. 
With $\|P^{\chi_{E+1}}-P_0\|_{\L(\H)}=\O(\veps)$ this entails $\|P-P_0\|_{\L(\H)}=\O(\veps)$. Hence, $1-(P-P_0)^2$ is strictly positive and thus has a bounded inverse. Therefore $U:=\big(P_0P+(1-P_0)(1-P)\big)\,\big(1-(P-P_0)^2\big)^{-1/2}$ is also bounded independent of $\veps$ as an operator on $\H$ and satisfies
\begin{equation*}
U\;=\;U_0\,\big(P\,+\,\O(\veps^2)\big).
\end{equation*}
We set $S:=\big(1-(P-P_0)^2\big)^{-1/2}$. It is easy to verify that $[P,1-(P-P_0)^2]=0=[P_0,1-(P-P_0)^2]$ and thus $[P,S]=0=[P_0,S]$. The latter implies $\tilde U^*\tilde U=1=\tilde U\tilde U^*$. So $\tilde U$ maps $P\H$ unitarily to $P_0\H$. Since $U_0$ is unitary when restricted to $P_0\H$, we see that $U=U_0\tilde U$ is unitary when restricted to $P\H$.

\smallskip

The combination of (\ref{almostPro}) with Steps \ref{step4} and \ref{step6} immediately yields
\[\|(P^{\chi_{E+1}})^2-P^{\chi_{E+1}}\|_{\L(\D(H_\veps^m))}=\O(\veps).\]  
for all $m\in\NNN_0$. So for $\veps<1/2C$ and $z\in\partial B_{1/2}(1)$ the resolvent $\big(P^{\chi_{E+1}}-z\big)^{-1}$ is an operator bounded independent of $\veps$ even on $\D(H_\veps^m)$. In view of $P$'s definition, this implies $\|P\|_{\L(\D(H_\veps^m))}\lesssim1$  for all $m\in\NNN$ . Then we obtain that $\|P-P_0\|_{\L(\D(H_\veps^m))}=\O(\veps)$ in the same way we did for $m=0$.  

\bigskip

\begin{step}\label{step8}
$\|[H_\veps,P]\|_{\L(\D(H_\veps^{m+1}),\D(H_\veps^m))}=\O(\veps) \ \,\&\ \,\|[H_\veps,P]\,\chi_E\|_{\L(\H,\D(H_\veps^m))}=\O(\veps^3)$ for all $m\in\NNN_0$. 
\end{step}

We observe that
\[[H_\veps,P]\;=\;\frac{\I}{2\pi}\oint_{\Gamma}\big(P^{\chi_{E+1}}-z\big)^{-1}[H_\veps,P^{\chi_{E+1}}]\big(P^{\chi_{E+1}}-z\big)^{-1}\,dz.\]
Since we saw that $\|\big(P^{\chi_{E+1}}-z\big)^{-1}\|_{\L(\D(H_\veps^m))}\lesssim1$ in the preceding step, the first estimate we claimed follows by inserting the result from Step \ref{step6}. To deduce the second one we set $R_{P^{\chi_{E+1}}}(z):=\big(P^{\chi_{E+1}}-z\big)^{-1}$ and use $\chi=\chi_E\chi$ to compute
\begin{eqnarray}\label{hierrein}
[H_\veps,P]\,\chi
&=& \frac{\I}{2\pi}\oint_{\Gamma}R_{P^{\chi_{E+1}}}(z)\,[H_\veps,P^{\chi_{E+1}}]\,R_{P^{\chi_{E+1}}}(z)\,\chi_E\,\chi\,dz\nonumber\\
&=& \frac{\I}{2\pi}\oint_{\Gamma}R_{P^{\chi_{E+1}}}(z)\,[H_\veps,P^{\chi_{E+1}}]\chi_E\,R_{P^{\chi_{E+1}}}(z)\,\chi\nonumber\\
&& \qquad\quad\;+\; R_{P^{\chi_{E+1}}}(z)\,[H_\veps,P^{\chi_{E+1}}]\big[R_{P^{\chi_{E+1}}}(z),\chi_E\big]\,\chi\,dz.\quad
\end{eqnarray}
Furthermore,
\begin{eqnarray*}
\big[R_{P^{\chi_{E+1}}}(z),\chi_E\,\big]\,\chi
&=& R_{P^{\chi_{E+1}}}(z)\,[P^{\chi_{E+1}},\chi_E]\,R_{P^{\chi_{E+1}}}(z)\,\chi_E\,\chi\\
&=& R_{P^{\chi_{E+1}}}(z)\,[P^{\chi_{E+1}},\chi_E]\,\chi_E\,R_{P^{\chi_{E+1}}}(z)\,\chi\\
&& \quad\,+\  R_{P^{\chi_{E+1}}}(z)\,[P^{\chi_{E+1}},\chi_E]\,\big[R_{P^{\chi_{E+1}}}(z),\chi_E\,\big]\,\chi\\
&=& R_{P^{\chi_{E+1}}}(z)\,[P^{\chi_{E+1}},\chi_E]\,\chi_E\,R_{P^{\chi_{E+1}}}(z)\,\chi\\
&& \quad\,+\  \Big(R_{P^{\chi_{E+1}}}(z)\,[P^{\chi_{E+1}},\chi_E]\Big)^2\,R_{P^{\chi_{E+1}}}(z)\,\chi.
\end{eqnarray*}

Since due to Step \ref{step6} we have $\big\|[P^{\chi_{E+1}},H_\veps]\big\|_{\L(\D(H_\veps^{m+1}),\D(H_\veps^m))}=\O(\veps)$ and $\|[P^{\chi_{E+1}},H_\veps]\chi_E\|_{\L(\H,\D(H_\veps^m))}=\O(\veps^3)$, Lemma \ref{microlocal} yields
\[\big\|[P^{\chi_{E+1}},\chi_E]\big\|_{\L(\D(H_\veps^{m}),\D(H_\veps^{m+1}))}=\O(\veps),\, \|[P^{\chi_{E+1}},\chi_E]\chi_E\|_{\L(\H,\D(H_\veps^m))}=\O(\veps^3).\] 

Applying these estimates, $\|R_{P^{\chi_{E+1}}}(z)\|_{\L(\D(H_\veps^m))}\lesssim1$, and Step \ref{step6} to (\ref{hierrein}) we obtain $\|[H_\veps,P]\,\chi(H_\veps)\|_{\L(\H,\D(H_\veps^m))}=\O(\veps^3)$.

\bigskip

\begin{step}\label{step9}
$\forall\,j,l\in\NNN,m\in\{0,1\}:\ \|\langle\nu\rangle^l \,P\,\langle\nu\rangle^j\|_{\L(\D(H_\veps^m))}\,\lesssim\,1.$
\end{step}

This can be seen by applying the spectral calculus to $P^{\chi_{E+1}}$ which we know to be bounded and symmetric. Let $f:\CCC\to\CCC$ be defined by $f(z):=z$ and let $g:\CCC\to\{0,1\}$ be the characteristic function of $B_{2/3}(1)$. Then due to (\ref{almostPro}) the spectral calculus implies that for $\veps$ small enough
\begin{eqnarray}\label{cheap}
P\;=\;g(P^{\chi_{E+1}}) &=& f(P^{\chi_{E+1}})\,(g/f^2)(P^{\chi_{E+1}})\,f(P^{\chi_{E+1}})\nonumber\\ 
&=&P^{\chi_{E+1}}\,(g/f^2)(P^{\chi_{E+1}})\,P^{\chi_{E+1}}.
\end{eqnarray}
We note that $(g/f^2)(P^{\chi_{E+1}})\in\L(\H)$  because $g\equiv0$ in a neighborhood of zero. Since $g/f^2$ is holomorphic on $B_{1/2}(1)$, it holds
\[(g/f^2)(P^{\chi_{E+1}})\;=\;\frac{\I}{2\pi}\oint_{\partial B_{1/2}(1)}(g/f^2)(z)R_{P^{\chi_{E+1}}}(z)\,dz\]
by the Cauchy integral formula for bounded operators (see e.g.\ \cite{DS}). In the proof of Step \ref{step7} we saw that $\|R_{P^{\chi_{E+1}}}(z)\|_{\D(H_\veps)}\lesssim1$ for $z\in\partial B_{1/2}(1)$,
which implies that also $\|(g/f^2)(P^{\chi_{E+1}})\|_{\L(\D(H_\veps))}\lesssim1$. Then applying the result of Step~\ref{step4} to (\ref{cheap}) yields the claim.

\bigskip

\begin{step}\label{step10}
$\forall\;m\in\NNN_0:\ \big\|(P-P^{\chi_{E+1}})\chi\big\|_{\L(\H,\D(H_\veps^m))} \,=\, \O(\veps^3)$
\end{step}

By construction we have $T_1=-T_1^*$ and $T_2+T_2^*=-T_1T_1^*$ as well as $P_0T_1P_0=0$. With this it is straight forward to verify that $P^{(2)}=P_0+\tilde P$ satisfies
\begin{equation}\label{inverse}
\big\|\chi_E\big(P^{(2)}P^{(2)}-P^{(2)}\big)\chi\big\|_{\L(\H,\D(H_\veps^m))}\;=\; \O(\veps^3).
\end{equation}

Since $\|[P^{\chi_{E+1}},H_\veps]\,\chi\|_{\L(\H,\D(H_\veps^{m-1}))}=\O(\veps^3)$ by Step \ref{step6}, Lemma \ref{microlocal} yields
\[\|[P^{\chi_{E+1}},\chi_E]\,\chi\|_{\L(\H,\D(H_\veps^m))}\,=\,\O(\veps^3).\]
Recalling that $\|P^{\chi_{E+1}}\|_{\L(\D(H_\veps^m))}\lesssim1$ due to Step \ref{step4} we have that in the norm of $\L(\H,\D(H_\veps^m))$
\begin{eqnarray*}
\lefteqn{\big((P^{\chi_{E+1}})^2-P^{\chi_{E+1}}\big)\,\chi}\\
&& \;=\; (P^{\chi_{E+1}}-1)P^{\chi_{E+1}}\,\chi_E\,\chi\\
&& \;=\; (P^{\chi_{E+1}}-1)\chi_E P^{\chi_{E+1}}\,\chi \,+\,(P^{\chi_{E+1}}-1)[P^{\chi_{E+1}},\chi_E]\,\chi\\ 
&& \;=\; \chi_E\,(P^{\chi_{E+1}}-1)P^{\chi_{E+1}}\,\chi \,+\,[P^{\chi_{E+1}},\chi_E]P^{\chi_{E+1}}\,\chi\,+\,\O(\veps^3)\\
&& \;=\; \chi_E\,\big(P^{(2)}-1\big)P^{(2)}\,\chi\,+\,\O(\veps^3)\\
&& \;=\; \chi_E\,\big(P^{(2)}P^{(2)}-P^{(2)}\big)\,\chi\,+\,\O(\veps^3) \qquad
\;\stackrel{(\ref{inverse})}{=}\; \O(\veps^3).
\end{eqnarray*}

Since we know from the proof of Step \ref{step7} that $\|R_{P^{\chi_{E+1}}}(z)\|_{\L(\D(H_\veps^m))}\lesssim1$ for $z$ away from $0$ and $1$, the formula
\begin{equation}\label{Nenciu}
P-P^{\chi_{E+1}} \;=\; \frac{\I}{2\pi\I}\oint_{\Gamma}\frac{R_{P^{\chi_{E+1}}}(z)+R_{P^{\chi_{E+1}}}(1-z)}{1-z}\,dz\,\big((P^{\chi_{E+1}})^2-P^{\chi_{E+1}}\big),
\end{equation}
which was proved by Nenciu in \cite{N}, implies that
\begin{equation}\label{Nen}
\big\|(P-P^{\chi_{E+1}})\chi(H_\veps)\big\|_{\L(\H,\D(H_\veps^m))} \;=\; \O(\veps^3).
\end{equation}

\bigskip

\begin{step}\label{step11}
There are $U_1,U_2\in\L(\H)\,\cap\,\L(\D(H_\veps))$ with norms bounded independently of $\veps$ satisfying $P_0U_1 P_0=0$ and $U_2 P_0=P_0U_2 P_0=P_0U_2$ such that $\tilde U= 1+\veps U_1+\veps^2 U_2$. 
In addition, 
$\|P_0U_1 \langle\nu\rangle^l\|_{\L(\D(H_\veps^m))}\lesssim1$ for all $l\in\NNN_0$ and $m\in\{0,1\}$.
\end{step}

We define 
\[U_1:=\veps^{-1}\big(P_0(\tilde U-1)(1-P_0)+(1-P_0)(\tilde U-1)P_0\big)\] 
and 
\[U_2:=\veps^{-2}\big(P_0(\tilde U-1)P_0+(1-P_0)(\tilde U-1)(1-P_0)\big).\] 
Then $\tilde U= 1+\veps U_1+\veps^2 U_2$, $P_0U_1P_0=0$, and $P_0U_2=P_0U_2P_0=U_2P_0$ are clear. Next we fix $m\in\NNN_0$ and prove that $U_1\in\L(\D(H_\veps^m))$ with norm bounded independent of $\veps$. The proof for $U_2$ is similar and will be omitted. We recall that
\[\tilde U\;=\; \big(P_0 P\,+\,(1-P_0)(1-P)\big)\,S\] 
with $S:=\big(1-(P-P_0)^2\big)^{-1/2}$ and that we showed $[P,S]=0=[P_0,S]$ in Step \ref{step7}. Therefore
\begin{eqnarray}\label{U1full}
U_1 &=& \veps^{-1}\big(P_0\tilde U(1-P_0)+(1-P_0)\tilde U P_0\big)\nonumber\\
&=& \veps^{-1}S\big(P_0P(1-P_0)\,+\,(1-P_0)(1-P)P_0\big)\nonumber\\
&=& \veps^{-1}S\big(P_0(P-P_0)(1-P_0)\,-\,(1-P_0)(P-P_0)P_0\big).
\end{eqnarray}
By Taylor expansion it holds
\begin{equation}\label{Taylor}
1-S\,=\,\int_0^1\textstyle{\frac{1}{2}}(1-s)\big(1-s(P-P_0)^2\big)^{-\frac{3}{2}}\,ds\;(P-P_0)^2.
\end{equation}
Let $h(x):=(1-sx^2\big)^{-3/2}$ with $s\in[0,1]$. $h$ is holomorphic in $B_{1/2}(0)$. Due to Step \ref{step7} the spectrum of $P-P_0$ as an operator on $\L(\D(H_\veps^m))$ is contained in $B_{1/4}(0)$ for $\veps$ small enough. Therefore $\|R_{P-P_0}(z)\|_{\L(\D(H_\veps^m))}\lesssim1$ for $z\in\partial B_{1/2}(0)$ and
$h(P-P_0)=\frac{\I}{2\pi}\oint_{\partial B_{1/2}(0)}h(z)R_{P-P_0}(z)\,dz$. This allows us to conclude that the integral on the right hand side of (\ref{Taylor}) is an operator bounded independent of $\veps$ on $\D(H_\veps^m)$. This implies that the whole right hand side is of order $\veps^2$ in $\L(\D(H_\veps^m))$ because $\|(P-P_0)^2\|_{\L(\D(H_\veps^m))}=\O(\veps^2)$ by Step~\ref{step7}. So we get
\begin{eqnarray}
U_1 &=& \veps^{-1}\big(P_0(P-P_0)(1-P_0)\,-\,(1-P_0)(P-P_0)P_0\big)\,+\,\O(\veps).\label{U1lo}
\end{eqnarray}
This yields the desired bound because $\|P-P_0\|_{\L(\D(H_\veps^m))}=\O(\veps)$.
We now turn to the claim that $\|P_0U_1 \langle\nu\rangle^l\|_{\L(\D(H_\veps^m))}\lesssim1$ for $m\in\{0,1\}$: Using $[S,P_0]=0$ and $\|P_0 \langle\nu\rangle^l\|_{\L(\D(H_\veps^m))}\lesssim1$ due to Step \ref{step1} we obtain from (\ref{U1full}) that
\begin{eqnarray*}
\|P_0U_1 \langle\nu\rangle^l\|_{\L(\D(H_\veps^m))} &=& \|\veps^{-1}SP_0(P-P_0)(1-P_0) \langle\nu\rangle^l\|_{\L(\D(H_\veps^m))}\\ 
&\lesssim& \|\veps^{-1}(P-P_0) \langle\nu\rangle^l\|_{\L(\D(H_\veps^m))} 
\end{eqnarray*}
We note that the decay properties of $P$ and $P_0$ themselves are not enough. Because of the $\veps^{-1}$ we really need to consider the difference. However, it holds $P-P_0 = (P-P^{\chi_{E+1}}) +(P^{\chi_{E+1}}-P_0)$ and via (\ref{Nenciu}) the first difference can be expressed by $(P^{\chi_{E+1}})^2-P^{\chi_{E+1}}$. Looking at the proof of Step \ref{step4} we see that both differences consist only of terms that carry an $\veps$ with them and have the desired decay property.

\bigskip

\begin{step}\label{step11b}
For $B:=P_0\tilde U\chi(H_\veps)$ and every $u\in\{1,U_1^*,U_2^*\}$
\begin{equation*}
\big\|\,[-\veps^2\Delta_{\rm h}+E_{\rm f},uP_0]B\,\big\|_{\L(\H)}\;=\;\O(\veps).
\end{equation*} 
\end{step}

\vspace{-0.8cm}
Again we restrict ourselves to the case $u=U_1^*$. It is obvious from the definition of $U_1$ in Step \ref{step11} that $[E_{\rm f},U_1^*P_0]=0$. In view of (\ref{U1lo}), $U_1$ (and thus also $U_1^*$) contains, up to terms of order $\veps$,  a factor $P-P_0$ . As long as we commute $(-\veps^2\Delta_{\rm h})P_0$ with the other factors, $P-P_0$ cancels the $\veps^{-1}$ in the definition of $U_1$ and the commutation yields the desired $\veps$ by Step \ref{step1}. Using that $B=P_0\tilde U\chi=P_0\chi+\O(\veps)$ we have
\begin{eqnarray*}
[-\veps^2\Delta_{\rm h},U_1^*P_0]B &=& [-\veps^2\Delta_{\rm h},U_1^*P_0]P_0\chi\,+\,\O(\veps)\\
&\stackrel{(\ref{U1lo})}{=}& [-\veps^2\Delta_{\rm h},\veps^{-1}(1-P_0)(P-P_0)P_0]P_0\chi\,+\,\O(\veps)\\
&=& (1-P_0)[-\veps^2\Delta_{\rm h},\veps^{-1}(P-P_0)]P_0\chi_E\chi\,+\,\O(\veps)\\
&=& (1-P_0)[-\veps^2\Delta_{\rm h},\veps^{-1}(P-P_0)\chi_E]P_0\chi\,+\,\O(\veps),
\end{eqnarray*}
The last step follows from $[(-\veps^2\Delta_{\rm h})P_0,\chi_E]\chi=\O(\veps)$, which is implied by Lemma \ref{microlocal} because $(-\veps^2\Delta_{\rm h})P_0$ satisfies the assumption on $A$ in Lemma \ref{expH} and thus
\begin{eqnarray*}
[H_\veps,(-\veps^2\Delta_{\rm h})P_0]\,\chi  
&=& [-\veps^2\Delta_{\rm h}+H_{\rm f},(-\veps^2\Delta_{\rm h})P_0]\,\chi\,+\,\O(\veps)\\ 
&=& 
[V_0,-\veps^2\Delta_{\rm h}]P_0\,\chi \,-\, 
\veps^2\Delta_{\rm h}[-\veps^2\Delta_{\rm h},P_0]\,\chi\,+\,\O(\veps)\\
&=& \O(\veps)
\end{eqnarray*}
as in Step \ref{step1}. Furthermore, due to Step \ref{step10}
\begin{eqnarray*}
\lefteqn{(1-P_0)[-\veps^2\Delta_{\rm h},\veps^{-1}(P-P_0)\chi_E]P_0\chi}\\
&& \quad=\ \;(1-P_0)[-\veps^2\Delta_{\rm h},\veps^{-1}(P^{\chi_{E+1}}-P_0)\chi_E]P_0\chi\,+\,\O(\veps^2)\\
&& \quad=\ \;(1-P_0)[-\veps^2\Delta_{\rm h},\big(P_1\chi_{E+1}+\chi_{E+1}P_1(1-\chi_{E+1})\big)\chi_E]P_0\chi\,+\,\O(\veps)\\
&& \quad=\ \;(1-P_0)[-\veps^2\Delta_{\rm h},(T_1^*P_0+P_0T_1)\chi_E]P_0\chi\,+\,\O(\veps).
\end{eqnarray*}
On the one hand,
\begin{eqnarray*}
(1-P_0)[-\veps^2\Delta_{\rm h},P_0T_1\chi_E]
&=& (1-P_0)[-\veps^2\Delta_{\rm h},P_0]P_0T_1\chi_E\ \,=\ \, \O(\veps)
\end{eqnarray*}
by Step \ref{step1} and Step \ref{step2}. On the other hand,
\begin{eqnarray*}
(1-P_0)[-\veps^2\Delta_{\rm h},T_1^*P_0\chi_E]P_0\chi
&=& (1-P_0)T_1^*P_0[(-\veps^2\Delta_{\rm h}),\chi_E]P_0\chi\\
&& \,+\,(1-P_0)[-\veps^2\Delta_{\rm h},T_1^*P_0]\chi_EP_0\chi \\
&=& (1-P_0)T_1^*P_0[(-\veps^2\Delta_{\rm h})P_0,\chi_E]\chi\,+\,\O(\veps)\\
&& \,+\,(1-P_0)[-\veps^2\Delta_{\rm h},T_1^*P_0]\chi_EP_0\chi\\
&=& \O(\veps)
\end{eqnarray*}
due to Step \ref{step5} and the above argument that $[(-\veps^2\Delta_{\rm h})P_0,\chi_E]\chi=\O(\veps)$.

\bigskip

\begin{step}\label{step12}
$\big\|\big(U_1^{*} \,+\,T_1^*P_0\big)\,B\big\|_{\L(\H,\D(H_\veps^m))} \,=\, \O(\veps)$ for all $m\in\NNN_0$.
\end{step}

All the following estimates will be in the norm of $\L\big(\H,\D(H_\veps)\big)$. It is easy to prove $[P_0,\chi_E]\chi=\O(\veps)$ in the same way we proved $[(-\veps^2\Delta_{\rm h})P_0,\chi_E]\chi=\O(\veps)$ in Step \ref{step11}. Using again that $B=P_0\tilde U\chi=P_0\chi+\O(\veps)$, $\chi=\chi_E\chi$, as well as  $P-P_0=\O(\veps)$ we obtain that 
\begin{eqnarray*}
U_1^*\,B &=& U_1^*\,P_0\chi_E\chi\,+\,\O(\veps)\\
&\stackrel{(\ref{U1lo})}{=}& \veps^{-1}(1-P_0)(P-P_0)P_0\chi_E\chi\,+\,\O(\veps)\\
&=& \veps^{-1}(1-P_0)(P-P_0)\chi_EP_0\chi\,+\,\O(\veps)\\
&\stackrel{(\ref{Nen})}{=}& \veps^{-1}(1-P_0)(P^{\chi_{E+1}}-P_0)\chi_EP_0\chi\,+\,\O(\veps)\\
&=& (1-P_0)\big(P_1\chi_{E+1}+(1-\chi_{E+1})P_1\chi_{E+1}\big)\chi_EP_0\chi\,+\,\O(\veps)\\
&=& (1-P_0)(T_1^*P_0+P_0T_1)\chi_EP_0\chi\,+\,\O(\veps)\\
&=& (1-P_0)T_1^*P_0\chi\,+\,\O(\veps)\\
&=& T_1^*P_0B\,+\,\O(\veps)
\end{eqnarray*}
because $(1-P_0)T_1^*P_0=T_1^*P_0$ by definition and $P_0\chi=B+\O(\veps)$. 

\bigskip

\begin{step}\label{step13}
It holds $E_{\rm f}\in C^\infty_{\rm b}(\C)$. 
If $\varphi_{\rm f}\in C^\infty_{\rm b}(\C,\H_{\rm f})$, then 
\[\|U_0\|_{\L(\D(H_\veps),\D(-\veps^2\Delta_\C+E_{\rm f}))}\,\lesssim\,1,\quad \|U_0^*\|_{\L(\D(-\veps^2\Delta_\C+E_{\rm f}),\D(H_\veps))}\,\lesssim\,1,\]  
and there is $\lambda_0\gtrsim1$ with $\sup_{q}\|\e^{\lambda_0\langle\nu\rangle}\varphi_{\rm f}(q)\|_{\H_{\rm f}(q)}\lesssim1$ and
\[\sup_{q}\|\e^{\lambda_0\langle\nu\rangle}\nabla^{\rm v}_{\nu_1,\dots,\nu_l}\nabla^{\rm h}_{\tau_1,\dots,\tau_m}\varphi_{\rm f}(q)\|_{\H_{\rm f}(q)}\lesssim1\]
for all $\nu_1,\dots,\nu_l\in\Gamma_{\rm b}(N\C)$ and $\tau_1,\dots,\tau_m\in\Gamma_{\rm b}(T\C)$.
\end{step}

We recall that $U_0\psi=\langle\varphi_{\rm f}|\psi\rangle_{\H_{\rm f}}$ and $U_0^*\psi=\varphi_{\rm f}\psi$. Using Lemma \ref{expdecay} ii) we easily obtain $\|(-\veps^2\Delta_\C+E_{\rm f})U_0\psi\|\lesssim\|\e^{-\lambda_0\langle\nu\rangle/2}(\nabla^{\rm h})^2\psi\|$ for all $\psi\in\D(H_\veps)$ and $\|H^\veps U_0^*\psi\|\lesssim\|\veps^2\nabla{\rm d}\psi\|$ for all $\psi\in\D(-\veps^2\Delta_\C+E_{\rm f})$. 
By (\ref{raus}) the former estimate implies
$\|U_0\|_{\L(\D(H_\veps),\D(-\veps^2\Delta_\C+E_{\rm f}))}\lesssim1$. Due to the bounded geometry of $\C$  any differential operator of second order with coefficients in $C^\infty_{\rm b}$ is operator-bounded by the elliptic $-\Delta_\C$. So the latter estimate implies $\|U_0^*\|_{\L(\D(-\veps^2\Delta_\C+E_{\rm f}),\D(H_\veps))}\lesssim1$.
The other statements are true by Lemma \ref{expdecay} i) and ii). 

\bigskip

The results of Step \ref{step1} and Steps \ref{step7} to \ref{step13} together form Proposition \ref{unitary}.\qed


\bigskip

{\sc Proof of Lemma \ref{expdecay}:}
\newline 
Because of $V_0\in C^\infty_{\rm b}(\C,C^\infty_{\rm b}(N_q\C))$ and $[\nabla^{\rm h}_\tau,\Delta_{\rm v}]=0$ for all $\tau$ due to Lemma~\ref{opequations} the mapping $q\mapsto (H_{\rm f}(q)-z)^{-1}$ is in $C^\infty_{\rm b}(\C,\L(\H_{\rm f}))$. Since $E_{\rm f}$ is a constraint energy band and thus separated, the projection $P_0(q)$ associated with $E_{\rm f}(q)$ is given via the Riesz formula: 
\begin{eqnarray*}
P_0(q) &=& \frac{\I}{2\pi}\oint_{\gamma(q)} \big(H_{\rm f}(q)-z\big)^{-1}\,dz,
\end{eqnarray*} 
where $\gamma(q)$ is positively oriented closed curve encircling $E_{\rm f}(q)$ once. It can be chosen independent of $q\in\C$ locally because the gap condition is uniform. Therefore $(H_{\rm f}(\cdot)-z)^{-1}\in C^\infty_{\rm b}(\C,\L(\H_{\rm f}))$ entails $P_0\in C^\infty_{\rm b}(\C,\L(\H_{\rm f}))$. This means in particular that $P_0\H$ is a smooth subbundle. Therefore locally it is spanned by a smooth section $\varphi_{\rm f}$ of normalized eigenfunctions. By
\begin{eqnarray*}
E_{\rm f}(q)P_0(q) \ \;=\ \;H_{\rm f}(q)P_0(q) &=& \frac{\I}{2\pi}\oint_{\gamma(q)} z\big(H_{\rm f}(q)-z\big)^{-1}\,dz
\end{eqnarray*} 
we see that also $E_{\rm f}P_0\in C^\infty_{\rm b}(\C,\L(\H_{\rm f}))$. Then $E_{\rm f}={\rm tr}_{\H_{\rm f}(\cdot)}\big(E_{\rm f}P_0\big)\in C^\infty_{\rm b}(\C)$ because covariant derivatives commute with taking the trace over smooth subbundles and derivatives of $E_{\rm f} P_0$ are trace-class operators. For example
\begin{eqnarray*}
\nabla_\tau\, {\rm tr}\big(E_{\rm f}P_0\big) &=& \nabla_\tau\, {\rm tr}\big((E_{\rm f}P_0)P_0\big)\\ 
&=&  {\rm tr}\big((\nabla^{\rm h}_\tau E_{\rm f}P_0)P_0\,+\, (E_{\rm f}P_0)\nabla^{\rm h}_\tau P_0\big)\\
&=&  {\rm tr}\big((\nabla^{\rm h}_\tau E_{\rm f}P_0)P_0\big) \,+\, {\rm tr}\big((E_{\rm f}P_0)\nabla^{\rm h}_\tau P_0\big)  \ \;<\ \; \infty
\end{eqnarray*} 
for all $\tau\in\Gamma_{\rm b}(T\C)$ because $P_0$ and $E_{\rm f} P_0$ are trace-class operators and the product of a trace-class operator and a bounded operator is again a trace-class operator (see e.g.\ \cite{RS1}, Theorem VI.19). The argument that higher derivatives of $E_{\rm f}P_0$ are trace-class operators is very similar.

\smallskip

Next we will prove the statement about invariance of exponential decay under the application of $R_{H_{\rm f}}(E_{\rm f}):=(1-P_0)(H_{\rm f}-E_{\rm f})^{-1}(1-P_0)$. So let $\Psi\in\H_{\rm f}$ be arbitrary. The claim is equivalent to showing that there is $\lambda_0>0$ such that for all $\lambda\in[-\lambda_0,\lambda_0]$
\[\Phi\,:=\;\e^{\lambda\langle\nu\rangle}R_{H_{\rm f}}(E_{\rm f})\e^{-\lambda\langle\nu\rangle}\Psi\]
satisfies $\Phi\|_{\H}\;\lesssim\;\|\Psi\|_{\H}$. The latter immediately follows from
\begin{equation}\label{crucial}
\|\Phi\|_{\H}\;\lesssim\;\|\e^{\lambda\langle\nu\rangle}(H_{\rm f}-E_{\rm f})\e^{-\lambda\langle\nu\rangle}\Phi\|_{\H}
\end{equation}
because
\begin{eqnarray*}
\|\e^{\lambda\langle\nu\rangle}(H_{\rm f}-E_{\rm f})\e^{-\lambda\langle\nu\rangle}\Phi\|_{\H} &=& \|\e^{\lambda\langle\nu\rangle}(1-P_0)\e^{-\lambda\langle\nu\rangle}\Psi\|_{\H}\\
&\leq& \|\Psi\|_{\H}\,+\,\sup_{q\in\C}\|\e^{\lambda\langle\nu\rangle}P_0\e^{-\lambda\langle\nu\rangle}\|_{\L(\H_{\rm f}(q))}\,\|\Psi\|_{\H}\\
&\lesssim& \|\Psi\|_{\H},
\end{eqnarray*}
where we used that $E_{\rm f}$ is a constraint energy band by assumption. We now turn to (\ref{crucial}). We note that by the Cauchy-Schwarz inequality it suffices to find a $\lambda_0>0$ such that for all $\lambda\in[-\lambda_0,\lambda_0]$
\begin{equation}\label{enough}
\langle\Phi|\Phi\rangle_\H \;\lesssim\; \big|{\rm Re}\,\big\langle\Phi\,\big|\,\e^{\lambda\langle\nu\rangle}(H_{\rm f}-E_{\rm f})\e^{-\lambda\langle\nu\rangle}\Phi\big\rangle_\H\big|
\end{equation}
To derive (\ref{enough}) we start with the following useful estimate, which is easily obtained by commuting $H_{\rm f}-E_{\rm f}$ with $\e^{-\lambda\langle\nu\rangle}$.
\begin{eqnarray}\label{IMS}
\big|{\rm Re}\,\big\langle\Phi\,\big|\,\e^{\lambda\langle\nu\rangle}(H_{\rm f}-E_{\rm f})\e^{-\lambda\langle\nu\rangle}\Phi\big\rangle\big| &=& \big|\langle\Phi|(H_{\rm f}-E_{\rm f})\Phi\rangle-\lambda^2\langle\Phi|(|\nu|^2/\langle\nu\rangle^2)\Phi\rangle\big|\nonumber\\
&\geq& \big|\langle\Phi|(H_{\rm f}-E_{\rm f})\Phi\rangle\big| \,-\, \lambda^2\langle\Phi|\Phi\rangle.\nonumber
\end{eqnarray}
Since $E_{\rm f}$ is assumed to be a constraint energy band and thus separated by a gap, we have
\begin{eqnarray*}
\big|\langle\Phi|(H_{\rm f}-E_{\rm f})\Phi\rangle \big|
&=& \big|\,\big\langle(1-P_0)\Phi\,\big|\,(H_{\rm f}-E_{\rm f})(1-P_0)\Phi\big\rangle\,\big|\\
&\geq& c_{\rm gap}\big\langle(1-P_0)\Phi\,\big|\,(1-P_0)\Phi\big\rangle\\
&=& c_{\rm gap}\big(\langle\Phi|\Phi\rangle-\langle\Phi|P_0\Phi\rangle\big).
\end{eqnarray*}
Since $\lambda_0$ can be chosen arbitrary small, we are left to show that $\langle\Phi|P_0\Phi\rangle$ is strictly smaller than $\langle\Phi|\Phi\rangle$ independent of $\lambda\in[-\lambda_0,\lambda_0]$. Since $E_{\rm f}$ is a constraint energy band by assumption, we know that there are $\Lambda_0>0$ and $C<\infty$ independent of $q\in\C$ such that $\|\e^{\Lambda_0\langle\nu\rangle}P_0(q)\e^{\Lambda_0\langle\nu\rangle}\|_{\H_{\rm f}(q)}\leq C$. Hence,
\begin{eqnarray*}
1\;=\;{\rm tr}_{\H_{\rm f}(q)}\big(P_0^2(q)\big) &=&  {\rm tr}_{\H_{\rm f}(q)}\big(\e^{\Lambda_0\langle\nu\rangle}P_0(q)\e^{\Lambda_0\langle\nu\rangle}\e^{-\Lambda_0\langle\nu\rangle} P_0(q)\e^{-\Lambda_0\langle\nu\rangle}\big) \\
&\leq& \|\e^{\Lambda_0\langle\nu\rangle}P_0(q)\e^{\Lambda_0\langle\nu\rangle}\|_{\H_{\rm f}(q)}\,{\rm tr}_{\H_{\rm f}(q)}\big(\e^{-\Lambda_0\langle\nu\rangle} P_0\e^{-\Lambda_0\langle\nu\rangle}\big)\\
&\leq& C\,{\rm tr}_{\H_{\rm f}(q)}\big(\e^{-\Lambda_0\langle\nu\rangle} P_0\e^{-\Lambda_0\langle\nu\rangle}\big).
\end{eqnarray*}
So we have that for any $\lambda$ with $\lambda\in[-\Lambda_0,\Lambda_0]$
\begin{eqnarray*}
\inf_q\,{\rm tr}_{\H_{\rm f}(q)}\big(\e^{-\lambda\langle\nu\rangle} P_0(q)\e^{-\lambda\langle\nu\rangle}\big) &\geq& \inf_q\,{\rm tr}_{\H_{\rm f}(q)}\big(\e^{-\Lambda_0\langle\nu\rangle} P_0(q)\e^{-\Lambda_0\langle\nu\rangle}\big)
\;\geq\;C^{-1}.
\end{eqnarray*}
Since $P_0\e^{-\lambda\langle\nu\rangle}\Phi=P_0R_{H_{\rm f}}\e^{-\lambda\langle\nu\rangle}\Psi=0$  by definition of $\Phi$, we have
\begin{eqnarray*}\label{clue}
\langle\Phi|P_0\Phi\rangle &=& \langle\Phi|(P_0-\e^{-\lambda\langle\nu\rangle}\ P_0\e^{-\lambda\langle\nu\rangle})\Phi\rangle\\
&\leq& \langle\Phi|\Phi\rangle\,\sup_q\,{\rm tr}_{\H_{\rm f}(q)}\big(P_0-\e^{-\lambda\langle\nu\rangle} P_0(q)\e^{-\lambda\langle\nu\rangle}\big)\\
&\leq& \langle\Phi|\Phi\rangle\,\Big(\sup_q{\rm tr}_{\H_{\rm f}(q)}(P_0)\,-\,\inf_q{\rm tr}_{\H_{\rm f}(q)}\big(\e^{-\lambda\langle\nu\rangle} P_0(q)\e^{-\lambda\langle\nu\rangle}\big)\Big)\\
&\leq& (1-C^{-1})\,\langle\Phi|\Phi\rangle,
\end{eqnarray*}
which finishes the proof of (\ref{enough}).

\smallskip

For i) it remains to show that the derivatives of $P_0$ produce exponential decay. By definition $P_0$ satisfies
\begin{equation}\label{projeqn}
0\;=\;(H_{\rm f}-E_{\rm f})P_0\;=\;-\Delta_{\rm v}P_0\,+\,V_0P_0\,-\,E_{\rm f}P_0.
\end{equation}
Let $\tau_1,...\tau_m\in\Gamma_{\rm b}(T\C)$ be arbitrary. To show that the derivatives of $P_0$ decay exponentially, we consider equations obtained by commutating the operator identity (\ref{projeqn}) with $\nabla^{\rm h}_{\tau_1,...,\tau_m}$. Since $\Delta_{\rm v}$ commutes with $\nabla^{\rm h}$ by Lemma \ref{opequations}, this yields the following hierachy of equations:
\begin{eqnarray*}
(H_{\rm f}-E_{\rm f})(\nabla^{\rm h}_{\tau_1}P_0) &=& (\nabla_{\tau_1}E_{\rm f}-\nabla^{\rm h}_{\tau_1}V_0)P_0,\\
(H_{\rm f}-E_{\rm f})(\nabla^{\rm h}_{\tau_1,\tau_2}P_0)
&=& (\nabla_{\tau_1,\tau_2}E_{\rm f}-\nabla^{\rm h}_{\tau_1,\tau_2}V_0)P_0 +(\nabla_{\tau_2}E_{\rm f}-\nabla^{\rm h}_{\tau_2}V_0)(\nabla^{\rm h}_{\tau_1}P_0)\\
&&\;+\,(\nabla_{\tau_1}E_{\rm f}-\nabla^{\rm h}_{\tau_1}V_0)(\nabla^{\rm h}_{\tau_2}P_0),
\end{eqnarray*}
and analogous equations for higher and mixed derivatives. Applying the reduced resolvent $R_{H_{\rm f}}(E_{\rm f})$ to both sides of the first equation we obtain that
\begin{eqnarray*}
(1-P_0)(\nabla^{\rm h}_{\tau_1}P_0) &=& R_{H_{\rm f}}(E_{\rm f})(\nabla^{\rm h}_{\tau_1}E_{\rm f}-\nabla^{\rm h}_{\tau_1}V_0)P_0.
\end{eqnarray*}
From $\big\|\,\e^{\lambda_0\langle\nu\rangle}P_0\e^{\lambda_0\langle\nu\rangle}\,\big\|_{\L(\H)}\lesssim 1$ we conclude that
\[\big\|\,\e^{\lambda_0\langle\nu\rangle}(1-P_0)\big(\nabla^{\rm h}_{\tau_1}P_0\big)\e^{\lambda_0\langle\nu\rangle}\,\big\|_{\L(\H)}\lesssim 1\] because the derivatives of $V_0$ and $E_{\rm f}$ are globally bounded and application of $R_{H_{\rm f}}(E_{\rm f})$ preserves exponential decay as we have shown above. Inductively, we obtain that 
\[\big\|\,\e^{\lambda_0\langle\nu\rangle}(1-P_0)\big(\nabla^{\rm v}_{\nu_1,\dots,\nu_l}\nabla^{\rm h}_{\tau_1,\dots,\tau_m}P_0\big)\e^{\lambda_0\langle\nu\rangle}\,\big\|_{\L(\H)}\;\lesssim\;1.\] 
The same arguments yield $\big\|\,\e^{\lambda_0\langle\nu\rangle}\big(\nabla^{\rm v}_{\nu_1,\dots,\nu_l}\nabla^{\rm h}_{\tau_1,\dots,\tau_m}P_0\big)(1-P_0)\e^{\lambda_0\langle\nu\rangle}\,\big\|_{\L(\H)}\lesssim 1$ when we start with $0=P_0(H_{\rm f}-E_{\rm f})$. The assumption
$\big\|\e^{\lambda_0\langle\nu\rangle}P_0\e^{\lambda_0\langle\nu\rangle}\big\|_{\L(\H)}\lesssim 1$ immediately implies $\big\|\,\e^{\lambda_0\langle\nu\rangle}P_0\big(\nabla^{\rm v}_{\nu_1,\dots,\nu_l}\nabla^{\rm h}_{\tau_1,\dots,\tau_m}P_0\big)P_0\e^{\lambda_0\langle\nu\rangle}\,\big\|_{\L(\H)}\lesssim 1$. These three statements together result in
\[\big\|\,\e^{\lambda_0\langle\nu\rangle}\big(\nabla^{\rm v}_{\nu_1,\dots,\nu_l}\nabla^{\rm h}_{\tau_1,\dots,\tau_m}\big)\e^{\lambda_0\langle\nu\rangle}\,\big\|_{\L(\H)}\;\lesssim\;1.\] 

\smallskip

We now turn to ii). So we assume that $\varphi_{\rm f}\in C^m_{\rm b}(\C,\H_{\rm f}(q))$ for some $m\in\NNN_0$. By definition $\varphi_{\rm f}$ satisfies
\begin{equation}\label{noreigenfct}
0\;=\;(H_{\rm f}-E_{\rm f})\varphi_{\rm f}\;=\;-\Delta_{\rm v}\varphi_{\rm f}\,+\,V_0\varphi_{\rm f}\,-\,E_{\rm f}\varphi_{\rm f}.
\end{equation}
for all $q\in\C$. Because of $V_0\in C^\infty_{\rm b}(\C,C^\infty_{\rm b}(N_q\C))$ and $E_{\rm f}\in C^\infty_{\rm b}(\C)$ this is an elliptic equation with coefficients in $C^0_{\rm b}(\C,C^\infty_{\rm b}(N_q\C))$ on each fibre. Therefore $\varphi_{\rm f}\in C^0_{\rm b}(\C,C^\infty_{\rm b}(N_q\C))$ follows from $\varphi_{\rm f}\in C^0_{\rm b}(\C,\H_{\rm f}(q))$ and standard elliptic theory immediately. 
Due to $\varphi_{\rm f}\in C^m_{\rm b}(\C,\H_{\rm f}(q))$
we may take horizontal derivatives of (\ref{noreigenfct}). Using that $[\Delta_{\rm v},\nabla^{\rm h}_{\tau}]$ for all $\tau$ by Lemma \ref{opequations} ii), we end up with the following equations 
\begin{eqnarray}\label{hierachy}
(H_{\rm f}-E_{\rm f})\nabla^{\rm h}_{\tau_1}\varphi_{\rm f} &=& (\nabla_{\tau_1}E_{\rm f}-\nabla^{\rm h}_{\tau}V_0)\varphi_{\rm f},\\
(H_{\rm f}-E_{\rm f})\nabla^{\rm h}_{\tau_1,\tau_2}\varphi_{\rm f} &=& (\nabla_{\tau_1,\tau_2}E_{\rm f}-\nabla^{\rm h}_{\tau_1,\tau_2}V_0)\varphi_{\rm f} \,+\, (\nabla_{\tau_1}E_{\rm f}-\nabla^{\rm h}_{\tau_1}V_0)(\nabla^{\rm h}_{\tau_2}\varphi_{\rm f}) \nonumber\\
&&\,+\,(\nabla_{\tau_2}E_{\rm f}-\nabla^{\rm h}_{\tau_2}V_0)(\nabla^{\rm h}_{\tau_1}\varphi_{\rm f}),\nonumber
\end{eqnarray}
and analogous equations up to order $m$. Iteratively, we see that these are all elliptic equations with coefficients in $C^0_{\rm b}(\C,C^\infty_{\rm b}(N_q\C))$ on each fibre. Hence, we obtain $\varphi_{\rm f}\in C^m_{\rm b}(\C,C^\infty_{\rm b}(N_q\C))$. So we may take also vertical derivatives of the above hierachy:
\begin{eqnarray}\label{hierachy2}
(H_{\rm f}-E_{\rm f})\nabla^{\rm v}_{\nu_1}\varphi_{\rm f} &=& -\,(\nabla^{\rm v}_{\nu_1}V_0)\,\varphi_{\rm f},\\
(H_{\rm f}-E_{\rm f})\nabla^{\rm v}_{\nu_1}\nabla^{\rm h}_{\tau_1}\varphi_{\rm f} &=& -\,(\nabla^{\rm v}_{\nu_1}\nabla^{\rm h}_{\tau_1}V_0)\varphi_{\rm f} \,-\, (\nabla^{\rm v}_{\nu_1}V_0)(\nabla^{\rm h}_{\tau_1}\varphi_{\rm f})\nonumber\\
&& \,+\,(\nabla_{\tau_1}E_{\rm f}-\nabla^{\rm h}_{\tau_1}V_0)\nabla^{\rm v}_{\nu_1}\varphi_{\rm f})\nonumber
\end{eqnarray}
and so on.
Since $E_{\rm f}$ is assumed to be a constraint energy band, we have that 
\[\big\|\e^{\Lambda_0\langle\nu\rangle}\varphi_{\rm f}\,\langle\e^{\Lambda_0\langle\nu\rangle}\varphi_{\rm f}|\psi\rangle_{\H_{\rm f}(q)} \big\|_{\H_{\rm f}(q)} \;=\;\|\e^{\Lambda_0\langle\nu\rangle}P_0\e^{\Lambda_0\langle\nu\rangle}\psi\|_{\H_{\rm f}(q)}\;\lesssim\;\|\psi\|_{\H_{\rm f}(q)}\] 
with a constant independent of $q$. Choosing $\psi=\e^{-\Lambda_0\langle\nu\rangle}\varphi_{\rm f}$ and taking the supremum over $q\in\C$ we obtain the desired exponential decay of $\varphi_{\rm f}$. 
Because of $V_0\in C^\infty_{\rm b}(\C,C^\infty_{\rm b}(N_q\C))$ and $E_{\rm f}\in C^\infty_{\rm b}(\C)$ also the right-hand sides of (\ref{hierachy}) and (\ref{hierachy2}) decay exponentially. 
By i) an application of $R_{H_{\rm f}}(E_{\rm f})$ preserves exponential decay. So we may conclude that the $\varphi_{\rm f}$-orthogonal parts of $\nabla^{\rm h}_{\tau_1}\varphi_{\rm f}$ and $\nabla^{\rm v}_{\nu_1}\varphi_{\rm f}$ decay exponentially. Together with the exponential decay of $\varphi_{\rm f}$ this entails the desired exponential decay of $\nabla^{\rm h}_{\tau_1}\varphi_{\rm f}$ and $\nabla^{\rm v}_{\nu_1}\varphi_{\rm f}$. This argument can now easily be iterated for the higher derivatives. 

\medskip

Finally, we turn to iii). We consider a normalized trivializing section $\varphi_{\rm f}$, in particular $\sup_{q\in\C}\|\varphi_{\rm f}\|_{\H_{\rm f}}$ is globally bounded. By assumption $\varphi_{\rm f}$ is smooth as a section of $P_0\H$. In order to see that it is also smooth in $(1-P_0)\H$, one applies $R_{H_{\rm f}}(E_{\rm f})$ to the equations (\ref{hierachy}), which can be justified by an approximation argument. Hence, we only need to show boundedness of all the derivatives. If $\C$ is compact, this is clear. 

\smallskip

We recall that the eigenfunction $\varphi_{\rm f}(q)$ can be chosen real-valued for any $q\in\C$.
If $\C$ is contractible, all bundles over $\C$ are trivializable. In particular, already the real eigenspace bundle $P_0\H$ has a global smooth trivializing section $\varphi_{\rm f}$. We choose a covering of $\C$ by geodesic balls of fixed diameter and take an arbitrary one of them called $\Omega$. We choose geodesic coordinates $(x^i)_{i=1,\ldots,d}$ and bundle coordinates $(n^\alpha)_{\alpha=1,\ldots,k}$ with respect to an orthonormal trivializing frame $(\nu_\alpha)_\alpha$ over $\Omega$ as in Remark \ref{coordinates}. Since $\varphi_{\rm f}$ is the only normalized element of the real $P_0\H$, we have that 
\begin{equation}\label{unique}
\varphi_{\rm f}(q) \;=\; \frac{P_0(x)\varphi_{\rm f}(x_0)}{\|P_0(x)\varphi_{\rm f}(x_0)\|}
\end{equation}
for any fixed $x_0\in\Omega$ and $x$ close to it. In view of the coordinate expression $\nabla^{\rm h}_{\partial_{x^i}}=\partial_{x^i}-\Gamma^{\alpha}_{i\beta}n^\beta\partial_{n^\alpha}$, we can split up $\nabla^{\rm h}_{\partial_{x^i}}\varphi_{\rm f}$ into terms depending on $\nabla^{\rm h}_{\partial_{x^i}} P_0$, which are bounded due to i), and $-\Gamma^{\alpha}_{i\beta}n^\beta\partial_{n^\alpha}\varphi_{\rm f}(x_0)$. We already know that $\varphi_{\rm f}\in C^0_{\rm b}\big(\C,\H_{\rm f}(q)\big)$. By ii) this implies $\varphi_{\rm f}\in C^0_{\rm b}\big(\C,C^\infty_{\rm b}(N_q\C)\big)$ with $\sup_q\|\e^{\lambda_0\langle\nu\rangle}\varphi_{\rm f}\|\lesssim1$. Recalling that $\langle\nu\rangle=\sqrt{1+\delta_{\alpha\beta}n^\alpha n^\beta}$ we have that $-\Gamma^{\alpha}_{i\beta}n^\beta\partial_{n^\alpha}\varphi_{\rm f}(x_0)$ is bounded. Noticing that all the bounds are independent of $\Omega$ due to (\ref{bndcurv2}) (as was explained in Remark~\ref{coordinates}) we obtain that $\varphi_{\rm f}\in C^1_{\rm b}\big(\C,\H_{\rm f}(q)\big)$. Now we can inductively make use of (\ref{unique}) and ii) to obtain $\varphi_{\rm f}\in C^\infty_{\rm b}\big(\C,\H_{\rm f}(q)\big)$.

\smallskip

If $E_{\rm f}=\inf\sigma(H_{\rm f}(q))$ for all $q\in\C$, again the real eigenspace bundle is already trivializable. To see this we note that the groundstate of a Schr\"odinger operator with a bounded potential can always be chosen strictly positive (see \cite{RS4}), which defines an orientation on the real eigenspace bundle. A real line bundle with an orientation is trivializable. So we may argue as in the case of a contractable $\C$ that the derivatives are globally bounded. \qed

\bigskip

{\sc Proof of Lemma \ref{notspoiled}:}

Let the assumption (\ref{asslambda}) be true for $l\in\NNN_0$ and $m\in\NNN$. The proof for $-l\in\NNN$ is very similar. We fix $z_1,\dots,z_m\in(\CCC\setminus\RRR)\,\cap\,({\rm supp}\,\chi\times[-1,1])$ and claim that there is a $c>0$ independent of the $z_i$ such that
\begin{equation}\label{gapdecay}
\Big\|\,\prod_{i=1}^{m}(H-z_i)\,\langle\lambda\nu\rangle^l\,\prod_{j=1}^{m}R_{H}(z_j)\,\langle\lambda\nu\rangle^{-l}\,\Big\|_{\L(\H)}
\;\leq\;2
\end{equation}
for $\lambda:=\min\Big\{1,C_1^{-1}\frac{c\,\prod_{i=1}^{m}|{\rm Im}z_i|}{1+\prod_{j=1}^{m}(|z_j|+|{\rm Im}z_j|)}\Big\}>0$. 

\medskip

To prove this we set $\Phi:=\prod_{i=1}^{m}(H-z_i)\,\langle\lambda\nu\rangle^l\prod_{j=1}^{m}R_{H}(z_j)\langle\lambda\nu\rangle^{-l}\Psi$ for $\Psi\in\H$ and aim to show that $\|\Psi\| \geq \|\Phi\|/2$. 
We have that
\begin{eqnarray*}
\|\Psi\| 
&=& \Big\|\langle\lambda\nu\rangle^l\,\prod_{j=1}^{m}(H-z_i)\,\langle\lambda\nu\rangle^{-l}\,\prod_{i=1}^{m}R_{H}(z_j)\,\Phi\Big\|\\
&\geq& \|\Phi\|\,-\,\Big\|\langle\lambda\nu\rangle^l\,\Big[\prod_{j=1}^{m}(H-z_i),\langle\lambda\nu\rangle^{-l}\Big]\,\prod_{i=1}^{m}R_{H}(z_j)\,\Phi\,\Big\|
\end{eqnarray*}
Using the assumption (\ref{asslambda}) and that $|z_i|\leq1$ for all $i$ we have that 
there is a $C<\infty$ independent of $\lambda$ and the $z_i$'s with
\begin{eqnarray*}
\|\Psi\|
&\geq& \|\Phi\|\,-\,C\,C_1\lambda\,\Big(\Big\|H^m\,\prod_{j=1}^{m}R_{H}(z_j)\,\Phi\Big\|
+\Big\|\prod_{j=1}^{m}R_{H}(z_j)\,\Phi\Big\|\Big)\\
&=& \|\Phi\|\,-\,C\,C_1\lambda\,\Big\|\prod_{j=1}^{m}H\,R_{H_\veps}(z_j)\,\Phi\Big\|
\,-\,C\,C_1\lambda\,\Big\|\prod_{j=1}^{m}R_{H}(z_j)\,\Phi\Big\|\\
&\geq& \|\Phi\|\,-\,C\,C_1\lambda\,\prod_{j=1}^{m}\Big(1+\frac{|z_j|}{|{\rm Im}z_j|}\Big)\,\|\Phi\|
\,-\,C\,C_1\lambda\,\prod_{j=1}^{m}|{\rm Im}z_j|^{-1}\,\big\|\Phi\big\|\\
&\geq& \|\Phi\|\,-\,C\,C_1\lambda\,\frac{1+\prod_{j=1}^{m}(|z_j|+|{\rm Im}z_j|)}{\prod_{i=1}^{m}|{\rm Im}z_i|}\,\|\Phi\|\\
&\geq& \|\Phi\|/2
\end{eqnarray*}
for $\lambda\leq C_1^{-1}\frac{(2C)^{-1}\,\prod_{i=1}^{m}|{\rm Im}z_i|}{1+\prod_{j=1}^{m}(|z_j|+|{\rm Im}z_j|)}$. This yields (\ref{gapdecay}).

\medskip

Now we make use of the Helffer-Sj\"ostrand formula. We recall from the proof of Lemma \ref{microlocal} that it says that 
\[f(H_\veps)\;=\;\frac{1}{\pi}\,\int_\CCC \partial_{\overline{z}}\tilde f(z)\,R_{H_\veps}(z)\,dz,\]
where $\tilde f$ is an arbitrary almost analytic extension of $f$. Here by $dz$ we mean again the usual volume measure on $\CCC$. By assumption $\chi$ is non-negative. So by the spectral theorem we have $\chi(H)=\prod_{i=1}^{m}\chi^{1/m}(H)$.
We choose an almost analytic extension of $\chi^{1/m}$ such that $K:={\rm supp}\,\widetilde{\chi^{1/m}}\subset{\rm supp}\,\chi\times[-1,1]$ (in particular the volume of $K$ is finite) and
\begin{equation}\label{almostanal2}
|\partial_{\overline{z}}\widetilde{\chi^{1/m}}(z)|\;=\;\O(|{\rm Im}z|^{l+1}).
\end{equation}
Then by the Helffer-Sj\"ostrand formula
\[\chi(H)\;=\;\frac{1}{\pi^m}\,\int_{\CCC^m} \prod_{i=1}^m\partial_{\overline{z}}\widetilde{\chi^{1/m}}(z_i)\,\prod_{i=1}^m R_{H}(z_i)\,dz_1\dots dz_m.\]
We will now combine (\ref{gapdecay}) and (\ref{almostanal2}) to obtain the claimed estimate. In the following, we use $\lesssim$ for 'bounded by a constant independent of $H$'.
\begin{eqnarray*}
\lefteqn{\big|\langle\nu\rangle^l\chi(H)\,\langle\nu\rangle^{-l}\,\Psi\big|}\\
&=& \Big|\frac{1}{\pi^m}\int_{\CCC^m} \prod_{i=1}^m\partial_{\overline{z}}\widetilde{\chi^{1/m}}(z_i)\langle\nu\rangle^l\langle\lambda\nu\rangle^{-l}\,\langle\lambda\nu\rangle^{l}\prod_{i=1}^m R_{H}(z_i)\langle\nu\rangle^{-l}\Psi\,dz_1\dots dz_m\,\Big|\\ 
&\stackrel{(\ref{almostanal2})}{\lesssim}& C_1^l\int_{K^m} \,\prod_{i=1}^m|{\rm Im}z_i|\,\Big|\langle\lambda\nu\rangle^l\,\prod_{i=1}^m R_{H}(z_i)\langle\nu\rangle^{-l}\,\Psi\Big|\,dz_1\dots dz_m
\end{eqnarray*} 
where we used that $\langle\nu\rangle^l\langle\lambda\nu\rangle^{-l}\leq\lambda^{-l}\sim C_1^l\prod_{i=1}^m|{\rm Im}\,z_i|^{-l}$ for small $|{\rm Im}\,z_i|$.
So
\begin{eqnarray*}
\lefteqn{\big\|\langle\nu\rangle^l\chi(H)\,\langle\nu\rangle^{-l}\,\Psi\big\|_{\D(H^m)} }\\
&\lesssim& C_1^l\,\Big\|\int_{K^m} \,\prod_{i=1}^m|{\rm Im}z_i|\,\Big|\langle\lambda\nu\rangle^l\,\prod_{i=1}^m R_{H}(z_i)\,\langle\nu\rangle^{-l}\,\Psi\Big|\,dz_1\dots dz_m\Big\|_{\D(H^m)} \\ 
&=& C_1^l\,\Big\|\int_{K^m} \,\prod_{i=1}^m|{\rm Im}z_i|\,\Big|\prod_{i=1}^m R_{H}(z_i)\,\prod_{i=1}^m(H-z_i)\,\langle\lambda\nu\rangle^l\\
&& \quad\qquad\qquad\qquad\times\,\prod_{i=1}^m R_{H}(z_i)\langle\lambda\nu\rangle^{-l}\,\langle\lambda\nu\rangle^{l}\,\langle\nu\rangle^{-l}\,\Psi\Big|\,dz_1\dots dz_m\Big\|_{\D(H^m)} \\ 
&\leq& C_1^l\int_{K^m} \,\prod_{i=1}^m|{\rm Im}z_i|\,\prod_{i=1}^m\|R_{H}(z_i)\|_{\L(\D(H^{m-i}),\D(H^{m-i+1}))}\,\|\langle\lambda\nu\rangle^{l}\,\langle\nu\rangle^{-l}\,\Psi\|_\H\\
&& \qquad\qquad\times\,\Big\|\prod_{i=1}^m(H_\veps-z_i)\,\langle\lambda\nu\rangle^l\,\prod_{i=1}^m R_{H_\veps}(z_i)\langle\lambda\nu\rangle^{-l}\Big\|_{\L(\H)}\,dz_1\dots dz_m \\ 
&\stackrel{(\ref{gapdecay})}{\lesssim}& C_1^l\,\|\Psi\|_\H,
\end{eqnarray*}  
because of the resolvent estimate (\ref{resestimate}) and $\langle\lambda\nu\rangle^l\langle\nu\rangle^{-l}\leq1$ for $\lambda\leq1$. Hence, $\|\langle\nu\rangle^l\chi(H)\,\langle\nu\rangle^{-l}\|_{\L(\H,\D(H_\veps^m))}$ is bounded by $C_1^l$ times a constant independent of $H$. 
\qed 


\section*{Appendix}
\addcontentsline{toc}{section}{Appendix}

\subsection*{Manifolds of bounded geometry}
\addcontentsline{toc}{subsection}{Manifolds of bounded geometry}

Here we explain shortly the notion of bounded geometry, which provides the natural framework for this work. More on the subject can be found in \cite{Sh}.

\begin{definition}
Let $(\M,g)$ be a Riemannian manifold and let $r_q$ denote the injectivity radius at $q\in\M$. Set $r_\M:=\inf_{q\in\M}r_q$.
$(\M,g)$ is said to be \emph{of bounded geometry}, if $r_\M>0$ and every covariant derivative of the Riemann tensor $\R$ is bounded, i.e. 
\begin{equation}\label{bndcurv3}
\forall\;m\in\NNN\ \ \exists\;C_m<\infty:\quad g(\nabla^m \R,\nabla^m \R)\;\leq\; C_m.
\end{equation}
Here $\nabla$ is the Levi-Civita connection on $(\M,g)$ and $g$ is extended to the tensor bundles $T^l_m\M$ for all $l,m\in\NNN$ in the canonical way. An open subset $U\subset\M$ equipped with the induced metric $g|_U$ is called a \emph{subset of bounded geometry}, if $r_\M>0$ and (\ref{bndcurv3}) is satisfied on $U$. 
\end{definition}

The definition of the Riemann tensor is given below. We note that $r_\M>0$ implies completeness of $\M$. The second condition is equivalent to postulating that every transition function between an arbitrary pair of geodesic coordinate charts has bounded derivatives up to any order. Finally, we note that the closure of a subset of bounded geometry is obviously metrically complete. 

\subsection*{The geometry of submanifolds}
\addcontentsline{toc}{subsection}{The geometry of submanifolds}

We recall here some standard concepts from Riemannian geometry. 
For further information see e.g.\ \cite{L}.

First we give the definitions of the inner curvature tensors we use because they vary in the literature. We note that they contain statements about tensoriality and independence of basis that are not proved here! In the following, we denote by $\Gamma(\mathcal{E})$ the set of all smooth sections of a bundle $\mathcal{E}$ and by $\T^l_{\,\,m}(\M)$ the set of all smooth $(l,m)$-tensor fields 
over a manifold $\M$.
\begin{definition}
Let $(\A,\overline{g})$ be a Riemannian manifold with Levi-Civita connection $\overline{\nabla}$. Let $\tau_1,\tau_2,\tau_3,\tau_4\in\Gamma(T\A)$.  

i) The \emph{curvature mapping} $\overline{{\rm R}}:\,\Gamma(T\A)\times \Gamma(T\A)\to \T^1_{\,\,1}(\A)$ is given by
\[\overline{{\rm R}}(\tau_1,\tau_2)\,\tau_3\;:=\;\overline{\nabla}_{\tau_1}\overline{\nabla}_{\tau_2}\tau_3\,-\,\overline{\nabla}_{\tau_2}\overline{\nabla}_{\tau_1}\tau_3\,-\,\overline{\nabla}_{[\tau_1,\tau_2]}\tau_3.\]

ii) The \emph{Riemann tensor} $\overline{\R}\in\T^0_{\,\,4}(\A)$ is given by 
\[\overline{\R}(\tau_1,\tau_2,\tau_3,\tau_4)\;:=\;\overline{g}\big(\tau_1,\overline{{\rm R}}(\tau_3,\tau_4)\,\tau_2\big).\]

iii) The \emph{Ricci tensor} $\overline{{\rm Ric}}\in\T^0_{\,\,2}(\A)$ is given by
\[\overline{{\rm Ric}}(\tau_1,\tau_2)\;:=\;{\rm tr}_\A\,\overline{{\rm R}}(\,.\,,\tau_1)\tau_2.\]

iv) The \emph{scalar curvature} $\overline{\kappa}:\A\to\RRR$ is given by
\[\overline{\kappa}\;:=\;{\rm tr}_\A\,\overline{{\rm Ric}}.\]

Here ${\rm tr}_\A\,t$ means contracting the tensor $t$ at any point $q\in\A$ by an arbitrary orthonormal basis of $T_q\A$. 
\end{definition}

\begin{remark}
The dependence on vector fields of $\overline{{\rm R}}$, $\overline{\R}$, and $\overline{\rm Ric}$ can be lifted to the cotangent bundle $T\C^*$ via the metric $\overline{g}$. The resulting objects are denoted by the same letters throughout this work. The same holds for all the objects defined below.
\end{remark} 

\medskip

Of course, all these objects can also be defined for a submanifold once a connection has been chosen. There is a canonical choice given by the induced connection.   

\begin{definition}
Let $\C\subset\A$ be a submanifold with induced metric $g$. Denote by $T\C$ and $N\C$ the tangent and the normal bundle of $\C$. Let $\tau_1,\tau_2,\tau_3\in\Gamma(T\C)$.

\smallskip

i) We define $\nabla$ to be the \emph{induced connection on $\C$} given via
\begin{equation*}
\nabla_{\tau_1}\tau_2\;:=\;P_T\overline{\nabla}_{\tau_1}\tau_2,
\end{equation*}
where $\tau_1,\tau_2$ are canonically lifted to $T\A=T\C\times N\C$ and $P_T$ denotes the projection onto the first component of the decomposition. The projection onto the second component of the decomposition will be denoted by $P_\perp$.

\smallskip

ii) ${\rm R}$, ${\rm Ric}$, and $\kappa$ are defined analogously with $\overline{{\rm R}}$, $\overline{{\rm Ric}}$ and $\overline{\kappa}$ from the preceding definition. The \emph{partial trace of 
$\overline{{\rm R}}$ with respect to $\C$} is given by
\[{\rm tr}_\C\,\overline{{\rm R}}\;:=\;{\rm tr}_\C\,\overline{{\rm Ric}}_\C,\]
with $\overline{{\rm Ric}}_\C(\tau_1,\tau_2)\;:=\;{\rm tr}_\C\,\overline{{\rm R}}(\,.\,,\tau_1)\tau_2$. 

Here ${\rm tr}_\C\,t$ means contracting the tensor $t$ at any point $q\in\C$ by an arbitrary orthonormal basis of $T_q\C$. 
\end{definition}

We note that $\nabla$ coincides with the Levi-Civita connection associated with the induced metric $g$.
Now we turn to the basic objects related to the embedding of a submanifold of arbitrary codimension.

\begin{definition}
Let $\tau,\tau_1,\tau_2\in\Gamma(T\C),\nu\in\Gamma(N\C)$. 

\smallskip
 
i) The \emph{Weingarten mapping} $\W:\,\Gamma(N\C)\to \T^1_{\,\,1}(\C)$ is given by
\[\W(\nu)\,\tau\;:=\;-P_T\overline{\nabla}_\tau \nu.\]
 
ii) The \emph{second fundamental form} ${\rm II}(\,.\,):\Gamma(N\C)\to\T^0_{\,\,2}(\C)$ is defined by
\[{\rm II}(\nu)\big(\tau_1,\tau_2\big)\;:=\;\overline{g}(\overline{\nabla}_{\tau_1} \tau_2,\nu).\]

iii) The \emph{mean curvature normal} $\eta\in\Gamma(N\C)$ is defined to be the unique vector field that satisfies
\[\overline{g}(\eta,\nu)\;=\;{\rm tr}_\C\W(\nu)\qquad\forall\ \nu\in\Gamma(N\C).\]

iv) We define the \emph{normal connection} $\nabla^\perp$ to be the bundle connection on the normal bundle given via
\begin{equation*}
\nabla^\perp_\tau\nu\;:=\;P_\perp\overline{\nabla}_\tau\nu,
\end{equation*}
where $\nu$ and $\tau$ are canonically lifted to $T\A=T\C\times N\C$.

v) ${\rm R}^\perp:\Gamma(T\C)\times\Gamma(T\C)\times\Gamma(N\C)\to\Gamma(N\C)$
denotes the \emph{normal curvature mapping} defined by
\[{\rm R}^\perp(\tau_1,\tau_2)\nu\;:=\;\nabla^\perp_{\tau_1}\nabla^\perp_{\tau_2}\nu\,-\,\nabla^\perp_{\tau_2}\nabla^\perp_{\tau_1}\nu\,-\,\nabla^\perp_{[\tau_1,\tau_2]}\nu.\]
\end{definition}

\medskip

\begin{remark}
i) The usual relations and symmetry properties for $\W$ and ${\rm II}$ also hold for codimension greater than one:
\[{\rm II}(\nu)(\tau_1,\tau_2)=g\big(\tau_1, \W(\nu)\,\tau_2\big)=g\big(\tau_2, \W(\nu)\,\tau_1\big)={\rm II}(\nu)(\tau_2,\tau_1).\]

ii) A direct consequence of the definitions is the Weingarten equation:
\begin{equation*}
\nabla^\perp_\tau\nu\;=\;\overline{\nabla}_\tau\nu\,+\,\W(\nu)\tau.
\end{equation*} 

iii) The normal curvature mapping ${\rm R}^\perp$ is identically zero, when the dimension or the codimension of $\C$ is smaller than two. 
\end{remark}

\pagebreak

\section*{Acknowledgements}
\addcontentsline{toc}{section}{Acknowledgements}

We are grateful to David Krej$\check{{\rm c}}$i$\check{\rm r}$\'ik for providing several references and for a careful reading of an earlier version resulting in lots of useful comments. We are also grateful to Luca Tenuta for helpful remarks when we began with this work. Furthermore, we thank Christian Loeschcke, Frank Loose, Christian Lubich, Olaf Post, Hans-Michael Stiepan, and Olaf Wittich for inspiring discussions about the topic of this paper.


\end{document}